%% file: manuscript_allinone_R1.tex
\documentclass[preprint,12pt]{elsarticle}

\usepackage{etoolbox}
\newtoggle{ONEFILE}
\togglefalse{ONEFILE} 

\iftoggle{ONEFILE}{
}{
	\usepackage{xr}
	\externaldocument{supplemental_R1}
}

\usepackage{amssymb}

\usepackage{url}
\usepackage{adjustbox}
\usepackage{booktabs}
\usepackage{url}
\usepackage[dvipsnames]{xcolor}
\usepackage{xcolor}
\usepackage{color, colortbl}
\usepackage{soul}
\usepackage{subfig}
\usepackage{caption}

\begin{document}

\begin{frontmatter}



\title{Journalists' ego networks in Twitter:\\invariant and distinctive structural features}


\author{Mustafa Toprak\corref{cor1}}
\ead{mustafa.toprak@iit.cnr.it}
\author{Chiara Boldrini}
\author{Andrea Passarella}
\author{Marco Conti}

\cortext[cor1]{Corresponding author}

\address{IIT-CNR, Via G. Moruzzi 1, 56124, Pisa, Italy}

\begin{abstract}

Ego networks have proved to be a valuable tool for understanding the relationships that individuals establish with their peers, both in offline and online social networks. Particularly interesting are the \emph{cognitive constraints} associated with the interactions between the ego and the members of their ego network, which limit individuals to maintain meaningful interactions with no more than 150 people, on average, and to arrange such relationships along concentric circles of decreasing engagement.
In this work, we focus on the ego networks of journalists on Twitter, considering 17 different countries, and we investigate whether they feature the same characteristics observed for other relevant classes of Twitter users, like politicians and generic users. Our findings are that journalists are generally more active and interact with more people than generic users, regardless of their country. Their ego network structure is very aligned with reference models derived in anthropology and observed in general human ego networks. Remarkably, the similarity is even higher than the one of politicians and generic users ego networks. This may imply a greater cognitive involvement with Twitter for journalists than for other user categories. From a dynamic perspective, journalists have stable short-term relationships that do not change much over time. In the longer term, though, ego networks can be pretty dynamic, especially in the innermost circles. Moreover, the ego-alter ties of journalists are often information-driven, as they are mediated by hashtags both at their inception and during their lifetime. Finally, we found that relationships between journalists are assortative in popularity: journalists tend to engage with other journalists of similar popularity, in all layers but especially in their innermost ones. Instead, when journalists interact with generic users, this assortativity is only present in the innermost layers.
 \end{abstract}




\end{frontmatter}


\section{Introduction}
\label{sec:introduction}

Online Social Networks (OSNs) have been shaping and transforming our daily lives for more than a decade. Currently, there are 4.2 billion active social media users and 4.15 billion active mobile social media users\footnote{https://www.statista.com/statistics/617136/digital-population-worldwide/}. With easy access to the Internet, especially via mobile devices, we started to carry our new communication mediums and online social groups everywhere we go. We can thus participate in online communities more than ever, amplifying the phenomenon known as cyber-physical convergence, whereby our offline and online lives become tightly intertwined.  
The huge volume of information produced online across different OSN platforms gives researchers new opportunities to analyze human behavior on large-scale data, overcoming the limitations of survey-based traditional approaches. However, there is a two-way exchange between our offline and online social life. 
On the one hand, we are transferring and engaging with our offline relationships on online platforms. On the other hand, we are establishing new relationships, which may not have been even possible in the offline world (due, e.g., to long distances and different time zones). Due to the importance of our online social life, not only academic studies on human behavior in the offline world via OSN data but also the comparison between offline and online human behavior have gained popularity. Within this latter line of research, scientists have tried to understand if and to which extent our offline and online relationships are similar. Quite surprisingly, their findings showed that the structure of online relationships by and large mirrors that of offline ones~\cite{Dunbar2015}. This has been interpreted as a sign that the very same cognitive constraints shaping humans' offline relationships act also on our online social life. 

Despite these findings, OSNs have certainly shaped how we communicate socially with each other, and they have also redefined what is expected from professionals in terms of online presence, especially for jobs that have a wide audience, such as politicians, artists, and journalists. 
%
As for the latter, social platforms have become digital newsstands, where user can find news sources and get instant access to articles. Journalism has been under transition to keep up with this new communication medium. Twitter has become one of the main access points to news resources. According to a survey carried out by the American Press Institute in collaboration with Twitter\footnote{https://www.americanpressinstitute.org/publications/reports/survey-research/how-people-use-twitter-news}, 86\% of Twitter users engage with the platform for news. As a result, Twitter has become extremely popular among journalists to share the news articles they produce, which makes journalists an interesting professional group to analyze. 

According to the related literature, journalists use Twitter as a platform for establishing their personal brands~\cite{canter2015personalised-tweeting} and promote content from their news websites~\cite{russell2015sets}. Famous journalists are now individual brands, news and opinion hubs~\cite{brems2017}, sometimes even more popular than the media companies they work for.  There are many studies analyzing this transformation of the profession~\cite{russell2015sets, brems2017, lee2017journalists}, as well as several comparative studies on journalism. In~\cite{macgregor2011cross}, journalists from the UK and 10 other European countries are compared in terms of their role perception and their belief in the usefulness of the digital and social media. Another study~\cite{weiss2015digital} analyzes Latin American journalists' view on their professional role in today's digital media platforms and how they are organized to create a new way of content management. In~\cite{hanusch2017comparing}, cross-cultural research has been carried out to assess the development of journalism culture in 66 countries. 

\begin{figure}[ht]
\begin{center}
\includegraphics[scale=0.6]{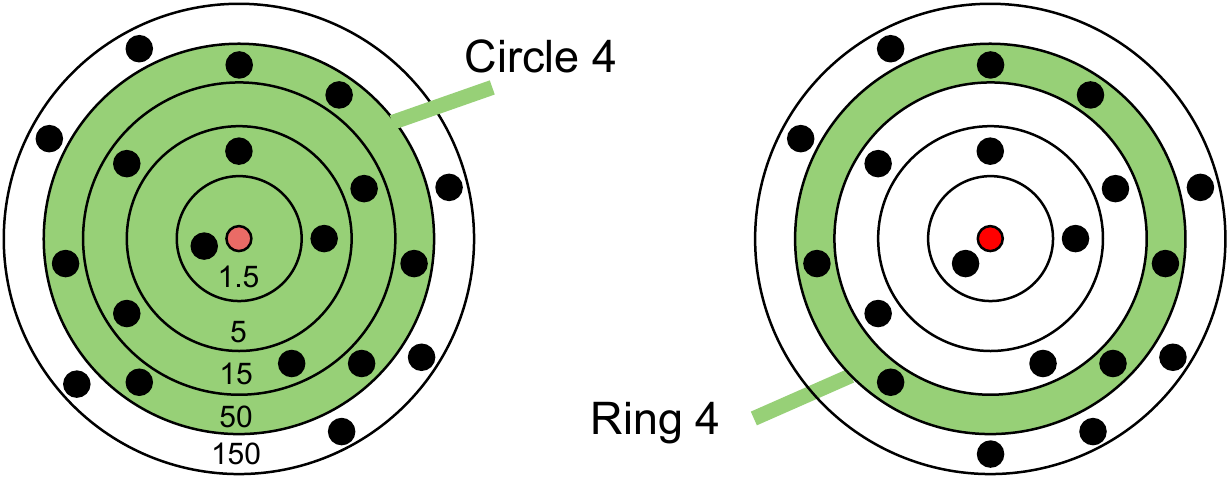}\vspace{-5pt}
\caption{Layered structure of human ego networks. The ego (red dot) is surrounded by alters (black dots), arranged in circles (left). The $i$-th circle contains all alters assigned to the $i$-th circle and the innermost ones. The $i$-th ring, instead, contains the alters in the $i$-th circle excluding those also in the innermost ones (right).}
\label{fig:egonet}\vspace{-10pt}
\end{center}
\end{figure}
    
The above studies are mainly carried out via traditional survey-based data collection and aim to understand journalism culture and journalists' views on the transformation of the profession.
In this work, we set out to investigate the nature and structure of the relationships journalists entertain on Twitter directly by studying their interactions on the platform. To this aim, we leverage a graph-based abstraction known as \emph{ego network}. 
The latter describes the relationships between an individual (ego) and its peers (alters)~\cite{everett2005ego,lin2001social,mccarty2002structure,hill2003social}. 
Grouping these relationships by their strength, a layered structure emerges in the ego network (Figure~\ref{fig:egonet}), with the inner circles containing the socially closest peers and the outer circles the more distant relationships. This structure originates from the limited cognitive resources that humans are able to allocate to \emph{meaningful} socialization (i.e., beyond the mere level of acquaintances), as discussed in detail in Section~\ref{sec:egonetworkmodel}.
In this model, five layers exist within the limit of the Dunbar's number, which is the maximum number of social relationships (around 150) that an individual, on average, can actively maintain~\cite{hill2003social, Zhou2005}. Beyond Dunbar's number, relationships are just acquaintances and their maintenance has a negligible effect on the cognitive resources of the ego.
The ego network is an important abstraction, as it is known that many traits of social behavior (resource sharing, collaboration, diffusion of information) are chiefly determined by its structural properties~\cite{sutcliffe2012relationships}.
While the ego network structure of general users in Online Social Networks have been analysed quite extensively recently, 
to the best of our knowledge, there are no studies on how journalists use their cognitive resources to maintain their relationships on Twitter, and no comparative analyses of ego network-based behavioral characteristics. As there are strong indications that Online Social Networks can be used as a reliable ``big data microscope" to investigate the key features of human social structures (also beyond the OSN environments), such a study can shed light on how a particularly critical community of users form their social relationships, and how they dynamically evolve over time.

In this paper, we study journalists from different countries and continents in order to highlight similarities and invariants in their social behavior online, and to understand how journalists allocate their cognitive resources among their colleagues and non-colleague alters. 
Specifically we address the following research questions:
\begin{description}
\item[RQ\#1:] How do journalists, as members of a specific community, organize their ego networks?
\item[RQ\#2] Does their geographical region affect how journalists organize their ego networks?
\item[RQ\#3] How do the ego networks of journalists change over time?
\item[RQ\#4] Are journalists topic-driven while establishing their relationships?
\item[RQ\#5] Does popularity play a role in how journalists engage with other users?
\end{description}
Our results about journalists on Twitter are compared, when possible, against those about politicians and generic users on Twitter, which have already been studied in the related literature. Specifically, the former have been investigated in~\cite{arnaboldi2017}, the latter in~\cite{Dunbar2015, arnaboldi2017}. 

The key findings presented in this study are the following:

\begin{itemize}
    \item Journalists establish numerous relationships, much more than generic Twitter users. However, when we consider only those relationships that are active (i.e., require a minimum cognitive effort), the ego network size (119 alters, on average) becomes very close to the Dunbar's number of 150. The social circle hierarchy of journalists also mirrors that observed for online networks, with an optimal number of social circles around 5.
    \item Journalists whose Twitter interactions mostly rely on replies (such as those from Finland, the Netherlands, and Denmark) have a below-average number of alters in their ego networks,  thus  suggesting  a  tendency  to  get  involved  with  the  same  group  of people. This might hint at the fact that replies are a more personal/intimate communication with respect to mention and retweets, hence they consume more cognitive resources on the ego side, who, in turn, is able to interact with fewer people. Vice versa, the journalists that interact via retweets and mentions tend to engage with an above-average number of distinct peers, suggesting the opposite effect. 
    \item From a dynamic perspective, journalists have stable short-term relationships that don't change a lot over time. They tend to keep their most and least intimate relationships as they are, while replacing the averagely intimate relationships much more. In the longer term, though, ego networks can be pretty dynamic, especially in the innermost circles. This is in contrast with the findings about generic Twitter users~\cite{Arnaboldi2013} and suggests that the ego networks of journalists, while structurally similar to that of generic users, may be affected by the information-driven nature of journalist engagement on the platform, thus yielding to much more variability in the composition of rings.
    \item Journalists are topic-driven users. They establish their relationships through hashtags. A relationship is hashtag-activated if the first contact of the relationship includes a hashtag. Journalists have more hashtag-activated relationships than politicians and generic Twitter users. In addition, they tend to utilize hashtags for hashtag-activated relationships more than for the relationships that are not activated by hashtags. The percentage of hashtag usage is higher in inner circles that include more intimate relationships.
    \item Highly popular journalists tend to engage with other journalists of similar popularity, and vice versa, creating an assortativity in popularity. In addition, journalist egos tend to keep their popular colleagues in the more intimate layers, hence allocating more cognitive resources for them. Instead, when journalist egos interact with non-journalist alters, the alter popularity i) does not seem to play a role for relationships in the outermost circles, ii) has a smaller effect than for journalist-journalist relationships even in the innermost circles.
\end{itemize}

Therefore, comparing journalists with generic users, we can highlight the following key aspects:
\begin{itemize}
    \item \emph{Structurally}, ego networks of all users are alike, meaning they are organised in a well-defined number of concentric circles of decreasing engagement with a well-defined limit in total size (the Dunbar's number).
    
    \item Also for journalists, as for generic users, \emph{the structure of the ego network} seems determined by the same cognitive constraints governing \emph{meaningful} human social relationships in general.
    
    \item \emph{Within these limits}, specific communities of users may ``use" the available social ties in distinct ways. For example, journalists show quite different patterns from generic users in the way they turn around alters, and use intimate relationships for mutual engagement based on popularity.
    
    \item Thus, while the well-established ego network model is confirmed to be a footprint of human microscopic social structures, the \emph{way alters are organised inside this model} can be a footprint of specific users communities, reflecting peculiar usages of Online Social Networks and Media.
\end{itemize}

The paper is organized as follows. In Section~\ref{sec:egonetworkmodel}, we describe the Dunbar's ego network model and how we apply it. In Section~\ref{sec:dataset}, we describe how we collected the datasets, how we cleaned and preprocessed them to determine which journalists are suitable for ego network analysis. In Section~\ref{sec:EGONETWORKANALYSIS}, we show the results of ego network analysis in four sub-sections with similarities and differences between journalists from different regions in addition to comparison with politicians and generic Twitter users. In Section~\ref{sec:conclusions} we conclude the paper. 
Given the large number of countries in our dataset, it will not be always possible to provide plots for all of them. In these cases, the plots can be found in the appendices.

\section{Background and methodology}
\label{sec:egonetworkmodel}

\subsection{The ego network model and related literature}

As discussed in Section~\ref{sec:introduction}, ego networks are graph-based abstractions used to study the relationships between a tagged individual (called ego) and its peers (alters). Specifically, an ego network is a local subgraph consisting of the ego (in the center) and ties/links that connect the ego to its alters. The strength of these links quantifies the emotional closeness/intimacy of the ego-alter relationship, and it is commonly measured in the related literature through the contact frequency~\cite{hill2003social, roberts2011costs, arnaboldi2013_2}.
When the ego-alter ties are grouped together~\cite{dunbar1998social, Dunbar2015} based on their strength,  \emph{intimacy layers} emerge, as shown in Figure~\ref{fig:egonet}.  The typical sizes of each layer are 1.5, 5, 15, 50, 150, respectively, where inner circles include alters with higher emotional closeness (higher contact frequency), and this closeness decreases through outer circles. The size (150 alters) of the outermost circle is known as \emph{Dunbar's number}. It represents the maximum number of relationships we can actively maintain~\cite{hill2003social, Zhou2005}. An active relationship is defined as one having at least one contact per year (see Sec.~\ref{sec:egonet_extraction} for more details). The relationships beyond the fifth circles are not active (just acquaintances), since we do not use our cognitive resources regularly to maintain them.
This finite-size layered structure is the result of our limited information-processing capacity, and this finding from anthropology goes under the name of \emph{social brain hypothesis}~\cite{dunbar1998social}. Since we have limited cognitive capacity and limited time, we try to optimize how we allocate our cognitive resources among our relationships~\cite{sutcliffe2012relationships}. The result of this optimisation is the ego network structure. An important structural invariant of ego networks is their \emph{scaling ratio}, defined as the ratio between the sizes of consecutive layers. Its value is typically around three in the offline and online networks studied in the related literature~\cite{Dunbar2015}~\cite{Zhou2005}.

The existence of Dunbar's ego network structures has been shown in different communication means in the offline world, including face-to-face interactions~\cite{roberts2009}, letters (Christmas cards)~\cite{hill2003social}, phone calls~\cite{miritello2013}. Recently, their occurence has been confirmed also for online interactions in OSNs~\cite{Dunbar2015,arnaboldi2017}, showing that the same cognitive capacity that limits our offline social interactions exists also in OSNs (Facebook, Twitter). In this sense, OSN become one of the outlets that is taking up the brain capacity of humans, and thus are subject to the same limitations that have been measured for more traditional social interactions, but are not capable of ``breaking'' the limits imposed by cognitive constraints to our social capacity.
Tie strengths and how they determine ego network structures have been the subject of several additional work. For example, in~\cite{gonccalves2011modeling} authors provide one of the first evidences of the existence of an ego network size comparable to the Dunbar's number in Twitter. The relationship between ego network structures and the role of users in Twitter was analysed in~\cite{quercia2012social}. In general, ego network structures are also known to impact significantly on the way information spreads in OSN, and the diversity of information that can be acquired by users~\cite{aral2011diversity}.
In~\cite{arnaboldi2017}, the ego network structure of politicians and how they allocate their cognitive resources among the alters have been studied. In the study, the authors showed the similarities between politicians' ego networks and Dunbar's model, and differences between generic users' ego networks by using Twitter data~\cite{Dunbar2015}. In addition to the characterization of the ego networks, the effect of ego network structure on information diffusion has also been investigated in the literature~\cite{Arnaboldi2017Facebook}.

\subsection{Social interactions on Twitter}
\label{sec:twitter_intro}

Twitter was launched in 2006 as a microblogging and social network platform. Users of the platform can share their thoughts in the form of \emph{tweets}, i.e., short messages of 280 characters or less, that are collected in the user's bulletin board (called \emph{timeline}).
Differently from Facebook posts, tweets are publicly visible by default, effectively creating a platform for engaging with and observing public discourse. Users can interact with each other in an information-driven way or in a socially-driven way. In the former, users place hashtags, i.e., words or phrases prefixed with a ``\#" sign, in their tweets, effectively marking the tweet as belonging to a certain topic (specified by the hashtag word). In the latter, users explicitly reply to other users' tweets, mention other users (via the ``@'' symbol) in their tweets, or retweet their content (i.e., share other users' tweets on their own timeline). In this work, we will mostly focus on Twitter social interactions, which will be used to build journalists' ego networks, as explained in the next section. 

\subsection{How to extract ego networks}
\label{sec:egonet_extraction}

Dunbar's ego network model captures how an individual allocates their cognitive resources among their relationships. We do not allocate our cognitive resources homogeneously across all the relationships that we establish.  Instead, we spend them mostly on the ones we try to keep ``active'' in our lives. In~\cite{hill2003social}, an active relationship has been defined as one where at least one contact per year occurs. This idea comes from the Christmas card exchange tradition of Western societies where people put effort to contact -- at least once a year -- the people they value the most. In the related studies about online ego networks~\cite{Arnaboldi2013, Dunbar2015, arnaboldi2017}, the concept of active relation has been adapted to the online domain by calculating the frequency of \emph{direct} contacts and labelling as \emph{inactive} those relationships with average contact frequency smaller than one direct tweet per year (in analogy with the Christmas card exchange in offline networks). 
In Twitter, a direct contact is a tweet (\emph{retweet}, \emph{reply}, \emph{mention}) where a user directly interacts with another user. Based on direct contacts, we calculate\footnote{We explain in Section~\ref{sec:dataset} how we have downloaded the necessary data from Twitter.} the contact frequency for a relationship between ego $i$ and alter $j$ as follows:
\begin{equation}
w_{ij} = \frac{N_{reply} + N_{mention} + N_{retweet}}{L_{R}},
\end{equation}
where $w_{ij}$ is contact frequency (which is a proxy for the intimacy between the ego and the alter~\cite{hill2003social, roberts2011costs, arnaboldi2013_2}), $N_{reply}$, $N_{mention}$, $N_{retweet}$ are the total number of replies, mentions, and retweets between the ego $i$ and the alter $j$, and $L_{R}$ is the observed duration (in years) of the relationship between $i$ and $j$, calculated as the time interval between their first direct tweet until the download date of the dataset. 
In agreement with the related literature, we filter out the relationships where $L_{R}$ is smaller than one year, since these recently established relationships may not have yet stabilized~\cite{Dunbar2015, arnaboldi2017, boldrini2018}. 

In order to extract the intimacy layers as in Figure~\ref{fig:egonet}, the active relationships must be clustered into groups. Each layer includes relationships that have similar intimacy with the ego. The number of layers, and the alters located in each of them, can be determined by means of clustering algorithms. Similarly to~\cite{quadri2018feature}, in this work we use the Mean Shift algorithm~\cite{comaniciu2002mean}, which is able to automatically identify the optimal number of clusters without additional interventions (e.g., $k$-means requires using the silhouette score or the elbow method to this purpose). Thus, for each ego in our datasets, starting from the active ego network, we derive the optimal number (typically around 5 in the related literature) of intimacy layers and the alters assigned to them. In this work, we denote the $i$-th circle with $\mathcal{C}_i$. Recall that circles are ``cumulative'' in the ego network model: circle $\mathcal{C}_i$ includes $\mathcal{C}_{n-1}$, which in turn includes $\mathcal{C}_{n-2}$, and so on. We use instead the term \emph{ring} to refer to the portion of a circle that excludes its inner circles. Thus, ring $\mathcal{R}_{i+1}$ is defined as $\mathcal{C}_{i+1} - \mathcal{C}_i$. Conventionally, $\mathcal{R}_1$ is equivalent to $\mathcal{C}_1$. The difference between circles and rings is illustrated in Figure~\ref{fig:egonet}.
Then, we can easily compute the ego network scaling ratios as the ratio between the sizes of consecutive layers (recall that this is typically an invariant, whose value is around three). Specifically, the scaling ratio $\rho_i$ is given by $\frac{|\mathcal{C}_{i}|}{|\mathcal{C}_{i-1}|}$, with $i \in \{2, \ldots, \tau\}$. 

The ego networks obtained as described above are typically referred to as \emph{static ego networks}, and they will be studied in Section~\ref{sec:egonets_analysis_static} for the journalists in our datasets. 
By modelling the ego-alter ties through a static ego network, we consider all the communications between the ego and the alter. This analysis provides an essential starting point for understanding the ego-alter interactions and for quantitative comparisons with other datasets studied in the related literature. 
At the opposite end of the spectrum, the \emph{dynamic} analysis of ego networks focuses on their evolution over time. Specifically, snapshots of one year are considered, each being shifted forward by a fixed amount of time with respect to the previous one. 
For each snapshot we focus on the alters that are members of each \emph{ring}, where we recall that a ring is defined as the portion of circle that excludes its inner circles.
Egos for whom we observe less than two years of tweets are filtered out, as they are not observed long enough to extract robust results. The dynamic analysis of the journalists' ego networks will be discussed in Section~\ref{sec:egonets_analysis_dynamic}.

With dynamic ego networks, the intimacy between the ego and alter may change between two consecutive time intervals, possibly resulting in the alter moving from one circle to another one. In order to capture the stability of rings (in terms of membership) over time, we use the Jaccard similarity, defined as the ratio between the cardinality of the intersection and that of the union set of the alters belonging to ring $i$ in consecutive time intervals. We calculate the Jaccard index $Jaccard_i$ of ring~$i$ as follows:
\begin{equation}
Jaccard_i = \frac{1}{T} \sum_{t=1}^{T-1} \frac{\mathcal{R}_i^{(t)}\cap \mathcal{R}_i^{(t+1)} }{
\mathcal{R}_i^{(t)} \cup \mathcal{R}_i^{(t+1)} } %
\end{equation}
where $T$ denotes the total number of consecutive snapshots. The closer this index to one, the higher the overlapping. In case of no overlap, the Jaccard index is equal to zero. 
The amount of movements between rings is measured through the Jump index, which simply counts (and then averages across all alters in the ring) the number of jumps between rings. This index can be computed as follows:
\begin{equation}
    Jump_i = \frac{1}{T} \sum_{t=1}^{T-1} \frac{1}{|\mathcal{R}_i^{(t+1)}|} \sum_{j \in \mathcal{R}_i^{(t+1)}} \Delta_j^{t,t+1},
\end{equation}
where we denote as $\Delta_j^{t,t+1}$ the number of jumps of alter $j$ between snapshot $t$ and $t+1$. For example, if the alter moves from $\mathcal{R}_1$ at $t$ to $\mathcal{R}_4$ at $t+1$, then $\Delta_j^{t,t+1}$ will be equal to three. Of course, it is possible that, at $t+1$, an alter enters the active network from the outside. In this case, we assume that it comes from the ``last plus one'' ring: e.g., for an ego with $n$ social circles, the alters is assumed to come from the $(n+1)$-th ring. Thus, each jump value $\Delta_j^{t,t+1}$ ranges between zero (no jump) and $n+1$ (e.g., for a jump from outside the ego network into $\mathcal{R}_1$). $Jump_i$ is then obtained as the average of these jump values across all alters and across all snapshots. Since Section~\ref{sec:egonets_analysis_dynamic} focuses on egos with five circles (i.e., $n=5$), $Jump_i$ will take values in $[0,6]$.

The notation we use in the paper is summarised in Table~\ref{tab:notation}.

\begin{table}[h!]
\centering
\caption{Summary of notation (for a tagged ego)}
\footnotesize
\begin{tabular}{@{}lcp{6cm}@{}}
\toprule
\textbf{Name} & \textbf{Notation} & \textbf{Definition/formula}\\
\midrule
Active network & $\mathcal{A}$ & ego-centered subgraph with weights greater than 1 \\
Optimal number of circles & $\tau$ & the results of the clustering on the edges in the active network $\mathcal{A}$ \\
Social circle (or layer) & $\mathcal{C}_i$ & $i$-th social circles of the tagged ego, with $i \in \{1, \ldots, \tau\}$ \\
Scaling ratio of layer $i$ & $\rho_i$ & $\frac{|\mathcal{C}_{i}|}{|\mathcal{C}_{i-1}|}$, with $i \in \{2, \ldots, \tau\}$ \\
Ring  & $\mathcal{R}_i$ & $\mathcal{C}_i - \mathcal{C}_{i-1}$ \\
Jaccard index of ring $i$ & $Jaccard_i$ & $\frac{1}{T} \sum_{t=1}^{T-1} \frac{\mathcal{R}_i^{(t)}\cap \mathcal{R}_i^{(t+1)} }{\mathcal{R}_i^{(t)} \cup \mathcal{R}_i^{(t+1)} }$ \\
Jump index of ring $i$ & $Jump_i$ & $\frac{1}{T} \sum_{t=1}^{T-1} \frac{1}{|\mathcal{R}_i^{(t+1)}|} \sum_{j \in \mathcal{R}_i^{(t+1)}} \Delta_j^{t,t+1}$\\
\bottomrule
\end{tabular}
\label{tab:notation}
\end{table}

\section{The dataset}
\label{sec:dataset}

We downloaded the Twitter timelines of journalists belonging to 17 different countries from 8 different continental regions in May 2018\footnote{Note that, for capturing the general ego network properties of journalists online, pre-pandemic datasets are to be preferred, as they capture ``baseline'' social behaviours in normal times. Note also that Twitter has undergone no major changes during these years, with the number of  monthly active users being quite stable since 2018 (\url{https://backlinko.com/twitter-users#usage-growth}).}. To this end, we used the Twitter REST API, which allows us to download, for a given user, the last 3200 tweets  since the time of the download. We classify geographical regions (continents and continental regions) by using UN M49 (standard country or area codes for statistical use\footnote{\url{https://unstats.un.org/unsd/methodology/m49/}}). The countries, which can be seen in Table~\ref{tab:DatasetSummary}, were selected based on the availability of journalists lists on Twitter. Twitter lists are collections of Twitter accounts that are included in a list by Twitter users with the aim of having a pool of accounts, generally quite popular, that are related to a specific topic of interest. For our study, we manually looked up lists including journalists from the same country\footnote{The Twitter lists that we used as seeds can be found in \url{https://github.com/mustafatoprak/TwitterJournalistLists}}.
For some of the countries, we found more than one relevant list, and in this case the lists are merged into one. Our dataset contains a different number of journalists for each country, as it can be seen in Table~\ref{tab:DatasetSummary} (under the \textit{before filtering} column). As a result of the Twitter API limitations, only the 3200 most recent tweets of each journalist (with a public profile) have been downloaded, and timelines without any tweeting activity were removed from the dataset.



\definecolor{NA}{rgb}{0.42, 0.43, 0.81}
\definecolor{SA}{rgb}{.32, 0.33, 0.64}
\definecolor{EA}{rgb}{0.84, 0.38, 0.42}
\definecolor{WA}{rgb}{0.68, 0.29, 0.29}
\definecolor{NE}{rgb}{0.71, 0.81, 0.42}
\definecolor{SE}{rgb}{0.55, 0.64, 0.32}
\definecolor{WE}{rgb}{0.39, 0.47, 0.22}
\definecolor{AN}{rgb}{1, 0.73, 0.47}
\definecolor{ALL}{rgb}{0.5, 0.5, 0.5}

\begin{table}[ht]
\centering
\caption{General statistics for the datasets}
\footnotesize
\begin{adjustbox}{width=1\textwidth}
\small
\begin{tabular}{@{}llllllll@{}} \toprule
 & & & \multicolumn{2}{l}{\textbf{before filtering}} & \multicolumn{3}{l}{\textbf{after filtering}} \\ \cmidrule(r){4-5}\cmidrule(lr){6-8}
\textbf{continent} & \textbf{region} & \textbf{dataset} & \textbf{profiles (\#)} & \textbf{tweets (\#)} & \textbf{profiles (\#)} & \textbf{tweets (\#)} &
\begin{tabular}{@{}cc@{}} \textbf{fully observed(\%) /} \\ \textbf{partially observed(\%)}
\end{tabular}\\ \midrule

\rowcolor{NA!30}
Americas & Northern America & USA & 1,722 & 4,671,922 & 617 & 1,927,183 & 10.86 \hspace{5mm}/\hspace{5mm} 89.14\\ 
\cmidrule{3-8}
\rowcolor{NA!30}
 & & Canada & 917 & 2,462,998 & 427 & 1,281,149 & 	20.14 \hspace{5mm}/\hspace{5mm} 79.86\\ \cmidrule{2-8}
 \rowcolor{SA!30}
 & South America & Brazil & 897 & 2,489,943 & 217 & 1,236,620 & 13.82 \hspace{5mm}/\hspace{5mm} 86.18\\ \midrule
 \rowcolor{EA!30}
Asia & Eastern Asia & Japan & 566 & 1,361,392 & 192 & 521,676 & 34.90 \hspace{5mm}/\hspace{5mm} 65.1\\ \cmidrule{2-8}
 \rowcolor{WA!30}
 & Western Asia & Turkey & 2,189 & 5,495,545 & 731 & 2,203,729 & 14.64 \hspace{5mm}/\hspace{5mm} 85.36\\ \midrule

\rowcolor{NE!30}
Europe & Northern Europe & UK & 512 & 1,330,467 & 235 & 717,341 & 14.47 \hspace{5mm}/\hspace{5mm} 85.53\\ \cmidrule{3-8}
\rowcolor{NE!30}
& & Denmark & 3,694 & 3,725,084 & 656 & 1,671,441 & 50.00 \hspace{5mm}/\hspace{5mm} 50.0\\ \cmidrule{3-8}
\rowcolor{NE!30}
& & Finland & 936 & 1,477,304 & 321 & 824,823 & 46.11 \hspace{5mm}/\hspace{5mm} 53.89\\ \cmidrule{3-8}
\rowcolor{NE!30}
& & Norway & 1,190 & 1,504,906 & 201 & 584,251 & 28.86 \hspace{5mm}/\hspace{5mm} 71.14\\ \cmidrule{3-8}
\rowcolor{NE!30}
& & Sweden & 761 & 1,792,973 & 275 & 847,196 & 12.36 \hspace{5mm}/\hspace{5mm} 87.64\\ \cmidrule{2-8}
\rowcolor{SE!30}
 & Southern Europe & Greece & 862 & 1,858,442 & 265 & 778,403 & 24.15 \hspace{5mm}/\hspace{5mm} 75.85\\ \cmidrule{3-8}
\rowcolor{SE!30}
& & Italy & 486 & 1,301,858 & 255 & 781,415 & 16.47 \hspace{5mm}/\hspace{5mm} 83.53\\ \cmidrule{3-8}
\rowcolor{SE!30}
& & Spain & 468 & 1,309,723 & 195 & 611,233 & 10.26 \hspace{5mm}/\hspace{5mm} 89.74\\ \cmidrule{2-8}

\rowcolor{WE!30}
 & Western Europe & France & 559 & 1,525,600 & 323 & 975,941 & 21.98 \hspace{5mm}/\hspace{5mm} 78.02\\ \cmidrule{3-8}
\rowcolor{WE!30}
& & Germany & 405 & 879,071 & 146 & 415,560 & 30.14 \hspace{5mm}/\hspace{5mm} 69.86\\ \cmidrule{3-8}
\rowcolor{WE!30}
& & Netherlands & 4,303 & 9,621,422 & 1,674 & 4,859,433 & 27.18 \hspace{5mm}/\hspace{5mm} 72.82\\ \midrule

\rowcolor{AN!30}
Oceania & \begin{tabular}{@{}cc@{}} Australia and \\ New Zealand \end{tabular} & Australia & 956 & 220,4974 & 400 & 1,153,010 & 29.50 \hspace{5mm}/\hspace{5mm} 70.5\\ \midrule
\rowcolor{ALL!30}
& \textbf{All journalists} & \begin{tabular}{@{}ccc@{}} \textbf{mean $\pm$ sd} \end{tabular} & \begin{tabular}{@{}ccc@{}} \textbf{1,260.18} \\ \textbf{$\pm$} \\ \textbf{1,134.95} \end{tabular} & \begin{tabular}{@{}ccc@{}} \textbf{2,647,860.24} \\ \textbf{$\pm$} \\ \textbf{2,203,177.69} \end{tabular} & \begin{tabular}{@{}ccc@{}} \textbf{419.41 } \\ \textbf{$\pm$} \\ \textbf{367.15} \end{tabular} & \begin{tabular}{@{}ccc@{}} \textbf{1,224,831 } \\ \textbf{$\pm$} \\ \textbf{1,064,588} \end{tabular} & \textbf{24.87} \hspace{5mm}/\hspace{5mm} \textbf{75.13}\\
\hline
\end{tabular}
\end{adjustbox}
\label{tab:DatasetSummary}
\end{table}

\subsection{Observability}

Since the Twitter API only discloses the last 3200 tweets of any public Twitter account, we were not able to download the full timelines of journalists having more than 3200 tweets. We refer to these accounts as \emph{partially observed}, because we can only observe a portion of their timelines instead of their full Twitter activity since their registration to the platform.  Across our datasets, on average $75.13\%$ of the users are partially observed (i.e., they posted more than 3200 tweets, see Table~\ref{tab:DatasetSummary}). We were able to fully observe only $23.7\%$ of the users (Table~\ref{tab:DatasetSummary}). 
Please note that being partially or fully observed is an indirect measure of the user’s tweeting frequency: the more the tweets posted, the quicker the 3200 slots are saturated. Indeed, the average daily tweeting frequency, across all journalists datasets, is 0.5 tweets/day for the fully observed users and 6.1 tweets/day for the partially observed ones (note that this number is fully compatible with human activity). The observability of the journalists' timelines changes among the different countries: most of the datasets have at least $66\%$ of the users that are partially observed, except for the Danish and Finnish journalists where approximately half of the users have partially observed timelines. 
Note also that journalists that are partially observed are actually more suitable for our analysis than fully observed ones. Indeed, as it will become clear in Figure~\ref{fig:user_classifier}, fully observed journalists are fully observed because they are less active on Twitter, hence they do not provide enough data over time and most of them will be discarded from our analysis.

\begin{figure}[t]
\subfloat[American journalists
\label{fig:tweeting_volume_AmericanJournalists}]
{\includegraphics[width=0.49\textwidth]
{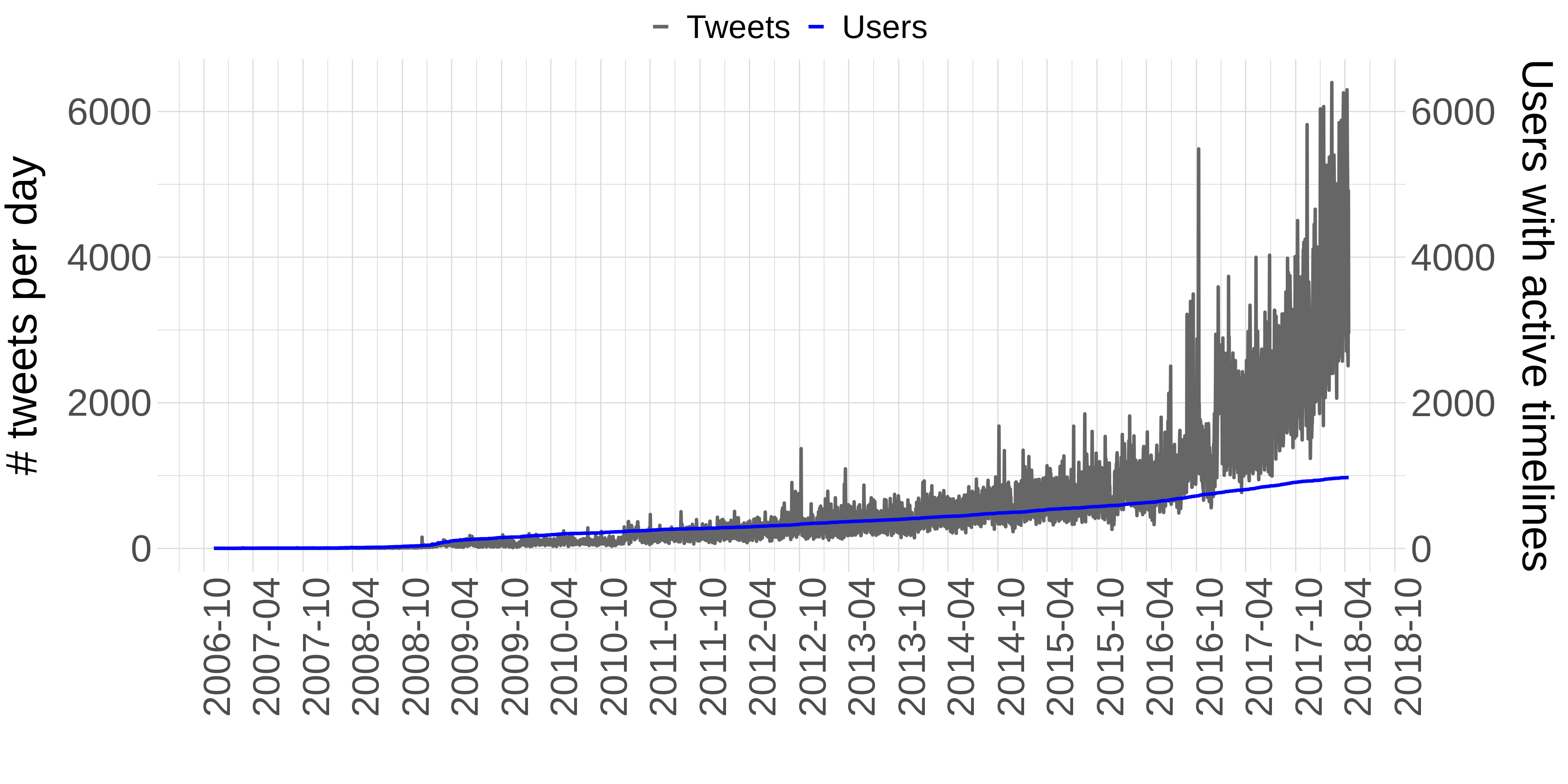}}
\hfill
\subfloat[Danish journalists
\label{fig:tweeting_volume_DanishJournalists}]
{\includegraphics[width=0.49\textwidth]
{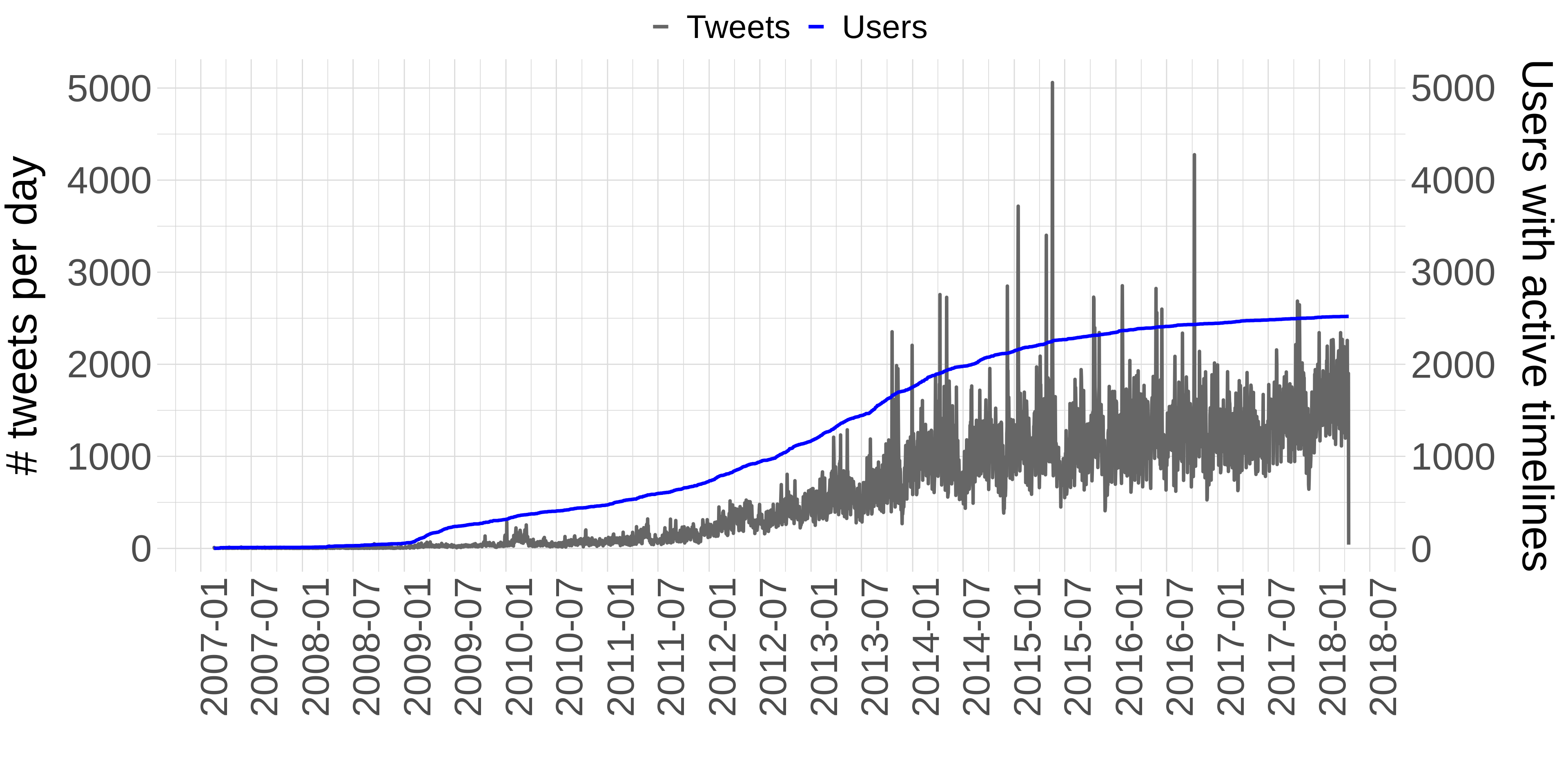}}
\hfill
\caption{Tweeting volume and active users over time for two representative countries}
\label{fig:tweet_volume_evolution}
\end{figure}

We define the \emph{observed timeline} as the portion of the user's timeline we can access via the Twitter API. If the 3200-tweets API limitation is not hit, the observed timeline is the same as the \emph{active timeline}, which instead covers the user activity starting from the date the user registered to Twitter until the dataset was downloaded. We provide the distribution of the observed timelines for all countries in~\ref{appendix_observed_timeline_legnth}\iftoggle{ONEFILE}{}{ of the Supplemental Material}.
%
Since the coverage of the timeline is based on the number of tweets published by the user, the observed portion of the timeline is user-based. As a result, if we measure the tweeting volume (total amount of daily tweets over time across all users in a given dataset) in the datasets, we observe a striking growth in the most recent weeks (Figure~\ref{fig:tweeting_volume_AmericanJournalists}, gray curve). This is a spurious effect, due to the fact that the observed timelines of the partially observed users fall in the latter period of the datasets timeframe. This is evident when plotting the number of active users who produced the corresponding tweeting volume (blue curve in Figure~\ref{fig:tweeting_volume_AmericanJournalists}).  
As we can see from the figure, the plots start from the early days of Twitter\footnote{Twitter has been established in 2006.} (since we have fully observed users whose observed timelines go back to the registration date). As we follow the timelines throughout the later dates, we notice that, while the number of active users increases (approximately linearly in most of the datasets), the total tweet volume per day increases exponentially. 
While the behaviour observed in Figure~\ref{fig:tweeting_volume_AmericanJournalists} well represents the general trend across several countries, there are also countries (like the Northern European ones, together with Germany and the Netherlands) where the observed behavior is different. As shown in Figure~\ref{fig:tweeting_volume_DanishJournalists} for a representative country in this pool, very prolific journalists do not seem to be part of the dataset in this case, and there is no exponential growth of the tweeting volume in the most recent days of the datasets. The plots for all the countries in our dataset can be found in~\ref{appendix_tweetingVolume}\iftoggle{ONEFILE}{}{ of the Supplemental Material}.

\subsection{Keeping only real journalists}
\label{sec:labellingusers}

\begin{figure}[t]
\subfloat[British journalists
\label{fig:mcc_british}]
{\includegraphics[width=0.49\textwidth]
{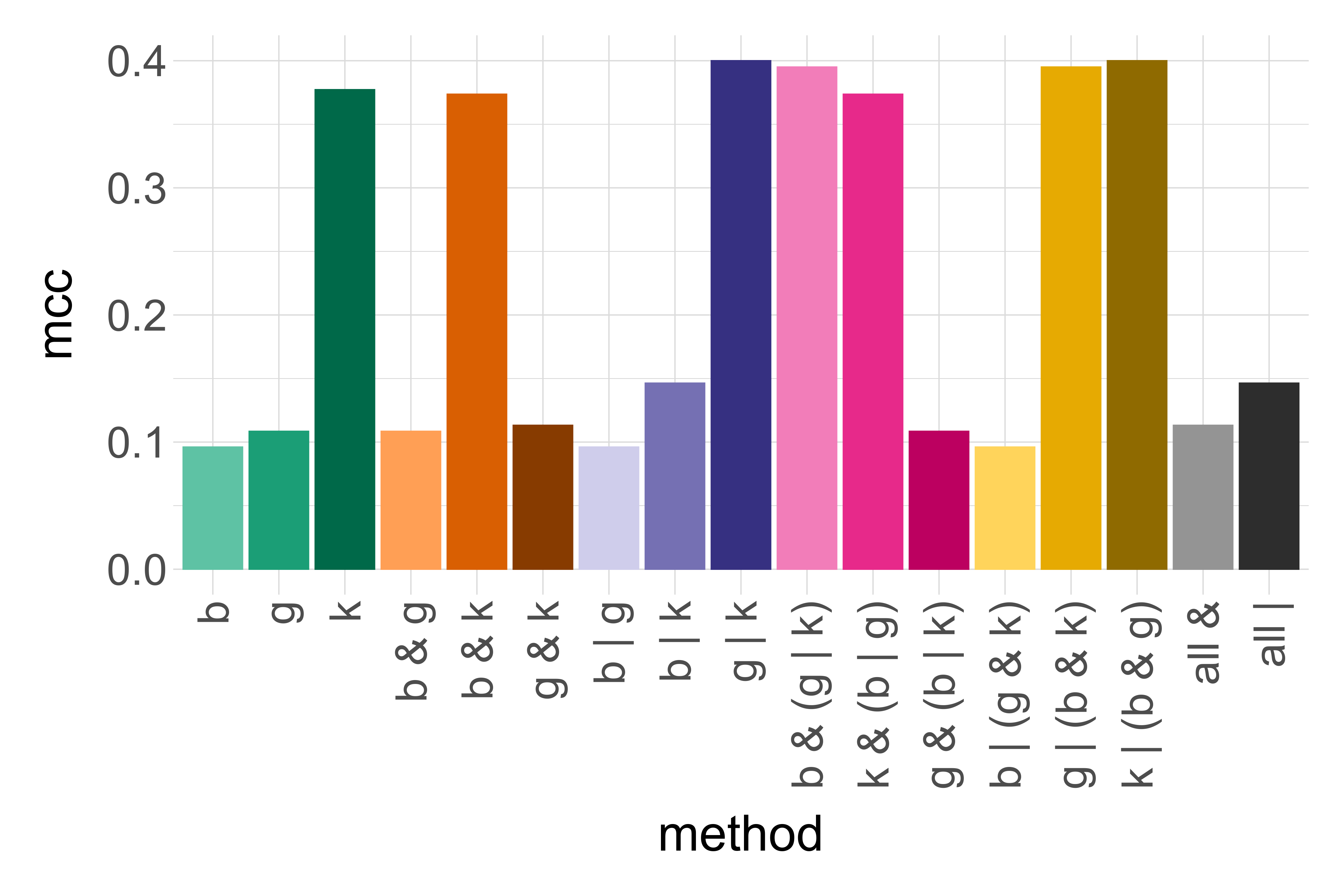}}
\hfill
\subfloat[Italian journalists
\label{fig:mcc_italian}]
{\includegraphics[width=0.49\textwidth]
{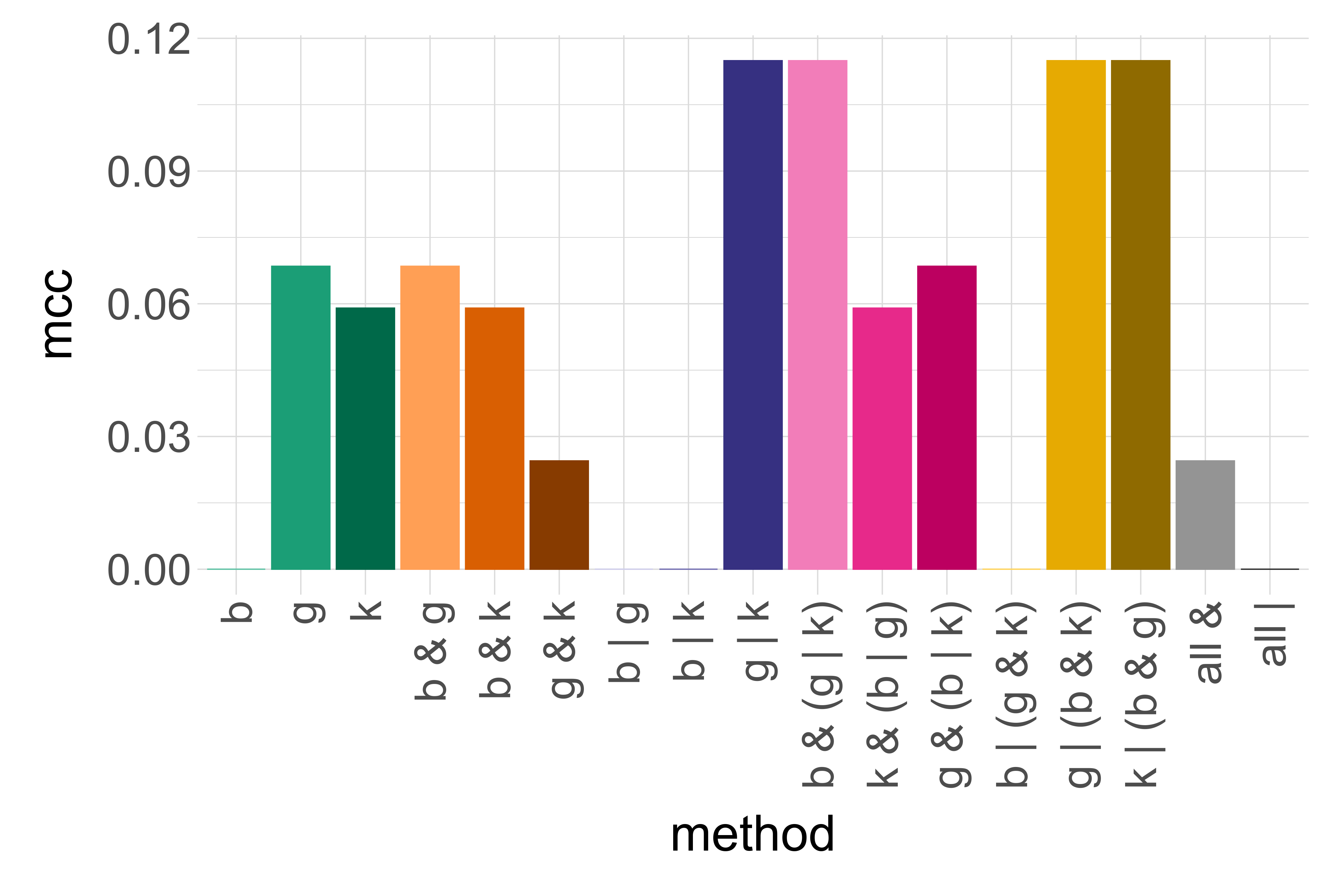}}
\hfill
\caption{MCC scores. In the plots, $g$ represents the classifier that leverages the Google Knowledge Graph, $k$ represents the keyword extraction method, $b$ represents the bot detection classifier. The symbol $\&$ represents the logical AND operation, symbol $|$ represents the logical OR. They are used to specify the test combinations of the three classifiers.}
\label{fig:mcc}
\end{figure}

The Twitter lists that we used to identify our reference pool of journalists are created by other Twitter users\footnote{Note that no \emph{official} Twitter list exists, hence relying on user-generated lists is the only way to collect this kind of thematic datasets at a scale. Manual selection of journalist accounts, beside behind time-consuming, would have been extremely difficult, given the language barrier for some countries.}. The general aim of these lists is to provide a curated collection of news/information sources generally related to a specific country or region. Therefore, these lists may include not only journalists but also news companies, anchormen, radio hosts, and so on.
Even though these accounts are sources of information, they are not journalists. Including these accounts may lead us to a characterization of ego networks that do not represent journalists accurately. Thus, we need to apply a classifier to decrease the noise in the datasets. In order to correctly identify ``true'' journalists, we have tested three different approaches (and their combinations). The first one is based on journalism-related keyword-matching in the user Twitter bio (where the match involves words such as \textit{columnist} or \textit{correspondent}). The second one entails searching for the person's job in the Google Knowledge Graph (GKG). The third one filters out users that are marked as bots by the Botometer bot detector. For more details, please refer to~\ref{appendix_labellingusers}\iftoggle{ONEFILE}{}{ of the Supplemental Material}. We have evaluated the performance of these three classifiers -- in isolation and combined -- on a subset of journalists and non-journalists (Italian and English ones) that we have manually labelled. Since the journalist/non-journalist classes are quite imbalanced in our labelled dataset, we used MCC (Matthews' Correlation Coefficient~\cite{Matthews1975}) for assessing which combination performs best. As explained in~\ref{appendix_labellingusers}\iftoggle{ONEFILE}{}{ of the Supplemental Material}, MCC ranges in [-1,1], and higher values correspond to better classification performance. The results are shown in Figure~\ref{fig:mcc}. For British journalists, the top performance is achieved by the \emph{g $|$ k} combination (i.e., a user is labelled as journalist if it passes the keyword-matching strategy \emph{or} if its name is found in the GKG) and by the \emph{k $|$ (b $\&$ g)} combination (meaning that a user is labelled as journalist if it passes the keyword-matching strategy or if it is in GKG and not a bot). The MCC performance being equal, it is convenient to reduce the number of methods used, since each method used in the combination takes up some time for preprocessing. From this standpoint, the classifier jointly considering the information in the GKG together with the keywords in the bio provides the best trade-off (high MCC, fewer computational steps) on the labelled dataset of British journalists. The same line of reasoning also holds for the labelled Italian journalists in Figure~\ref{fig:mcc_italian}. Thus, we use this strategy for filtering out non-journalists from the remaining datasets in our study.
In the end, we removed 7815 non-journalists, i.e., 38.2\% of our datasets. 

\subsection{Identifying the regular and active users}

Of all the journalist users remaining after the filtering in Section~\ref{sec:labellingusers}, we want to keep in our study only those that are regular and active Twitter users. 
The \emph{inactive life} of a user is defined as the time interval between the last tweet of the user and the download time of the dataset.
Long inactive periods are a sign of decreased engagement with the platform.
In~\cite{Arnaboldi2013}, a user is accepted as inactive if there is no activity for the last 6 months (mimicking Twitter 6-month inactivity criterion that labels users as inactive if they do not login at least once every six month\footnote{https://help.twitter.com/en/rules-and-policies/inactive-twitter-accounts}). In this paper, we use instead the \emph{intertweet time}, which we have previously introduced in~\cite{boldrini2018}. This method assumes that a user has abandoned the platform if the inactive period is longer than six months plus the longest observed intertweet time (which corresponds to the longest period between two consecutive tweets of the user). The intuition behind this formulation is to embed the user's personal regularity pattern (by considering their longest break in their tweeting activity) instead of fixing an a-priori period of absence equal for all users. Using this filter, 1169 users are marked as having abandoned Twitter at the time we downloaded the dataset.

\begin{figure}[ht]
\begin{center}
\includegraphics[scale=0.2]{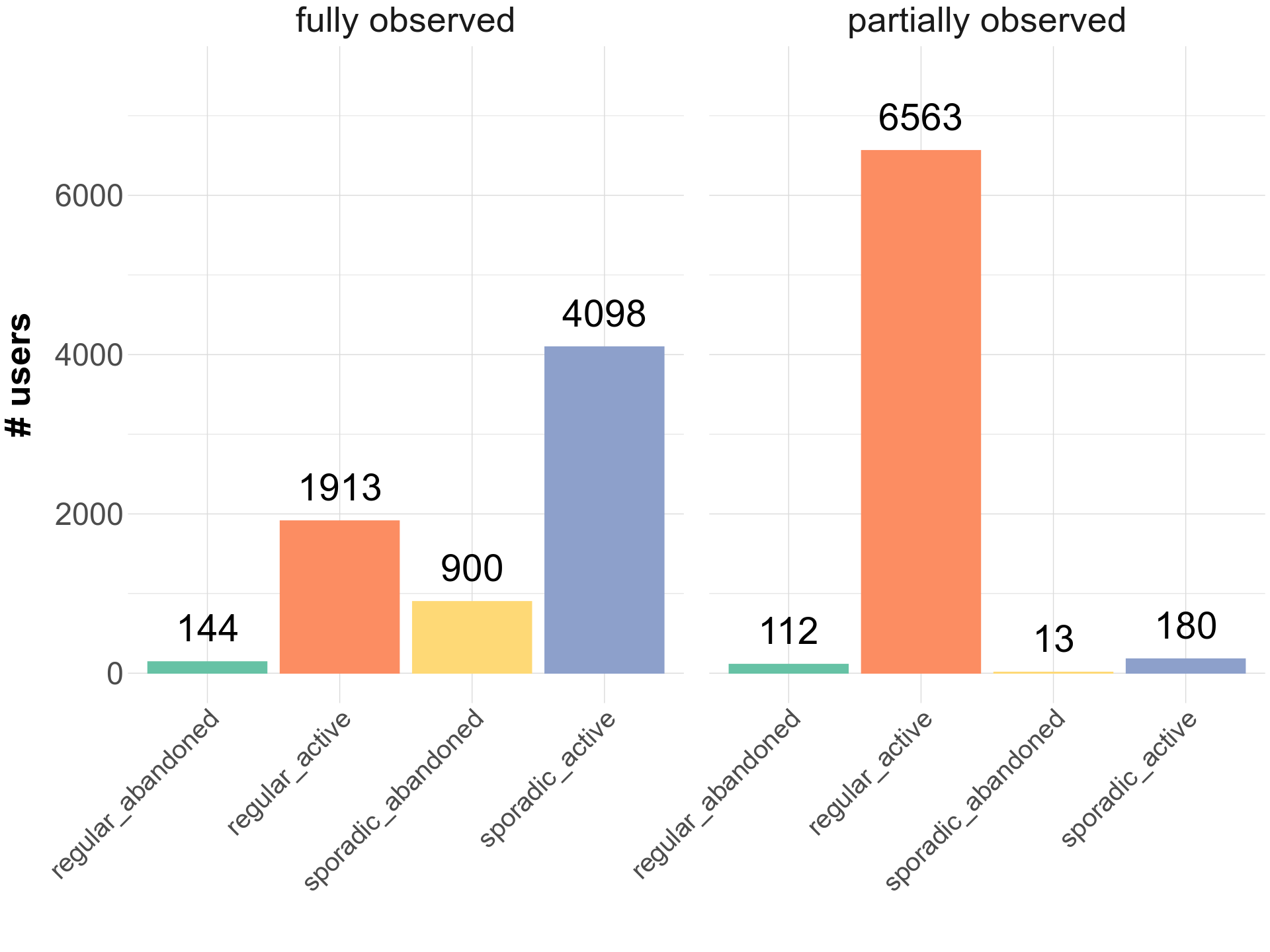}\vspace{-15pt}
\caption{User classification}
\label{fig:user_classifier}\vspace{-10pt}
\end{center}
\end{figure}

We now focus on how regular a user activity is. We assume that a Twitter user is regular if, as in~\cite{arnaboldi2017}, the user posts at least one tweet every 3 days as an average for at least half of the observed timeline. If this is not the case, the user is labelled as \emph{sporadic}. In Figure~\ref{fig:user_classifier}, we show the classification of users both based on the abandonment classifier (previous paragraph) and the regularity classifier. As expected, partially observed users are much more active than fully observed one. For our analysis, we focus on regular and active users, since they engage with the platform regularly and, thus, spend on it more cognitive resources than others.

At the opposite end of the spectrum of inactive users are users that tweet a lot. These users could have a significant impact on the ego network statistics, due to their intense Twitter activity, hence they should be studied separately. In order to detect these outliers, we have run DBSCAN (a standard clustering technique that automatically identifies outliers~\cite{ester1996density-based-algorithm}) on the journalists tweeting frequencies. For all the datasets, we found that all the outliers feature an observed timeline shorter than one year (\ref{appendix_twittingactivity}\iftoggle{ONEFILE}{}{ of the Supplemental Material}). This is due to the fact that, as these users produce numerous tweets per day, they saturate the Twitter API limitation fast, hence their observed timelines are too short. Given that their observed timelines are shorter than one year, their ego networks are discarded, as discussed in Section~\ref{sec:egonet_extraction}.

The last check we perform is about the stationarity of the Twitter activity of the users in our dataset. 
As previously shown in the literature (\cite{Arnaboldi2017Facebook}, \cite{miritello2013}, \cite{viswanath2009}), newly registered social network users add relationships into their network and interact with others at a higher rate than long-term users. After a while, their activity stabilizes. Newly registered users are thus outliers with respect to the general population of users, and they should be discarded from the analysis. Again, we use the approach described in~\cite{boldrini2018}. We do not observe a drastic change in tweeting activity for any of the datasets. Therefore, we do not discard any part of the timeline as a transient period. For more details on this analysis, please refer to~\ref{appendix_stationarity}\iftoggle{ONEFILE}{}{ of the Supplemental Material}.

\subsection{Dataset overview}

After the preprocessing steps described above, we can be reasonably confident that the accounts we use for our analysis belong to real journalists who are also regular and active Twitter users. Under this assumption, all the observed differences between countries that we highlight in Section~\ref{sec:EGONETWORKANALYSIS} can be attributed to different ways of using Twitter in different countries\footnote{Note that, with the current data, we cannot provide the exact reasons (e.g., Twitter less popular among journalists in some countries, existence of competing social networks where journalists share their opinions, etc.) for the differences between countries. Addressing this problem is left as future work.}. The number of journalists in our dataset at the end of the preprocessing phase is reported, per country, in Table~\ref{tab:DatasetSummary} (``after filtering" column). In Table~\ref{tab:summary} we provide some summary statistics regarding the length of their active life on Twitter, their average daily tweets, and their predominant type of tweet. With an average Twitter life between 6 and 9 years, the journalists in out datasets tend to be long-time Twitter users. These journalists tweet, on average, twice or thrice per day. $65\%$ of the tweets we observe are social tweets (\emph{reply}, \emph{retweet}\footnote{Quote tweets are treated as retweets.} and \emph{mention}), i.e., entail a direct communication between two Twitter users. 

\begin{table}[ht!]
\centering
\caption{Summary statistics}
\footnotesize
\begin{adjustbox}{width=1\textwidth}
\small
\begin{tabular}{@{}llrrrrr@{}}  \toprule

dataset & statistic type & active life [years] & total tweets & observed tweets & tweet/day & \% social\\ \midrule
\rowcolor{NA!30}
\textbf{US} & mean & 9.08 & 13199.12 & 3123.47 & 3.24 & 65.41\\
\rowcolor{NA!30}
 & sd & 0.97 & 13407.07 & 318.19 & 2.02 & 16.89\\
 \cline{1-7}
\rowcolor{NA!30}
\textbf{Canada} & mean & 8.04 & 10544.86 & 3000.35 & 3.30 & 61.35\\
\rowcolor{NA!30}
 & sd & 1.43 & 13027.79 & 511.59 & 2.19 & 19.47\\ \cmidrule{1-7}
\rowcolor{SA!30}
\textbf{Brazil} & mean & 8.58 & 14874.91 & 3079.88 & 3.14 & 56.09\\
\rowcolor{SA!30}
 & sd & 1.59 & 16040.28 & 394.39 & 2.05 & 18.97\\ \midrule

\rowcolor{WA!30}
\textbf{Japan} & mean & 6.42 & 7416.97 & 2717.06 & 2.81 & 48.37\\
\rowcolor{WA!30}
 & sd & 2.33 & 7156.52 & 817.50 & 1.99 & 19.14\\ \cmidrule{1-7}
\rowcolor{EA!30}
\textbf{Turkey} & mean & 7.49 & 11167.86 & 3014.68 & 2.94 & 58.67\\
\rowcolor{EA!30}
 & sd & 1.32 & 10180.01 & 465.07 & 1.80 & 19.73\\ \midrule

\rowcolor{NE!30}
\textbf{UK} & mean & 8.45 & 11369.24 & 3052.51 & 3.18 & 67.39\\
\rowcolor{NE!30}
 & sd & 1.07 & 10756.58 & 460.81 & 2.12 & 15.86\\ \cmidrule{2-7}
\rowcolor{NE!30}
\textbf{Denmark} & mean & 7.14 & 5085.30 & 2547.93 & 2.01 & 67.63\\
\rowcolor{NE!30}
 & sd & 1.83 & 5795.11 & 822.68 & 1.59 & 14.80\\ \cmidrule{2-7}
\rowcolor{NE!30}
\textbf{Finland} & mean & 6.36 & 5141.03 & 2569.54 & 2.28 & 62.18\\
\rowcolor{NE!30}
 & sd & 1.78 & 4791.04 & 840.97 & 1.77 & 16.93\\ \cmidrule{2-7}
\rowcolor{NE!30}
\textbf{Norway} & mean & 8.59 & 6993.00 & 2906.72 & 1.78 & 64.29\\
\rowcolor{NE!30}
 & sd & 1.26 & 7303.59 & 586.30 & 1.21 & 17.66\\ \cmidrule{2-7}
\rowcolor{NE!30}
\textbf{Sweden} & mean & 8.57 & 12047.46 & 3080.71 & 2.50 & 66.29\\
\rowcolor{NE!30}
 & sd & 1.08 & 12224.38 & 424.71 & 1.70 & 16.22\\ \cmidrule{1-7}

\rowcolor{SE!30}
\textbf{Greece} & mean & 7.33 & 10426.07 & 2937.37 & 2.45 & 53.10\\
\rowcolor{SE!30}
 & sd & 1.29 & 15919.70 & 576.30 & 1.74 & 20.90\\ \cmidrule{2-7}
\rowcolor{SE!30}
\textbf{Italy} & mean & 7.85 & 10536.59 & 3064.37 & 2.89 & 66.34\\
\rowcolor{SE!30}
 & sd & 1.30 & 8622.80 & 415.26 & 1.84 & 19.34\\ \cmidrule{2-7}
\rowcolor{SE!30}
\textbf{Spain} & mean & 8.43 & 17051.02 & 3134.53 & 3.45 & 65.80\\
\rowcolor{SE!30}
 & sd & 1.07 & 14219.35 & 274.32 & 2.06 & 16.03\\ \cmidrule{1-7}

\rowcolor{WE!30}
\textbf{France} & mean & 8.03 & 10269.97 & 3021.49 & 3.06 & 71.92\\
\rowcolor{WE!30}
 & sd & 1.30 & 9034.66 & 445.22 & 2.16 & 14.49\\ \cmidrule{2-7}
\rowcolor{WE!30}
\textbf{Germany} & mean & 8.21 & 7810.44 & 2846.30 & 2.17 & 67.05\\
\rowcolor{WE!30}
 & sd & 1.32 & 7969.40 & 667.03 & 1.62 & 14.80\\ \cmidrule{2-7}
\rowcolor{WE!30}
\textbf{Netherlands} & mean & 8.07 & 9889.23 & 2902.89 & 2.34 & 63.35\\
\rowcolor{WE!30}
 & sd & 1.52 & 12482.73 & 617.79 & 1.72 & 17.19\\ \midrule

\rowcolor{AN!30}
\textbf{Australia} & mean & 7.85 & 7146.85 & 2882.53 & 2.62 & 72.06\\
\rowcolor{AN!30}
 & sd & 1.43 & 5984.29 & 612.12 & 1.80 & 15.36\\
\midrule

\rowcolor{ALL!30}
\textbf{All journalists} & \textbf{mean} & \textbf{9.08} & \textbf{13199.12} & \textbf{3123.47} & \textbf{3.24} & \textbf{65.41}\\ 
\rowcolor{ALL!30}
& \textbf{sd} & \textbf{0.97} & \textbf{13396.81} & \textbf{317.94} & \textbf{2.02} & \textbf{16.88}\\
\hline

\end{tabular}
\end{adjustbox}
\label{tab:summary}
\end{table}

\begin{figure}[!h]
\begin{center}
\includegraphics[scale=0.55]{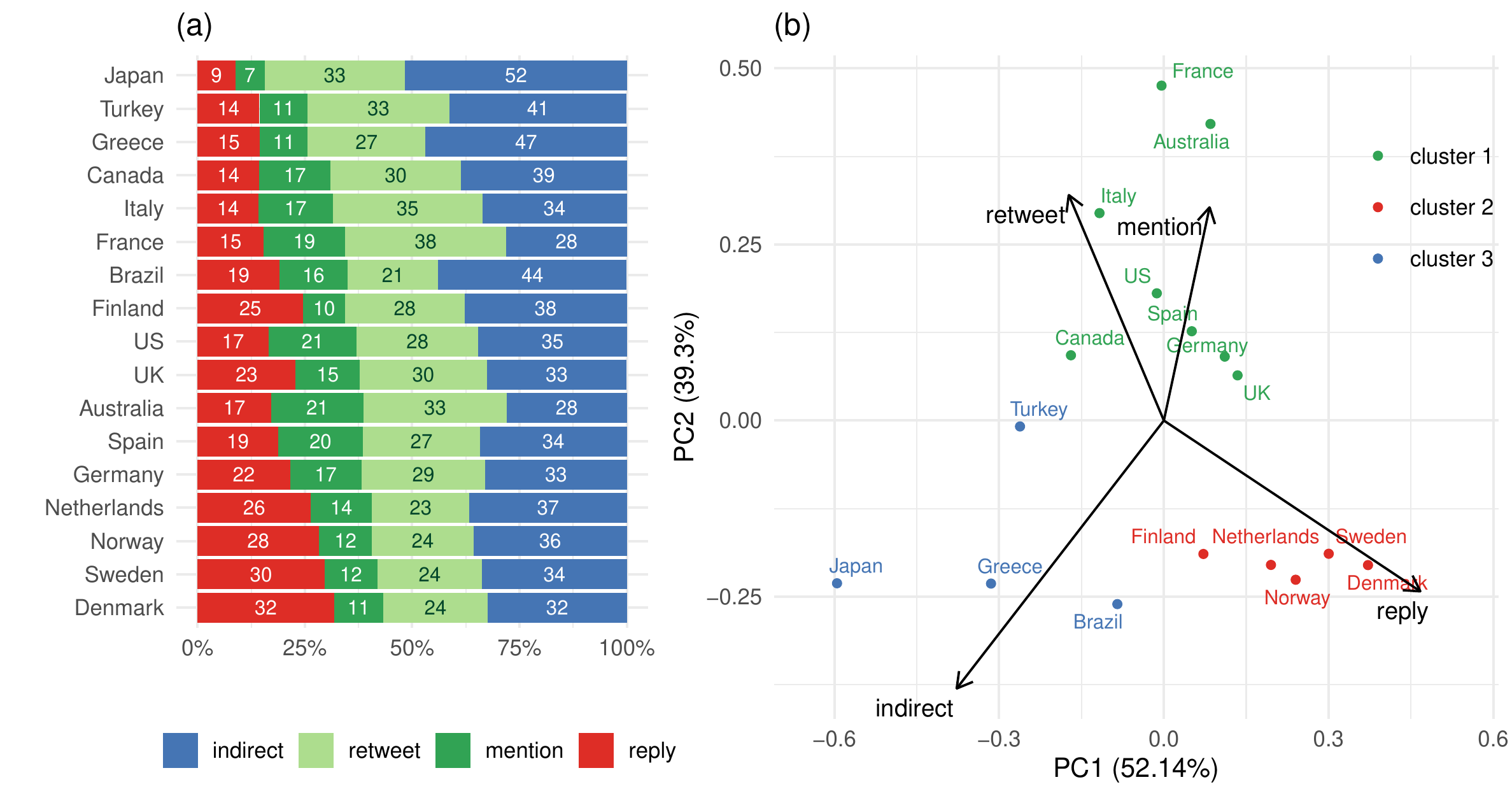}\vspace{-5pt}
\caption{Tweet types across countries. On the left, the average number of tweets in the different categories, computed by country. On the right, the corresponding PCA plot. The colors denote the 3 groups detected by the k-means clustering algorithm on the PCA components. The optimal number of groups has been obtained using the silhouette method. }\vspace{-10pt}
\label{fig:tweet_type}
\end{center}
\end{figure}

When considering the split between the different flavours of social tweets and indirect tweets, some trends emerge. Specifically, we can identify (Figure~\ref{fig:tweet_type}(b)) three distinct groups of countries. In the cluster denoted in blue in Figure~\ref{fig:tweet_type}(b) (comprising Japan, Greece, Turkey, Brazil), the dominant tweet type are indirect tweets. These journalists, thus, tend to post their tweets without interacting with others, not even mentioning their news outlet. At the opposite end of the spectrum, journalists in the red cluster predominantly engage on Twitter by means of replies. Interestingly, they all belong to the same geographical area (Northern Europe). Note that replies are reactive to what other people/news outlet have posted online. The third cluster (members colored in green) comprises all the English-speaking countries in our dataset plus the Central and Southern European states. For these countries the prevalent mode of interaction is through retweets and mentions. This is perhaps the most expected pattern of engagement for journalists on Twitter: sharing content from other users and tagging users (among which, their news outlets) in conversations.

The clusters in Figure~\ref{fig:tweet_type}(b) are obtained applying $k$-means clustering on their tweet type feature vector (shown in  Figure~\ref{fig:tweet_type}(a)) that consists of elements that represent the percentage of tweeting type (i.e., the tweeting type feature of Japanese journalists is $[9, 7, 33, 52]$ which corresponds to reply, mention, retweet, indirect tweeting types' percentages). We select the number of clusters based on the silhouette method. In Figure~\ref{fig:tweet_type}(b) we visualize these clusters by reducing the dimension of the data to the first two PCA components to be able to represent the data in 2D. Note that no geographical information has been used to obtain the clusters: the fact that nearby countries tend to belong to the same cluster shows the strong indirect effect of spatial correlation and similar cultures.

\section{Ego network analysis}
\label{sec:EGONETWORKANALYSIS}

In this section, we analyze the ego network structure of the journalists that have been filtered as described in Section~\ref{sec:dataset}, and we show the differences and invariants between different countries and regions. The frequency of \emph{direct tweets} (mentions, reply, or retweets) is used to represent the strength of the ties (that is, the intimacy or emotional closeness) between journalists (egos) and their alters as discussed in Section~\ref{sec:egonet_extraction}. Recall also that, similarly to the related literature, a relationship is considered \emph{active} if the ego and the alter have at least one contact (direct tweet) per year.

\subsection{Static properties of ego networks}
\label{sec:egonets_analysis_static}

The static view of an ego network is a single, aggregate snapshot of the ego and its ties. It is a representation of the ego network that includes the whole observed timeline of the journalist. This analysis allows us to understand the general distribution of cognitive resources through all the alters and to compare results with other findings from the related literature in a quantitative manner (since the analysis of static ego networks is the starting point in all related works). 

\begin{figure}[!h]
\begin{center}
\includegraphics[scale=0.2]{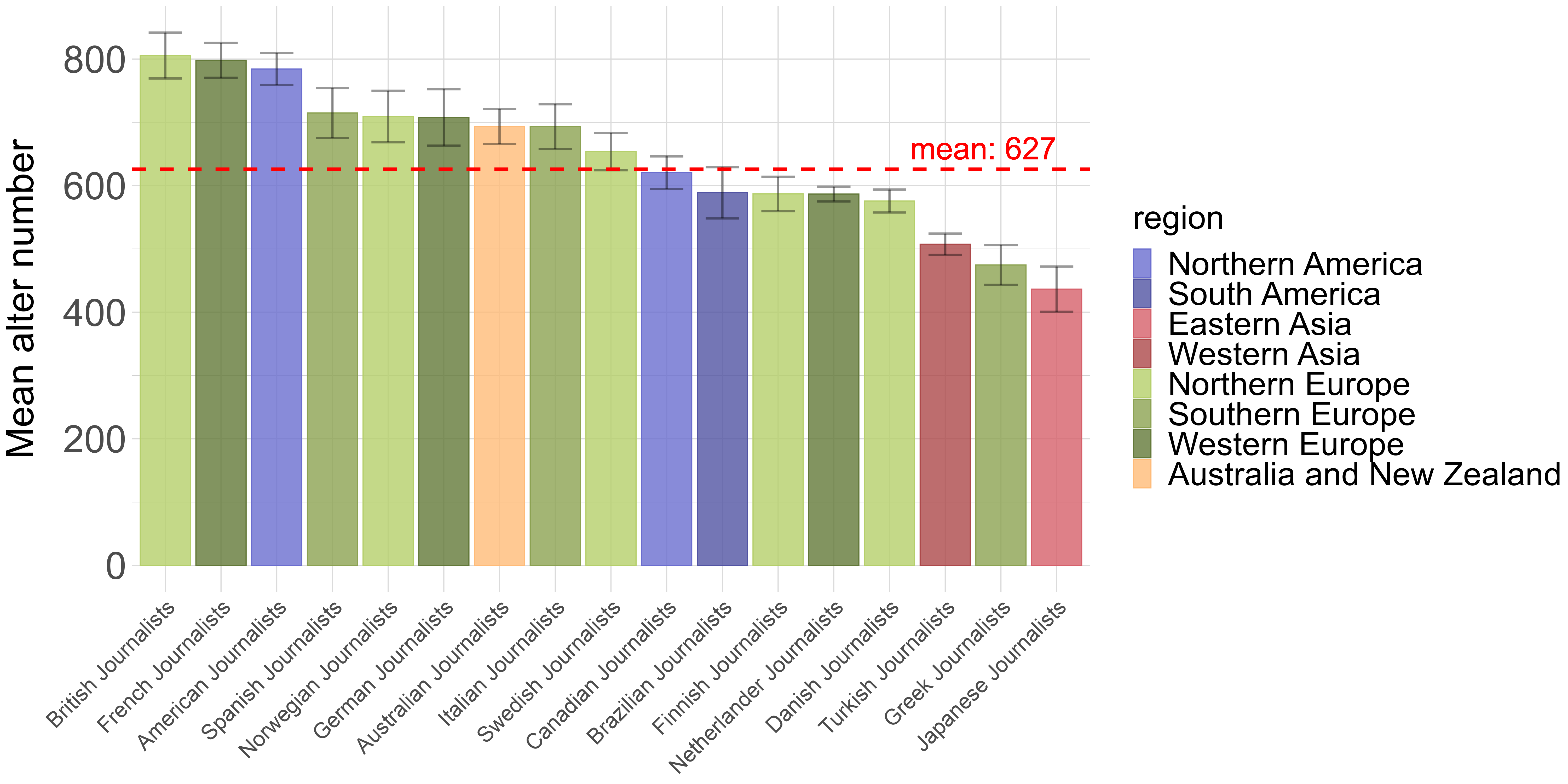}\vspace{-5pt}
\caption{Average number of alters per ego, grouped by country}
\label{fig:egonet_sizes}\vspace{-10pt}
\end{center}
\end{figure}

First, we show the average number of alters (with confidence intervals) of egos in different countries (Figure~\ref{fig:egonet_sizes}, the distribution of the number of alters per ego in each dataset, instead of dataset averages, can be found in Figure~\ref{fig_appendix:totalsize} of \ref{appendix_static_egonets}\iftoggle{ONEFILE}{}{ in the Supplemental Material}). This number captures how many peers the ego user interacts with during the last 3200 observed tweets, and it corresponds to the size of the ego network when both active and inactive relationships are considered (see Section~\ref{sec:egonetworkmodel}). As expected, the countries whose journalists engage on Twitter predominantly via indirect tweets (Figure~\ref{fig:tweet_type}(b)) tend to feature fewer alters than their more social counterparts. Figure~\ref{fig:egonet_sizes} also shows that journalists that mostly rely on replies (such as those from Finland, the Netherlands, and Denmark) have a below-average number of alters, thus suggesting a tendency to get involved with smaller groups of people. This is consistent with the fact that replies are a more personal/intimate communication with respect to mention and retweets, hence they consume more cognitive resources on the ego side, which, in turn, is able to interact with fewer people. Vice versa, the journalists in countries from the green cluster in Figure~\ref{fig:tweet_type}(b) tend to interact with an above-average number of distinct peers, hinting at the opposite effect. 
Even though there are differences for journalists from different regions and countries, still the average number of alters of journalists are much bigger than those of the generic Twitter users analyzed in~\cite{Dunbar2015}, where more than $90\%$ of users have less than 100 relationships. This supports the claim that journalists establish more relationships and they are prominent users on Twitter. 

The number of alters showed in Figure~\ref{fig:egonet_sizes} include relationships to which egos don't allocate their cognitive resources regularly (the typical case in the anthropology literature, which we also apply here, is at least once a year). We now consider the average number of \emph{active} relationships of egos in different countries (Figure~\ref{fig:active_egonet_sizes}, the distribution can be found in Figure~\ref{fig_appendix:activenet_size} of \ref{appendix_static_egonets}\iftoggle{ONEFILE}{}{ in the Supplemental Material}). As expected, the average number of active alters is significantly lower than in the previous case of Figure~\ref{fig:egonet_sizes}, with sizes varying between 75 alters (Japanese journalists) and 147 (Spanish journalists). Turkish, Greek and Japanese journalists still have the smallest network sizes. Also, the difference between journalists and generic Twitter users still exists: the active network size of generic Twitter users was 88 alters in~\cite{Dunbar2015}, while the average for journalists is 119.
Hence, we can conclude that the ego network size of journalists mimics offline ego network size predicted by Dunbar's number, and it is actually closer to the Dunbar's number than those of generic Twitter users were\footnote{While the initial filtering used in~\cite{Dunbar2015} is different from the one discussed in this paper (Section~\ref{sec:dataset}), the comparison between the two sets of results is still meaningful. In fact, the filtering in~\cite{Dunbar2015} (whereby only users with an average of more than 10 interactions per month are kept) tends to retain only very active users, thus it might overestimate the social interactions of generic users. However, our results indicate that journalists tend to be even more engaged than this active subset of generic users.}.

\begin{figure}[h]
\begin{center}
\includegraphics[scale=0.2]{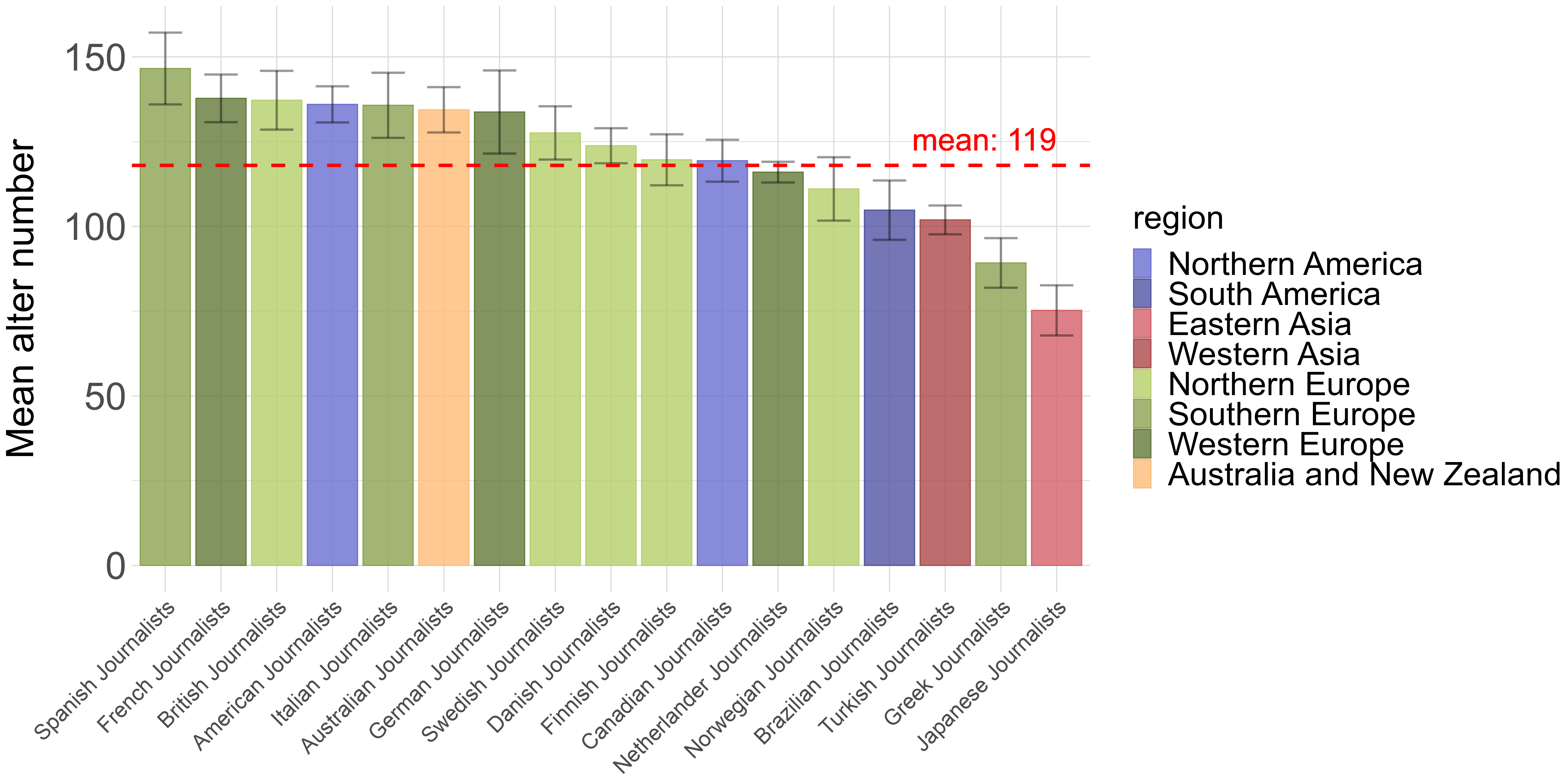}\vspace{-5pt}
\caption{Average active networks size for egos in each country}
\label{fig:active_egonet_sizes}\vspace{-10pt}
\end{center}
\end{figure}

\begin{figure}[h]
\begin{center}
\includegraphics[scale=0.2]{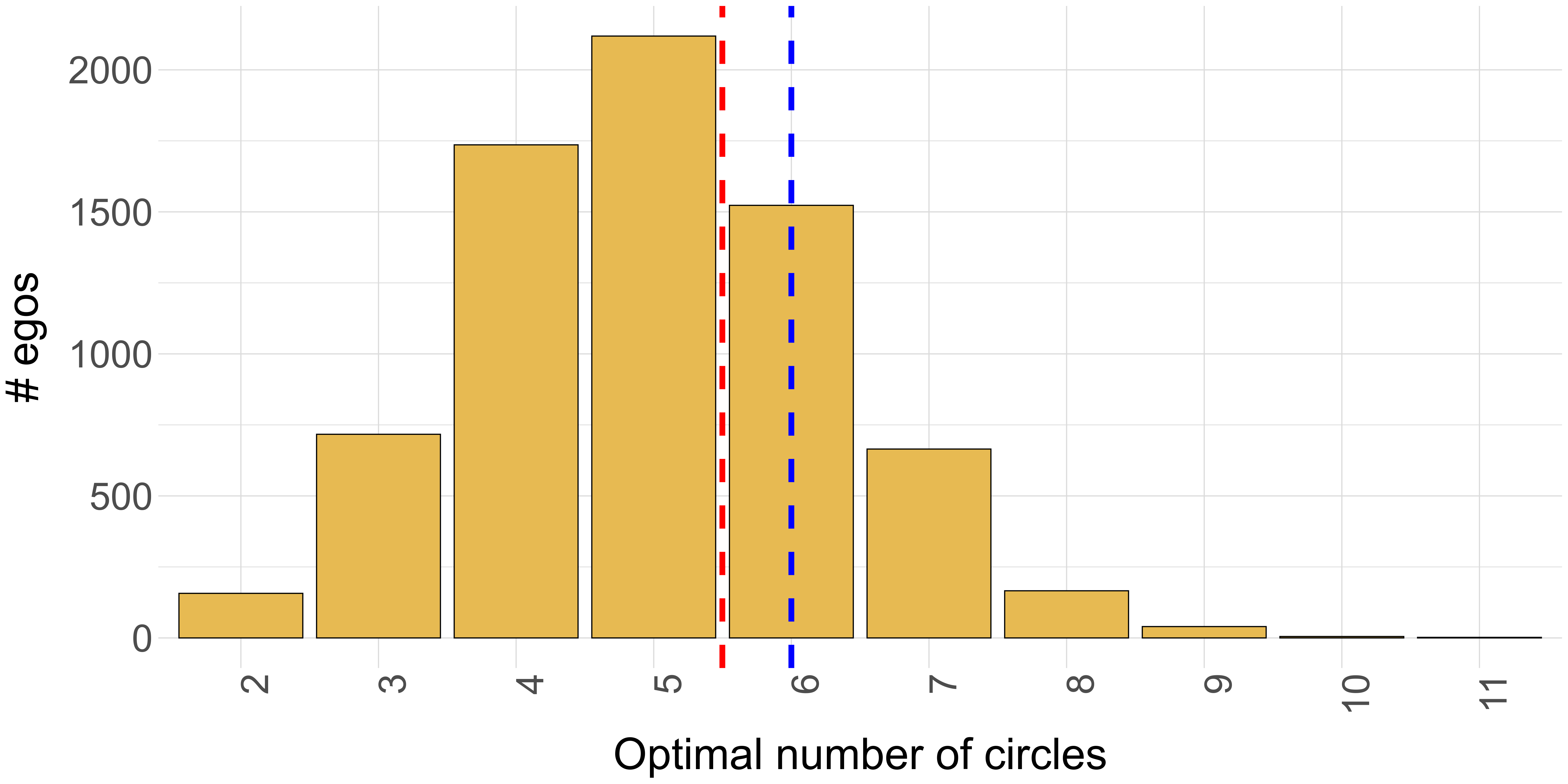}\vspace{-5pt}
\caption{Optimal number of circles per ego (red line: average, blue line: median)}
\label{fig:optimal_circles}\vspace{-10pt}
\end{center}
\end{figure}

The active relationships of egos can be grouped into layers (also known as Dunbar's \emph{circles}) based on their emotional closeness with the ego. It has been shown that there are four layers in offline human ego networks~\cite{Zhou2005}. This number is extended to five layers for OSNs~\cite{Dunbar2015}. The optimal number of circles can be determined by non-parametric clustering algorithms, as discussed in Section~\ref{sec:egonetworkmodel}. Figure~\ref{fig:optimal_circles} shows the distribution of the optimal circle number (i.e., the optimal number of relationship clusters, as found by Mean Shift) per ego for all journalists (the distribution of circle numbers per dataset is provided in Figure~\ref{fig_appendix:optimal_circles} of \ref{appendix_static_egonets}\iftoggle{ONEFILE}{}{ in the Supplemental Material}). The mode value of the optimal circle numbers is 5, its mean value is 5.6, and its median value is 6. All datasets feature a mean and median value for the optimal number of social circles around five, thus matching the Dunbar's model, with the exception of Japanese journalists (which is not surprising given that their active ego network sizes are smaller than for other countries). 
In the remaining of the paper, when comparing egos with each other, we will focus on egos with five circles, in order to rule out differences that are dependent on the different number of circles.

Table~\ref{tab:circle_sizes_opt_5} shows the circle sizes (number of alters in each Dunbar's circle). As we can see from the last row of the table (corresponding to the average of all journalists), the circle sizes of journalists are very close to the sizes predicted by the Dunbar's model (1.5, 5, 15, 50, and 150), with slight variations. Although generally the circle sizes of OSN users tend to be smaller than what Dunbar's model suggests~\cite{Dunbar2015, arnaboldi2017}, journalists are an exception to this rule, with more alters than expected in all circles except for the outermost one (5th circle). In Table~\ref{tab:circle_sizes_opt_5}, we also show the scaling ratio between consecutive circles, whose typical value in the related literature is around three~\cite{Zhou2005}. Again, journalists conforms quite accurately to the predictions of Dunbar's model.

\begin{table}[!t]
\centering
\caption{Circle sizes of egos with optimal cicle number 5}
\footnotesize
\begin{adjustbox}{width=1\textwidth}
\small
\begin{tabular}{@{}llllllllll@{}} \toprule
\textbf{dataset} & \textbf{C1} & \textbf{$ratio$} & \textbf{C2} & \textbf{$ratio$} & \textbf{C3} & \textbf{$ratio$} & \textbf{C4} & \textbf{$ratio$} & \textbf{C5}\\ \midrule
\rowcolor{NA!30}
\textbf{US} & 3.24 $\pm$ 0.21 &  & 8.3 $\pm$ 0.51 &  & 20.15 $\pm$ 1.2 &  & 48.42 $\pm$ 2.52 &  & 124.56 $\pm$ 5.78\\
\rowcolor{NA!30}
 &  & 2.96 $\pm$ 0.21 &  & 2.58 $\pm$ 0.11 &  & 2.56 $\pm$ 0.08 &  & 2.72 $\pm$ 0.1 & \\ \cmidrule{2-10}
 \rowcolor{NA!30}
\textbf{Canada} & 3.43 $\pm$ 0.27 &  & 8.29 $\pm$ 0.72 &  & 19.36 $\pm$ 1.68 &  & 44.68 $\pm$ 3.49 &  & 114.26 $\pm$ 7.53\\
\rowcolor{NA!30}
 &  & 2.63 $\pm$ 0.21 &  & 2.44 $\pm$ 0.12 &  & 2.46 $\pm$ 0.12 &  & 2.79 $\pm$ 0.18 & \\ \cmidrule{1-10}
 \rowcolor{SA!30}
\textbf{Brazil} & 3.14 $\pm$ 0.31 &  & 8.19 $\pm$ 0.85 &  & 19.32 $\pm$ 1.93 &  & 43.27 $\pm$ 4.1 &  & 105.3 $\pm$ 9.43\\
\rowcolor{SA!30}
 &  & 2.98 $\pm$ 0.33 &  & 2.45 $\pm$ 0.12 &  & 2.32 $\pm$ 0.1 &  & 2.61 $\pm$ 0.17 & \\ \midrule
 \rowcolor{EA!30}
\textbf{Japan} & 2.98 $\pm$ 0.31 &  & 7.14 $\pm$ 0.65 &  & 16.4 $\pm$ 1.56 &  & 37.56 $\pm$ 3.77 &  & 96.2 $\pm$ 9.32\\
\rowcolor{EA!30}
 &  & 2.72 $\pm$ 0.3 &  & 2.35 $\pm$ 0.13 &  & 2.31 $\pm$ 0.1 &  & 2.67 $\pm$ 0.17 & \\ \cmidrule{2-10}
 \rowcolor{WA!30}
\textbf{Turkey} & 3.1 $\pm$ 0.16 &  & 7.43 $\pm$ 0.43 &  & 17.46 $\pm$ 0.97 &  & 40.95 $\pm$ 2.05 &  & 106.54 $\pm$ 4.77\\
\rowcolor{WA!30}
 &  & 2.64 $\pm$ 0.15 &  & 2.46 $\pm$ 0.09 &  & 2.51 $\pm$ 0.07 &  & 2.79 $\pm$ 0.09 & \\ \midrule
 \rowcolor{NE!30}
\textbf{UK} & 3.5 $\pm$ 0.39 &  & 8.56 $\pm$ 0.94 &  & 22.08 $\pm$ 2.31 &  & 52.85 $\pm$ 4.71 &  & 134.07 $\pm$ 11.56\\
 \rowcolor{NE!30}
 &  & 2.74 $\pm$ 0.32 &  & 2.72 $\pm$ 0.21 &  & 2.72 $\pm$ 0.32 &  & 2.66 $\pm$ 0.19 & \\ \cmidrule{2-10}
 \rowcolor{NE!30}
\textbf{Denmark} & 3.63 $\pm$ 0.22 &  & 9.62 $\pm$ 0.58 &  & 22.69 $\pm$ 1.32 &  & 51.37 $\pm$ 2.9 &  & 119.87 $\pm$ 6.8\\
 \rowcolor{NE!30}
 &  & 2.92 $\pm$ 0.17 &  & 2.5 $\pm$ 0.09 &  & 2.33 $\pm$ 0.06 &  & 2.41 $\pm$ 0.08 & \\ \cmidrule{2-10}
 \rowcolor{NE!30}
\textbf{Finland} & 3.74 $\pm$ 0.37 &  & 9.29 $\pm$ 0.9 &  & 22.92 $\pm$ 2.11 &  & 53.57 $\pm$ 4.5 &  & 123.82 $\pm$ 10.05\\
 \rowcolor{NE!30}
 &  & 2.74 $\pm$ 0.25 &  & 2.61 $\pm$ 0.17 &  & 2.49 $\pm$ 0.14 &  & 2.4 $\pm$ 0.1 & \\ \cmidrule{2-10}
 \rowcolor{NE!30}
\textbf{Norway} & 3.35 $\pm$ 0.42 &  & 9.16 $\pm$ 1.09 &  & 21.74 $\pm$ 2.61 &  & 50.05 $\pm$ 5.29 &  & 115.56 $\pm$ 11.14\\
 \rowcolor{NE!30}
 &  & 3.12 $\pm$ 0.35 &  & 2.47 $\pm$ 0.16 &  & 2.45 $\pm$ 0.13 &  & 2.42 $\pm$ 0.12 & \\ \cmidrule{2-10}
 \rowcolor{NE!30}
\textbf{Sweden} & 3.38 $\pm$ 0.35 &  & 8.63 $\pm$ 0.86 &  & 20.86 $\pm$ 1.91 &  & 49.1 $\pm$ 3.97 &  & 122.77 $\pm$ 9.94\\
 \rowcolor{NE!30}
 &  & 2.93 $\pm$ 0.27 &  & 2.61 $\pm$ 0.17 &  & 2.49 $\pm$ 0.13 &  & 2.6 $\pm$ 0.14 & \\ \cmidrule{1-10}
 \rowcolor{SE!30}
\textbf{Greece} & 2.92 $\pm$ 0.28 &  & 6.66 $\pm$ 0.69 &  & 15.07 $\pm$ 1.68 &  & 34.5 $\pm$ 3.36 &  & 91.78 $\pm$ 7.68\\
\rowcolor{SE!30}
 &  & 2.49 $\pm$ 0.23 &  & 2.34 $\pm$ 0.16 &  & 2.44 $\pm$ 0.14 &  & 2.83 $\pm$ 0.16 & \\ \cmidrule{2-10}
  \rowcolor{SE!30}
\textbf{Italy} & 2.81 $\pm$ 0.29 &  & 7.26 $\pm$ 0.79 &  & 19.2 $\pm$ 2.23 &  & 46.72 $\pm$ 4.81 &  & 124.21 $\pm$ 10.06\\
\rowcolor{SE!30}
 &  & 2.97 $\pm$ 0.37 &  & 2.73 $\pm$ 0.19 &  & 2.65 $\pm$ 0.18 &  & 2.94 $\pm$ 0.23 & \\ \cmidrule{2-10}
 \rowcolor{SE!30}
\textbf{Spain} & 3.2 $\pm$ 0.3 &  & 8.34 $\pm$ 0.96 &  & 19.65 $\pm$ 2.39 &  & 46.97 $\pm$ 5.02 &  & 124.64 $\pm$ 10.65\\
\rowcolor{SE!30}
 &  & 2.74 $\pm$ 0.27 &  & 2.42 $\pm$ 0.16 &  & 2.59 $\pm$ 0.16 &  & 2.88 $\pm$ 0.2 & \\ \cmidrule{1-10}
  \rowcolor{WE!30}
\textbf{France} & 3.2 $\pm$ 0.32 &  & 8.17 $\pm$ 1.06 &  & 19.46 $\pm$ 2.09 &  & 49.56 $\pm$ 4.56 &  & 129.14 $\pm$ 10.29\\
\rowcolor{WE!30}
 &  & 2.73 $\pm$ 0.26 &  & 2.65 $\pm$ 0.22 &  & 2.7 $\pm$ 0.13 &  & 2.78 $\pm$ 0.17 & \\ \cmidrule{2-10}
 \rowcolor{WE!30}
\textbf{Germany} & 3.39 $\pm$ 0.54 &  & 8.99 $\pm$ 1.35 &  & 22.8 $\pm$ 2.97 &  & 52.99 $\pm$ 6.21 &  & 127.25 $\pm$ 13.74\\
\rowcolor{WE!30}
 &  & 3.1 $\pm$ 0.41 &  & 2.74 $\pm$ 0.23 &  & 2.45 $\pm$ 0.16 &  & 2.52 $\pm$ 0.23 & \\ \cmidrule{2-10}
\rowcolor{WE!30}
\textbf{Netherlands} & 3.28 $\pm$ 0.12 &  & 8.39 $\pm$ 0.31 &  & 19.9 $\pm$ 0.7 &  & 46.66 $\pm$ 1.53 &  & 114.79 $\pm$ 3.64\\
\rowcolor{WE!30}
  &  & 2.88 $\pm$ 0.11 &  & 2.49 $\pm$ 0.06 &  & 2.45 $\pm$ 0.05 &  & 2.55 $\pm$ 0.05 & \\ \midrule
\rowcolor{AN!30}
\textbf{Australia} & 3.19 $\pm$ 0.29 &  & 8.06 $\pm$ 0.75 &  & 20.23 $\pm$ 1.84 &  & 49.17 $\pm$ 3.83 &  & 128.7 $\pm$ 8.67\\
\rowcolor{AN!30}
 &  & 2.82 $\pm$ 0.23 &  & 2.63 $\pm$ 0.15 &  & 2.63 $\pm$ 0.15 &  & 2.86 $\pm$ 0.19 & \\ \hline
 
\rowcolor{ALL!30}
\textbf{All journalists} & \textbf{3.24 $\pm$ 0.05} &  & \textbf{8.3 $\pm$ 0.12} &  & \textbf{20.15 $\pm$ 0.28} &  & \textbf{48.42 $\pm$ 0.59} &  & \textbf{124.56 $\pm$ 1.35}\\
\rowcolor{ALL!30}
 &  & \textbf{2.96 $\pm$ 0.05} &  & \textbf{2.58 $\pm$ 0.03} &  & \textbf{2.56 $\pm$ 0.02} &  & \textbf{2.72 $\pm$ 0.02} & \\ \hline
\end{tabular}
\end{adjustbox}
\label{tab:circle_sizes_opt_5}
\end{table}

Wrapping up the results of the static ego networks analysis, we can provide answers to the first two research questions laid out in Section~\ref{sec:introduction}. 
\paragraph{RQ\#1} Overall, journalists' cognitive resource allocation to their alters mimics closely what was observed in offline human ego networks, with the circle sizes and the scaling ratio between the circle sizes matching Dunbar's model. Also, we have shown that journalists tend to have more active relationships than generic Twitter users and politicians, with ego network sizes closer to Dunbar's number than typical OSN users. 
\paragraph{RQ\#2} The country in which journalists are active have an effect on their ego network, but it does not alter significantly their structure. For example, we have observed that Turkish, Greek and Japanese journalists have smaller ego networks compared to other countries. British, French, and American journalists engage in more numerous relationships on Twitter, while Spanish journalists maintain more active relationships. However, by and large, Dunbar's model holds across countries.

\subsection{Dynamic properties of ego networks}
\label{sec:egonets_analysis_dynamic}

In this section, we analyze the evolution of ego networks over time, in order to capture changes in social relationships. If we divide the observed timeline of a user into equal time intervals, the intimacy of a relationship may change from interval to interval. That is, an ego may allocate more cognitive resources to an alter in the early time intervals and less in the latter ones, or vice versa. A varying intimacy entails a possible rearrangement of relationships in circles. In order to investigate both short-term and long-term dynamics, in this section we divide the observed timeline of an ego in one-year time intervals with 1-month step size (11 months overlap between two consecutive time windows) and 12-month step size (no overlap between consecutive time windows). Each time window provides a snapshot of the ego networks, and, for each snapshot, we focus on rings (the portion of circles excluding the inner circles) and how they change over time. Note that both in the short- and long-term case, we construct the ego network from one year’s worth of data. However, for the short-term analysis, there is a partial overlap between the one-year windows, while this overlap is not present in the long-term analysis.

\begin{figure}[p]
\begin{center}
\includegraphics[scale=0.32]{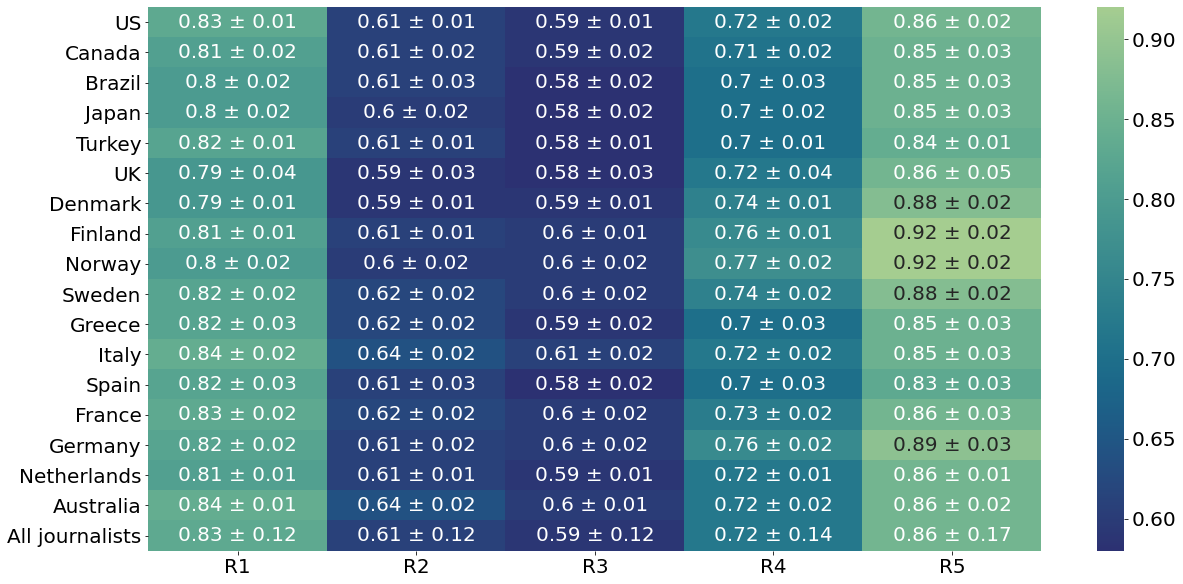}\vspace{-5pt}
\caption{Jaccard coefficients of egos with optimal circle number 5 with 1-month step size for 1-year intervals.}
\label{fig:jaccard_overlapping_opt_5}\vspace{-10pt}
\end{center}
\end{figure}

\begin{figure}[p]
\begin{center}
\includegraphics[scale=0.32]{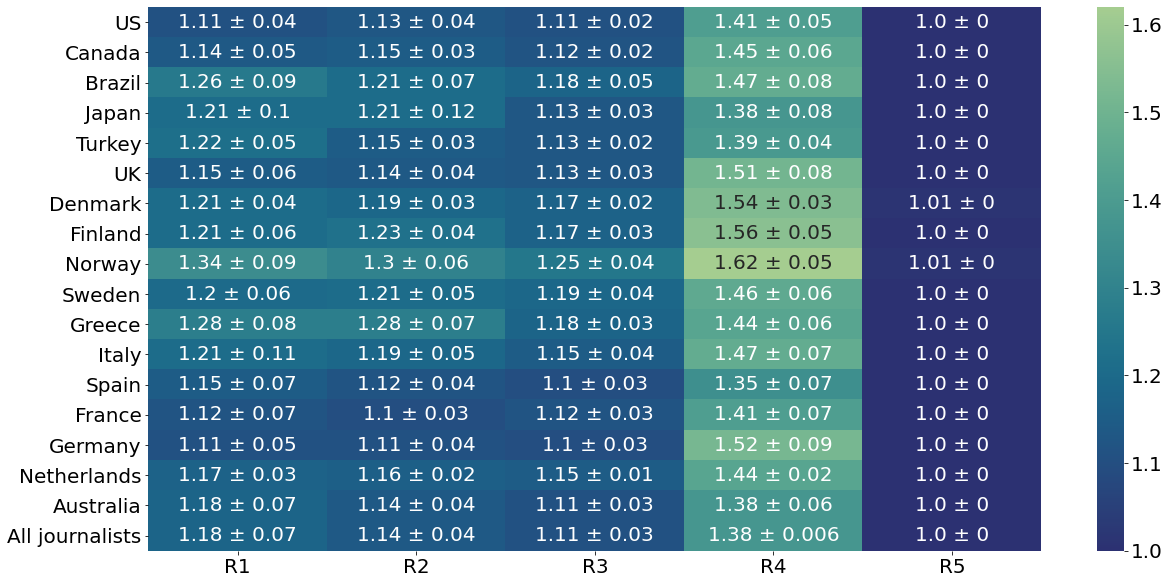}\vspace{-5pt}
\caption{Jump indices of egos with optimal circle number 5 with 1-month step size for 1-year intervals.}
\label{fig:jump_overlapping_opt_5_nonzero}\vspace{-10pt}
\end{center}
\end{figure}

As discussed in Section~\ref{sec:egonet_extraction}, we study the dynamic changes in ego networks by means of the Jaccard similarity and the Jump index. Considering the same circle in two consecutive time intervals as two separate sets, the Jaccard index is obtained dividing the cardinality of the intersection set by the cardinality of the union set. The closer this index to one, the more the overlapping. The amount of movements between rings is measured through the Jump index, which simply counts (and then averages across all alters in the ring) the number of ring jumps made by the alters that have just entered the ring at this temporal snapshot. We first describe the short-term changes of dynamic ego networks.
Figures~\ref{fig:jaccard_overlapping_opt_5} and~\ref{fig:jump_overlapping_opt_5_nonzero} show the values of Jaccard and jump indices with 1-year time interval and 1-month step size for the five rings. With this approach, we compare how ego networks change on a month-by-month basis. The same analysis was carried out in~\cite{arnaboldi2017} for politicians. We can see (Figure~\ref{fig:jaccard_overlapping_opt_5}) that the Jaccard indices are higher at the innermost (R1) and the outermost (R5) rings, while middle rings have lower index values (U-shaped pattern). This shows that the innermost and outermost rings don't change much over time, implying that the strongest and weakest active relations of journalists are pretty stable in the short terms. Vice versa, relationships with intermediate intimacy are more fluctuating. These results are extremely consistent across countries.
We provide Jump index in Figure~\ref{fig:jump_overlapping_opt_5_nonzero}. The Jump indices are generally close to one and stable across the rings, signalling that movements mostly occur between adjacent rings. The ring with slightly more jumps, on average, is the fourth. This may suggest that the fourth ring serves as a buffer zone, where mutating relationships transit during their movements from one ring to another one.

\begin{figure}[p]
\begin{center}
\includegraphics[scale=0.32]{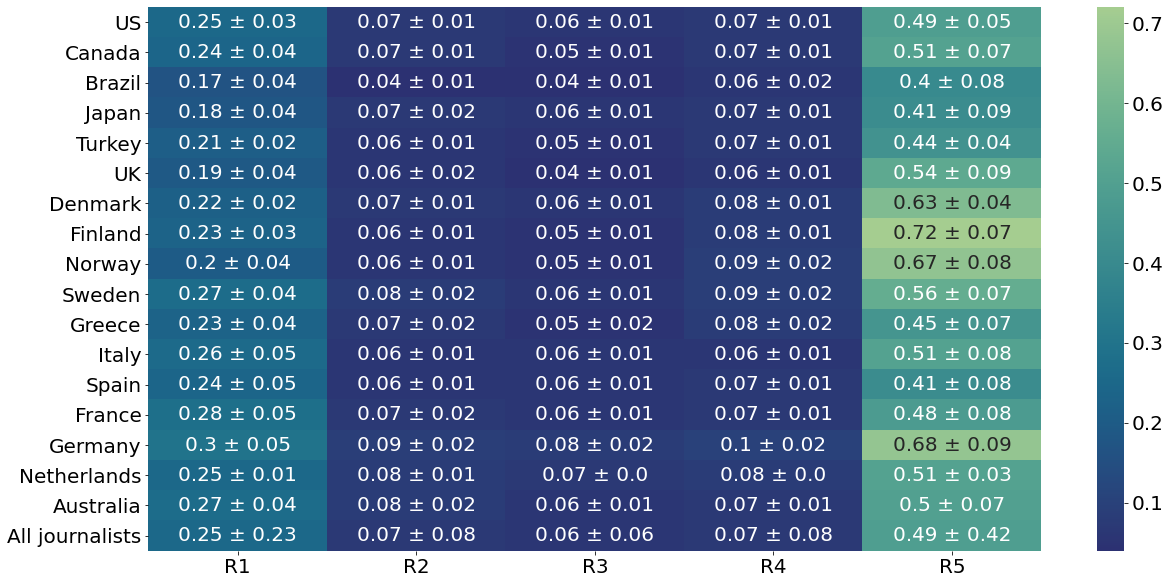}\vspace{-5pt}
\caption{Jaccard coefficients of egos with optimal cicle number 5 with 1-year step size for 1-year intervals.}
\label{fig:jaccard_nonoverlapping_opt_5}\vspace{-10pt}
\end{center}
\end{figure}

\begin{figure}[p]
\begin{center}
\includegraphics[scale=0.32]{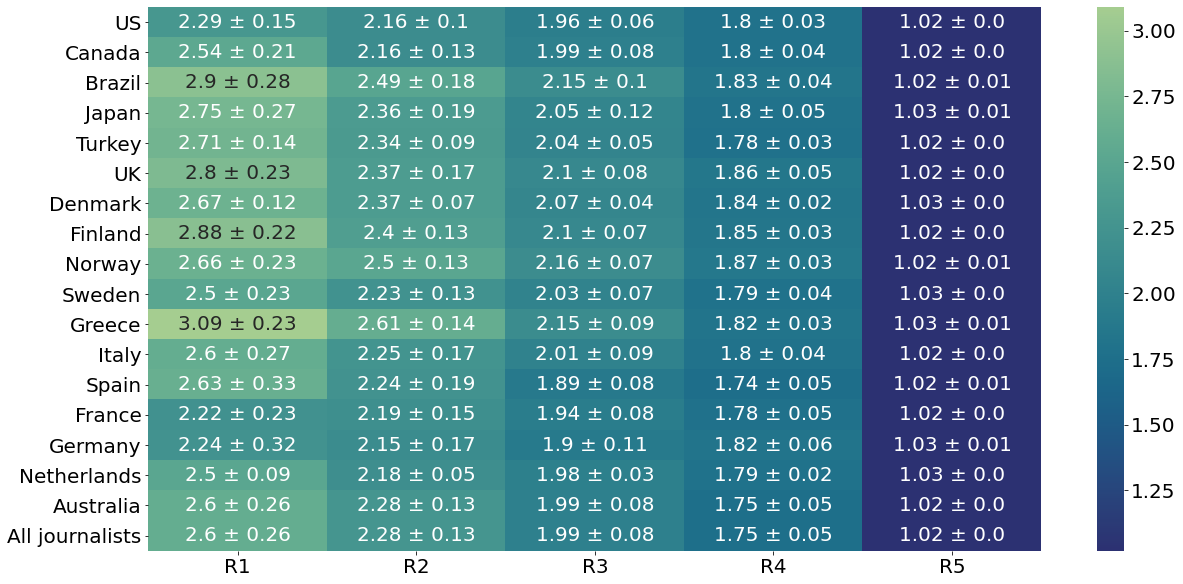}\vspace{-5pt}
\caption{Jump indices of egos with optimal cicle number 5 with 1-year step size for 1-year intervals.}
\label{fig:jump_nonoverlapping_opt_5_nonzero}\vspace{-10pt}
\end{center}
\end{figure}

Let us now study the long-term dynamics of ego networks. 
To this aim, we consider 1-year time intervals for observing ego networks, and 1-year step sizes (i.e., there is no intersection between two consecutive observation time intervals). Results are shown in Figures~\ref{fig:jaccard_nonoverlapping_opt_5} and~\ref{fig:jump_nonoverlapping_opt_5_nonzero}.
The U-shape pattern across rings for the Jaccard coefficient is still there, but the similarity values are much lower. This means that, on a yearly basis, the rings of ego networks tend to change significantly. In the central rings, especially, the turnover is almost complete, as indicated by the Jaccard index close to zero. Surprisingly, the most stable ring is R5, which contains the weakest active social links. We observe slightly more variability across countries with respect to the short-term results in Figure~\ref{fig:jaccard_overlapping_opt_5}. This suggests that, as expected, long-term dynamics tend to be more interesting than short-term ones.
We can compare the Jaccard index of journalists against that of politicians and generic users (whose dynamic ego networks are both studied in~\cite{arnaboldi2017}). We find the main differences in R5, where the Jaccard index of journalists is approximately twice as a large with respect to generic users and very similar to that of European political leaders. This implies that journalists and politicians have more stable weak social relationships than generic users. This finding hints at a distinctive feature of public figures on social media, suggesting a semantically different way in which social links are exploited on Twitter by these users.
The pattern observed for the Jump index, too, is completely different from the previous case (Figure~\ref{fig:jump_overlapping_opt_5_nonzero}): the index decreases as we move from R1 to R5, while previously it was stable across the rings and close to value of one. We see that alters come to the first ring after moving from non-adjacent circles, but this effect wanes as we move towards the outermost circles.


The results from this section allow us to provide an answer to the third research question we laid out in Section~\ref{sec:introduction}. 
\paragraph{RQ \#3} Journalists tend to have stable short-term relationships that don't change a lot over time. Specifically, they tend to keep unaltered their most and least intimate relationships while they replace the relationships in middle circles much more. In the longer term, though, ego networks can be pretty dynamic, especially in the innermost circles. While this is not surprising, the interesting finding is how journalists compare to generic users and politicians. From the ring stability standpoint, journalists resemble the politicians studied in~\cite{arnaboldi2017} and are different from generic users. This suggests that the ego networks of journalists and politicians, while structurally similar to that of generic users, may be affected by the information-driven nature of their engagement on the platform, thus yielding to distinctive ring dynamics.

\subsection{Social tweets and hashtags}
\label{sec:hashtags}

Considering that journalists are on Twitter to promote their work, establish their personal brands, and satisfy the user demand for news, we expect them to be more topic-driven than generic Twitter users. The typical way of searching for and publishing tweets about a specific topic is to use hashtags. Thus, we expect journalists to be using hashtags at a higher rate than regular users. To look into this aspect, we provide hashtag-related statistics about the journalist datasets. Due to space constraints, we will provide plots for selected countries (specifically, those representative of a group behavior), while the complete set of results can be found in~\ref{appendix_hashtags_egonets}\iftoggle{ONEFILE}{}{ of the Supplemental Material}.

In Figure~\ref{fig:hashtag_activated}, the percentage of hashtag-activated relationships are shown. A relationship is labelled \emph{hashtag-activated} if the first contact (direct tweet) of the relationship includes a hashtag. For all journalists, the average hashtag-activated relationship percentage is $23 \%$ while it is $6\%$ for generic Twitter users and approximately $15\%$ for politicians (both studied in~\cite{arnaboldi2017}). This confirms that journalists establish more topic-driven relationships. All of the journalist datasets have a higher hashtag-activated percentage than generic Twitter users, and, in most cases, also higher than the politicians. However, when we compare the different countries, there is a high variability in terms of hashtag-activated relationship percentages, with values ranging from $12\%$ in Japanese journalists to $36\%$ in Italian journalists.

\begin{figure}[h]
\begin{center}
\includegraphics[scale=0.22]{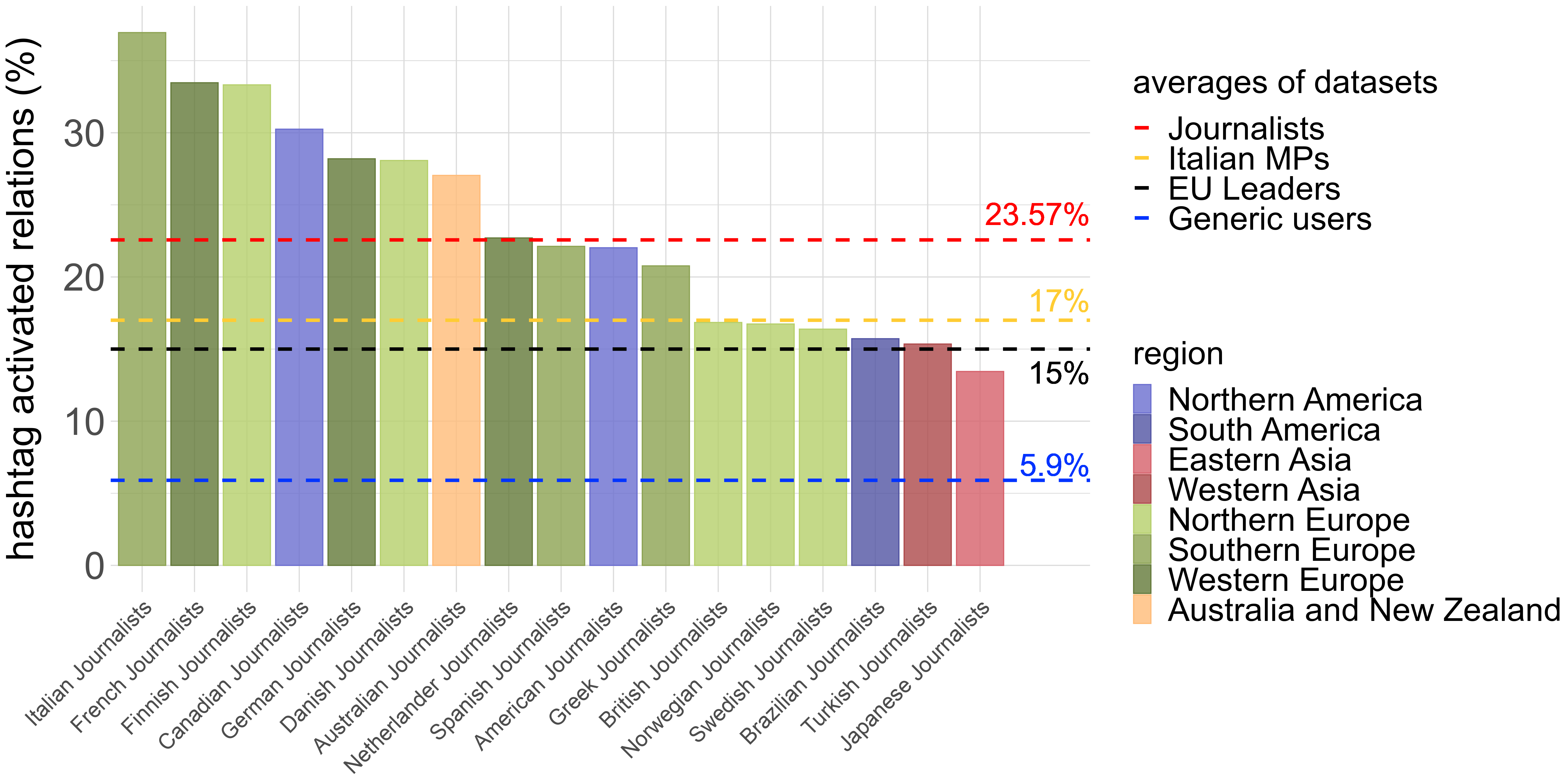}\vspace{-5pt}
\caption{Mean number of hashtags per alter in the different countries}
\label{fig:hashtag_activated}\vspace{-10pt}
\end{center}
\end{figure}

\begin{figure}[t]
\subfloat[American Journalists
\label{fig:relations_hashtags_rings_AmericanJournalists}]
{\includegraphics[width=0.33\textwidth]
{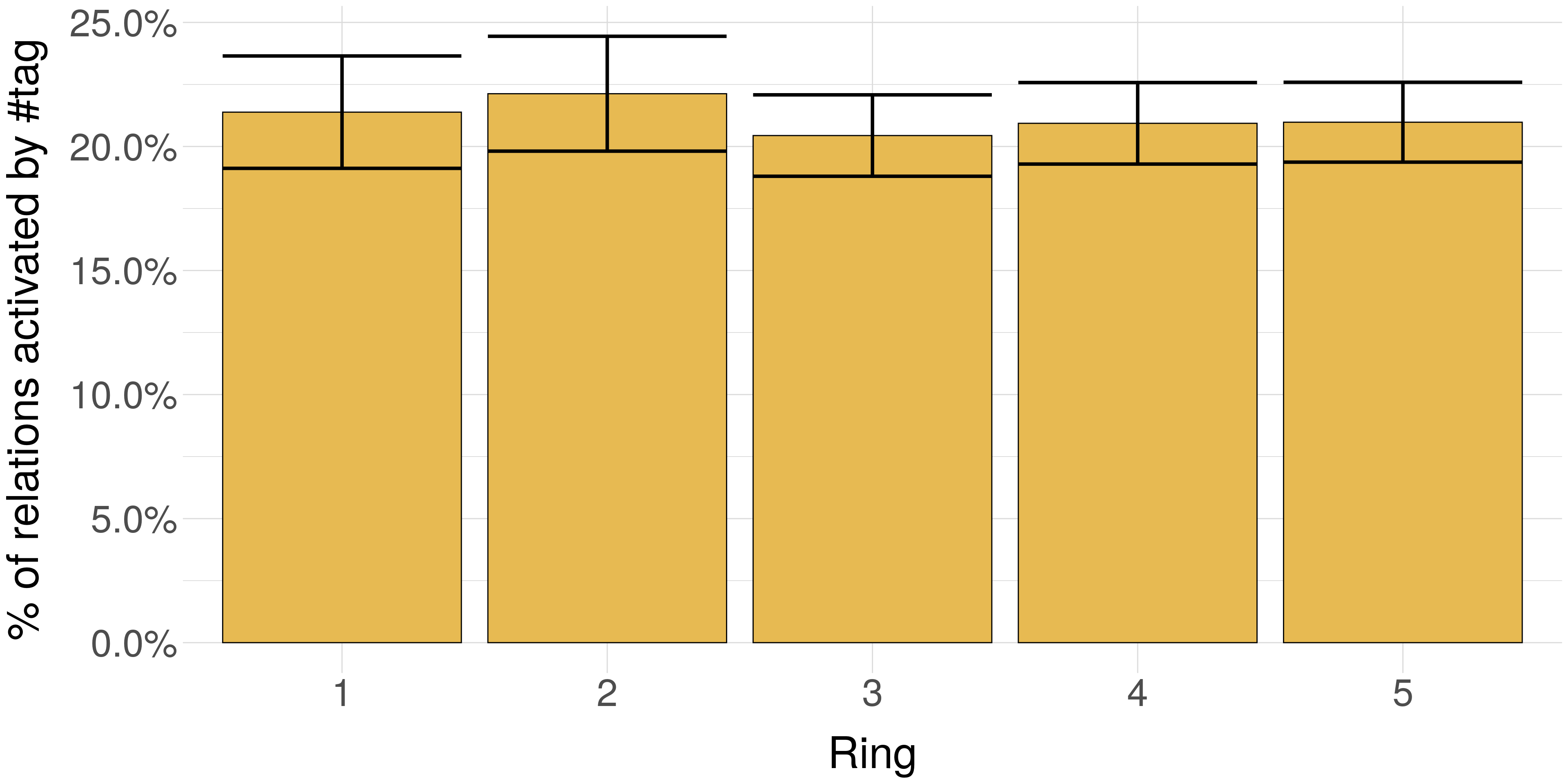}}
\hfill
\subfloat[Japanese Journalists
\label{fig:relations_hashtags_rings_JapaneseJournalists}]
{\includegraphics[width=0.33\textwidth]
{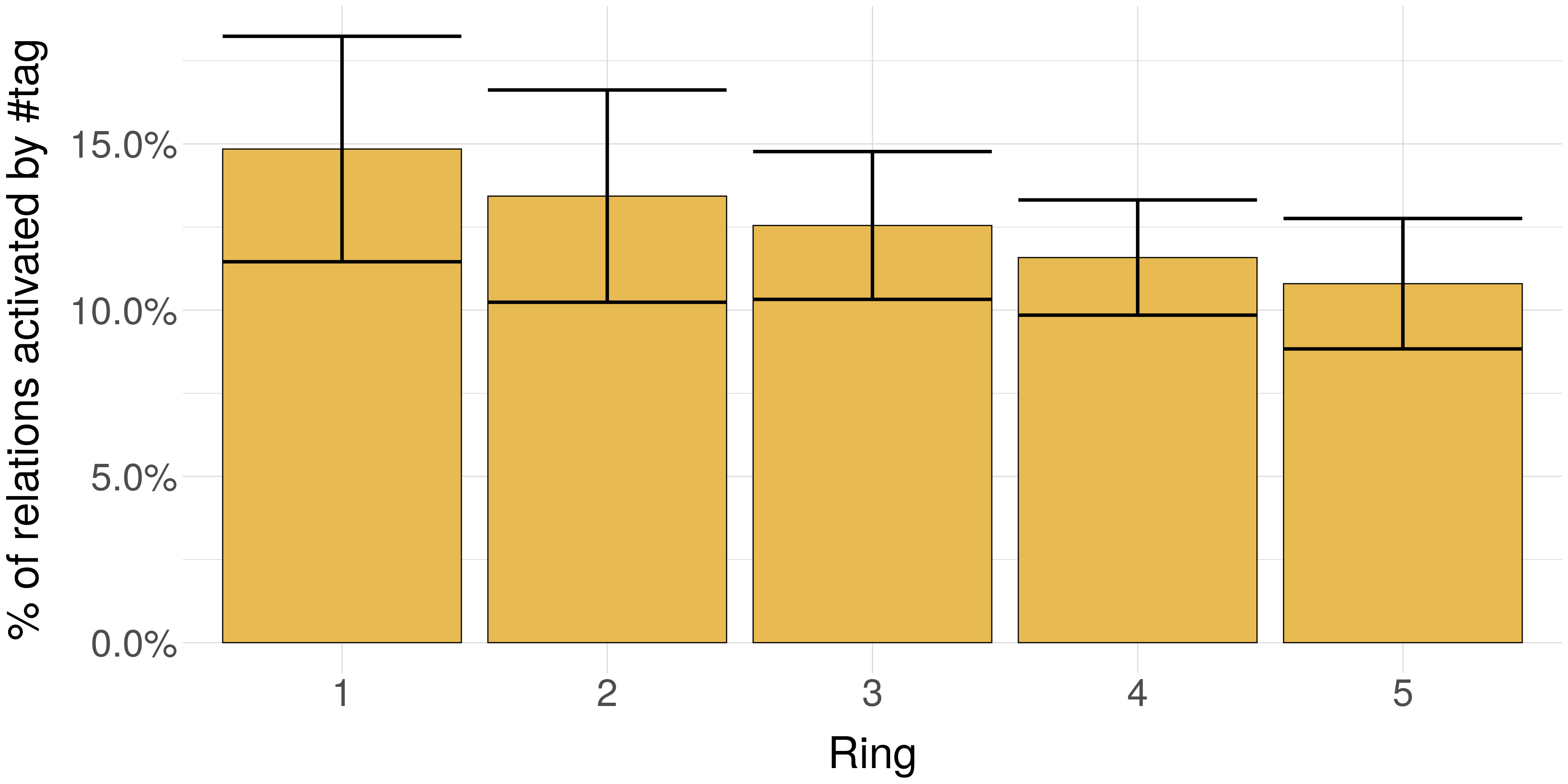}}
\hfill
\centering
\subfloat[Norwegian Journalists
\label{fig:relations_hashtags_rings_NorwegianJournalists}]
{\includegraphics[width=0.33\textwidth]
{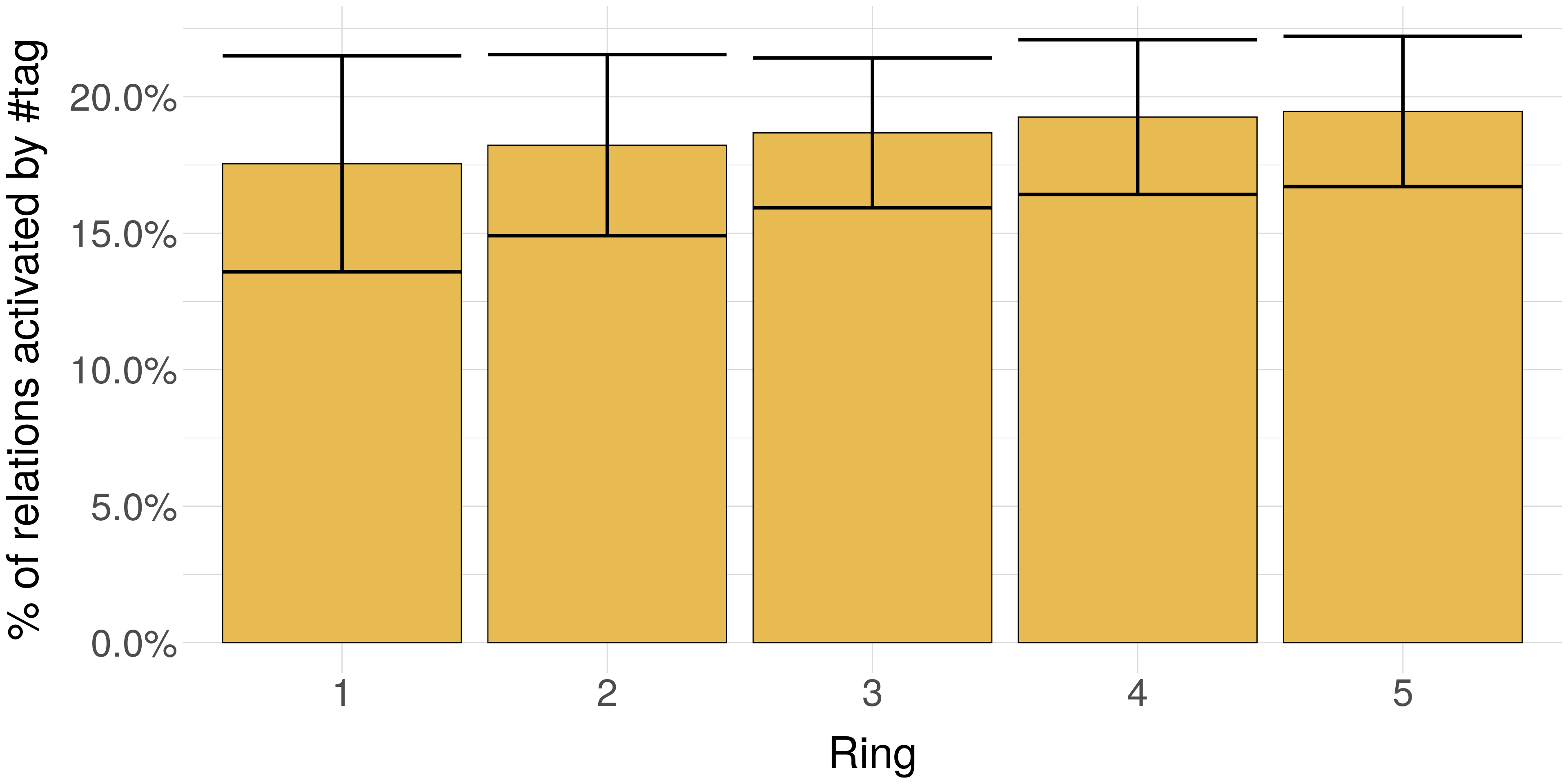}}
\hfill
\caption{Average percentage of relations activated by hashtags per ring, with confidence intervals}
\label{fig:relations_hashtags_rings}
\end{figure}

Only showing the percentage of relationships activated by hashtags does not give us information about the usage of hashtags across rings, i.e., about the dependency on intimacy. In Figure~\ref{fig:relations_hashtags_rings}, the percentage of hashtag-activated relationships is shown for selected countries, in order to highlight three different characteristics featured in the datasets. American journalists (together with Finnish, French, and Netherlander ones, not shown in the figure) use the same amount of hashtags across the rings. Hence, the percentage of their relationships activated by hashtags is independent of the intimacy level of relationships. Norwegian journalists (together with German and Greek ones) activate their less intimate relationships (outer rings) by hashtags more than more intimate layers (inner rings). The journalists from the remaining countries have higher hashtag-activated relationship percentage in the inner rings.

\begin{figure}[t]
	\begin{minipage}[t]{0.5\linewidth}
		\includegraphics[width=\textwidth]{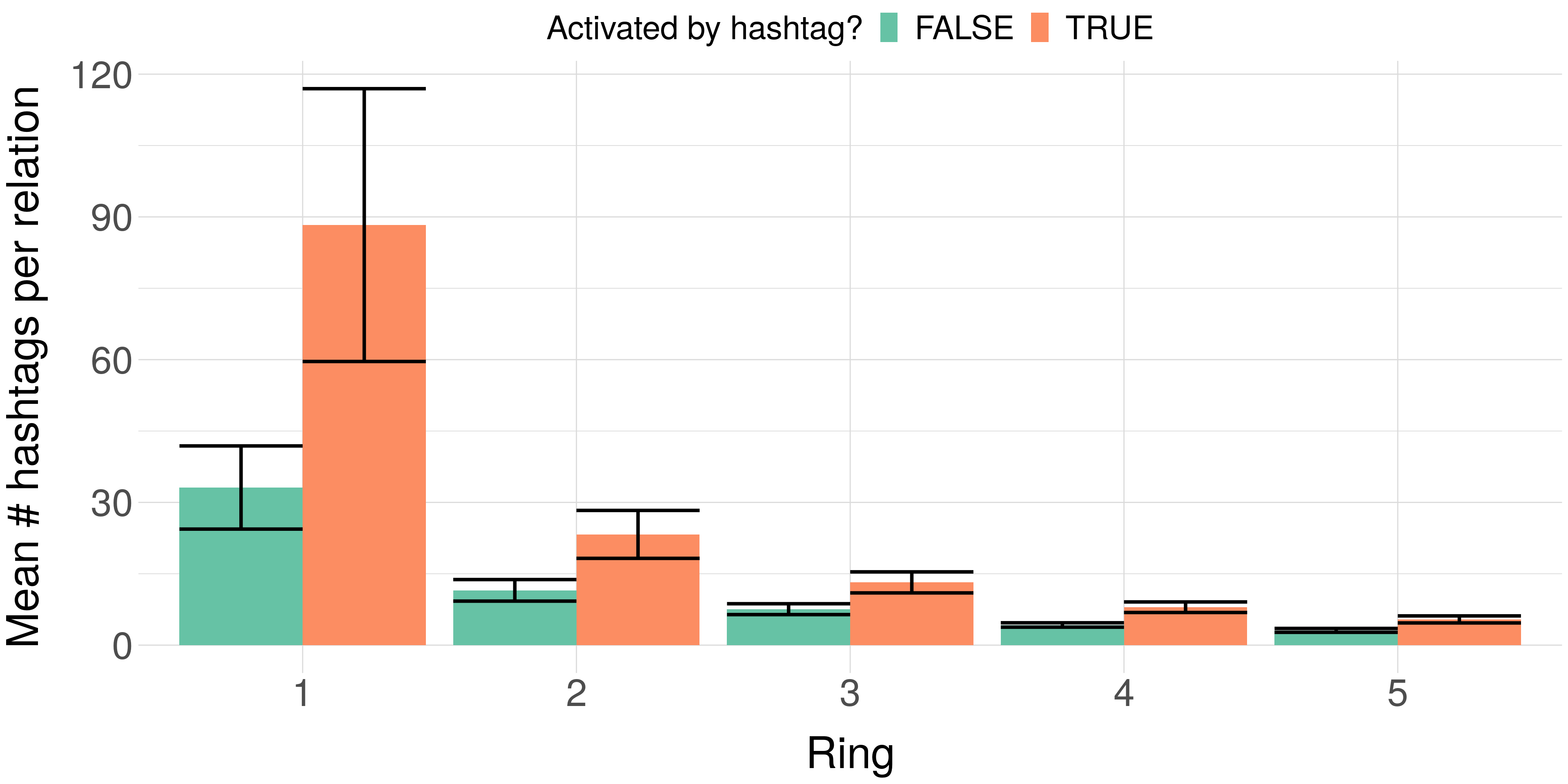}\vspace{-5pt}
		\caption{Mean number of hashtags per alter in ring (American journalists)}
		\label{fig:mean_hashtags_per_relation}\vspace{-10pt}
	\end{minipage} \hspace{10pt}
	\begin{minipage}[t]{0.5\linewidth}
		\includegraphics[width=\textwidth]{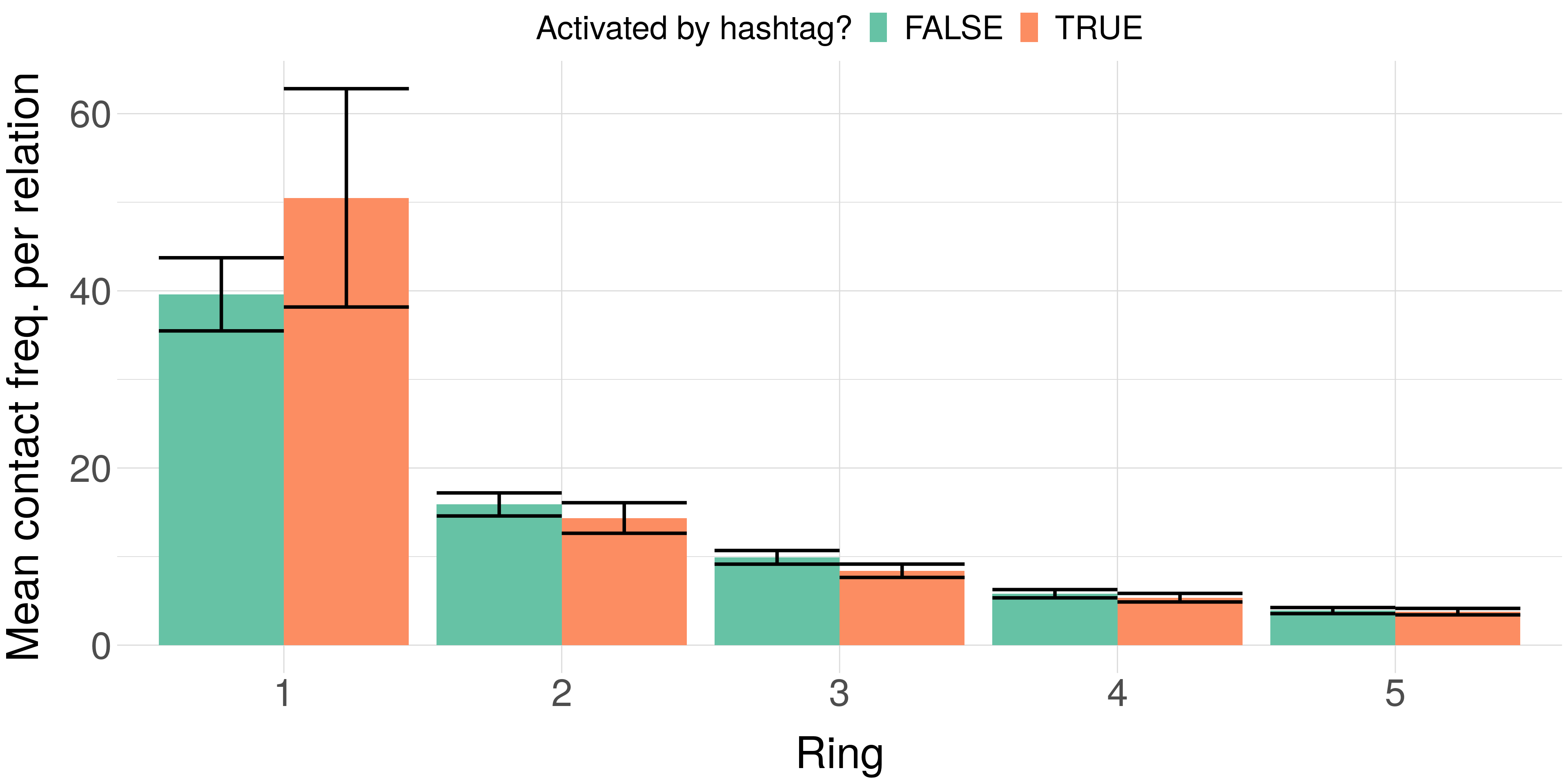}
		\caption{Contact frequency per ring (American journalists)}
		\label{fig:contact_freq_rings_hashtags}
	\end{minipage}
\end{figure}

We now shift the focus from the use of hashtags at the beginning of the relationship to the use of hashtags during its full duration. The number of hashtags used per relationship is much higher for the relationships that are activated by hashtag (i.e., their first interaction contains a hashtag) as can be seen in Figure~\ref{fig:mean_hashtags_per_relation} for a selected dataset (the vast majority of datasets feature the same behavior). 
However, this doesn't translate into an increased cognitive investment in those relationships. In fact, we don't observe statistically significant difference in terms of cognitive resources allocated (measured in terms of contact frequency) per ego, as shown in Figure~\ref{fig:contact_freq_rings_hashtags} (again, this finding is common to all datasets).

We summarise the main take-home message for this section below.
\paragraph{RQ \#4} Journalists are topic-driven users, as their relationships are initiated with a hashtag much more frequently than for politicians or generic users. Hashtags are not only used to establish a relationship (i.e., at the first contact) but they continue to be used during subsequent interactions. We also observed that hashtags are used more often for more intimate relationships, once again confirming the topic-driven nature of journalists' interactions on Twitter.

\subsection{Popularity of the alters and intimacy level of the relationships}
\label{sec:correlation}

Our goal in this section is to investigate if popular journalists interact with similarly popular journalists (a sort of assortativity in popularity) on Twitter and if there is a correlation between the popularity of alters and the intimacy level of the corresponding relationship. Similar analyses are not available for politicians and generic users, hence we will provide no comparison. 

In the following, we measure popularity in terms of the number of followers of the account, similarly to~\cite{imamori2016predicting}.
To understand the effect of the popularity of the alters in different ego network layers, we carry out a correlation analysis between the popularity of egos and that of their alters. Recall that correlation values vary between $-1$ and $+1$, where $+1$ represents the perfect positive association, $-1$ represents the perfect negative association and $0$ represents the independence of the variables. Since we are interested in assessing monotonic relationships, we will use Kendall's correlation~\cite{chok2010,hauke2011}. 
We also want to distinguish between journalist and non-journalist alters (because some effects might be dependent on the professional nature of the interaction). Thus, we first label the alters of the journalists using the same method discussed in Section~\ref{sec:labellingusers} (which entails leveraging the Google Knowledge graph together with the keywords in the Twitter bio). Then, the correlation scores between the ego's popularity (its follower number) and the average alter popularity (mean of follower number of alters) are calculated. 

\begin{figure}[h]
\centering
\includegraphics[width=0.8\textwidth]{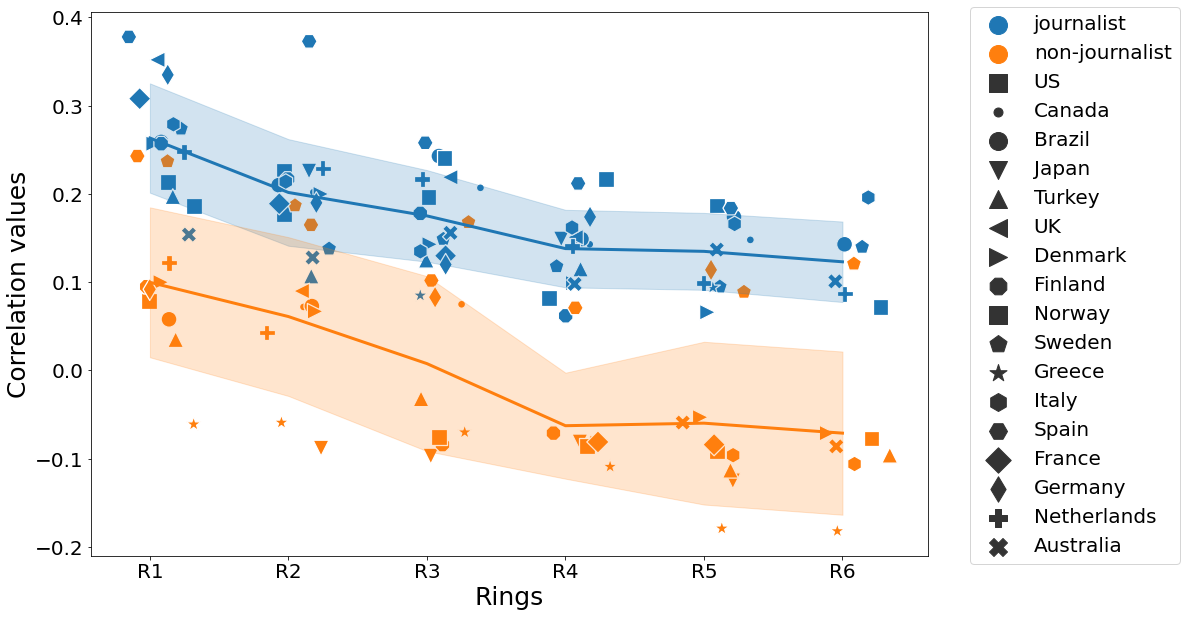}
\caption{Correlation coefficients between the popularity of ego and the journalist (in blue) vs non-journalist (in orange) alters per ring, per country. Only correlation values whose $p$-value is smaller than $0.1$ are reported.}
\label{fig:correlations_countries}
\end{figure}

In Figure~\ref{fig:correlations_countries}, we provide the Kendall's correlation between the popularity of the egos and their alters (which can be either journalists or non-journalists) in each ring. Blue dots represent journalist alters, orange dots represent non-journalist alters. The lines represent the averages of the groups with standard deviation bands. Note that the plot includes the correlation values whose corresponding $p$-value is low enough (that is, smaller than $0.1$) to draw reliable conclusions (therefore, for some countries, some of the dots are missing).
We visualize the results for the first six rings, since the number of egos with more than 6 circles decreases drastically, and we don't have enough samples to measure correlation.
From Figure~\ref{fig:correlations_countries}, we observe that the popularity of egos and their journalist alters is correlated, while that of egos and non-journalist alters much less so. In terms of absolute values of correlation, while ego-journalist relationships exhibit a positive correlation regarding their popularity at all rings, the correlation for relationships involving egos and non-journalists is greater than zero only in the innermost rings. This suggests that what we called ``assortativity in popularity'' is a phenomenon common to all ring for journalist-to-journalist relationships, while it is only relevant in the innermost ring for the other relationships. In other words, journalist-to-journalist relationships seem to be popularity-driven even when interactions are more sporadic. Vice versa, when journalists interact with non-journalists, the popularity of the alter seems to have a role only for strong connections. 

\begin{figure}[h]
\centering
\includegraphics[width=0.65\textwidth]{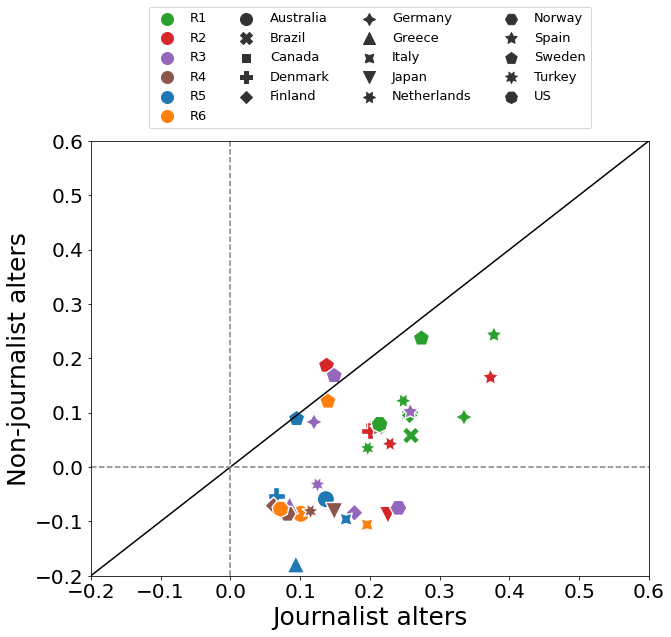}
\caption{Scatterplot of the correlation coefficients between the popularity of ego and of the journalist vs. non-journalist alters per ring, per country.  Only correlation values whose $p$-value is smaller than $0.1$ are reported.}
\label{fig:correlations_journalists_vs_nonjournalists}
\end{figure}

In Figure~\ref{fig:correlations_journalists_vs_nonjournalists} we provide a direct comparison between the correlation in popularity of journalist-journalist relationships and the others. Specifically, we provide a scatterplot where each point corresponds to a country-ring combination and its $x,y$ coordinates are the correlations in the non-journalists vs journalists groups. 
It can be seen from the plot that journalist alters never feature negative correlation values, while non-journalist alters do. 
In the plot we can identify three groups of country-ring combinations. In the first group, the popularity of non-journalist alters has negative correlation with the popularity of their egos, while the journalist alters exhibit positive correlation values (lower-right quadrant). In the second group, the popularity of both journalist and non-journalist alters is positively correlated with the of the ego, but since the values are under the identity line, the correlation values of journalist alters are larger than the non-journalist ones. These two groups cover almost all of the country-ring combinations. There is one exception: Sweden. The correlation values of Sweden journalists are located near the identity line, which implies that the correlation values of journalists and non-journalist are close to each other. And for the second and third rings, Swedish non-journalist alters have higher correlation values than their journalist counterparts. 

We can now provide an answer to the fifth research question we discussed in Section~\ref{sec:introduction}.

\paragraph{RQ \#5} Highly popular journalists tend to engage with other journalists of similar popularity, and vice versa. In addition, journalist egos tend to keep their popular colleagues in the more intimate layers, hence allocate more cognitive resources for them.
Instead, when egos interact with non-journalist users, the alter popularity i) does not seem to play a role for relationship in the outermost circles, ii) has a smaller effect than for journalist-journalist relationships even in the innermost circles.

\section{Conclusions}
\label{sec:conclusions}

In this paper, we have studied the ego network structure of journalists from 17 different countries belonging to 8 different continental regions and 4 continents. We have demonstrated that how they allocate their cognitive resources among their relationships is mostly independent of the countries they are from. In general, the ego networks of journalists on Twitter all mirror the same offline human ego network structure where, according to the Dunbar's model, our finite social cognitive capacity imposes limits to how many people we meaningfully interact with and to the intensity of these relationships.
Journalists show a different dynamic ``use" of their Twitter relationships with respect to general users, and much more similar to what has been observed for politicians. Specifically, \emph{most} and \emph{least} intense relationships (in terms of interaction frequency) are significantly more stable than \emph{intermediately} intense one, whose turnover is almost 100\% each year. This confirms a different user of Twitter relationships for public figures (such as politicians and journalists) with respect to the general population. In addition, differently from general users, journalists establish and maintain their relationships in a topic-oriented way, even more than politicians do. This suggests that journalists' engagement on Twitter is mostly information-driven. Journalists also tend to have stable relationships and to maintain their relationships longer than politicians. As a final step, we have shown that journalists are not randomly allocating their cognitive resources among their colleagues. They establish more intimate relationships with their counterparts with similar popularity. However, popularity plays a much less significant role when they interact with non-journalists.
 
\section*{Acknowledgements}

This work was partially funded by the SoBigData++, HumaneAI-Net, MARVEL, OK-INSAID, and SAI projects. The SoBigData++, HumaneAI-Net, and MARVEL projects have received funding from the European Union's Horizon 2020 research and innovation programme under grant agreements No 871042, No 952026, No 957337, respectively. The OK-INSAID project has received funding from the Italian PON-MISE program under grant agreement ARS01 00917. The SAI project is supported by the CHIST-ERA grant CHIST-ERA-19-XAI-010, funded by by MUR (grant No. not yet available), FWF (grant No. I 5205), EPSRC (grant No. EP/V055712/1), NCN (grant No. 2020/02/Y/ST6/00064), ETAg (grant No. SLTAT21096), BNSF (grant No. KP-06-DOO2/5).

\bibliographystyle{elsarticle-num}
\bibliography{CITATIONS}


\clearpage

\iftoggle{ONEFILE}{
	\appendix
	\input{appendices}
}{
}

\end{document}

%% file: appendices.tex
\renewcommand\thefigure{\Alph{section}.\arabic{figure}} 
\setcounter{figure}{0}

\renewcommand\thetable{\Alph{section}.\arabic{table}} 
\setcounter{table}{0}

\section{Preprocessing the dataset}
\label{appendix_dataset}

\vspace{-5pt}

\subsection{Observed timeline lengths - all datasets}
\label{appendix_observed_timeline_legnth}

\vspace{-15pt}

\begin{figure}[!h]
\begin{adjustbox}{minipage=0.97\linewidth}
\begin{center}
\centering
\subfloat[USA
\label{fig_appendix:observed_timelines_length}]
{\includegraphics[width=0.28\textwidth]
{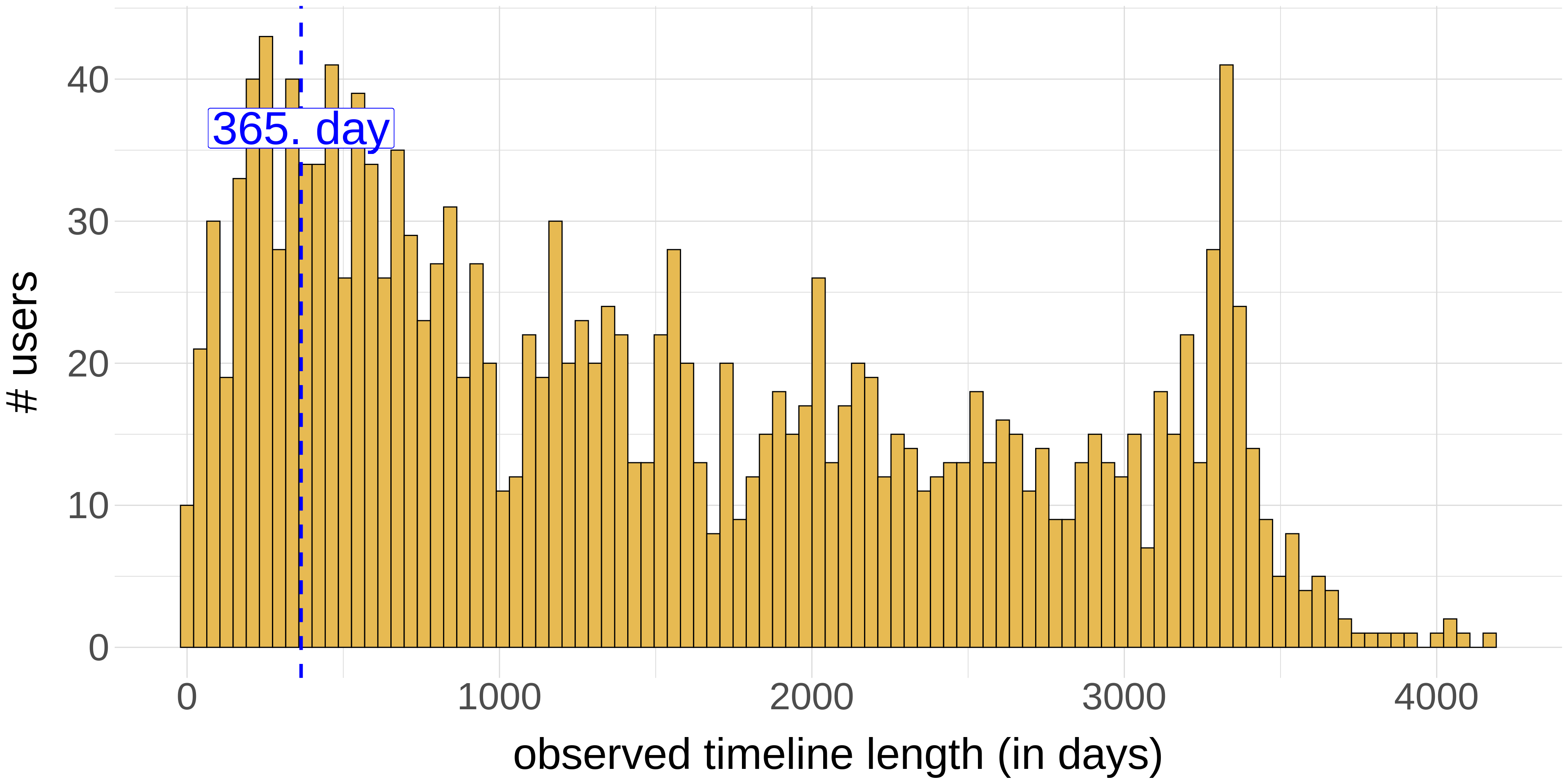}}
\hfill
\subfloat[Canada
\label{fig_appendix:observed_timelines_length_CanadianJournalists}]
{\includegraphics[width=0.28\textwidth]
{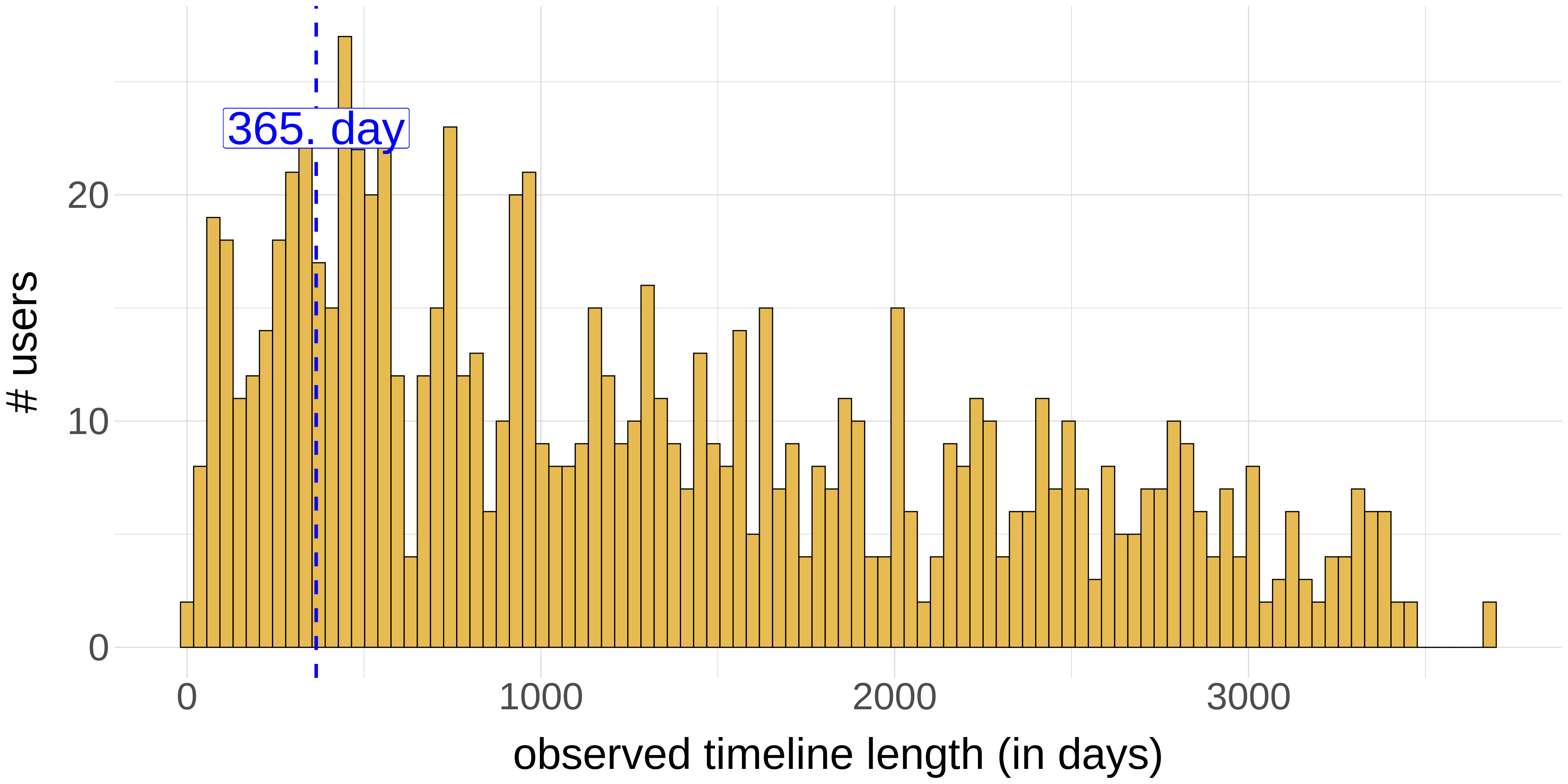}}
\hfill
\subfloat[Brasil
\label{fig_appendix:observed_timelines_length_BrazilianJournalists}]
{\includegraphics[width=0.28\textwidth]
{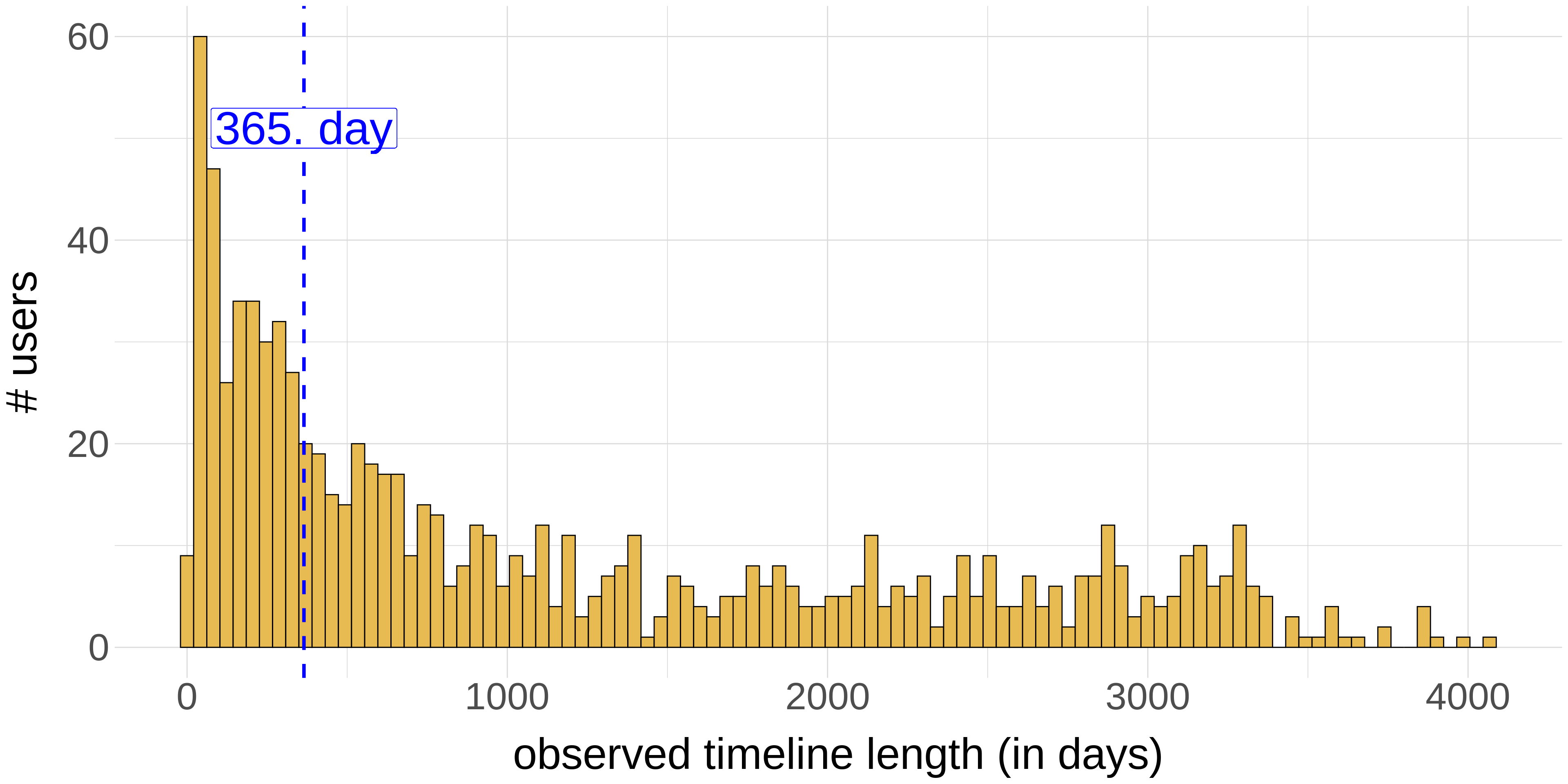}}
\hfill
\subfloat[Japan
\label{fig_appendix:observed_timelines_length_JapaneseJournalists}]
{\includegraphics[width=0.28\textwidth]
{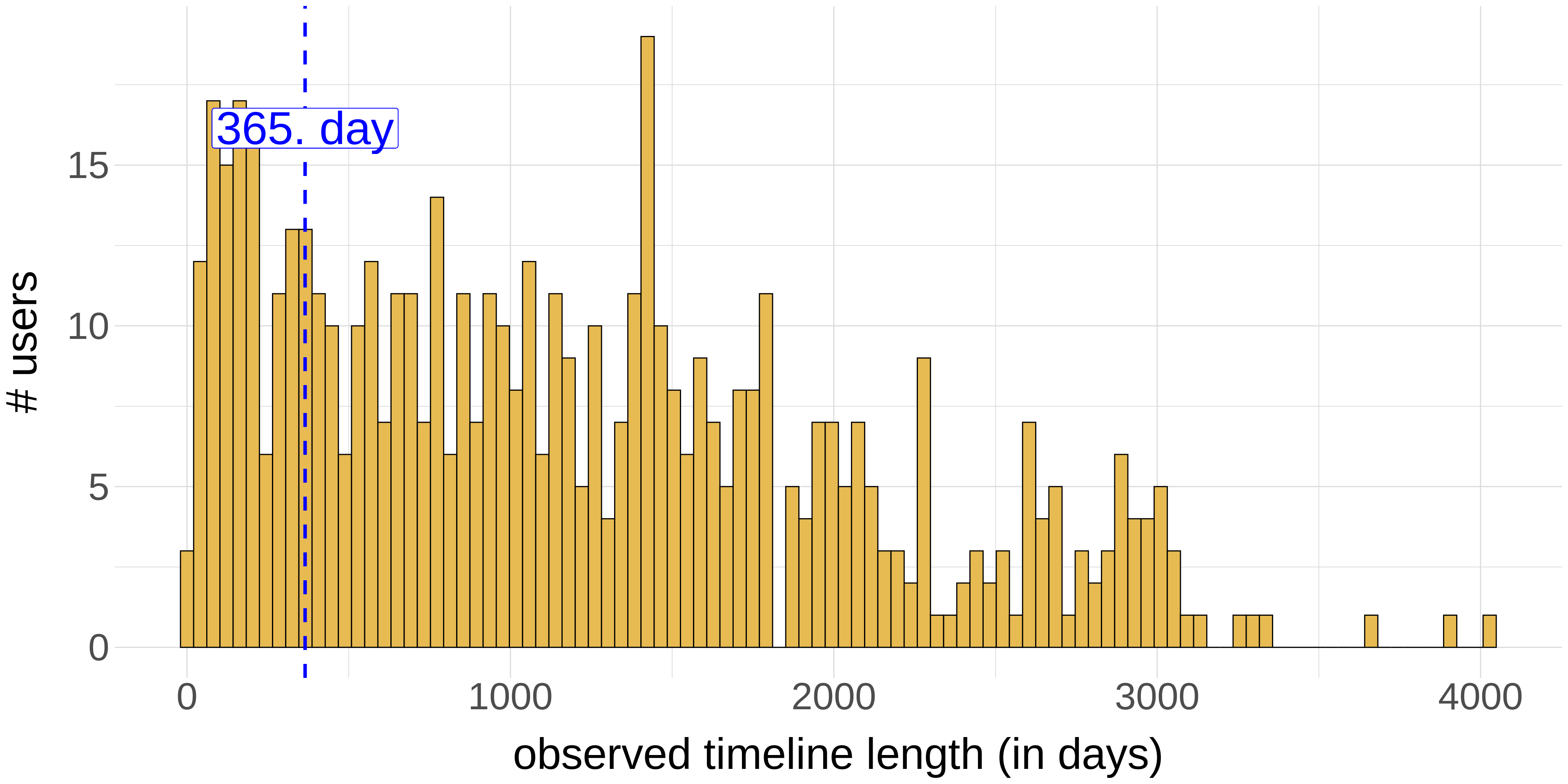}}
\hfill
\subfloat[Turkey
\label{fig_appendix:observed_timelines_length_TrukishJournalists}]
{\includegraphics[width=0.28\textwidth]
{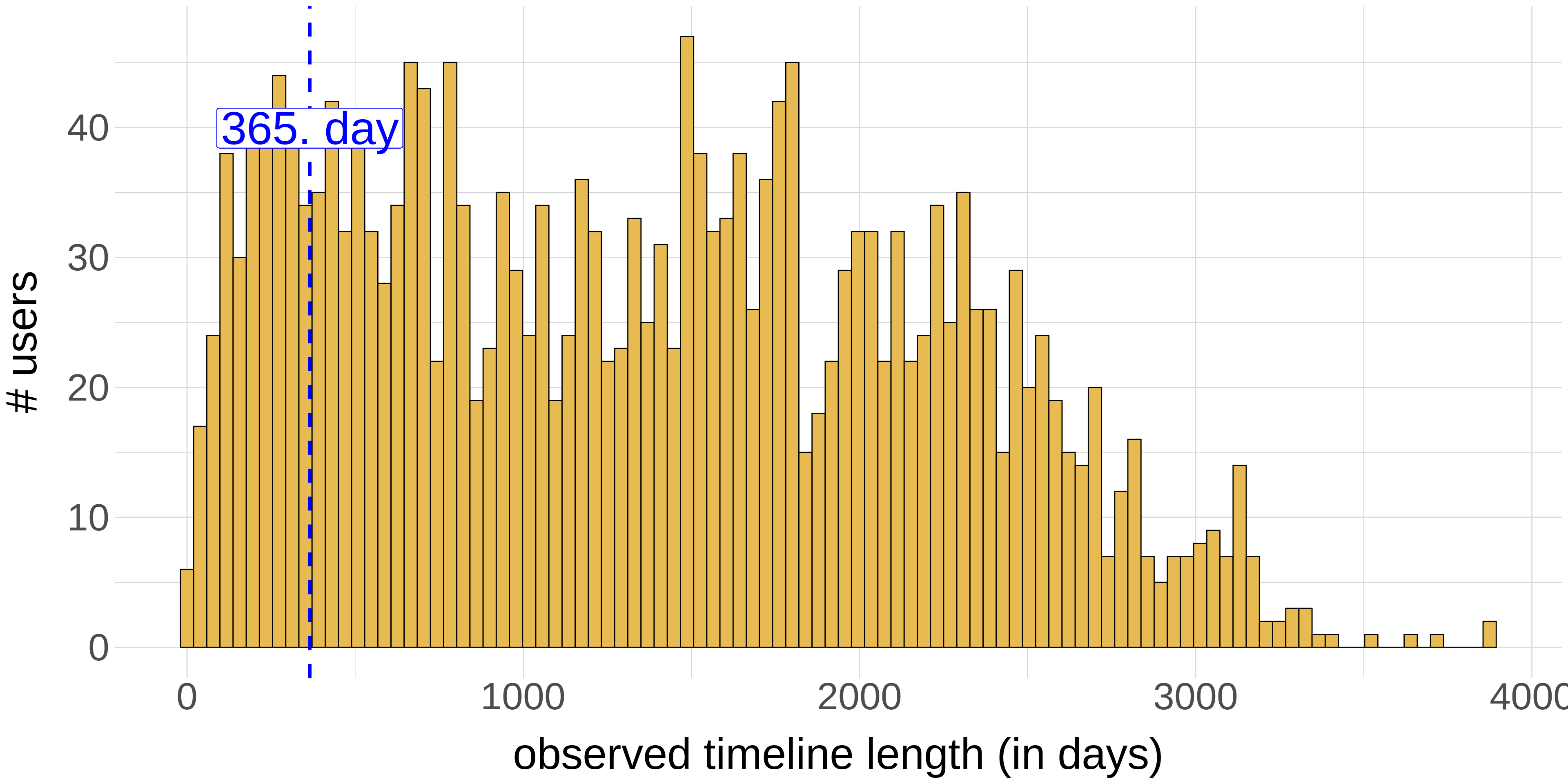}}
\hfill
\subfloat[UK
\label{fig_appendix:observed_timelines_length_BritishJournalists}]
{\includegraphics[width=0.28\textwidth]
{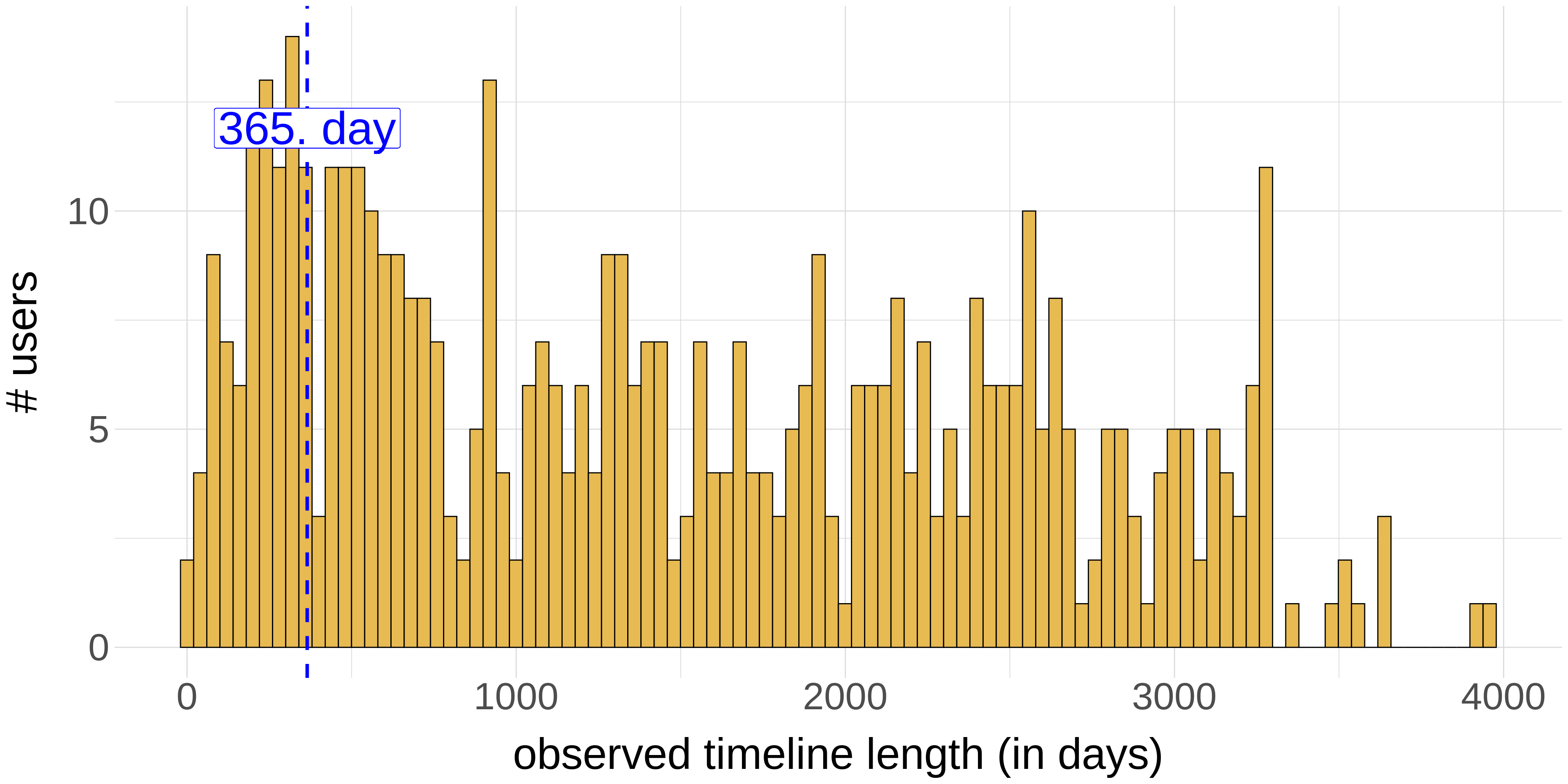}}
\hfill
\subfloat[Denmark
\label{fig_appendix:observed_timelines_length_DanishJournalists}]
{\includegraphics[width=0.28\textwidth]
{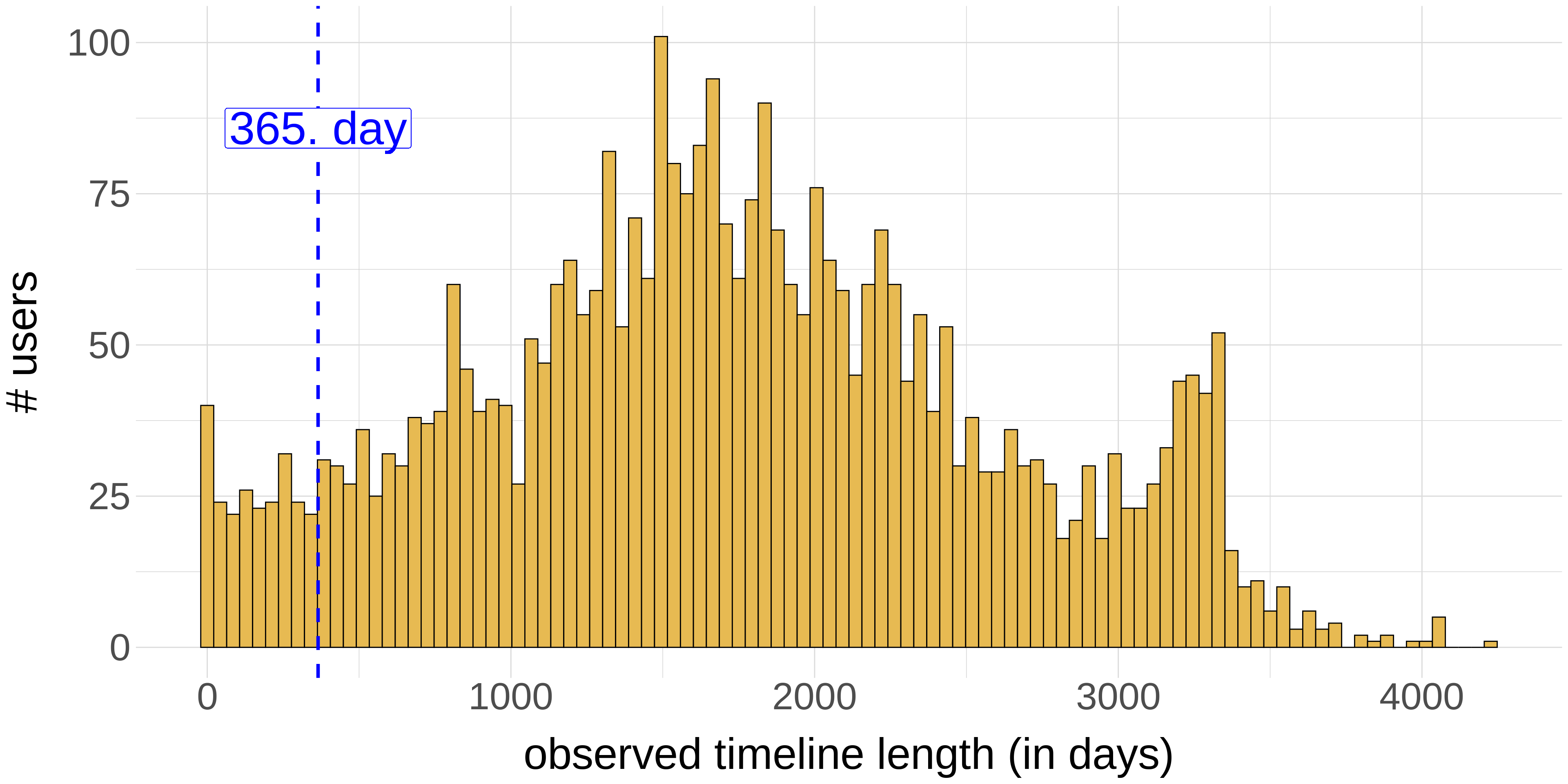}}
\hfill
\subfloat[Finland
\label{fig_appendix:observed_timelines_length_FinnishJournalists}]
{\includegraphics[width=0.28\textwidth]
{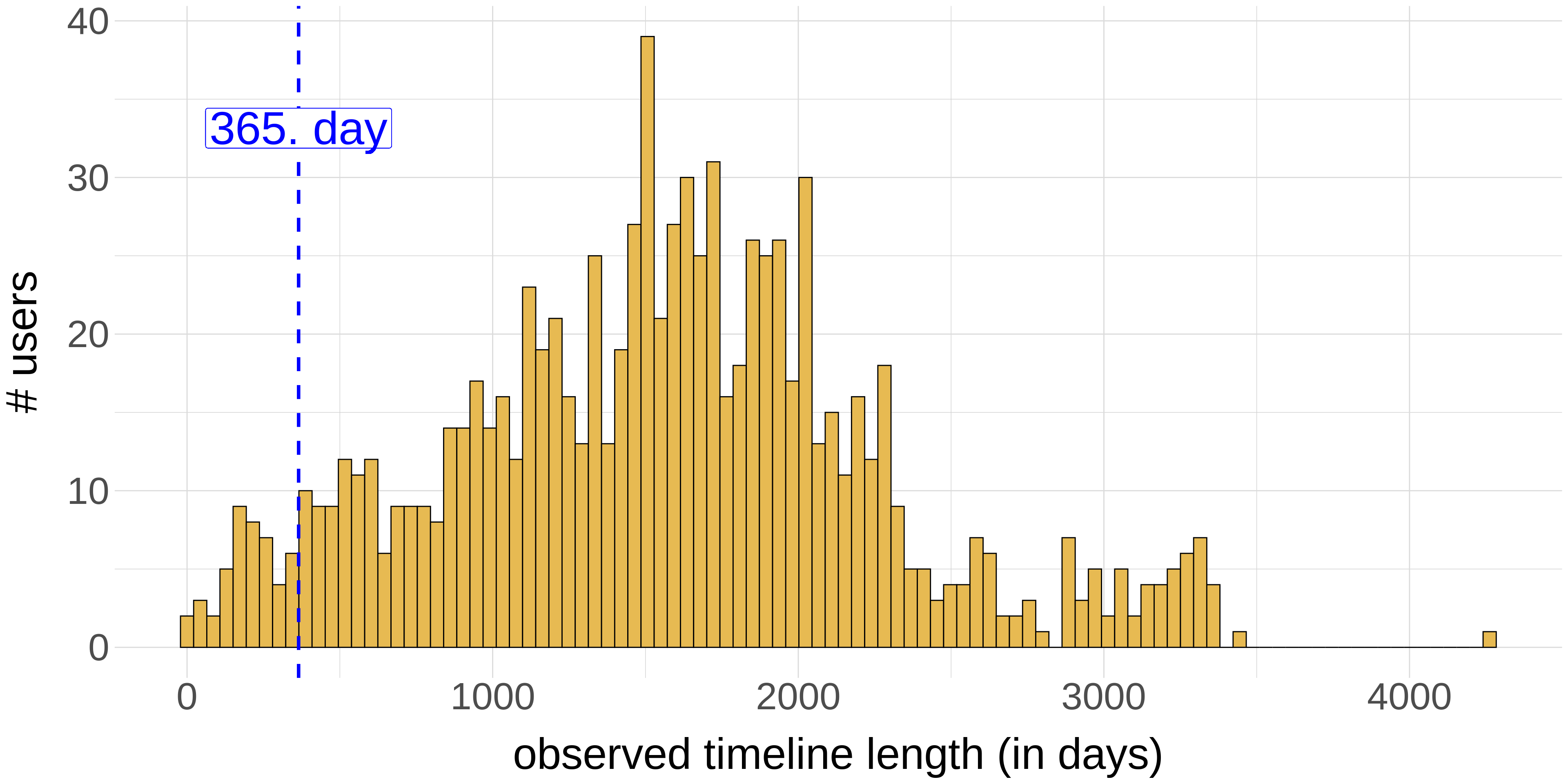}}
\hfill
\subfloat[Norway
\label{fig_appendix:observed_timelines_length_NorwegianJournalists}]
{\includegraphics[width=0.28\textwidth]
{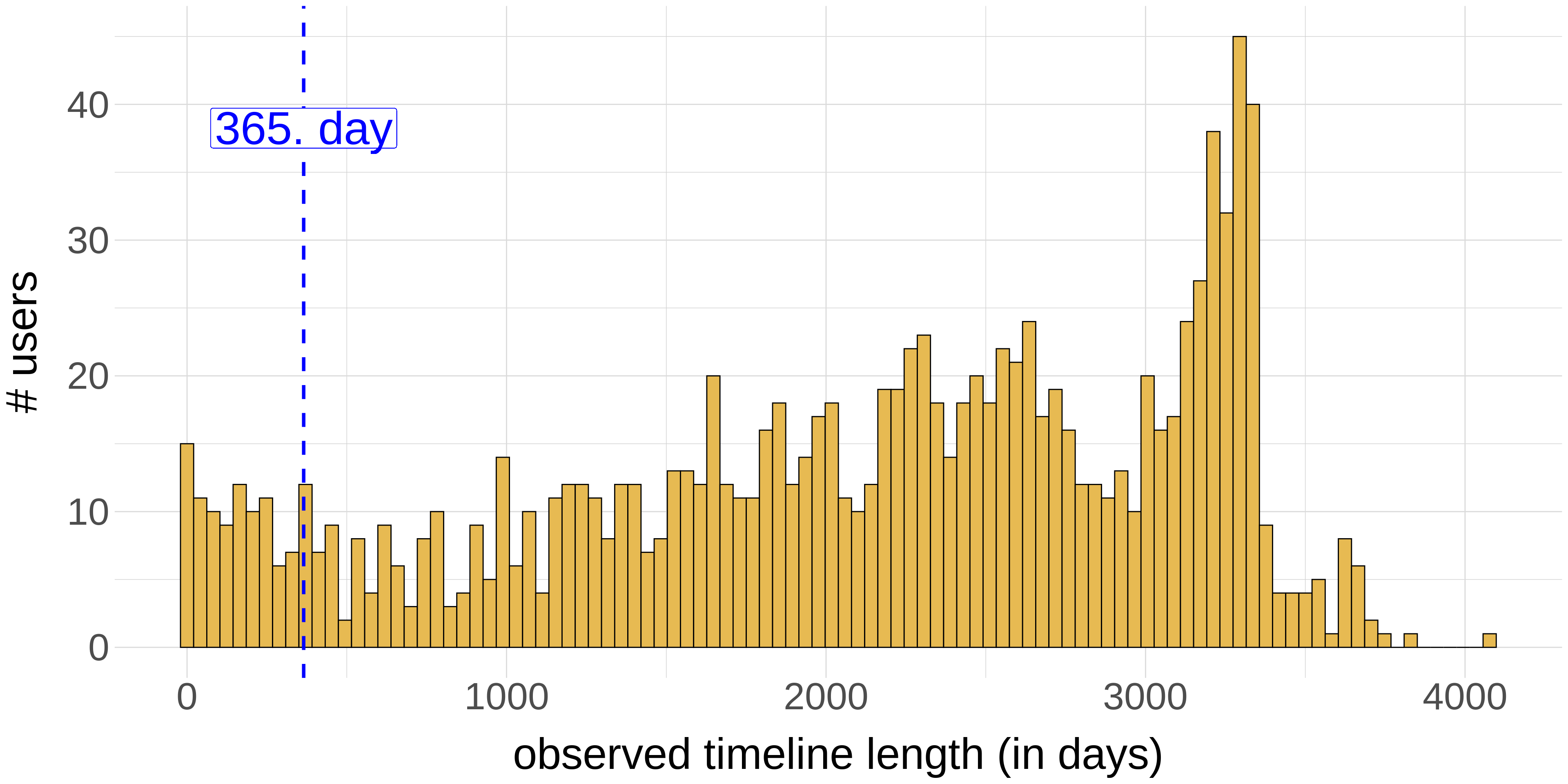}}
\hfill
\subfloat[Sweden
\label{fig_appendix:observed_timelines_length_SwedishJournalists}]
{\includegraphics[width=0.28\textwidth]
{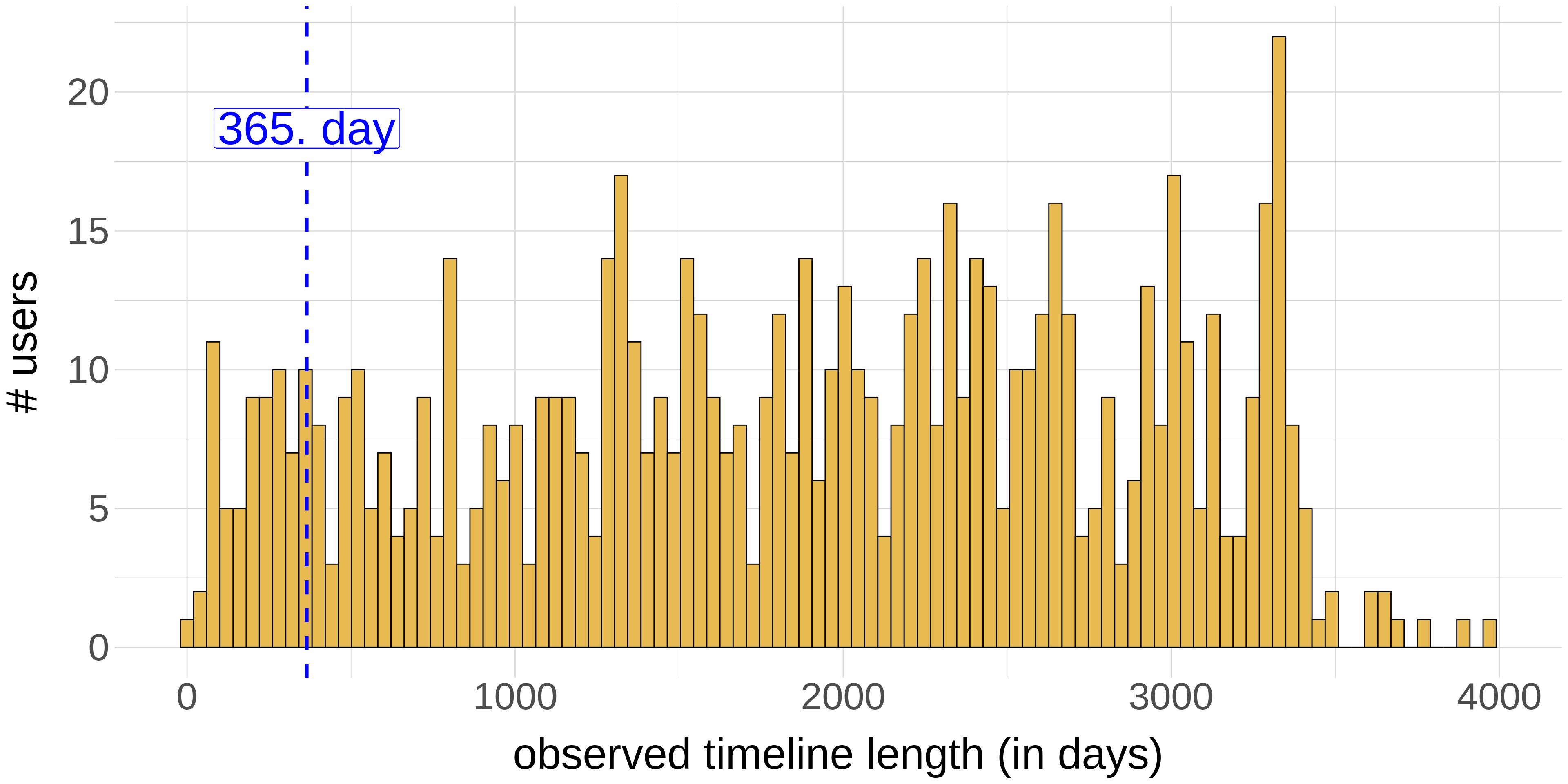}}
\hfill
\subfloat[Greece
\label{fig_appendix:observed_timelines_length_GreekJournalists}]
{\includegraphics[width=0.28\textwidth]
{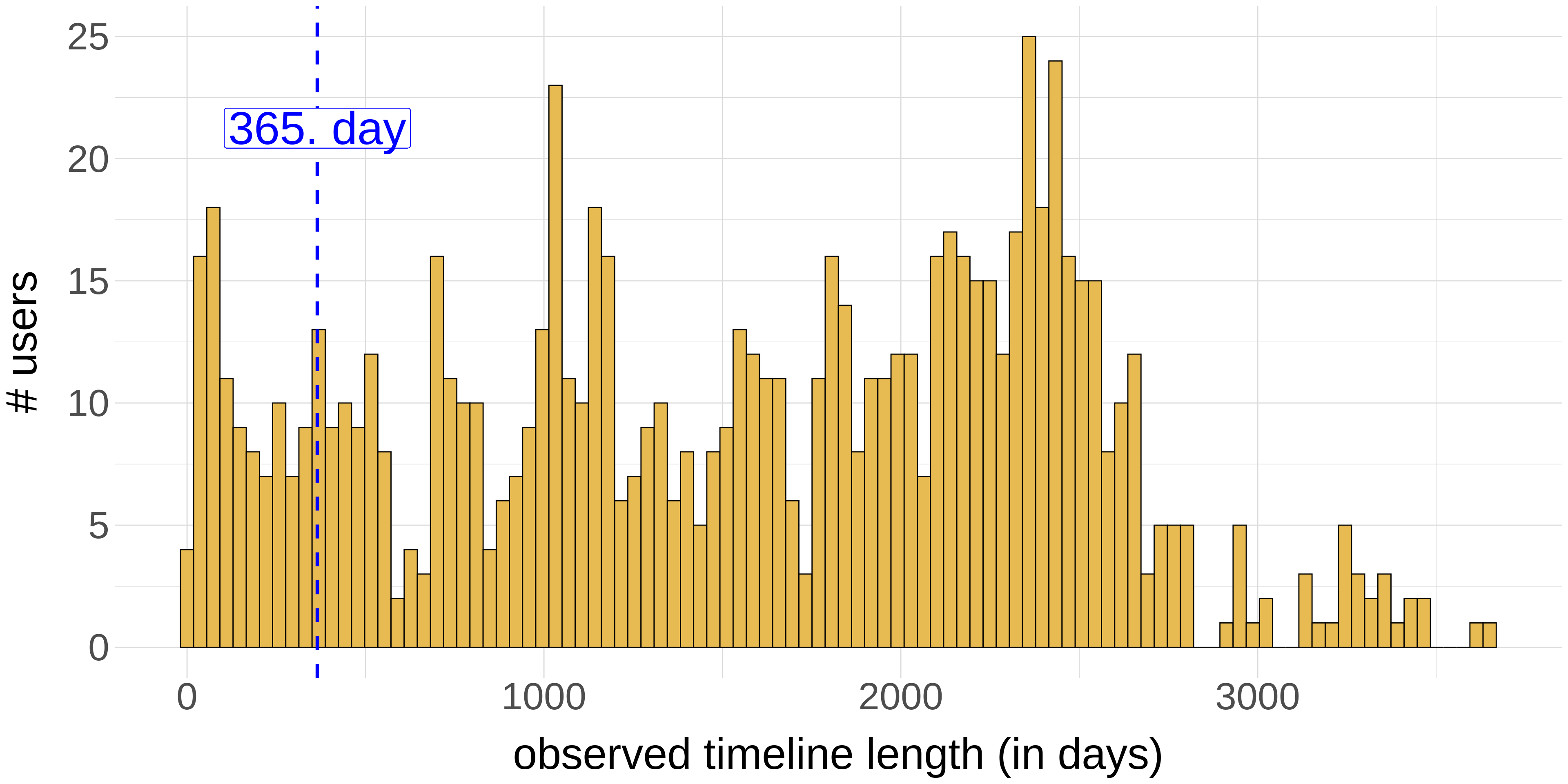}}
\hfill
\subfloat[Italy
\label{fig_appendix:observed_timelines_length_ItalianJournalists}]
{\includegraphics[width=0.28\textwidth]
{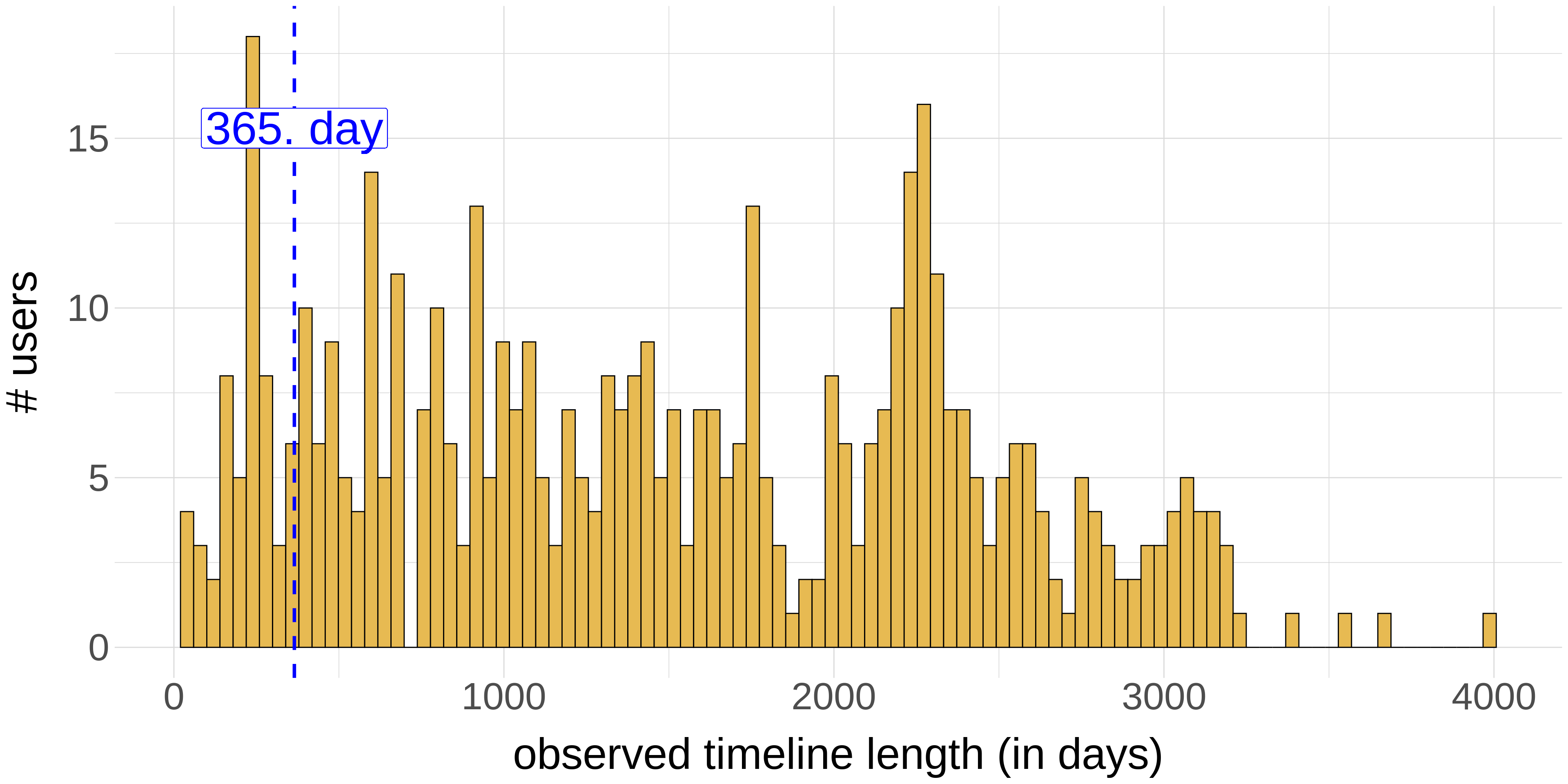}}
\hfill
\subfloat[Spain
\label{fig_appendix:observed_timelines_length_SpanishJournalists}]
{\includegraphics[width=0.28\textwidth]
{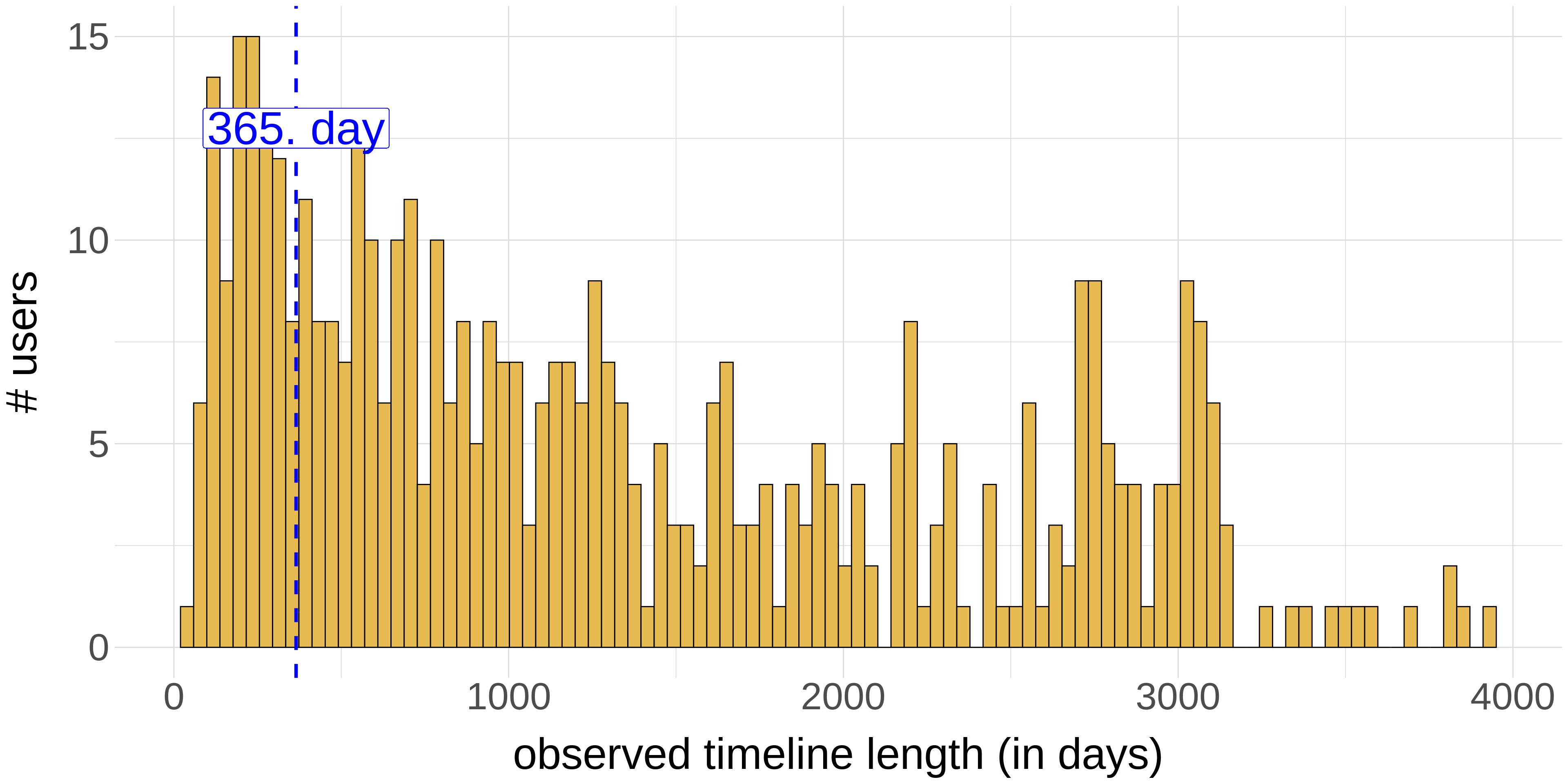}}
\hfill
\subfloat[France
\label{fig_appendix:observed_timelines_length_FrenchJournalists}]
{\includegraphics[width=0.28\textwidth]
{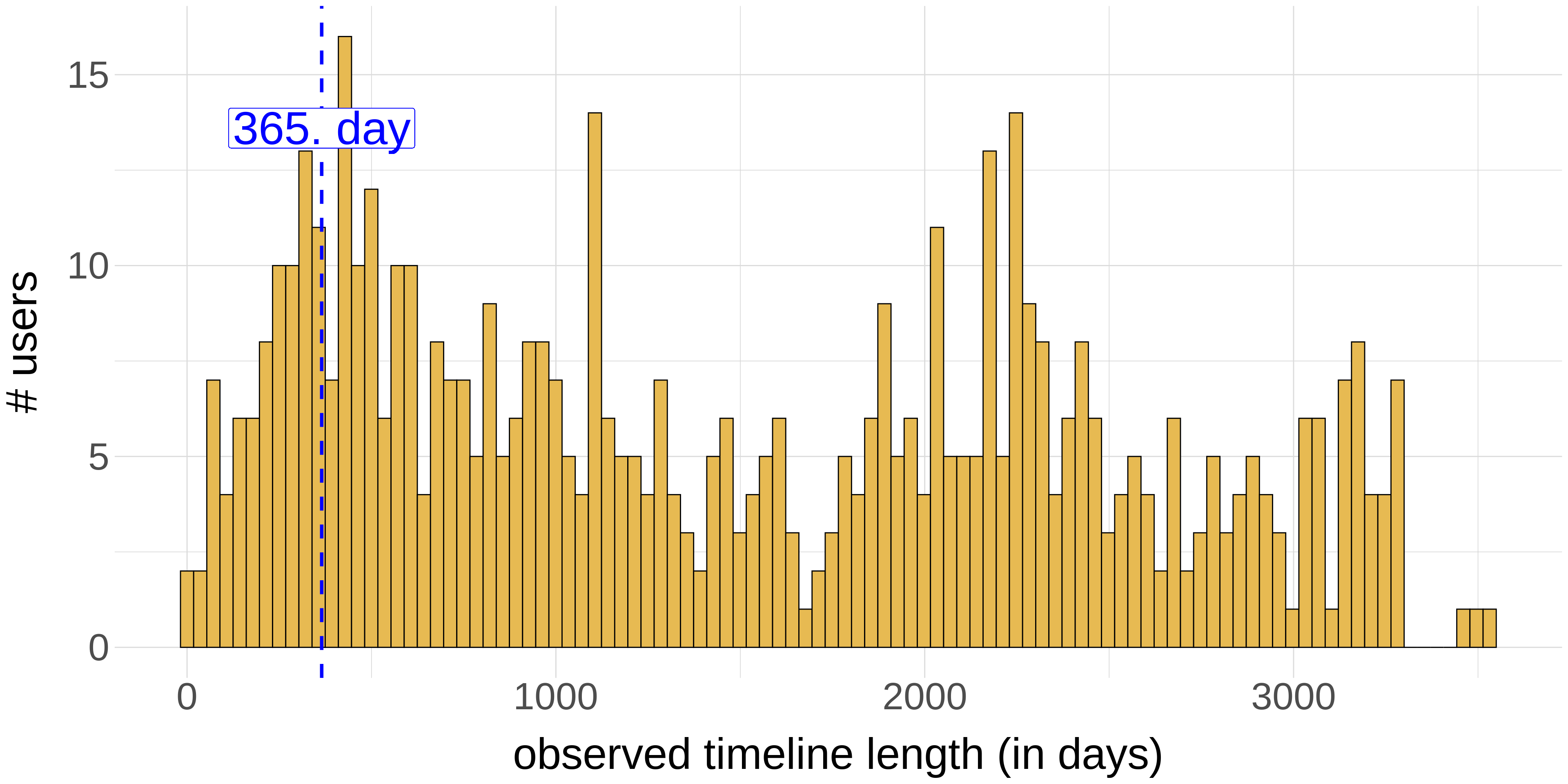}}
\hfill
\subfloat[Germany
\label{fig_appendix:observed_timelines_length_GermanJournalists}]
{\includegraphics[width=0.28\textwidth]
{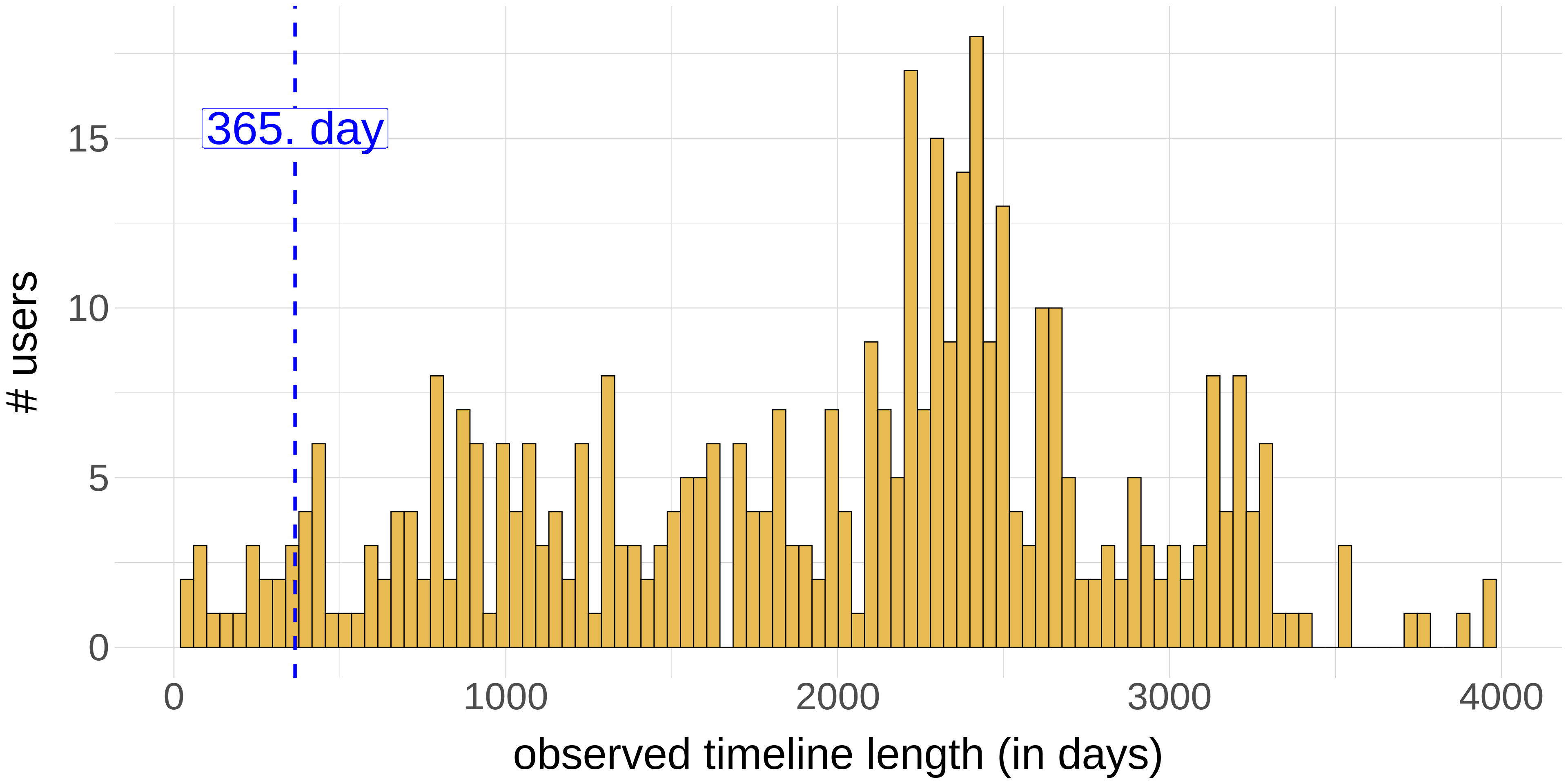}}
\hfill
\subfloat[Netherland
\label{fig_appendix:observed_timelines_length_NetherlanderJournalists}]
{\includegraphics[width=0.28\textwidth]
{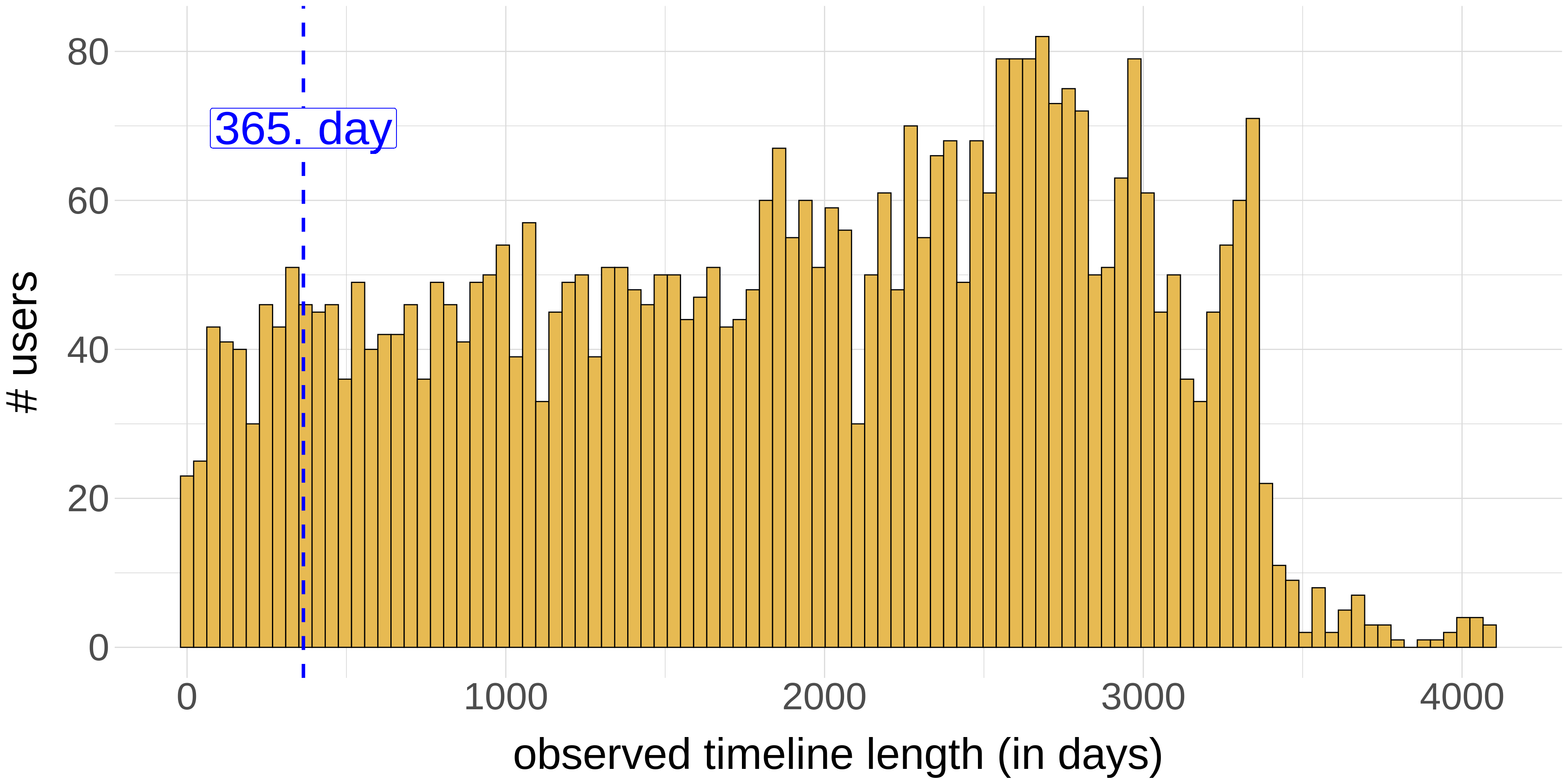}}
\hspace{1pt}
 \subfloat[Australia
\label{fig_appendix:observed_timelines_length_AustralianJournalists}]
{\includegraphics[width=0.28\textwidth]
{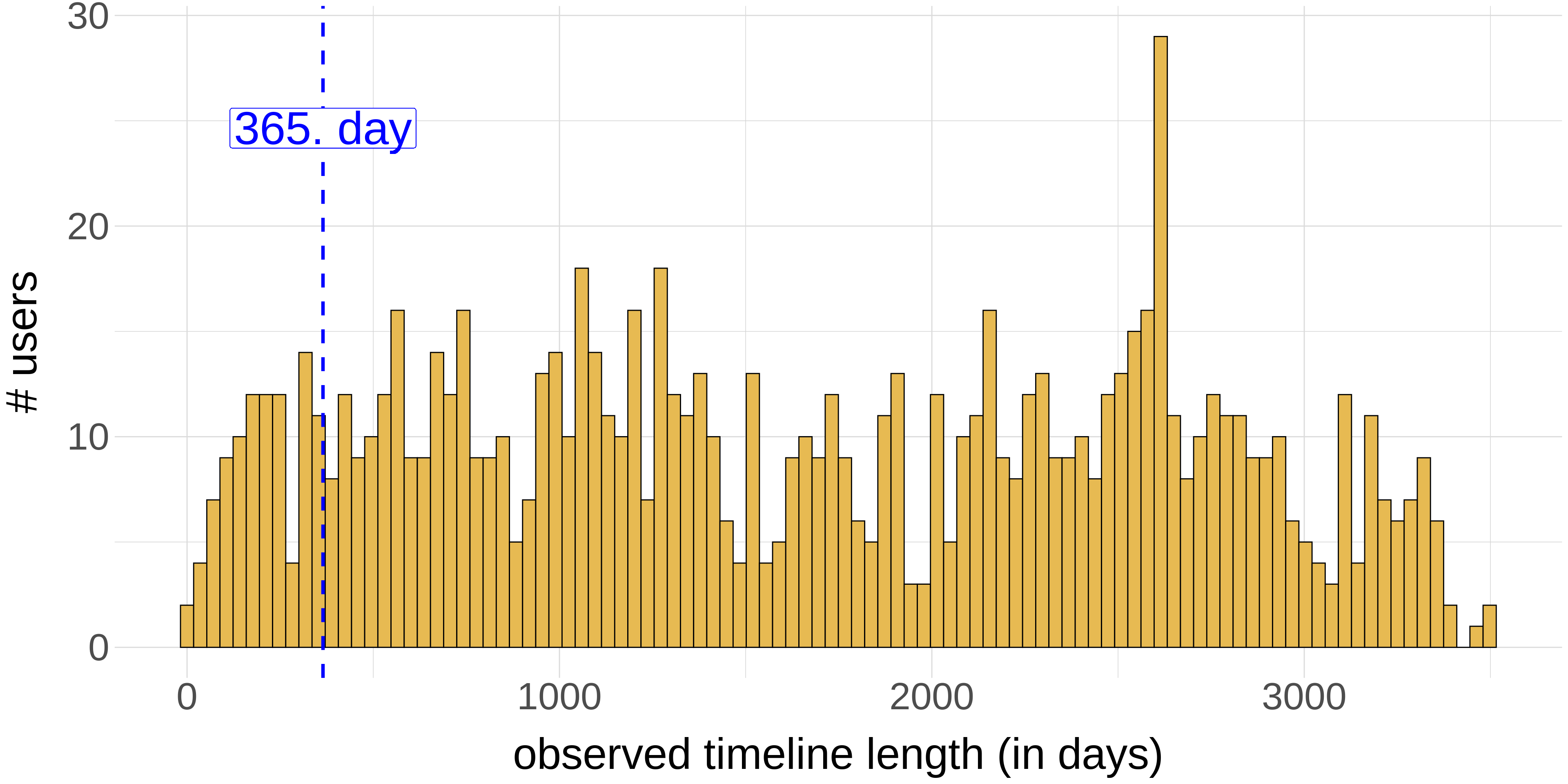}}
\end{center}
\end{adjustbox}
\caption{Distribution of observed timeline lengths. The blue line corresponds to the minimum threshold (365 days) for extracting active ego networks.}
\label{fig_appendix:observed_timelines}
\end{figure}


\subsection{Tweeting volume and active users over time - all datasets}
\label{appendix_tweetingVolume}

\vspace{-10pt}

\begin{figure}[!h]
\begin{adjustbox}{minipage=0.98\linewidth}
\begin{center}
\subfloat[USA
\label{fig_appendix:tweeting_volume_AmericanJournalists}]
{\includegraphics[width=0.3\textwidth]
{./figures_new/tweeting_volume/AmericanJournalists_tweet_vol_vs_active_user_before_filtering.png}}
\hfill
\subfloat[Canada
\label{fig_appendix:tweeting_volume_CanadianJournalists}]
{\includegraphics[width=0.3\textwidth]
{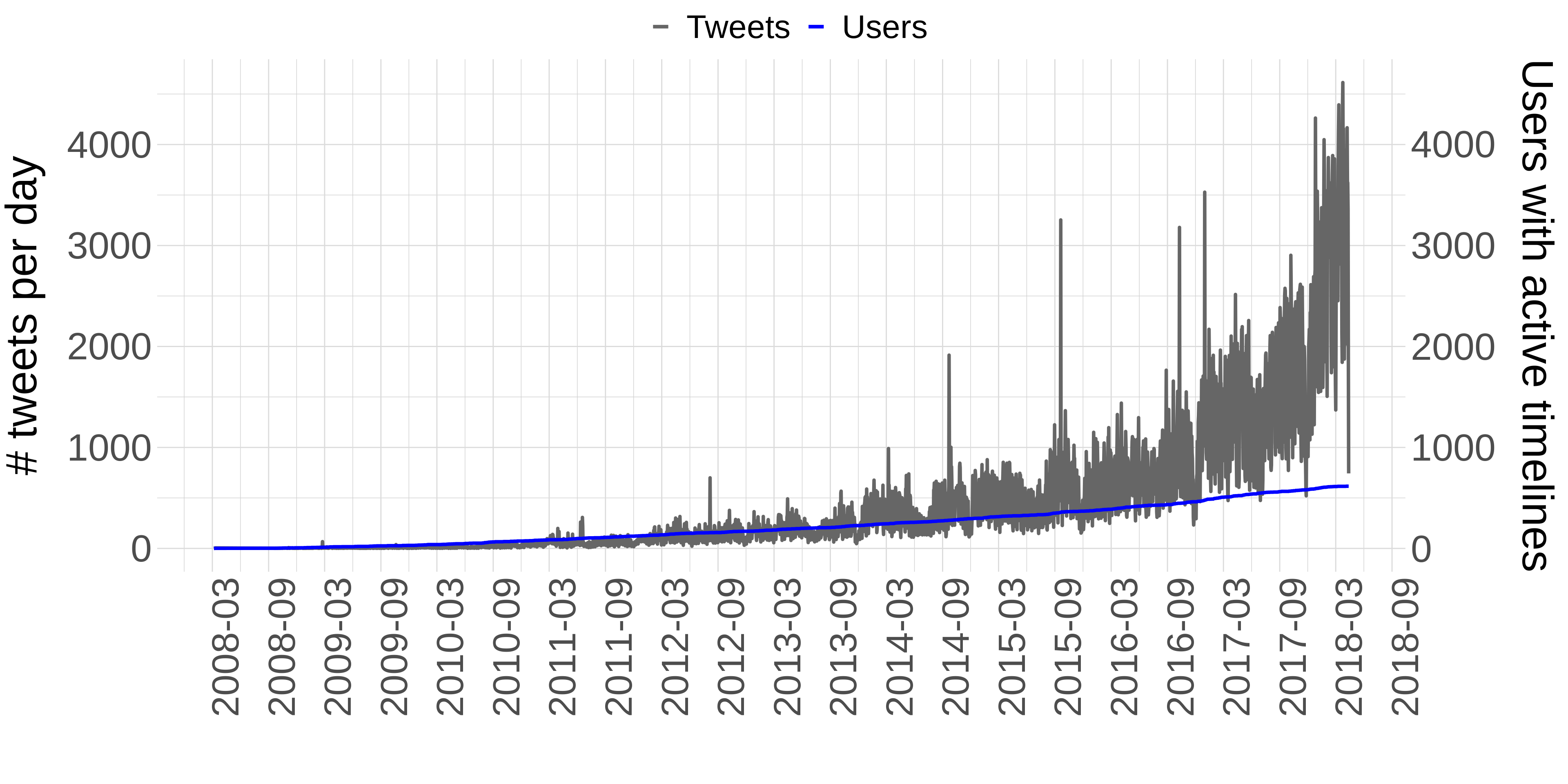}}
\hfill
\subfloat[Brasil
\label{fig_appendix:tweeting_volume_BrazilianJournalists}]
{\includegraphics[width=0.3\textwidth]
{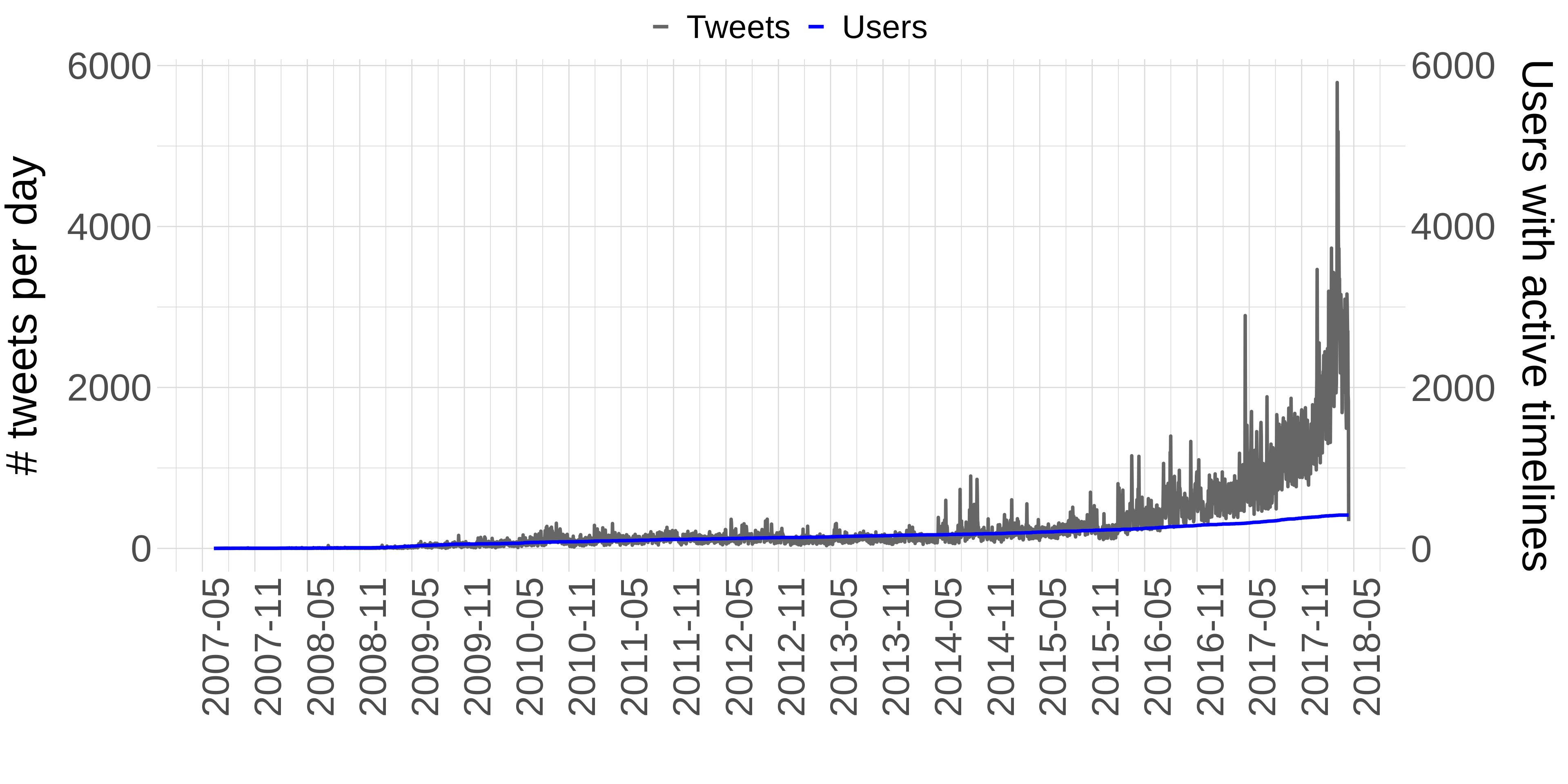}}
\hfill
\subfloat[Japan
\label{fig_appendix:tweeting_volume_JapaneseJournalists}]
{\includegraphics[width=0.3\textwidth]
{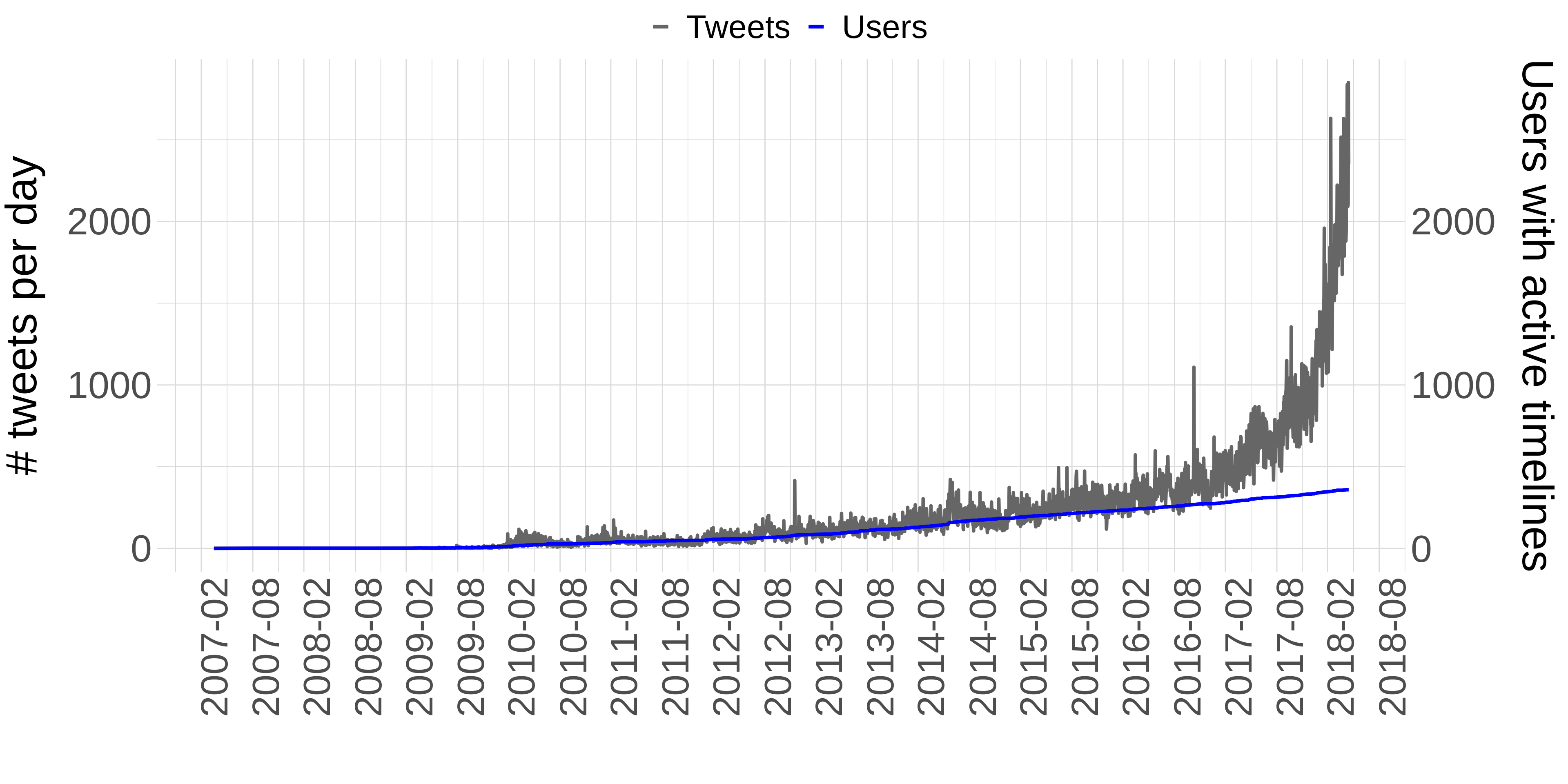}}
\hfill
\subfloat[Turkey
\label{fig_appendix:tweeting_volume_TrukishJournalists}]
{\includegraphics[width=0.3\textwidth]
{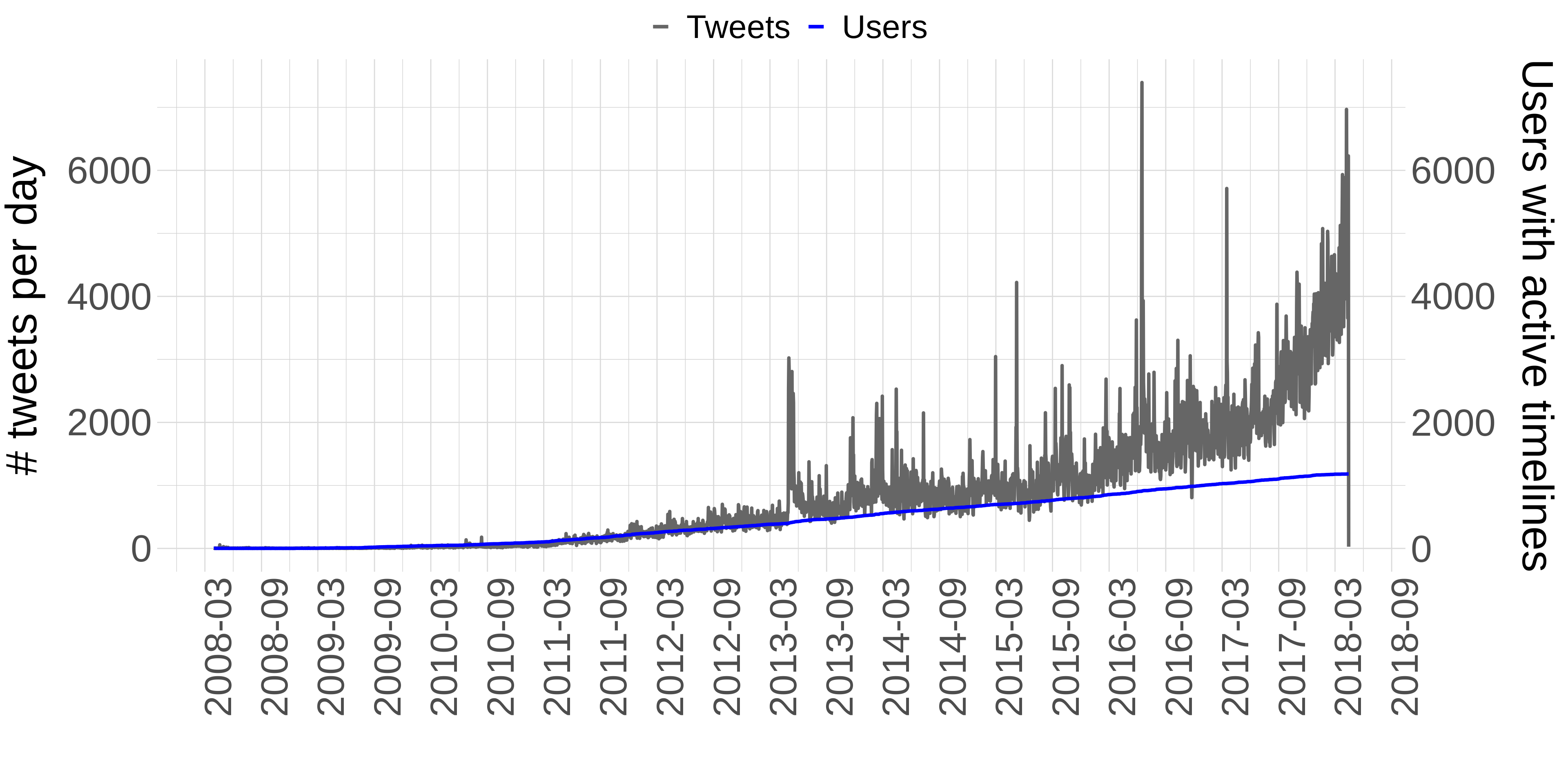}}
\hfill
\subfloat[UK
\label{fig_appendix:tweeting_volume_BritishJournalists}]
{\includegraphics[width=0.3\textwidth]
{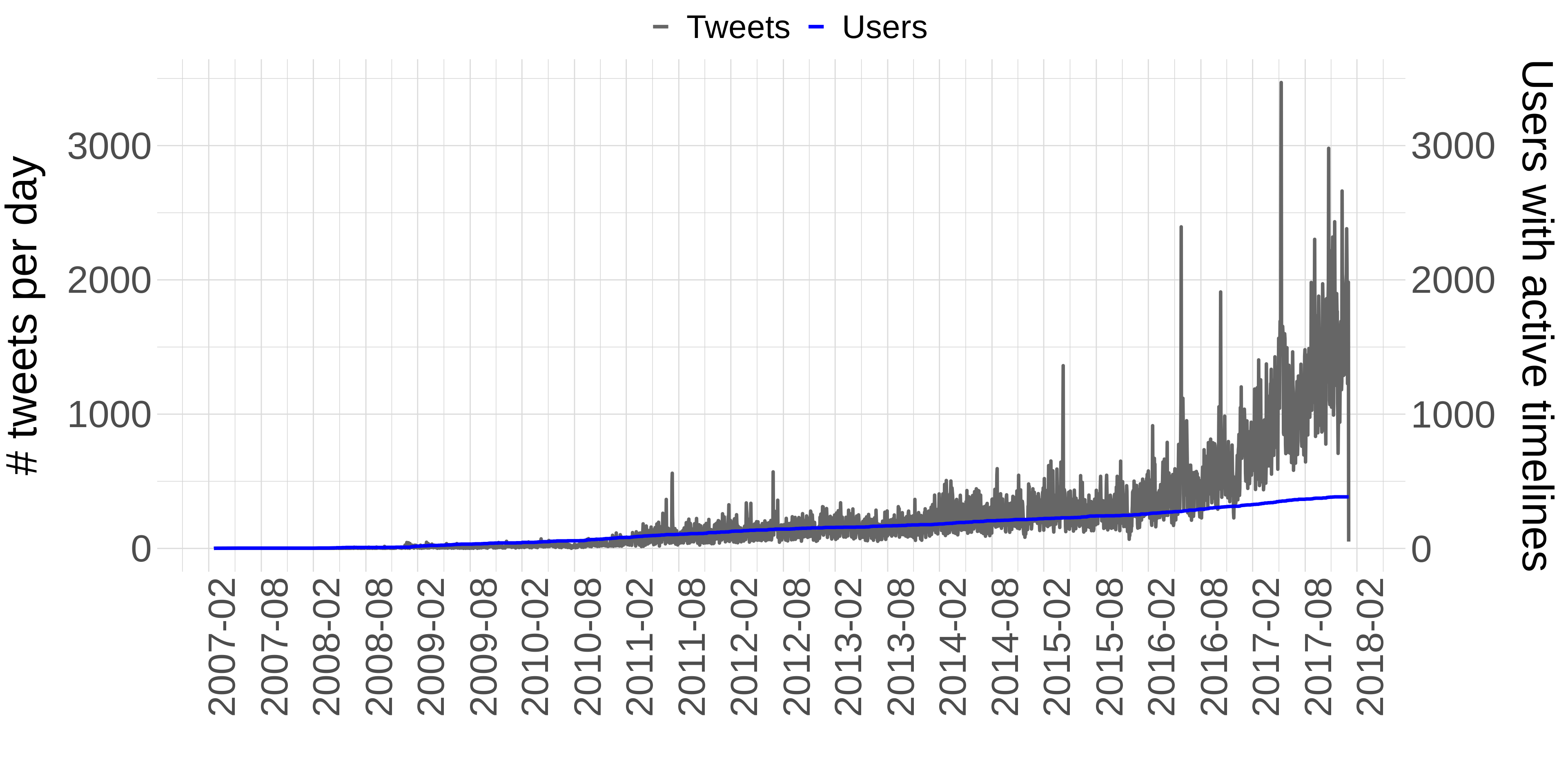}}
\hfill
\subfloat[Denmark
\label{fig_appendix:tweeting_volume_DanishJournalists}]
{\includegraphics[width=0.3\textwidth]
{./figures_new/tweeting_volume/DanishJournalists_tweet_vol_vs_active_user_before_filtering.png}}
\hfill
\subfloat[Finland
\label{fig_appendix:tweeting_volume_FinnishJournalists}]
{\includegraphics[width=0.3\textwidth]
{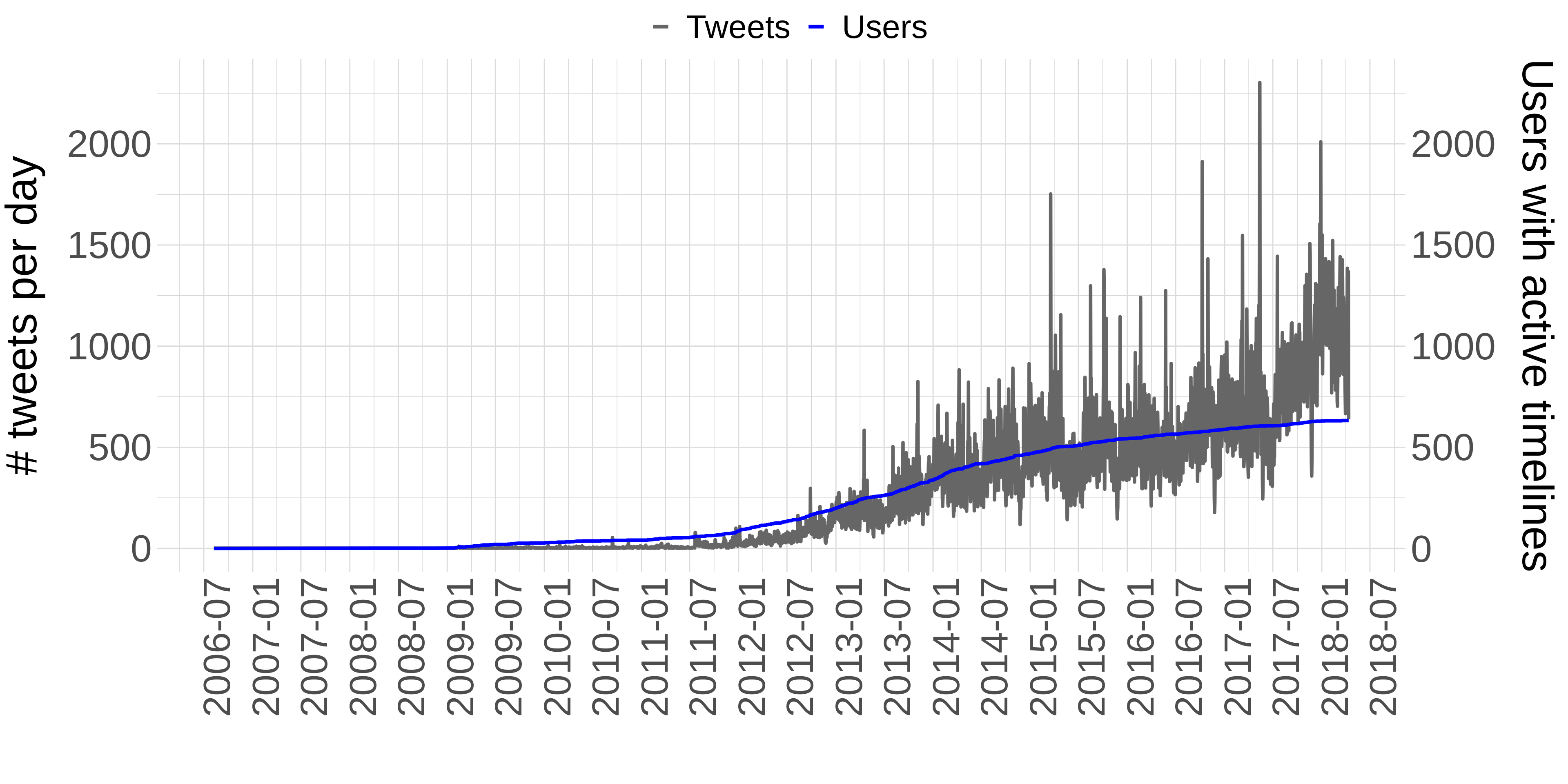}}
\hfill
\subfloat[Norway
\label{fig_appendix:tweeting_volume_NorwegianJournalists}]
{\includegraphics[width=0.3\textwidth]
{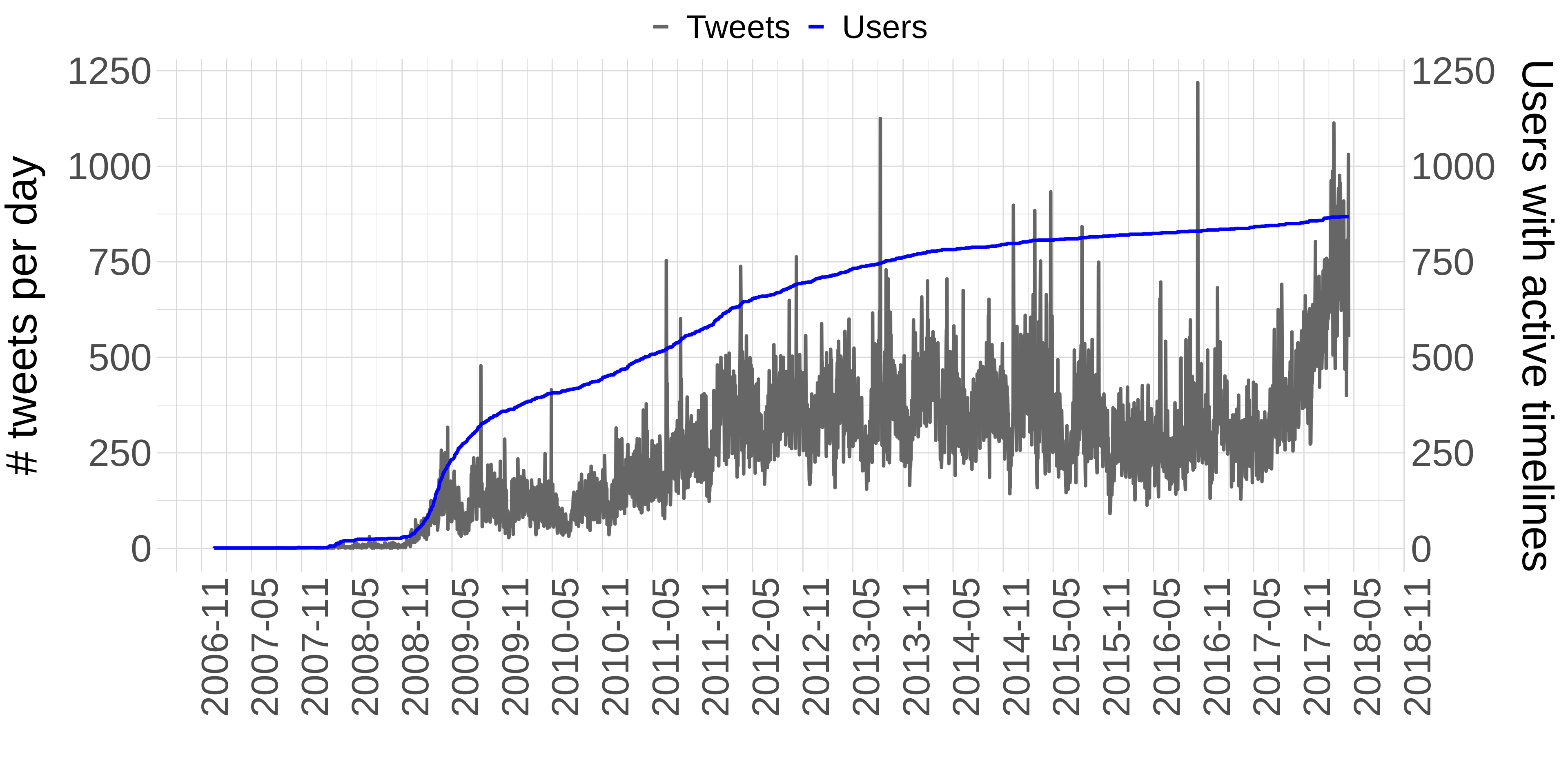}}
\hfill
\subfloat[Sweden
\label{fig_appendix:tweeting_volume_SwedishJournalists}]
{\includegraphics[width=0.3\textwidth]
{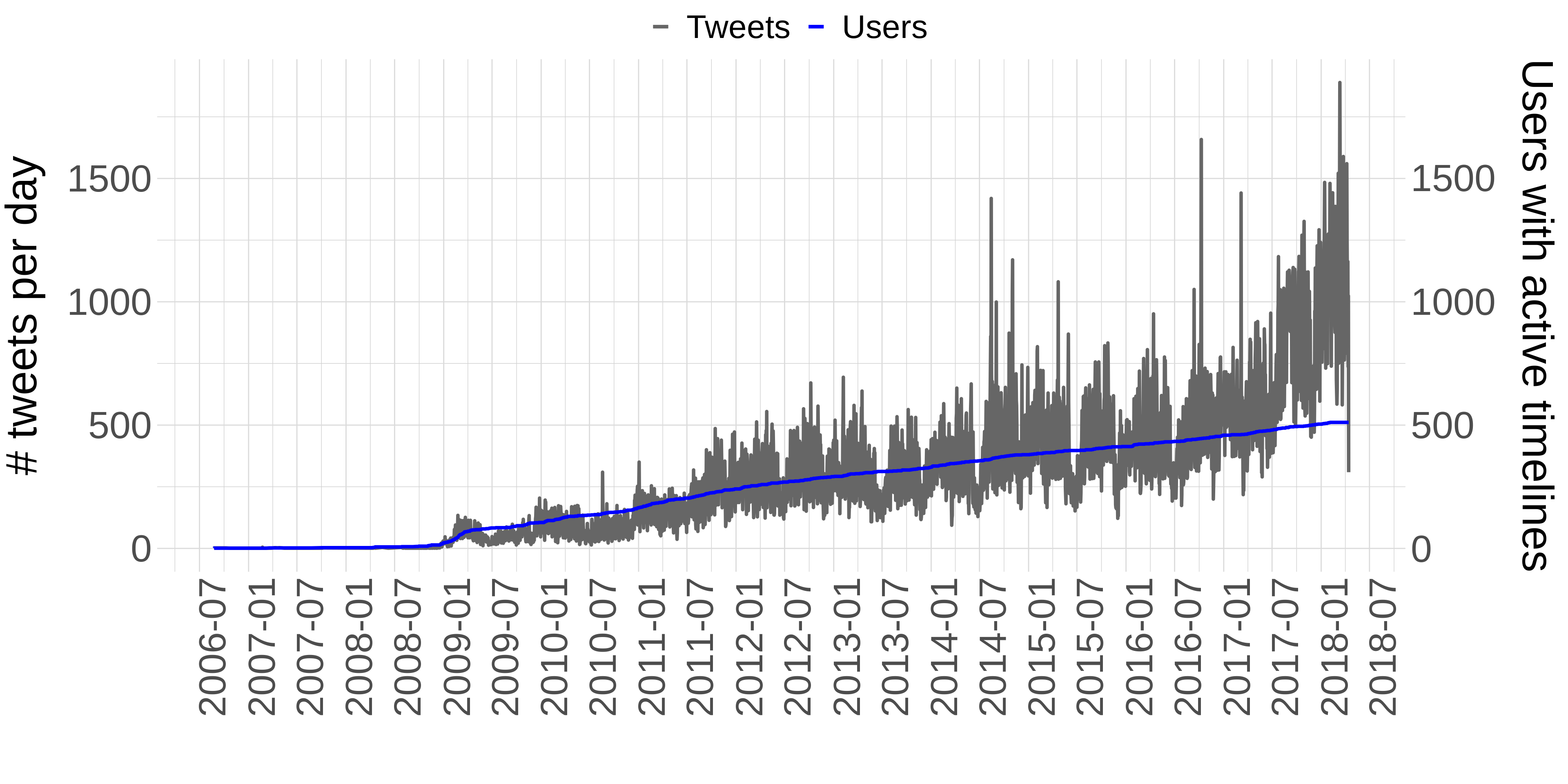}}
\hfill
\subfloat[Greece
\label{fig_appendix:tweeting_volume_GreekJournalists}]
{\includegraphics[width=0.3\textwidth]
{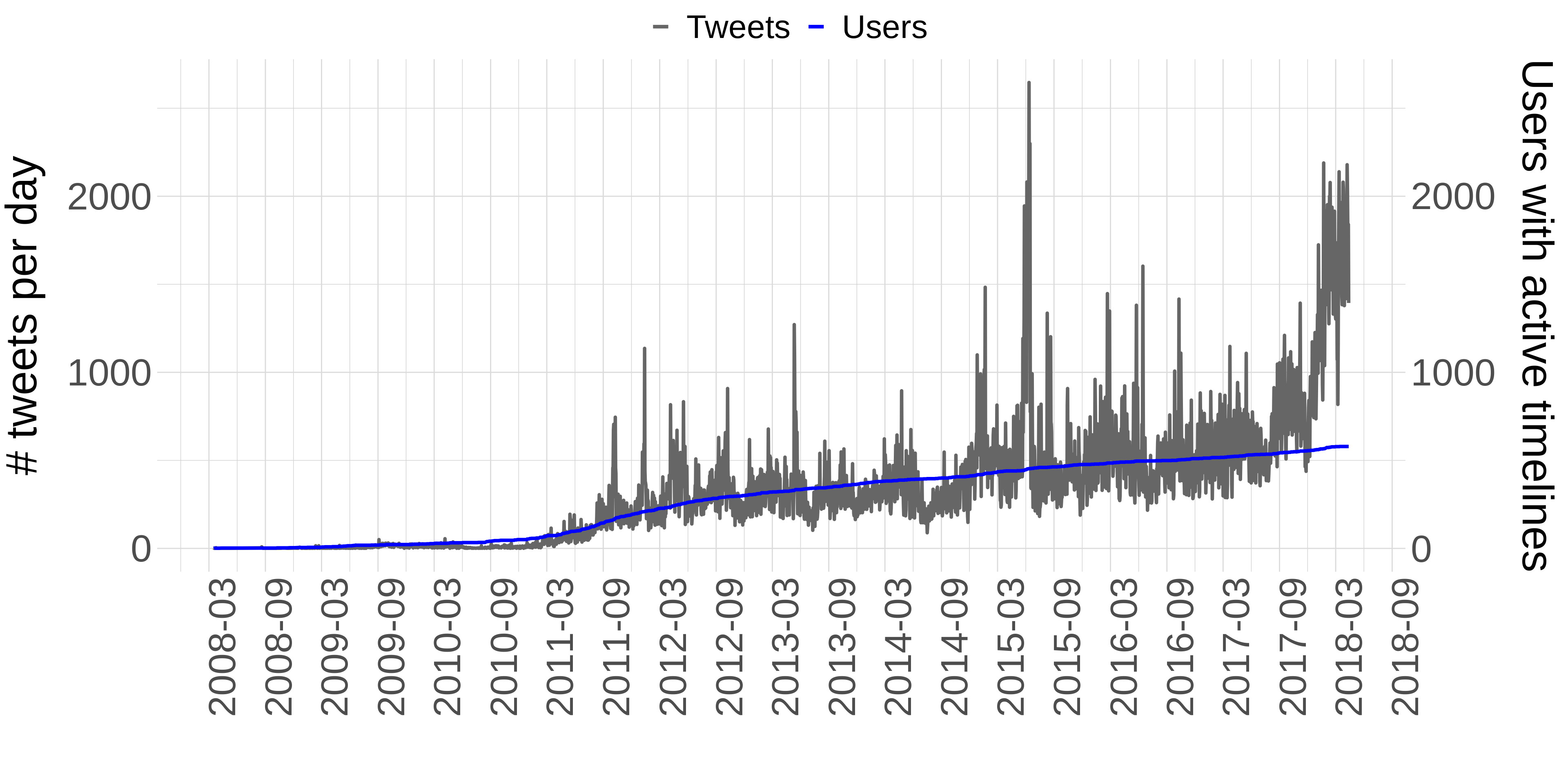}}
\hfill
\subfloat[Italy
\label{fig_appendix:tweeting_volume_ItalianJournalists}]
{\includegraphics[width=0.3\textwidth]
{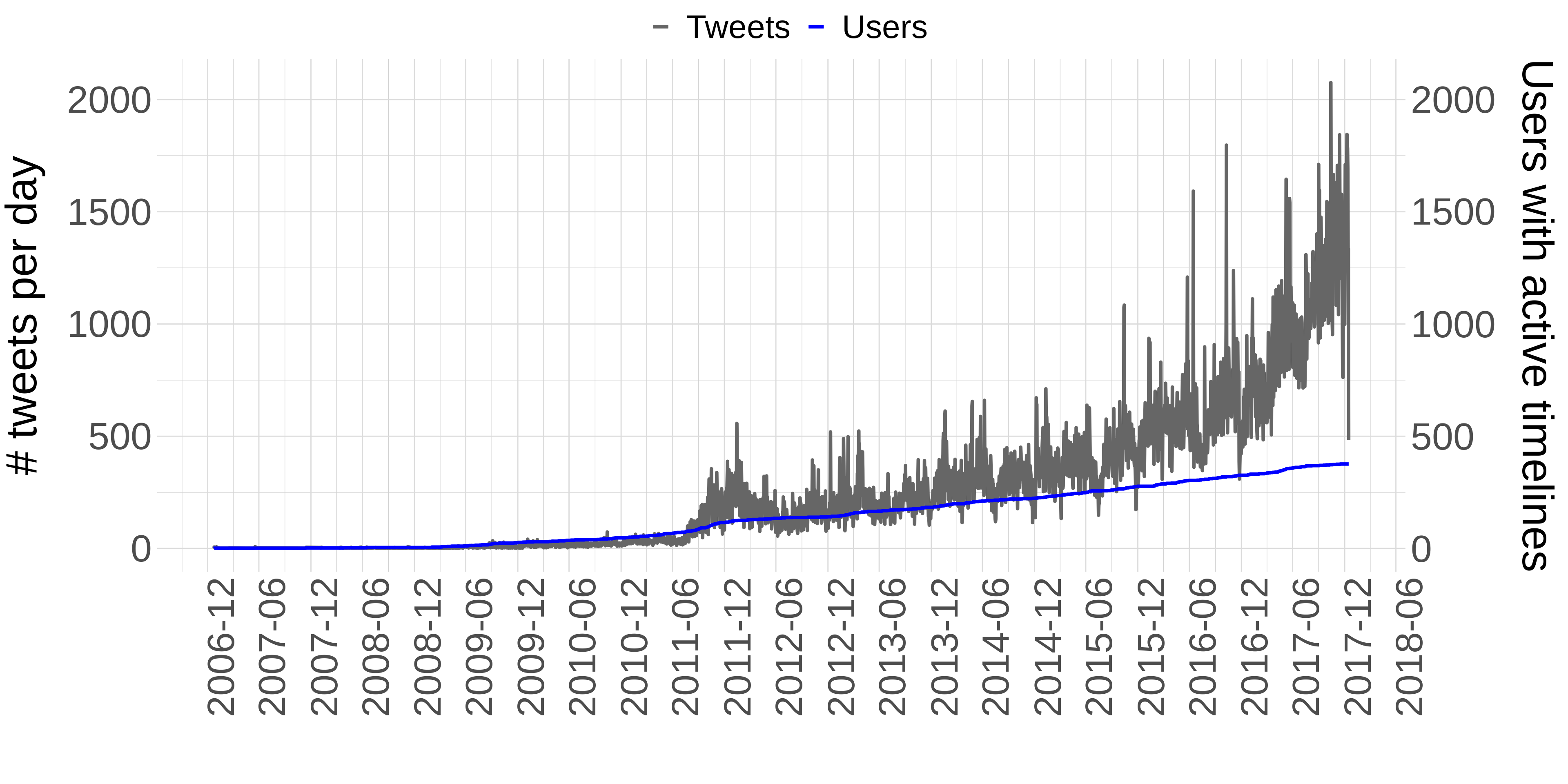}}
\hfill
\subfloat[Spain
\label{fig_appendix:tweeting_volume_SpanishJournalists}]
{\includegraphics[width=0.3\textwidth]
{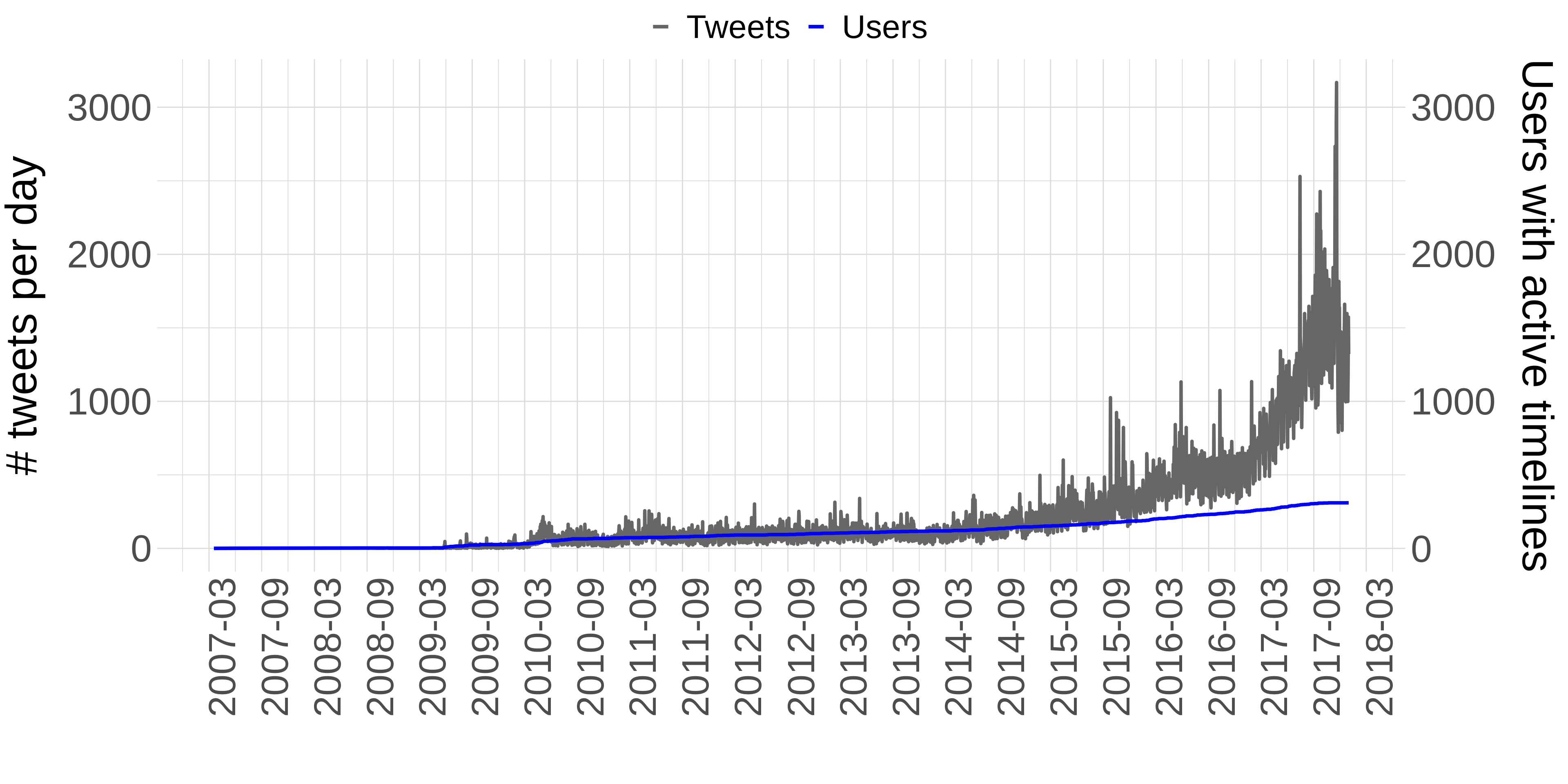}}
\hfill
\subfloat[France
\label{fig_appendix:tweeting_volume_FrenchJournalists}]
{\includegraphics[width=0.3\textwidth]
{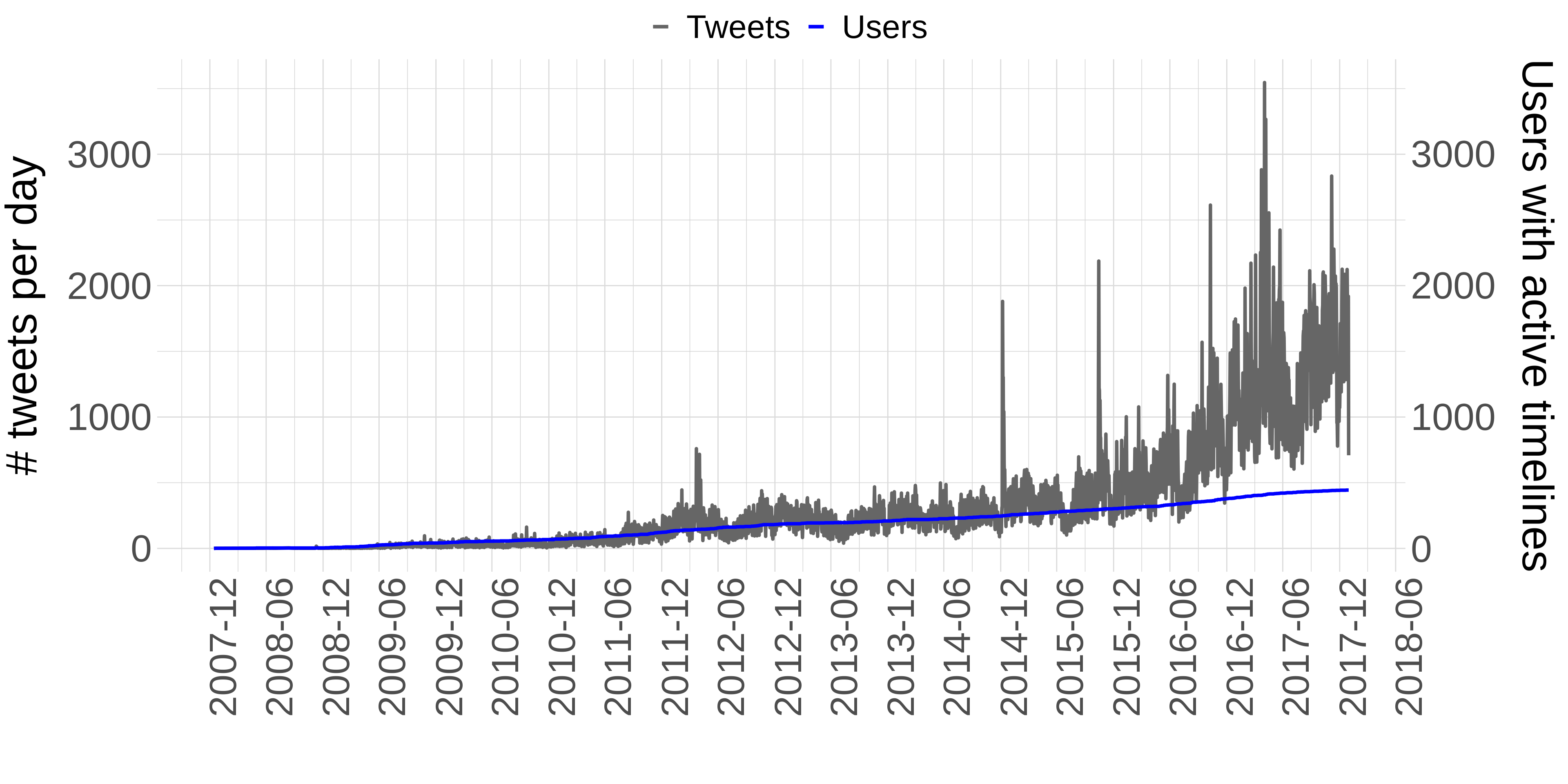}}
\hfill
\subfloat[Germany
\label{fig_appendix:tweeting_volume_GermanJournalists}]
{\includegraphics[width=0.3\textwidth]
{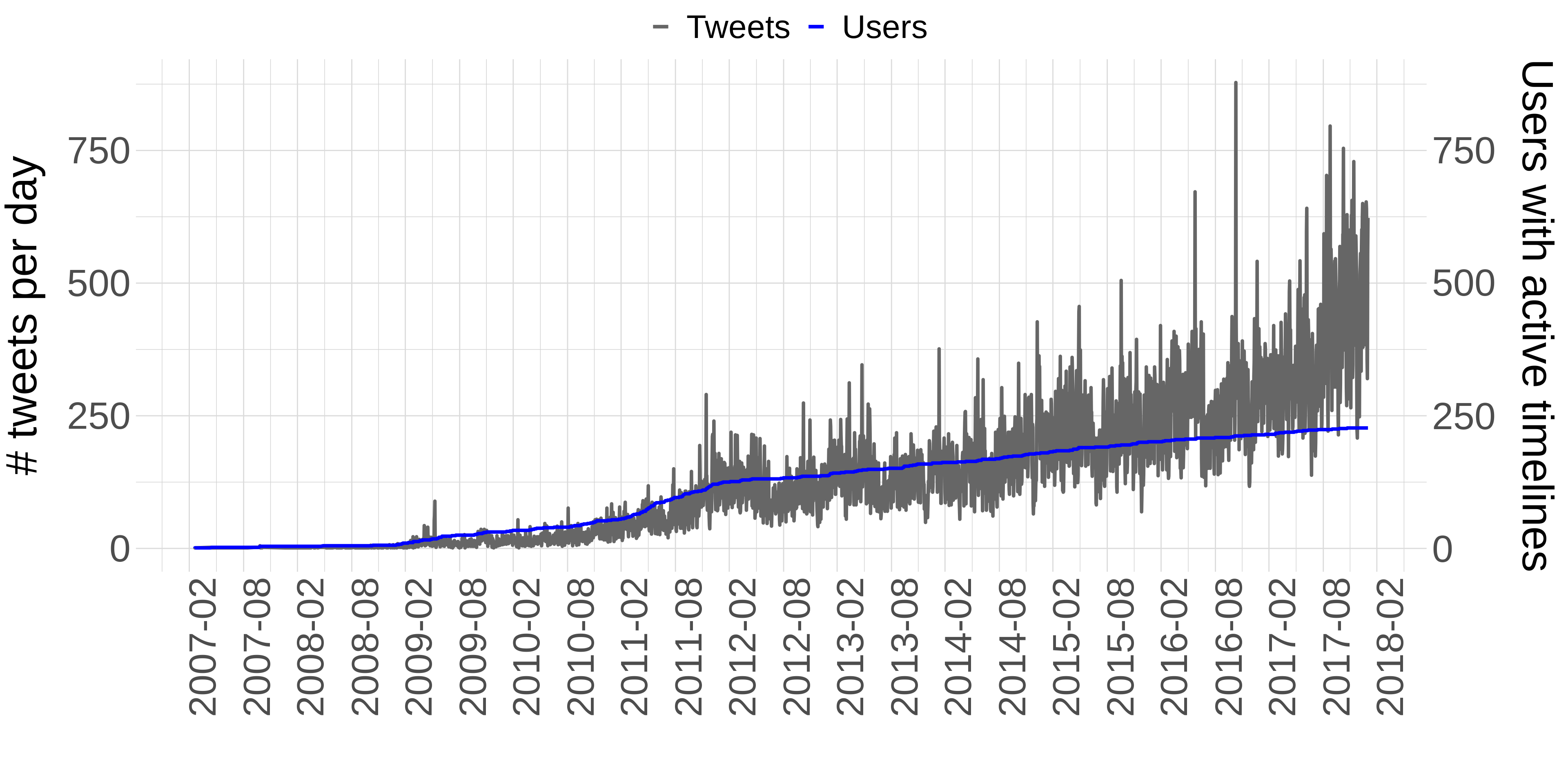}}
\hfill
\subfloat[Netherland
\label{fig_appendix:tweeting_volume_NetherlanderJournalists}]
{\includegraphics[width=0.3\textwidth]
{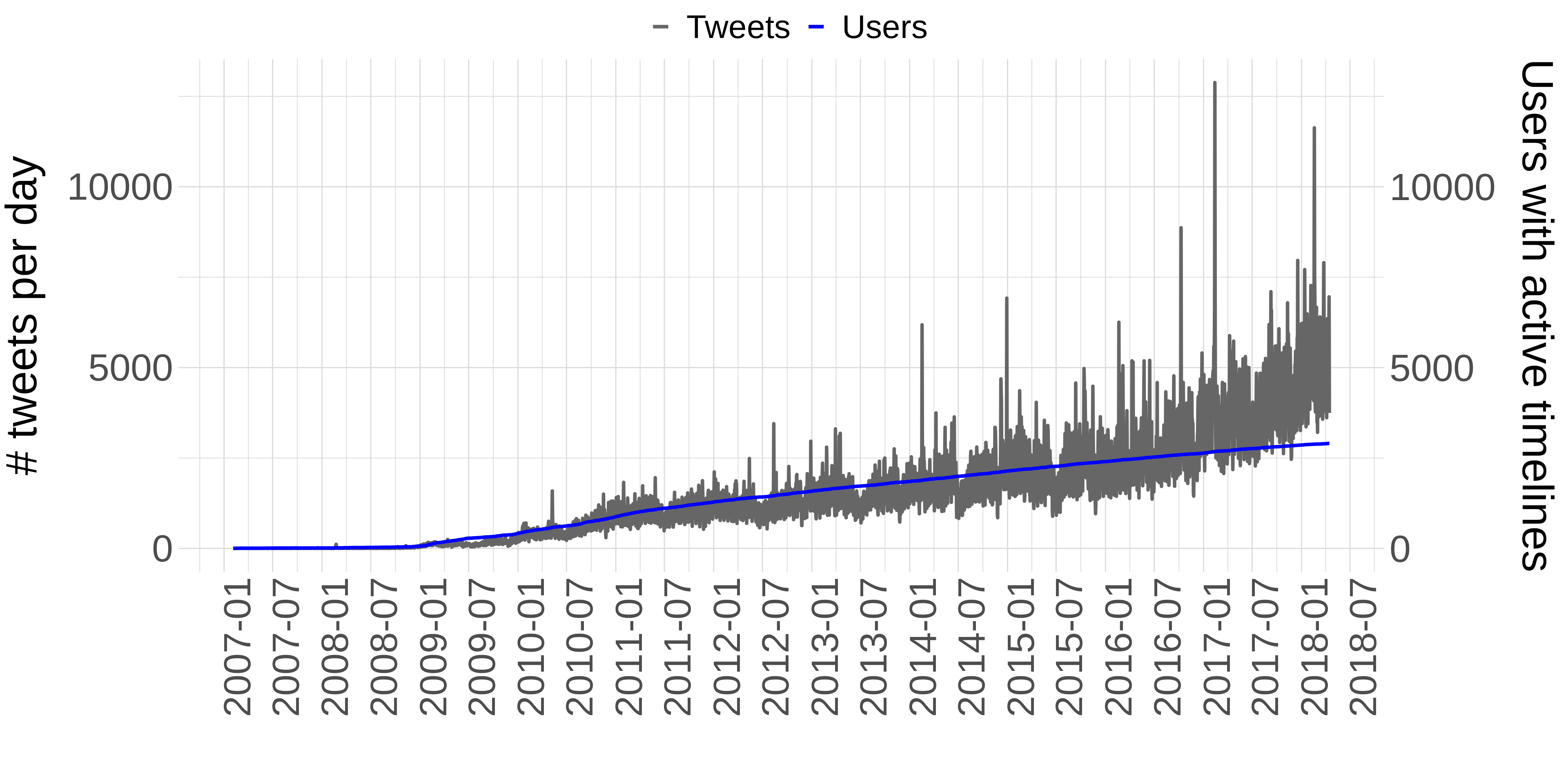}}
\hspace{1pt}
 \subfloat[Australia
\label{fig_appendix:tweeting_volume_AustralianJournalists}]
{\includegraphics[width=0.3\textwidth]
{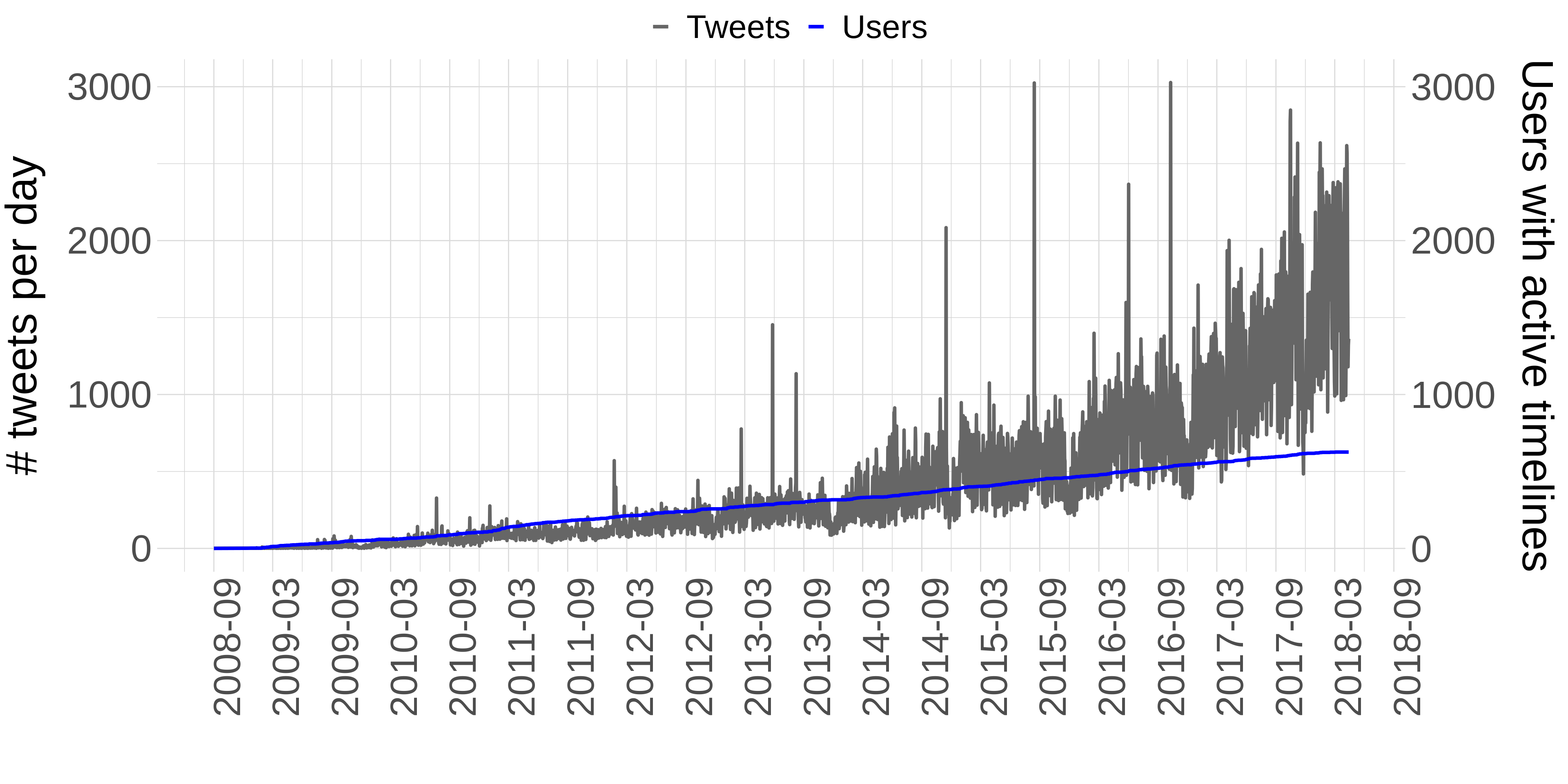}}
\end{center}
\end{adjustbox}
\caption{Tweeting volume and active users over time}
\label{fig_appendix:tweet_volume_evolution_all}
\end{figure}


\subsection{Labelling users as journalists}
\label{appendix_labellingusers}

Since Twitter lists are used-generated, they may contain some spurious or heterogeneous entries. In our case, we have to make sure that they indeed contain journalists exclusively. 
First of all, we have manually filtered the lists to remove news company accounts. However, there may still be accounts that are not journalist even though they are sources of information or instead some accounts may belong to former journalists that have switched jobs in the meanwhile. In order to address this problem, we have applied an automated labelling methodology to decrease noise in the dataset. As anticipated in Section~\ref{sec:labellingusers}, we have evaluated three approaches: keyword matching, Google's Knowledge Graph (GKG), bot detection, and their combinations. We will explain them in detail in the next paragraphs. In order to have a reference ground truth, we have manually labelled 513 British and 329 Italian accounts in our dataset as journalists/non-journalists. Table~\ref{tab_appendix:manuallylabbelled} provides some summary statistics. As expected, given the selection of thematic lists, most of the accounts belong indeed to real journalists.

\begin{table}[ht]
\centering
\caption{Manually labelled journalists}
\footnotesize
\begin{adjustbox}{width=0.8\textwidth}
\begin{tabular}{@{}lllll@{}}  \toprule
dataset & journalist & not journalist & imbalance ratio \\ \midrule
UK & 434 & 79 & 0.846 \\ \midrule
Italy & 316 & 13 & 0.960 \\
\hline
All & 750 & 92 & 0.890 \\
\hline
\end{tabular}
\end{adjustbox}
\label{tab_appendix:manuallylabbelled}
\end{table}

\begin{figure}
\subfloat[UK
\label{fig_appendix:wordcloud_british}]
{\includegraphics[width=0.49\textwidth]
{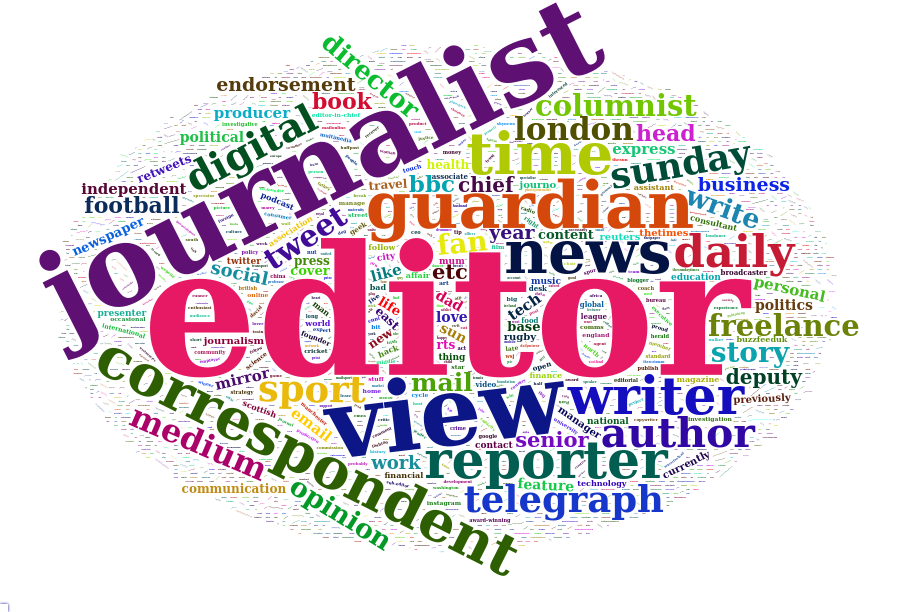}}
\hfill
\subfloat[Italy
\label{fig_appendix:wordcloud_italian}]
{\includegraphics[width=0.49\textwidth]
{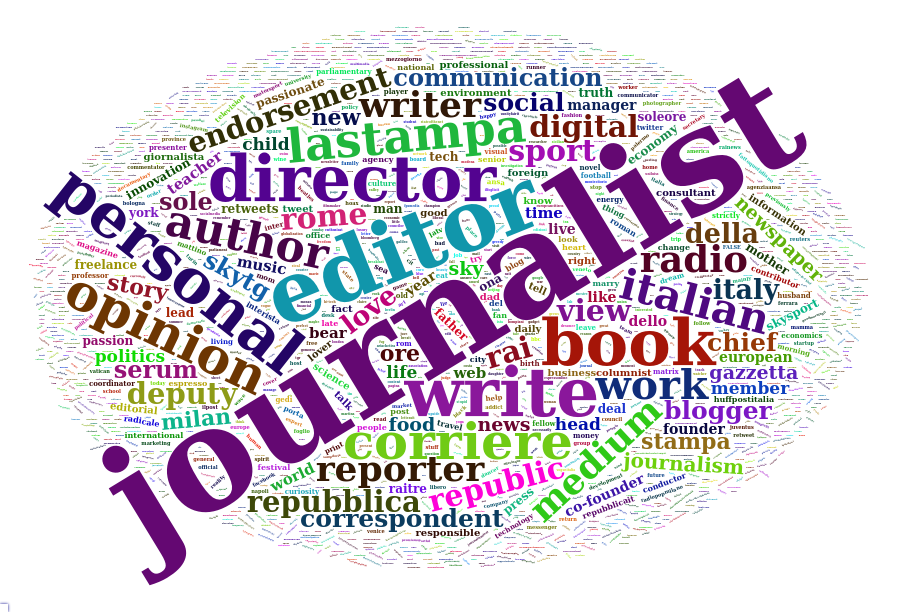}}
\hfill
\caption{Wordclouds of labelled dataset.}
\label{fig_appendix:wordclouds}
\end{figure}

\paragraph{Keyword matching} The Twitter API can be queried to extract the user profiles. We are particularly interested in the Twitter bios of users, the brief description about themselves that users can provide and that may contain information such as their profession, which football team they support, which video games they play, and so on.
Considering that the journalists use Twitter as a platform for personal branding and to promote content from their news websites, journalists are very likely to disclose their profession in their bio. Therefore, one of the approaches that we can use for automatizing the labelling process is to extract the keywords from the Twitter bios of the users in our dataset, then matching them to well-known journalism-related keywords. The main obstacle we had to face is the heterogeneity of languages in our dataset. In order to solve this problem, we decide to translate all the bios in English (which is the most convenient language for NLP, given the availability of a large amount of software tools). In order to facilitate the translation with standard offline tools (cloud-based solutions offering automatic language detection offer a limited number of free translations), we first detect the original language in which the bios is written. We used two offline Python libraries: \textit{pyplot}\footnote{\url{https://polyglot.readthedocs.io/en/latest/index.html}} and \textit{langdetect}\footnote{\url{https://pypi.org/project/langdetect/}}.
If both libraries detect the same language, the detected language is used to translate the bio into English.
Otherwise, the language is accepted as undetectable. After assigning the languages, English descriptions are directly used in keyword extraction while the others are translated into English first. The descriptions of whose language was undetectable or not English are translated into English by using a  Python library \textit{googletrans}\footnote{\url{https://pypi.org/project/googletrans/}} which uses \textit{Google Translate Ajax API} with the feature to detect the source language automatically. 
After translating the bios, their text is tokenized and stop words are removed by using the library \textit{spacy}\footnote{\url{https://spacy.io/usage/linguistic-features}}. As a final step, lemmatization is applied on the remaining words to find related lexical roots/dictionary forms of the words by using the \textit{WordNet}\footnote{\url{https://wordnet.princeton.edu/}} lexical database for English via  thePython library \textit{nltk}\footnote{\url{https://www.nltk.org/}}. 
The obtained lemmas are then compared against the lemmas of keywords that often appear in bios of real journalists. We took the manually labelled set as reference. Since there are approximately 2000 keywords for each manually labelled dataset, here we visualise it with word clouds instead of tables. The extracted keywords are shown in Figure~\ref{fig_appendix:wordclouds}. Based on this result, the selected keywords are: \textit{critic}, \textit{columnist}, \textit{correspondent}, \textit{editor}, \textit{journalist} and \textit{reporter}, as well as the bigrams \textit{staff writer} and \textit{senior writer}.

\paragraph{GKG} The term \textit{knowledge graph} was introduced in 2012 by Google with the motto ``things, not strings''\footnote{\url{https://googleblog.blogspot.com/2012/05/introducing-knowledge-graph-things-not.html}}. It was sold as the first step in the transition from information graph to knowledge graph, where the real-word objects are not just strings but entities with inter-relationships. The power of this representation can be leveraged by searching for real-world objects via the Google Knowledge Graph Search API\footnote{\url{https://developers.google.com/knowledge-graph}}. The API returns the top $n$ entities related to the keyword search as result. These entities could be different type of entities such as persons, sport teams, cities, and so on. For labelling the journalist using GKG, we search for users' Twitter display name and screen name. Among the returned entities, the first person entity is accepted as the target user's identity, and its profession is extracted from the entity details. If there is no person as an entity within top 10 result, the profession of the user is marked as not-journalist.

\paragraph{Bot elimination} A social bot is an automated social media account that interacts with other users of the social platform according to predefined algorithms~\cite{Allen2016}. The literature about bot detection is extensive. In this paper, we use bot detection to clean our dataset from outliers and non-human accounts. We used a bot detection tool named \emph{botometer} (formerly BotOrNot)\footnote{\url{https://botometer.iuni.iu.edu/#!/}}. Botometer is a publicly-available service to label Twitter accounts as either bot or not by using more than one thousand features~\cite{Allen2016}. Note that, since the language features are defined only for English texts, here we use them only for the accounts tweeting in English. Botometer returns two scores: the first one is the bot score (which is the result of the prediction algorithm) and the second one is the CAP (Complete Automation Probability) score,  which is calculated based on Bayes' theorem by considering an estimate of the overall prevalence of bots. Both scores are in the interval $[0, 1]$. In the scope of this study, similarly to the related literature, we mark an account as bot, therefore as non-journalist, if both the bot score and the CAP score are greater than $0.5$.

\begin{table}[p]
\centering
\caption{Prediction statistics.}
\footnotesize
\resizebox{!}{.34\paperheight}{
\small
\begin{tabular}{@{}llllllllll@{}} \toprule
\textbf{dataset} & \textbf{method name} & \textbf{TP} & \textbf{FP} & \textbf{TN} & \textbf{FN} & \textbf{precision} & \textbf{recall} & \textbf{accuracy} & \textbf{F1}\\ \midrule
UK &  $b$  & 428 & 77 & 2 & 6 & 0.84 & 0.98 & 0.83 & 0.91\\ \cmidrule{2-10}
 &  $g$  & 59 & 3 & 76 & 375 & 0.95 & 0.13 & 0.26 & 0.23\\ \cmidrule{2-10}
 &  $k$  & 342 & 25 & 54 & 92 & 0.93 & 0.78 & 0.77 & 0.85\\ \cmidrule{2-10}
 &  $b \;\&\; g$  & 59 & 3 & 76 & 375 & 0.95 & 0.13 & 0.26 & 0.23\\ \cmidrule{2-10}
 &  $b \;\&\; k$  & 337 & 24 & 55 & 97 & 0.93 & 0.77 & 0.76 & 0.84\\ \cmidrule{2-10}
 &  $g \;\&\; k$  & 44 & 1 & 78 & 390 & 0.97 & 0.1 & 0.23 & 0.18\\ \cmidrule{2-10}
 &  $b \;|\; g$  & 428 & 77 & 2 & 6 & 0.84 & 0.98 & 0.83 & 0.91\\ \cmidrule{2-10}
 &  $b \;|\; k$  & 433 & 78 & 1 & 1 & 0.84 & 0.99 & 0.84 & 0.91\\ \cmidrule{2-10}
 &  $g \;|\; k$  & 357 & 27 & 52 & 77 & 0.92 & 0.82 & 0.79 & 0.87\\ \cmidrule{2-10}
 &  $b \;\&\; (g \;|\; k)$  & 352 & 26 & 53 & 82 & 0.93 & 0.81 & 0.78 & 0.86\\ \cmidrule{2-10}
 &  $k \;\&\; (b \;|\; g)$  & 337 & 24 & 55 & 97 & 0.93 & 0.77 & 0.76 & 0.84\\ \cmidrule{2-10}
 &  $g \;\&\; (b \;|\; k)$  & 59 & 3 & 76 & 375 & 0.95 & 0.13 & 0.26 & 0.23\\ \cmidrule{2-10}
 &  $b \;|\; (g \;\&\; k)$  & 428 & 77 & 2 & 6 & 0.84 & 0.98 & 0.83 & 0.91\\ \cmidrule{2-10}
 &  $g \;|\; (b \;\&\; k)$  & 352 & 26 & 53 & 82 & 0.93 & 0.81 & 0.78 & 0.86\\ \cmidrule{2-10}
 &  $k \;|\; (b \;\&\; g)$  & 357 & 27 & 52 & 77 & 0.92 & 0.82 & 0.79 & 0.87\\ \cmidrule{2-10}
 &  $all \;\&\;$  & 44 & 1 & 78 & 390 & 0.97 & 0.1 & 0.23 & 0.18\\ \cmidrule{2-10}
 &  $all \;|\;$  & 433 & 78 & 1 & 1 & 0.84 & 0.99 & 0.84 & 0.91\\ \midrule
Italy &  $b$  & 316 & 13 & 0 & 0 & 0.96 & 1 & 0.96 & 0.97\\ \cmidrule{2-10}
 &  $g$  & 70 & 1 & 12 & 246 & 0.98 & 0.22 & 0.24 & 0.36\\ \cmidrule{2-10}
 &  $k$  & 215 & 7 & 6 & 101 & 0.96 & 0.68 & 0.67 & 0.79\\ \cmidrule{2-10}
 &  $b \;\&\; g$  & 70 & 1 & 12 & 246 & 0.98 & 0.22 & 0.24 & 0.36\\ \cmidrule{2-10}
 &  $b \;\&\; k$  & 215 & 7 & 6 & 101 & 0.96 & 0.68 & 0.67 & 0.79\\ \cmidrule{2-10}
 &  $g \;\&\; k$  & 37 & 1 & 12 & 279 & 0.97 & 0.11 & 0.14 & 0.2\\ \cmidrule{2-10}
 &  $b \;|\; g$  & 316 & 13 & 0 & 0 & 0.96 & 1 & 0.96 & 0.97\\ \cmidrule{2-10}
 &  $b \;|\; k$  & 316 & 13 & 0 & 0 & 0.96 & 1 & 0.96 & 0.97\\ \cmidrule{2-10}
 &  $g \;|\; k$  & 248 & 7 & 6 & 68 & 0.97 & 0.78 & 0.77 & 0.86\\ \cmidrule{2-10}
 &  $b \;\&\; (g \;|\; k)$  & 248 & 7 & 6 & 68 & 0.97 & 0.78 & 0.77 & 0.86\\ \cmidrule{2-10}
 &  $k \;\&\; (b \;|\; g)$  & 215 & 7 & 6 & 101 & 0.96 & 0.68 & 0.67 & 0.79\\ \cmidrule{2-10}
 &  $g \;\&\; (b \;|\; k)$  & 70 & 1 & 12 & 246 & 0.98 & 0.22 & 0.24 & 0.36\\ \cmidrule{2-10}
 &  $b \;|\; (g \;\&\; k)$  & 316 & 13 & 0 & 0 & 0.96 & 1 & 0.96 & 0.97\\ \cmidrule{2-10}
 &  $g \;|\; (b \;\&\; k)$  & 248 & 7 & 6 & 68 & 0.97 & 0.78 & 0.77 & 0.86\\ \cmidrule{2-10}
 &  $k \;|\; (b \;\&\; g)$  & 248 & 7 & 6 & 68 & 0.97 & 0.78 & 0.77 & 0.86\\ \cmidrule{2-10}
 &  $all \;\&\;$  & 37 & 1 & 12 & 279 & 0.97 & 0.11 & 0.14 & 0.2\\ \cmidrule{2-10}
 &  $all \;|\;$  & 316 & 13 & 0 & 0 & 0.96 & 1 & 0.96 & 0.97\\ \midrule
All &  $b$  & 744 & 90 & 2 & 6 & 0.89 & 0.99 & 0.88 & 0.93\\ \cmidrule{2-10}
 &  $g$  & 129 & 4 & 88 & 621 & 0.96 & 0.17 & 0.25 & 0.29\\ \cmidrule{2-10}
 &  $k$  & 557 & 32 & 60 & 193 & 0.94 & 0.74 & 0.73 & 0.83\\ \cmidrule{2-10}
 &  $b \;\&\; g$  & 129 & 4 & 88 & 621 & 0.96 & 0.17 & 0.25 & 0.29\\ \cmidrule{2-10}
 &  $b \;\&\; k$  & 552 & 31 & 61 & 198 & 0.94 & 0.73 & 0.72 & 0.82\\ \cmidrule{2-10}
 &  $g \;\&\; k$  & 81 & 2 & 90 & 669 & 0.97 & 0.1 & 0.2 & 0.19\\ \cmidrule{2-10}
 &  $b \;|\; g$  & 744 & 90 & 2 & 6 & 0.89 & 0.99 & 0.88 & 0.93\\ \cmidrule{2-10}
 &  $b \;|\; k$  & 749 & 91 & 1 & 1 & 0.89 & 0.99 & 0.89 & 0.94\\ \cmidrule{2-10}
 &  $g \;|\; k$  & 605 & 34 & 58 & 145 & 0.94 & 0.8 & 0.78 & 0.87\\ \cmidrule{2-10}
 &  $b \;\&\; (g \;|\; k)$  & 600 & 33 & 59 & 150 & 0.94 & 0.8 & 0.78 & 0.86\\ \cmidrule{2-10}
 &  $k \;\&\; (b \;|\; g)$  & 552 & 31 & 61 & 198 & 0.94 & 0.73 & 0.72 & 0.82\\ \cmidrule{2-10}
 &  $g \;\&\; (b \;|\; k)$  & 129 & 4 & 88 & 621 & 0.96 & 0.17 & 0.25 & 0.29\\ \cmidrule{2-10}
 &  $b \;|\; (g \;\&\; k)$  & 744 & 90 & 2 & 6 & 0.89 & 0.99 & 0.88 & 0.93\\ \cmidrule{2-10}
 &  $g \;|\; (b \;\&\; k)$  & 600 & 33 & 59 & 150 & 0.94 & 0.8 & 0.78 & 0.86\\ \cmidrule{2-10}
 &  $k \;|\; (b \;\&\; g)$  & 605 & 34 & 58 & 145 & 0.94 & 0.8 & 0.78 & 0.87\\ \cmidrule{2-10}
 &  $all \;\&\;$  & 81 & 2 & 90 & 669 & 0.97 & 0.1 & 0.2 & 0.19\\ \cmidrule{2-10}
 &  $all \;|\;$  & 749 & 91 & 1 & 1 & 0.89 & 0.99 & 0.89 & 0.94\\
\hline\end{tabular}
\label{tab_appendix:prediction_stats}
}
\end{table}

\begin{figure}[t]
\subfloat[UK
\label{fig_appendix:mcc_british}]
{\includegraphics[width=0.49\textwidth]
{./figures_new/mcc_BritishJournalists.png}}
\hfill
\subfloat[Italy
\label{fig_appendix:mcc_italian}]
{\includegraphics[width=0.49\textwidth]
{./figures_new/mcc_ItalianJournalists.png}}
\hfill
\centering
\subfloat[Both datasets together
\label{fig_appendix:mcc_all}]
{\includegraphics[width=0.49\textwidth]
{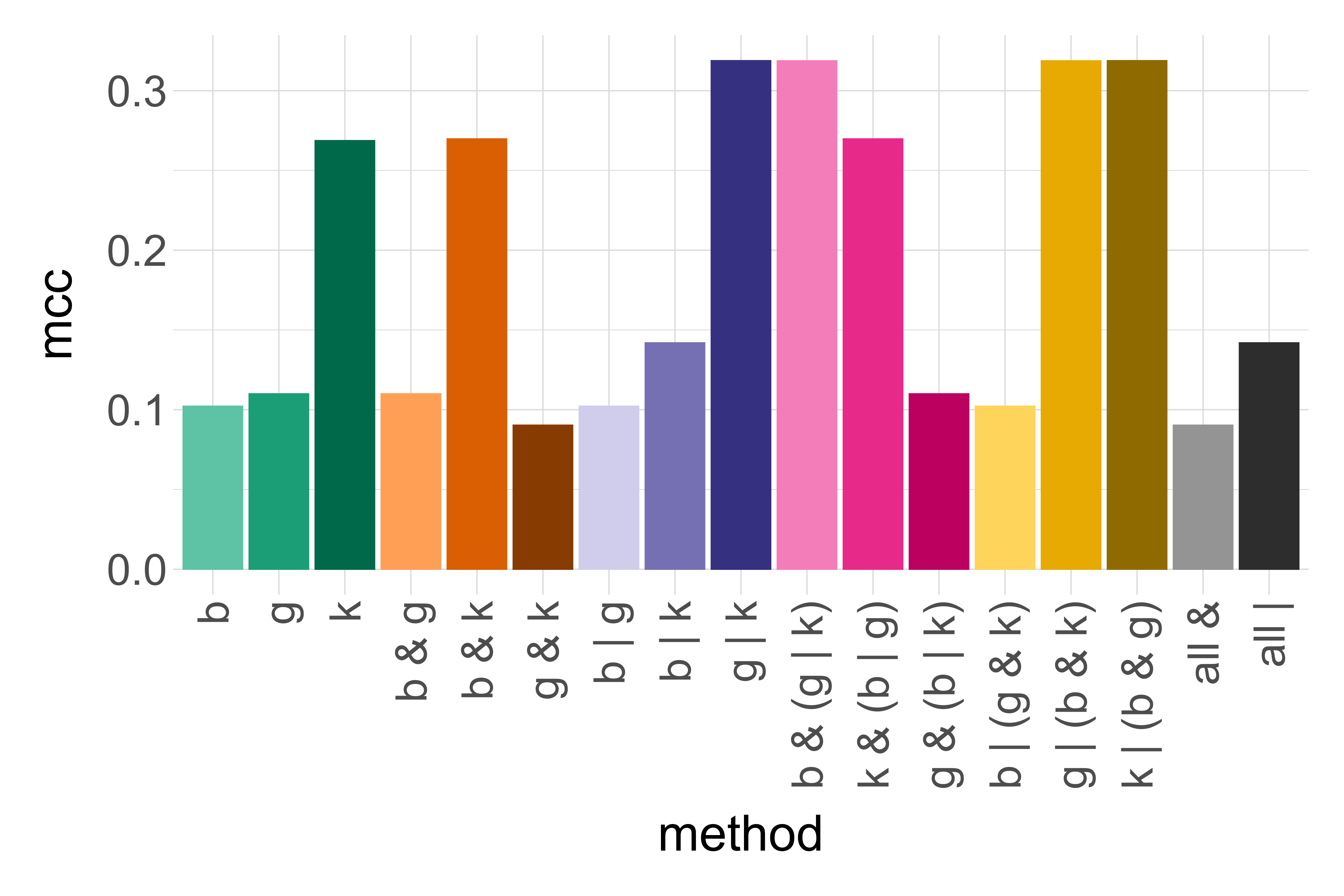}}
\hfill
\caption{MCC scores. In the plots, $g$ represents the classifier that leverages the Google Knowledge Graph, $k$ represents keyword extraction method, $b$ represents bot detection classifier. The symbol $\&$ represents the logical AND operation, symbol $|$ represents the logical OR. They are used to specify the test combinations of the three classifiers.}
\label{fig_appendix:mcc}
\end{figure}

As anticipated, we have tested the three methods above in isolation and by combining them. Their prediction performance on our labelled dataset is shown in Table~\ref{tab_appendix:prediction_stats}, using standard performance metrics. However, given that the two classes we are using for the classification are imbalanced and that we are interested in correctly classifying both the positive and negative class, we also provide the Matthews' Correlation Coefficient (MCC, \cite{Matthews1975}) in Figure~\ref{fig_appendix:mcc}.  The MCC is a Pearson product-moment correlation coefficient between actual and predicted classes that takes into account the balance ratios of the confusion matrix categories, as shown in Equation~\ref{eqn_appendix:MCC}. MCC values are in the interval $[-1, +1]$ where $-1$ is perfect misclassification and $+1$ is perfect classification. Since $g|k$ achieves the highest score by using less features (and it is faster since it doesn't use the bot detection API), we selected $g|k$ as the labelling method.  

\begin{equation}
\label{eqn_appendix:MCC}
MCC = \frac{TP \times TN - FP \times FN}{\sqrt{(TP + FP) \times (TP + FN) \times (TN + FP) \times (TN + FN)}}
\end{equation}

\clearpage

\subsection{Tweting activity - all datasets}
\label{appendix_twittingactivity}
\vspace{-10pt}

\begin{figure}[!h]
\begin{adjustbox}{minipage=\linewidth}
\begin{center}
\subfloat[USA
\label{fig_appendix:observedtimeline_AmericanJournalists}]
{\includegraphics[width=0.28\textwidth]
{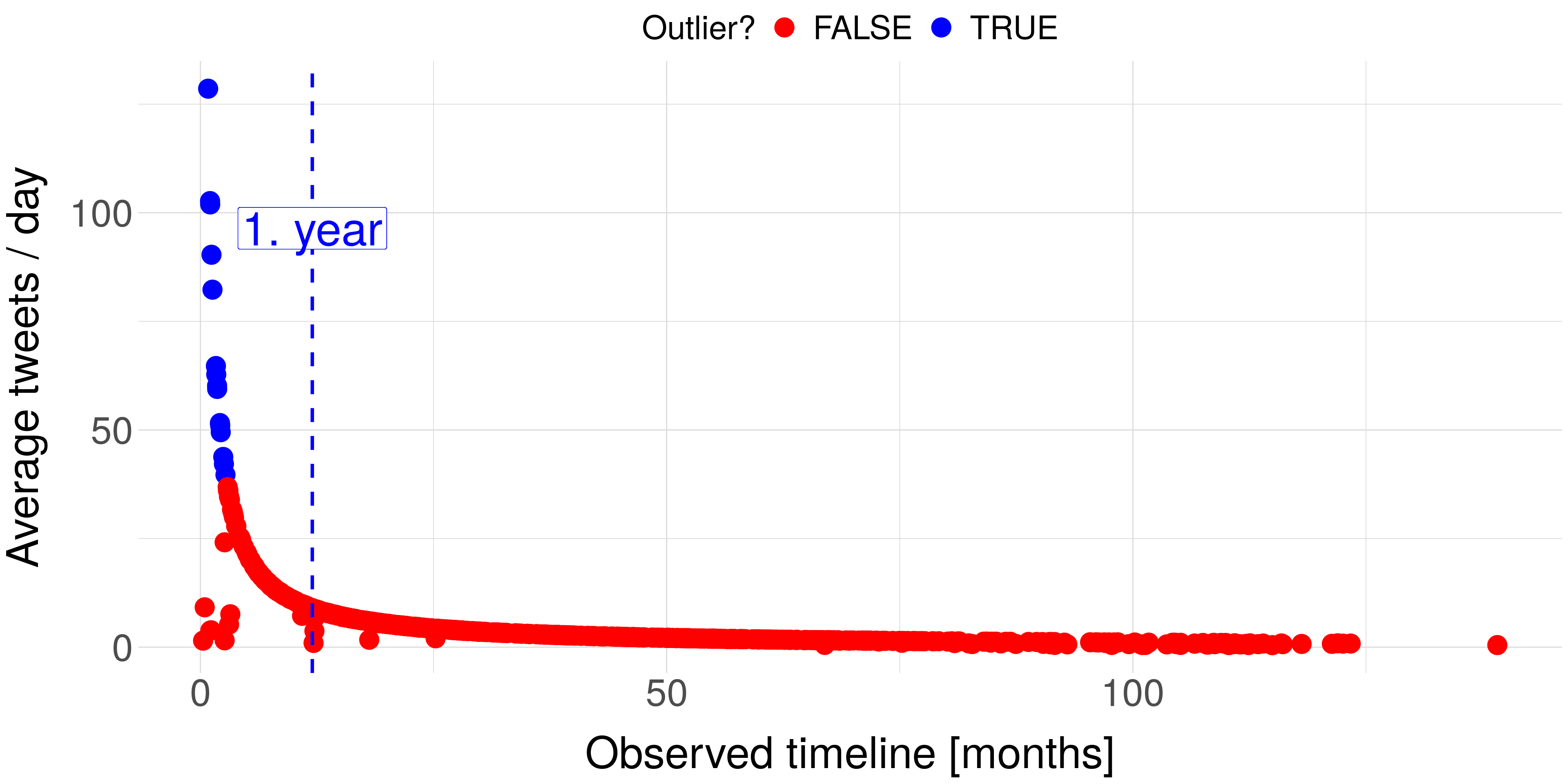}}
\hfill
\subfloat[Canada
\label{fig_appendix:observedtimeline_CanadianJournalists}]
{\includegraphics[width=0.28\textwidth]
{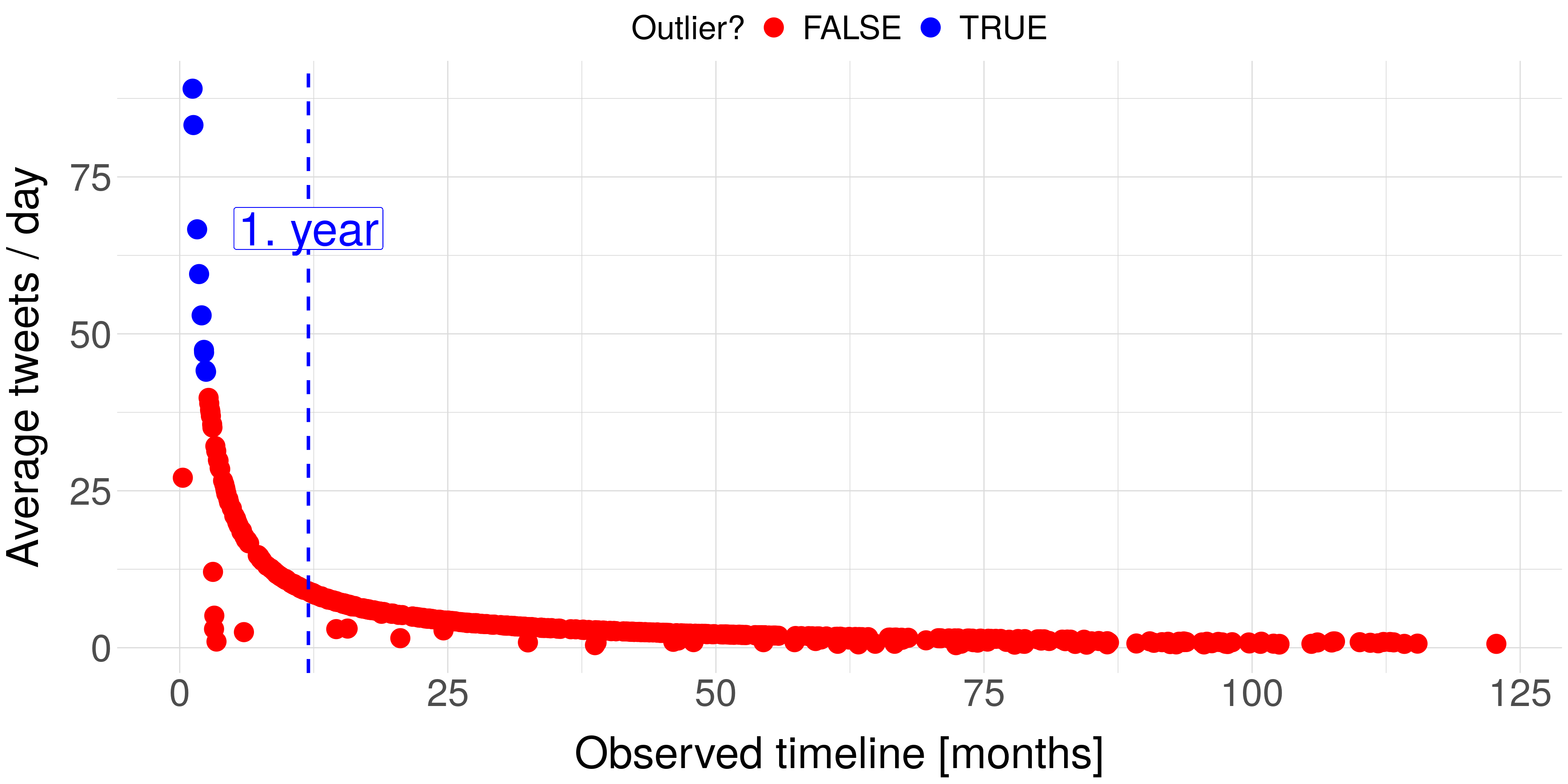}}
\hfill
\subfloat[Brasil
\label{fig_appendix:observedtimeline_BrazilianJournalists}]
{\includegraphics[width=0.28\textwidth]
{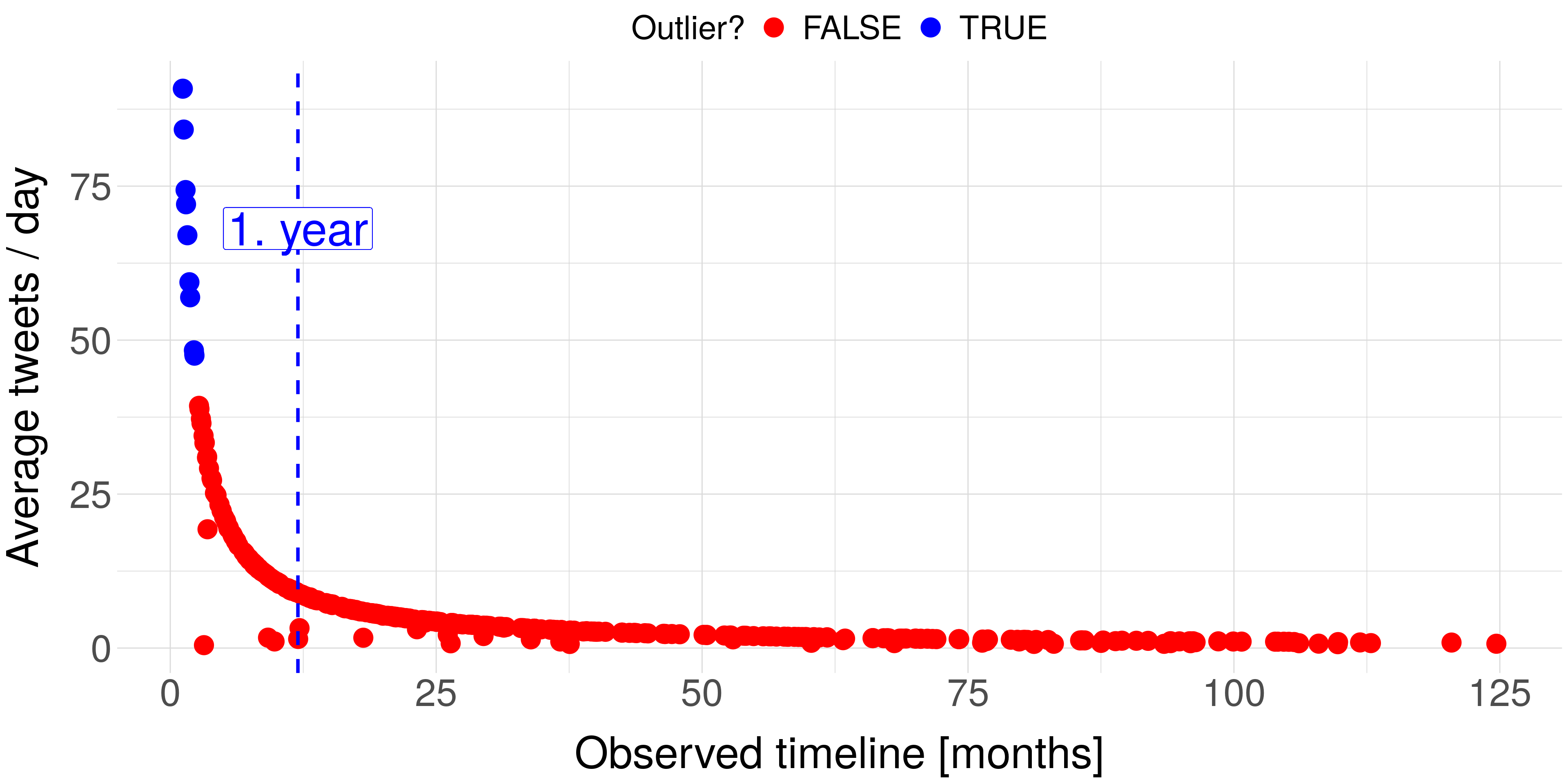}}
\hfill
\subfloat[Japan
\label{fig_appendix:observedtimeline_JapaneseJournalists}]
{\includegraphics[width=0.28\textwidth]
{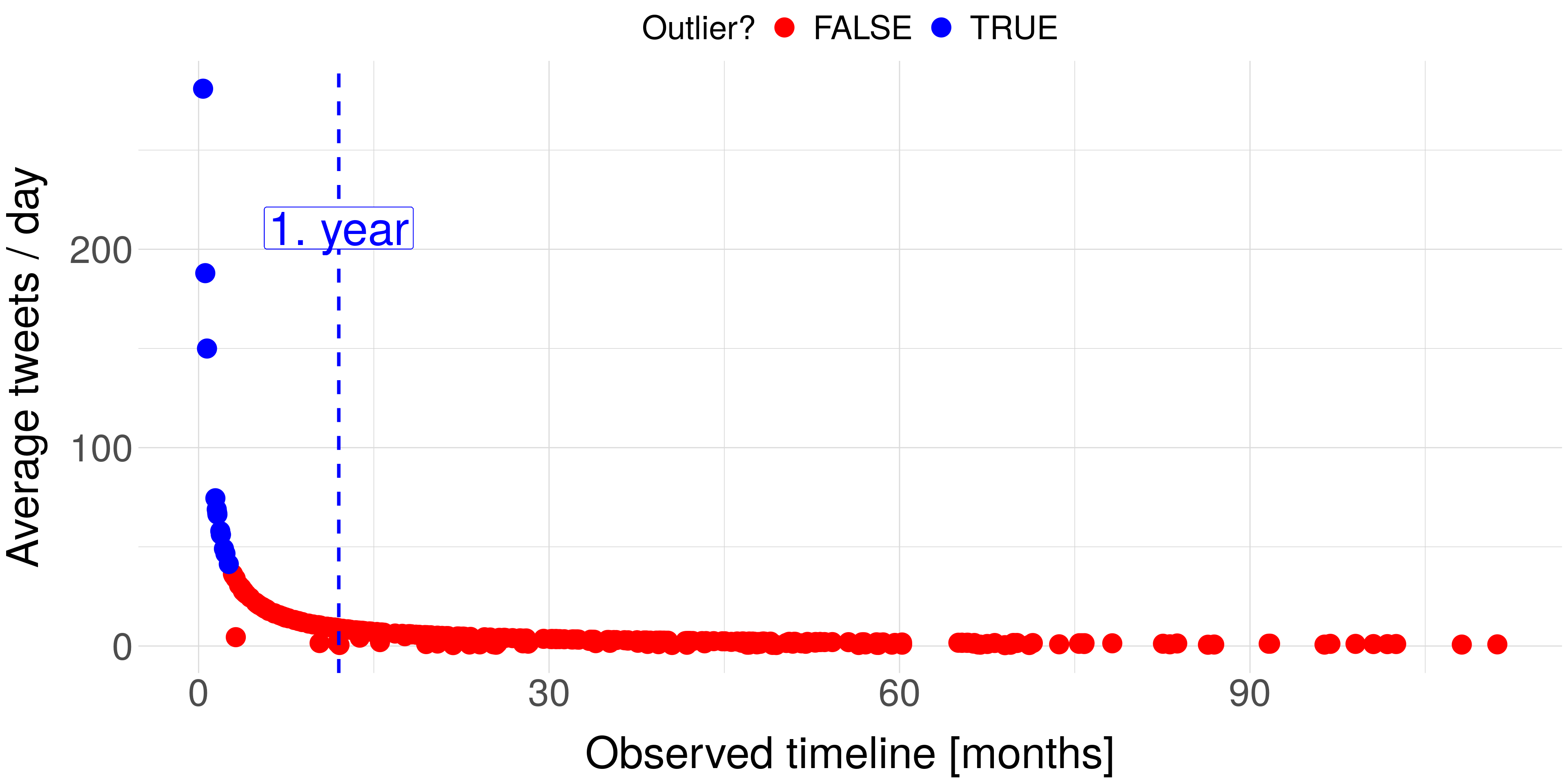}}
\hfill
\subfloat[Turkey
\label{fig_appendix:observedtimeline_TurkishJournalists}]
{\includegraphics[width=0.28\textwidth]
{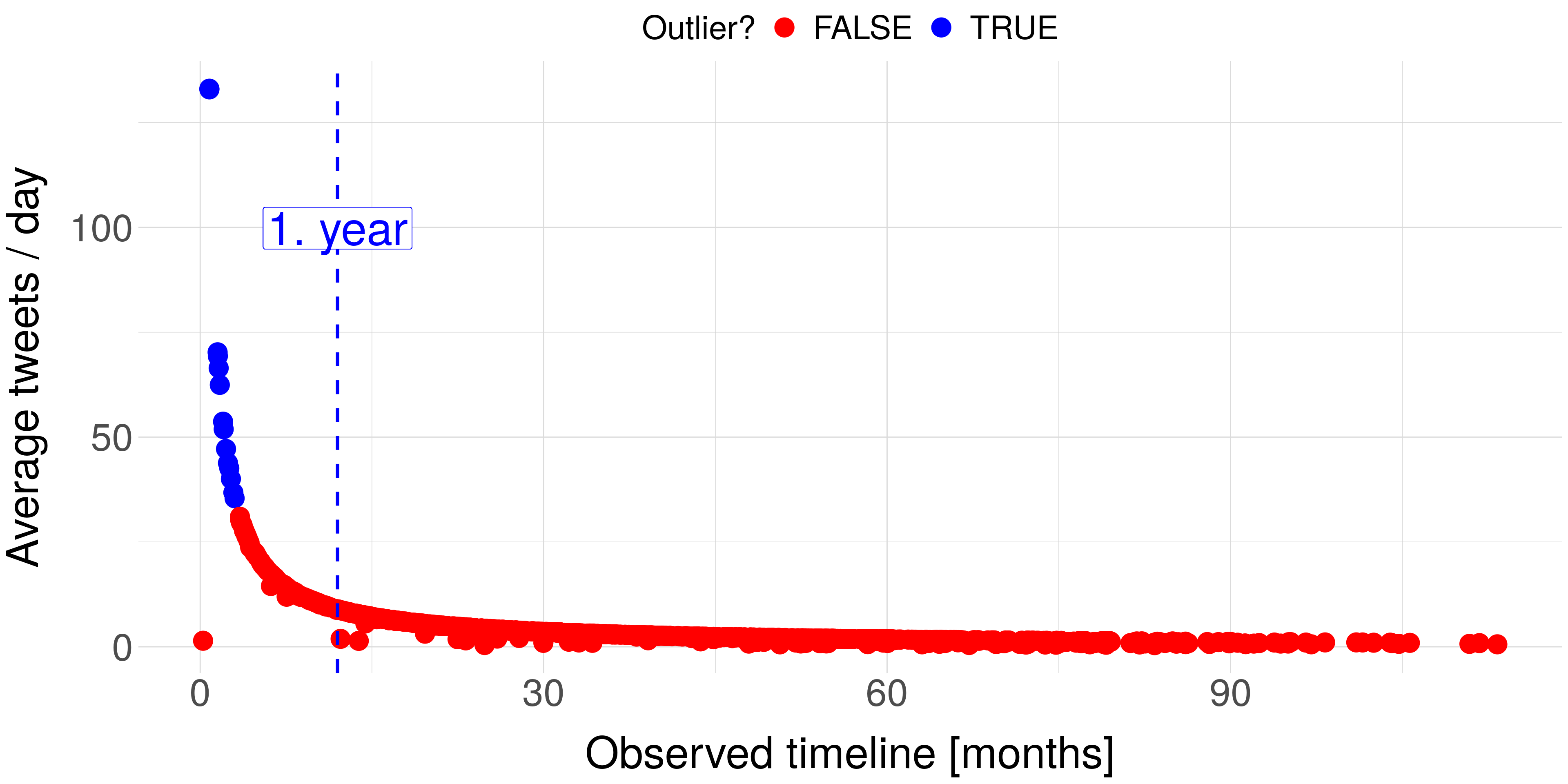}}
\hfill
\subfloat[UK
\label{fig_appendix:observedtimeline_BritishJournalists}]
{\includegraphics[width=0.28\textwidth]
{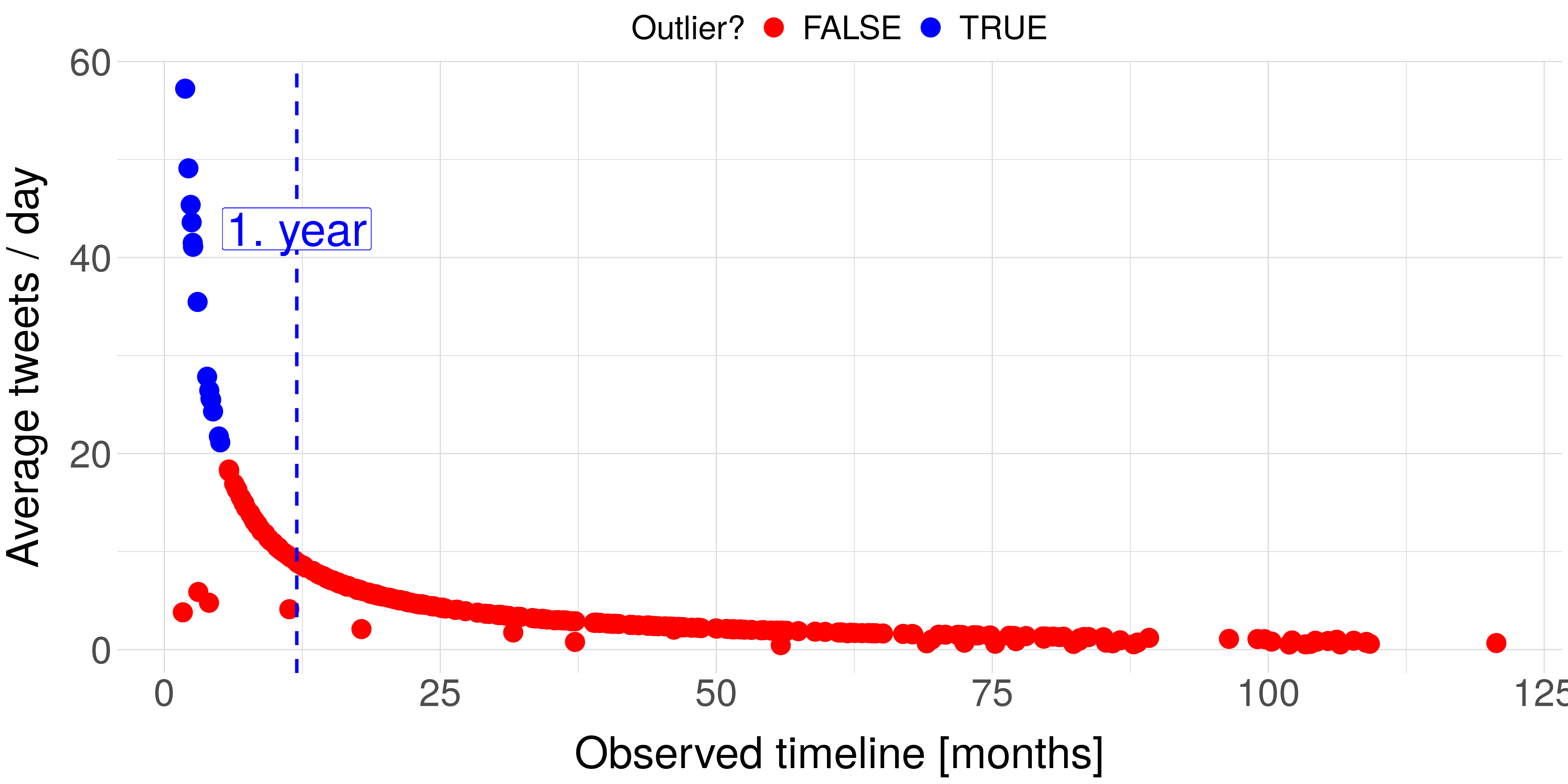}}
\hfill
\subfloat[Denmark
\label{fig_appendix:observedtimeline_DanishJournalists}]
{\includegraphics[width=0.28\textwidth]
{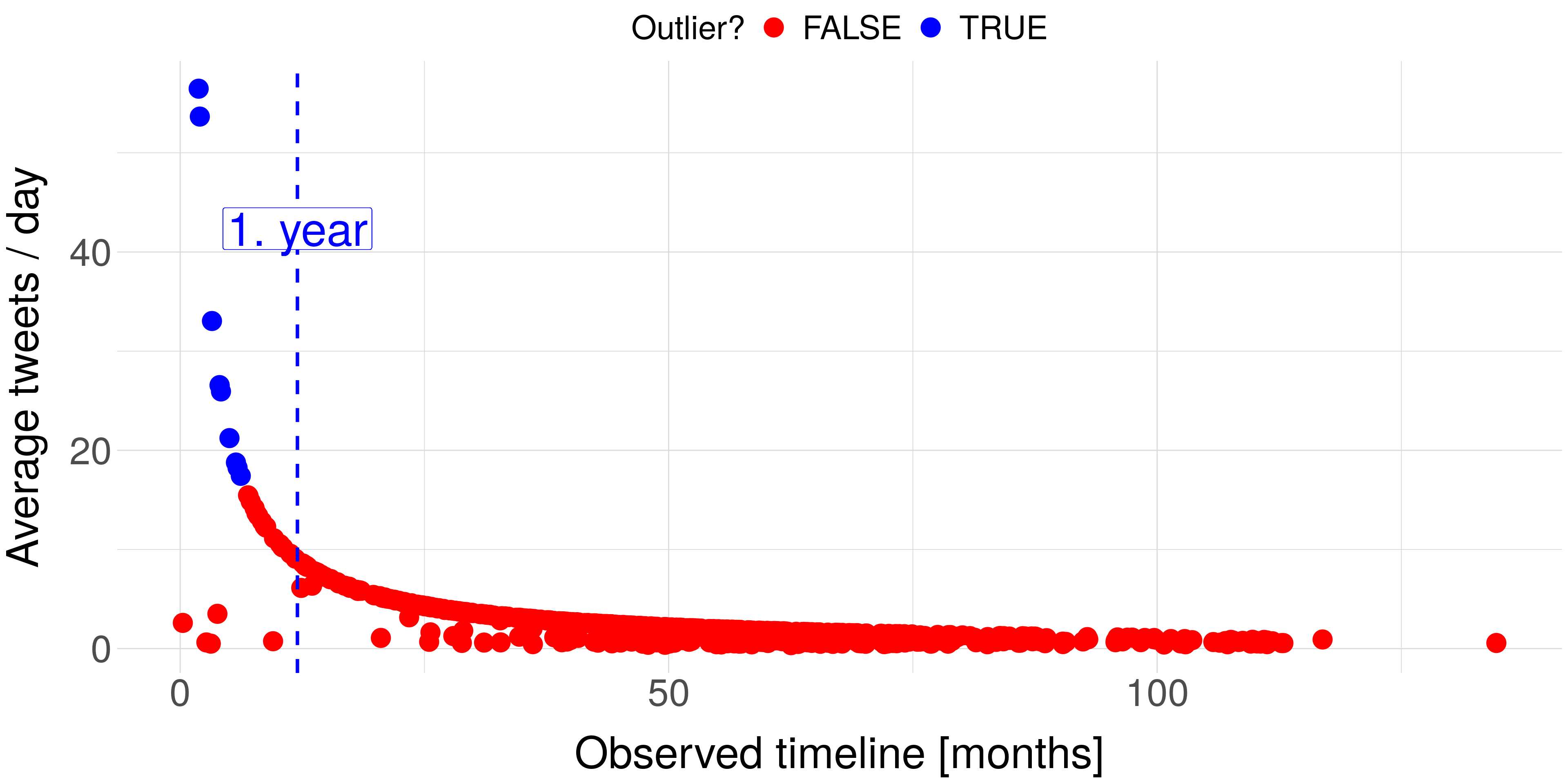}}
\hfill
\subfloat[Finland
\label{fig_appendix:observedtimeline_FinnishJournalists}]
{\includegraphics[width=0.28\textwidth]
{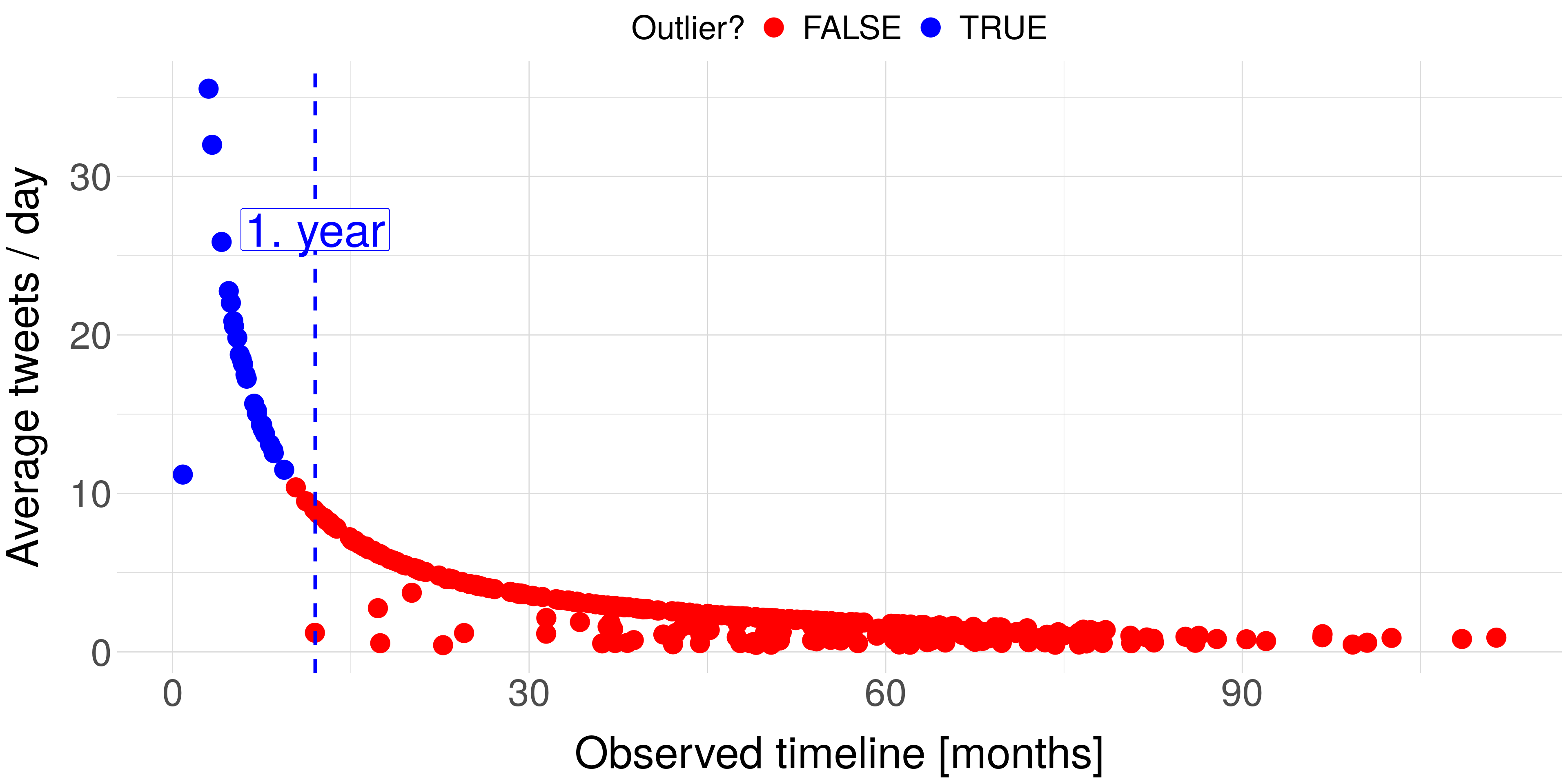}}
\hfill
\subfloat[Norway
\label{fig_appendix:observedtimeline_NorwegianJournalists}]
{\includegraphics[width=0.28\textwidth]
{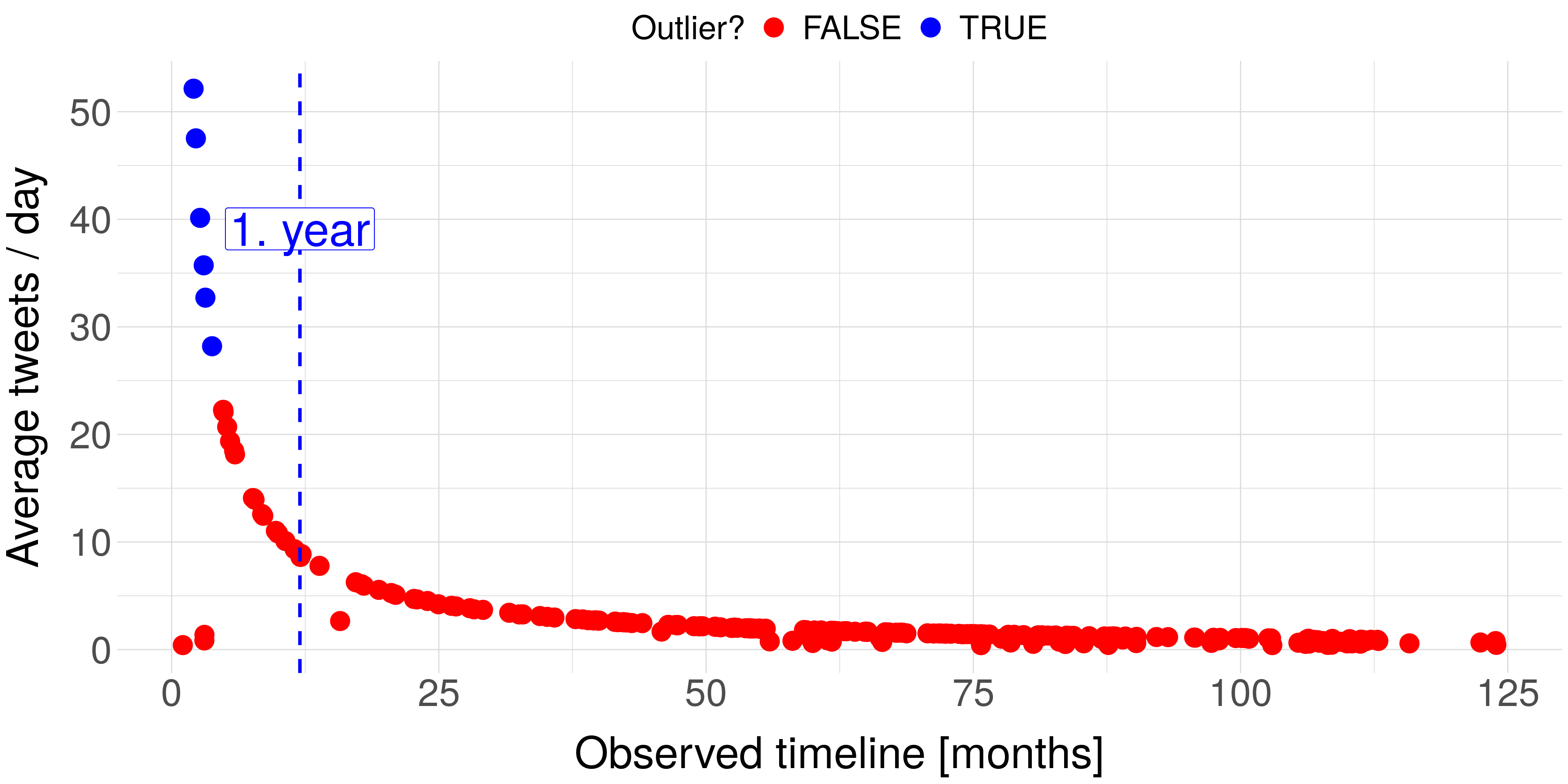}}
\hfill
\subfloat[Sweden
\label{fig_appendix:observedtimeline_SwedishJournalists}]
{\includegraphics[width=0.28\textwidth]
{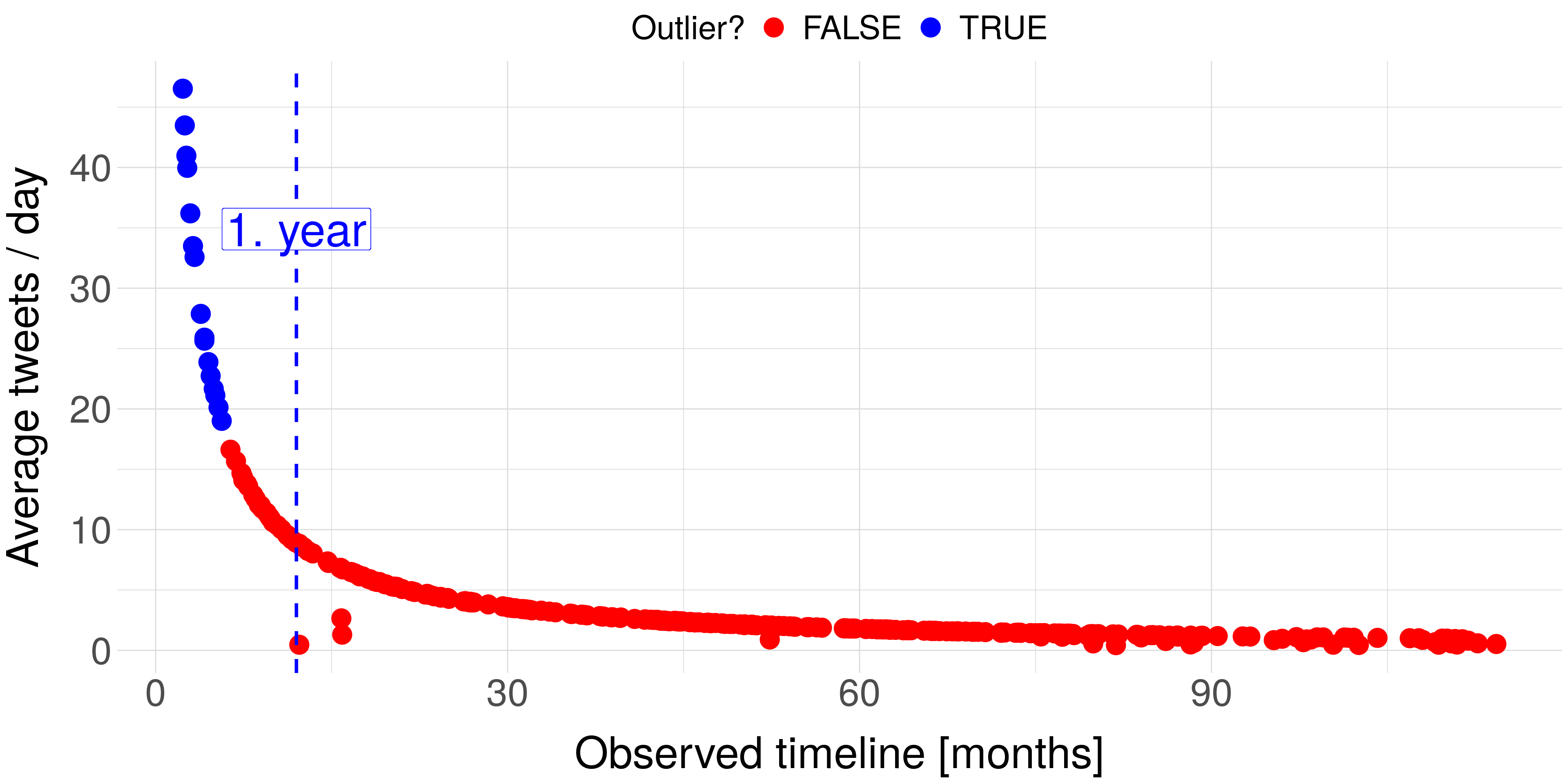}}
\hfill
\subfloat[Greece
\label{fig_appendix:observedtimeline_GreekJournalists}]
{\includegraphics[width=0.28\textwidth]
{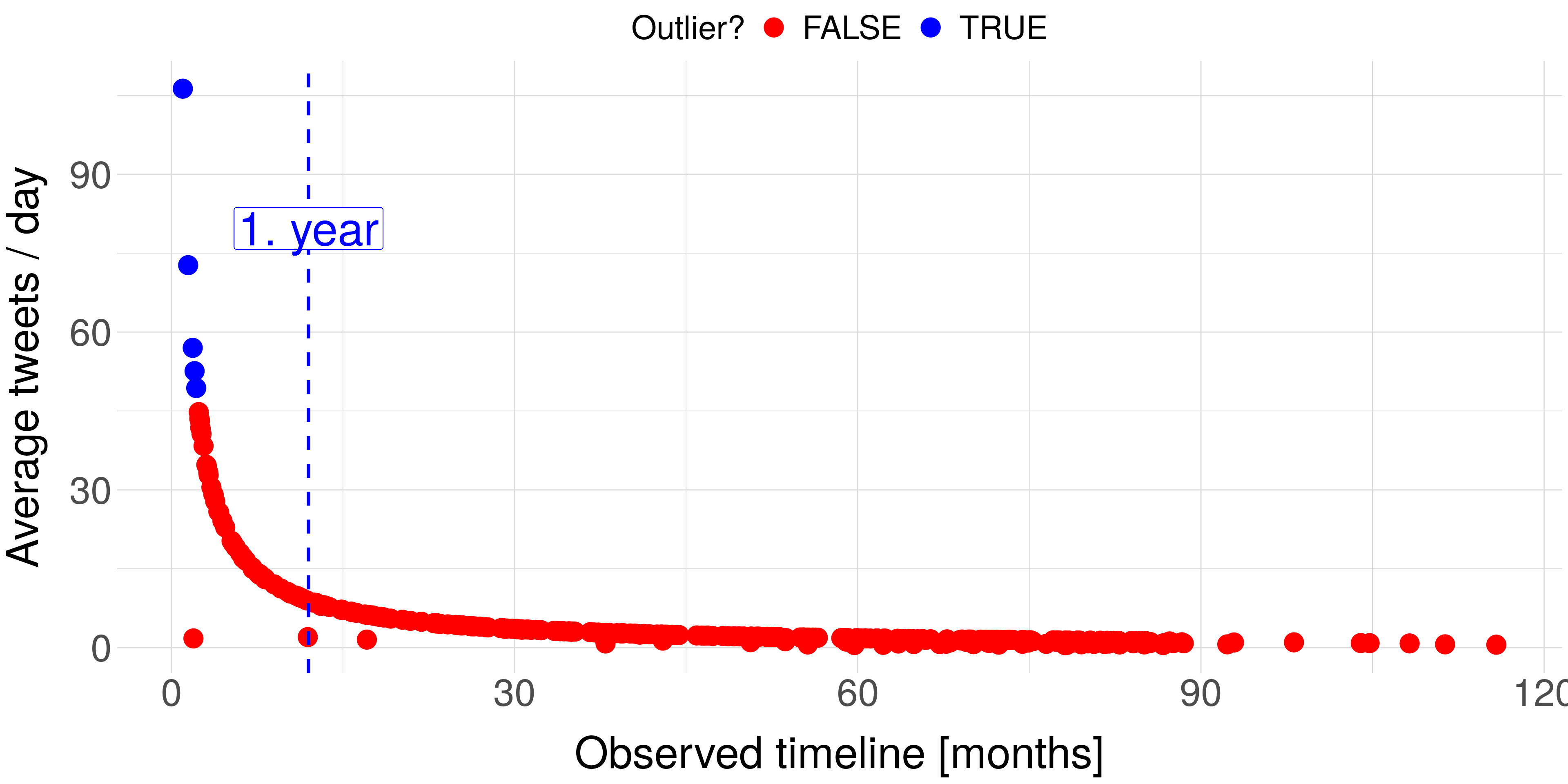}}
\hfill
\subfloat[Italy
\label{fig_appendix:observedtimeline_ItalianJournalists}]
{\includegraphics[width=0.28\textwidth]
{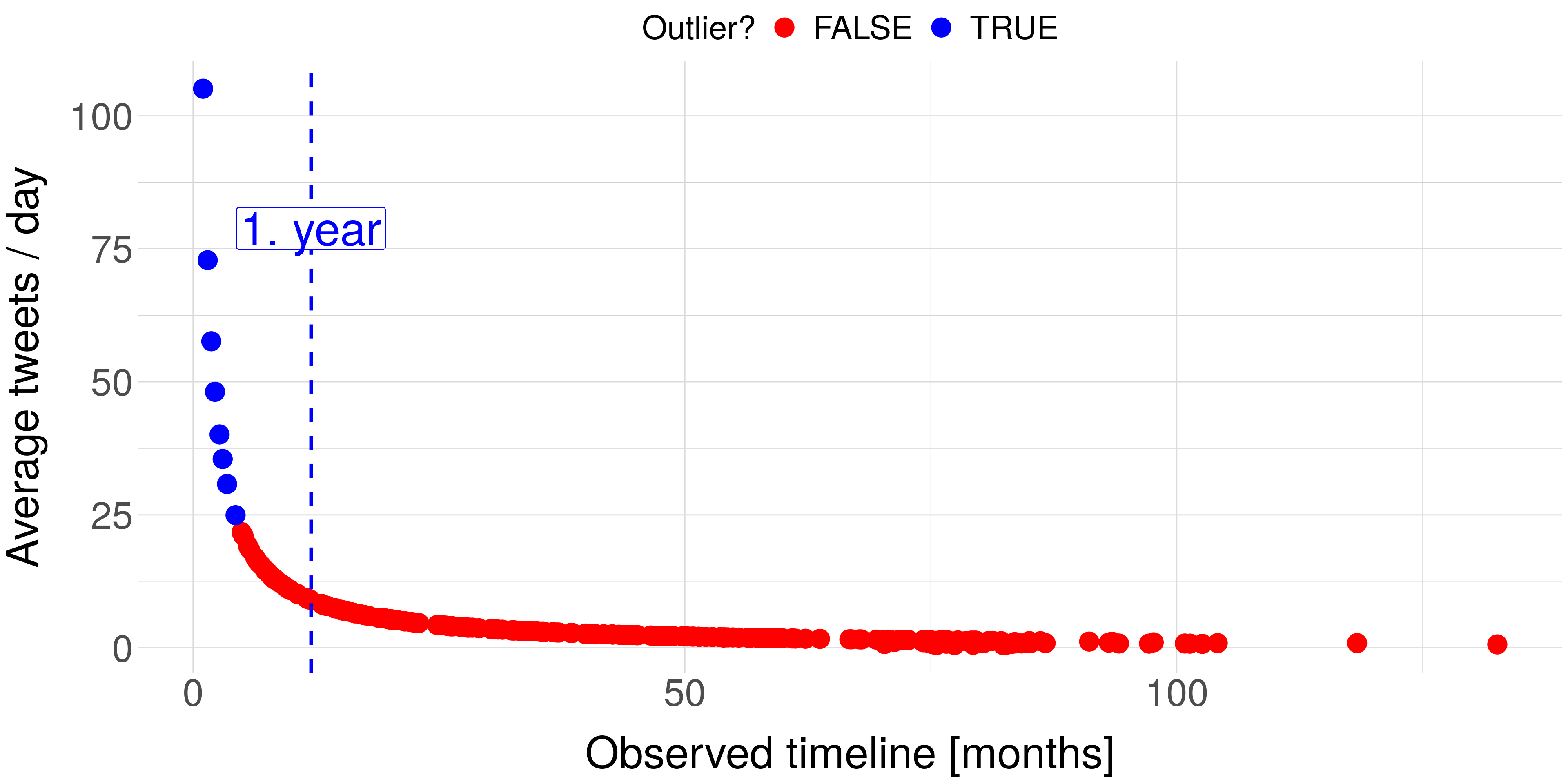}}
\hfill
\subfloat[Spain
\label{fig_appendix:observedtimeline_SpanishJournalists}]
{\includegraphics[width=0.28\textwidth]
{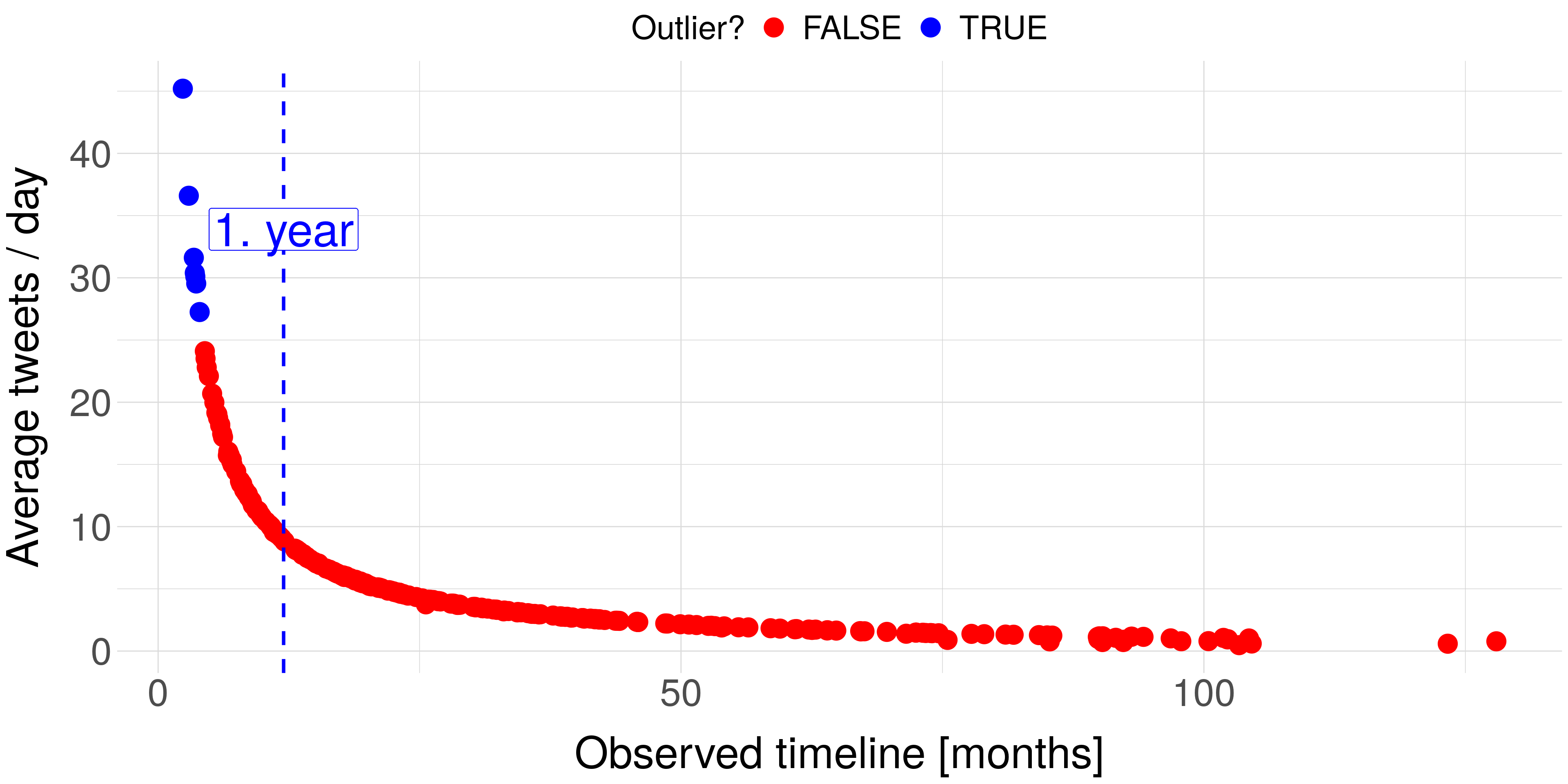}}
\hfill
\subfloat[France
\label{fig_appendix:observedtimeline_FrenchJournalists}]
{\includegraphics[width=0.28\textwidth]
{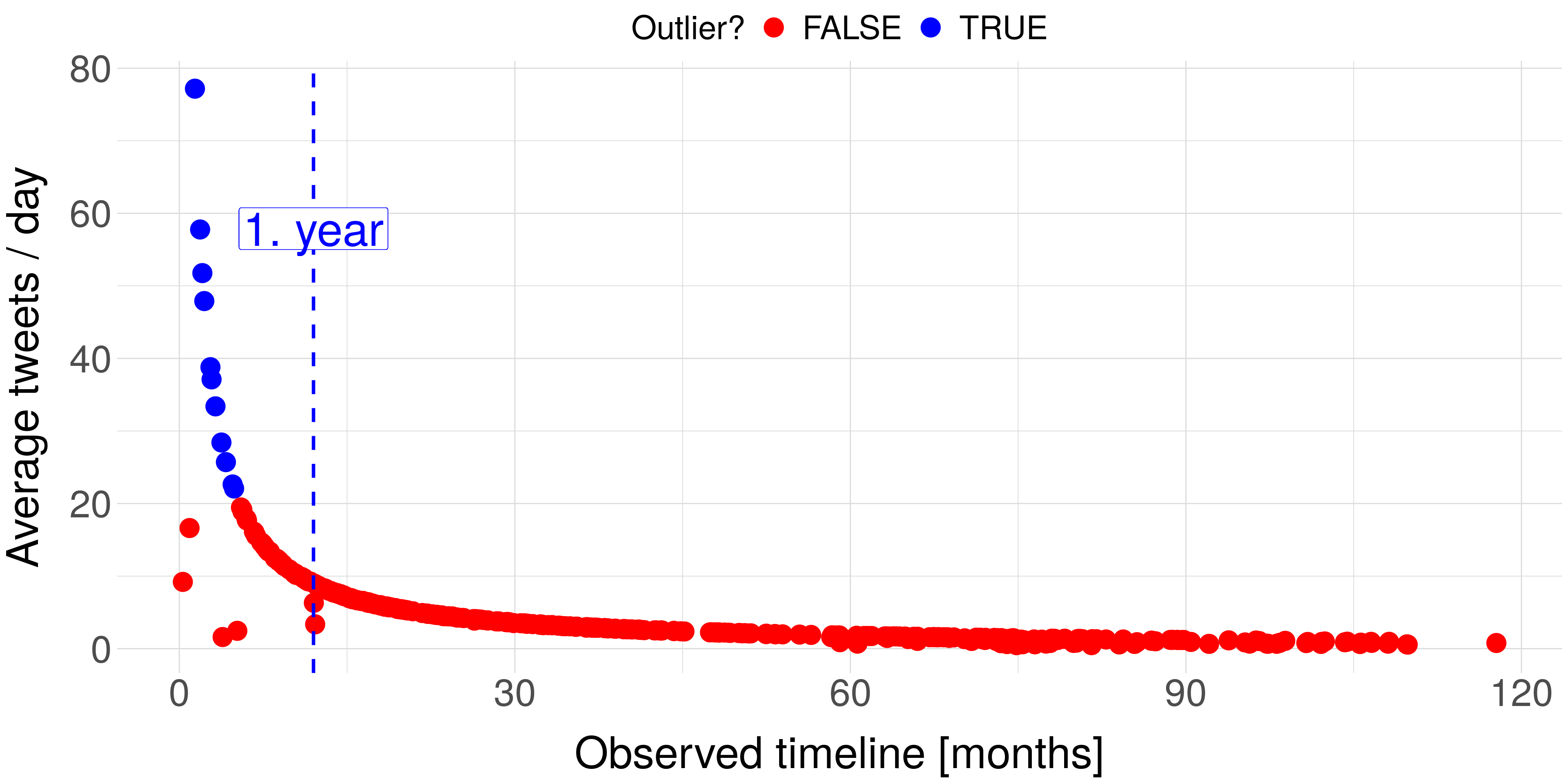}}
\hfill
\subfloat[Germany
\label{fig_appendix:observedtimeline_AGermanournalists}]
{\includegraphics[width=0.28\textwidth]
{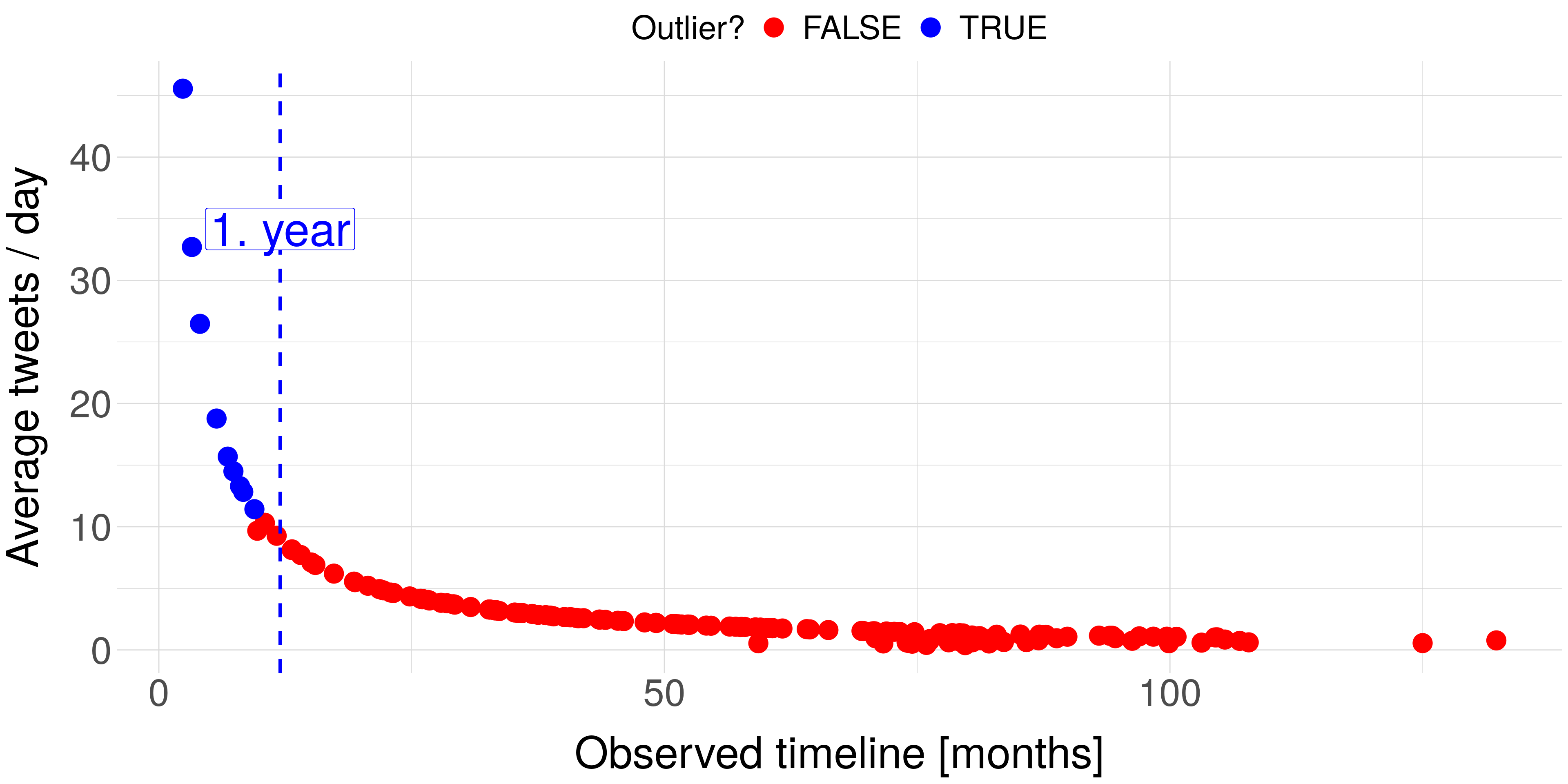}}
\hfill
\subfloat[Netherland
\label{fig_appendix:observedtimeline_NetherlanderJournalists}]
{\includegraphics[width=0.28\textwidth]
{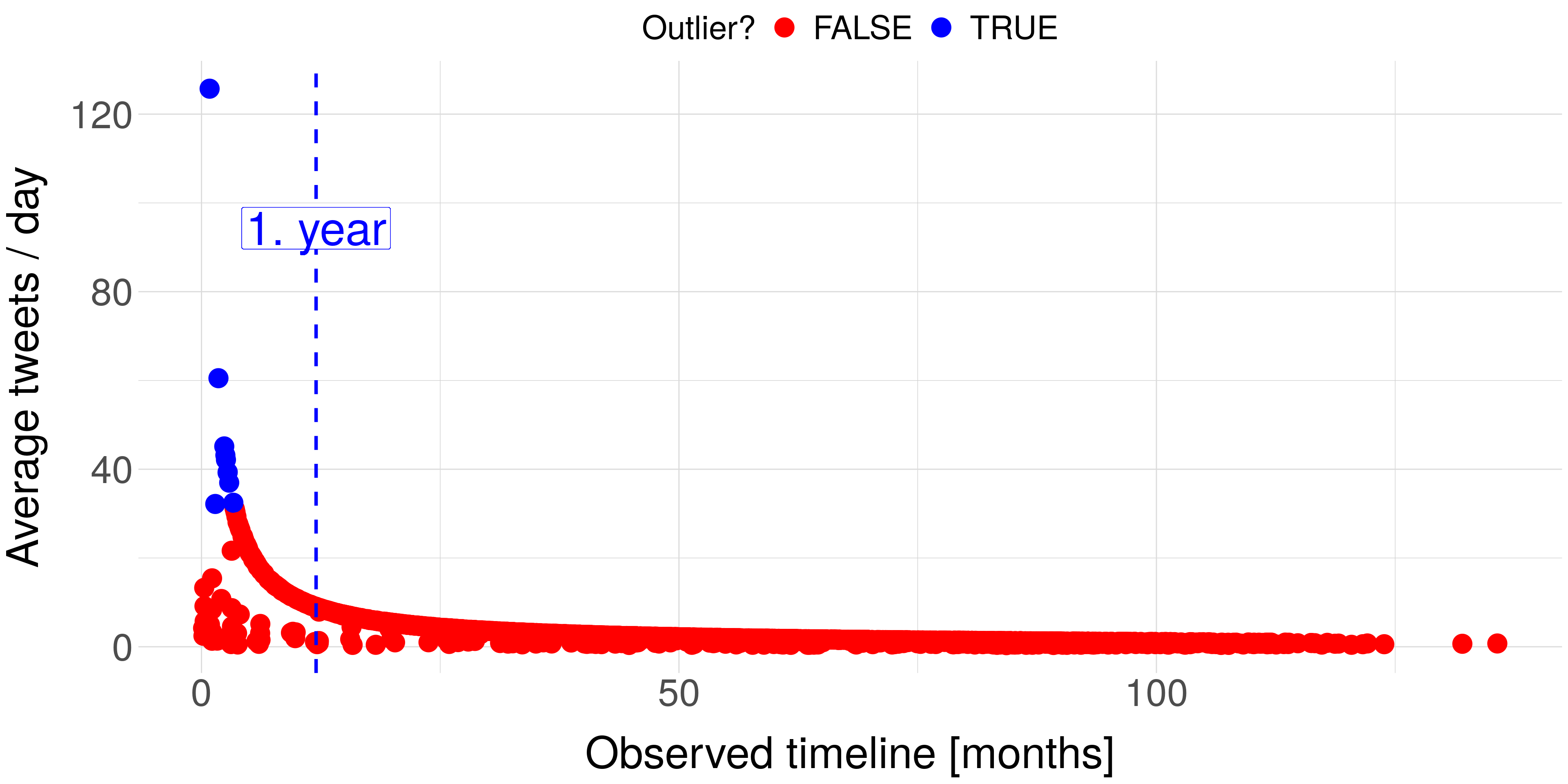}}
\hspace{1pt}
\subfloat[Australia
\label{fig_appendix:observedtimeline_AustralianJournalists}]
{\includegraphics[width=0.28\textwidth,height=0.1\hsize]
{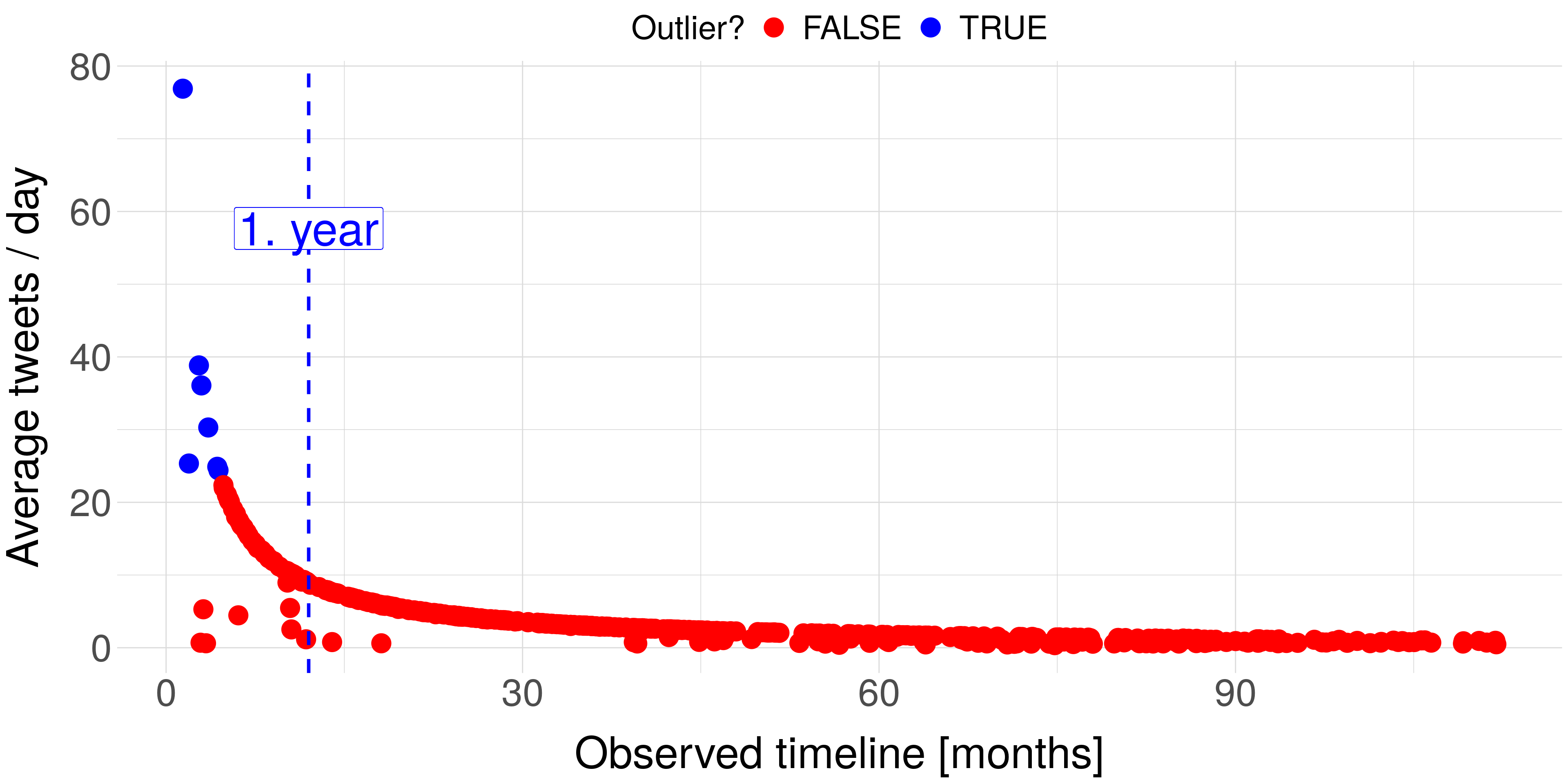}}
\end{center}
\end{adjustbox}
\caption{Observed timeline length vs average tweet frequency}
\label{fig_appendix:scatter_observedtl_vs_freq}
\end{figure}

\clearpage
\subsection{Stationarity}
\label{appendix_stationarity}

\begin{figure}[ht]
\begin{center}
\includegraphics[scale=0.2]{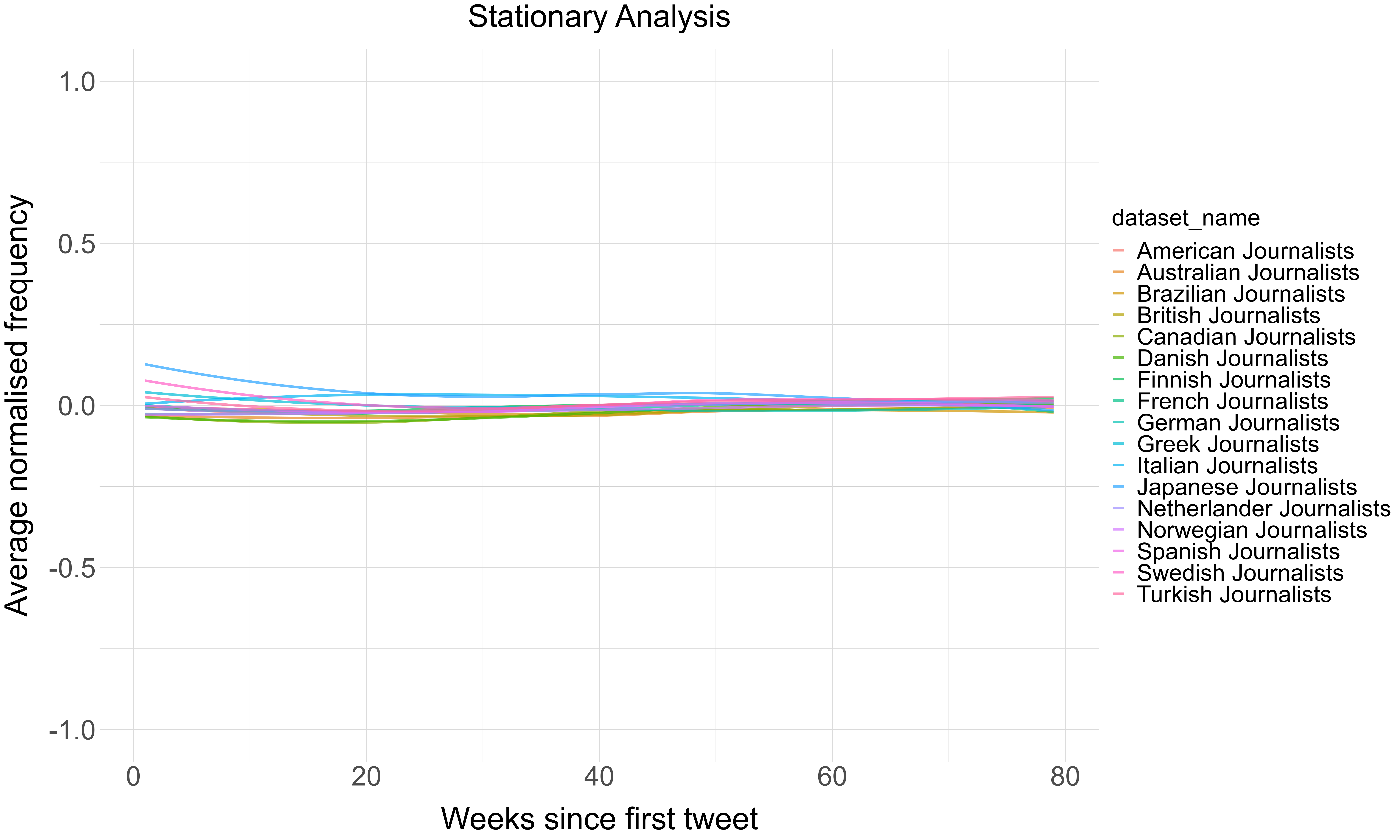}\vspace{-5pt}
\caption{Stationarity analysis}
\label{fig:stability}\vspace{-10pt}
\end{center}
\end{figure}

As previously shown in the literature (\cite{Arnaboldi2017Facebook,miritello2013,viswanath2009}), newly registered social network users add relationships into their network and interact with others at a higher rate than long-term users. After a while, though, their activity stabilizes. Newly registered users are thus outliers with respect to the general population of users, and they should be discarded from the analysis.
In order to verify whether this is the case also in our dataset, we compute, for each user, the average number of tweets generated each week. 
We align the first week of the observed timeline (starting from the first observed tweet) and get the average of the dataset. To be fair and keep this stability calculation at the individual user level, we use mean normalization for each user. As a result, the values are located in the range $[-1,1]$ where zero values are close to the mean value. 
In Figure~\ref{fig:stability}, we show the first 80 weeks of averaged normalized tweeting frequency per week, since the number of active users decreases drastically after 80th week. For the stability analysis, we only show fully observed users since we cannot observe the full timeline of the partially observed ones (for which the initial transient period is thus automatically discarded). As it can be seen from the figure, we do not have a drastic change in tweeting activity for any of the datasets. Therefore, we do not discard any part of the timeline of fully observed users.

\renewcommand\thefigure{\Alph{section}.\arabic{figure}} 
\setcounter{figure}{0}

\renewcommand\thetable{\Alph{section}.\arabic{table}} 
\setcounter{table}{0}

\section{Static ego network analysis - additional plots}
\label{appendix_static_egonets}

\vspace{-20pt}

\begin{figure}[!h]
\begin{adjustbox}{minipage=\linewidth}
\begin{center}
\subfloat[USA
\label{fig_appendix:totalsize_AmericanJournalists}]
{\includegraphics[width=0.28\textwidth]
{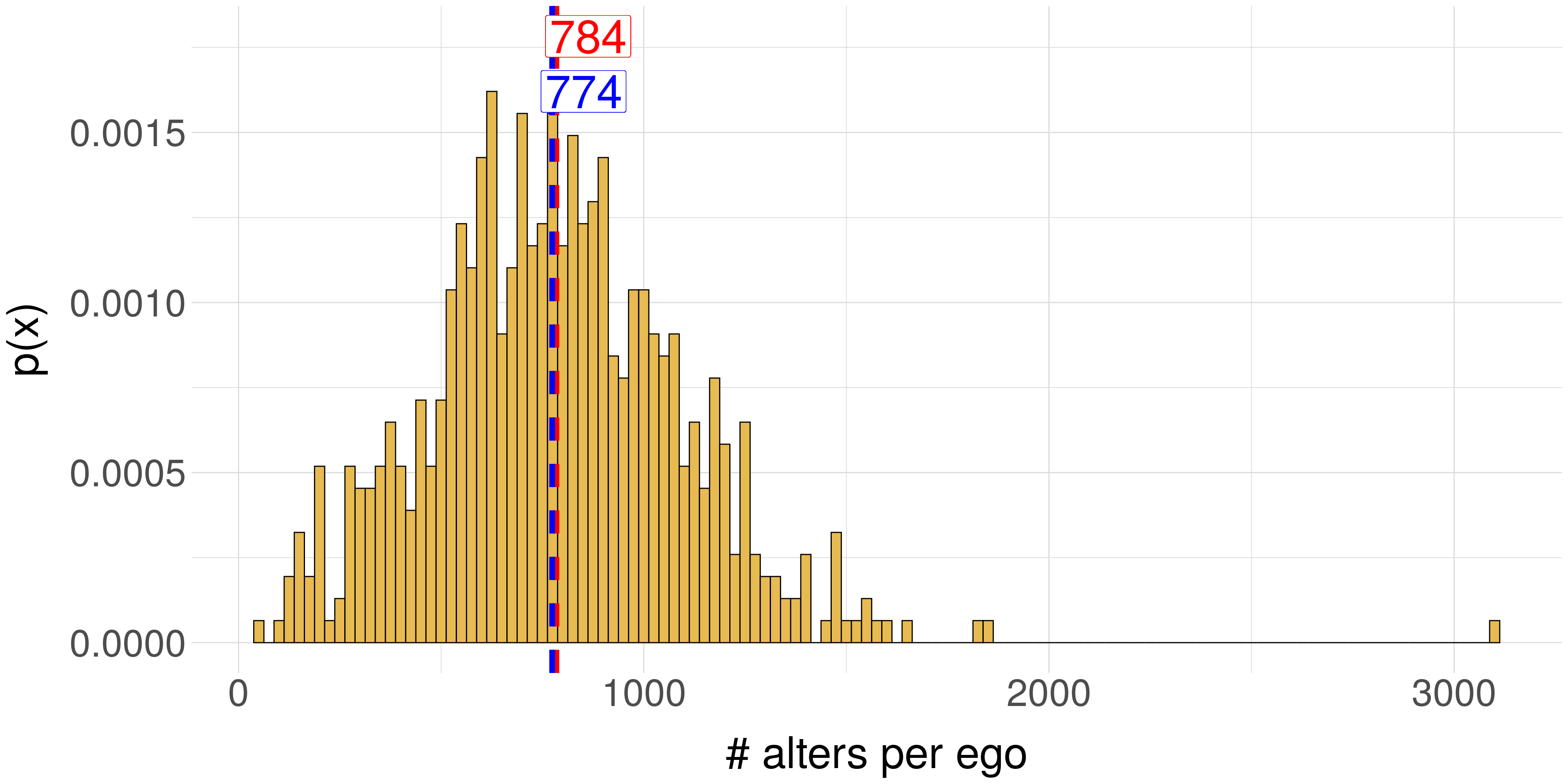}}
\hfill
\subfloat[Canada
\label{fig_appendix:totalsize_CanadianJournalists}]
{\includegraphics[width=0.28\textwidth]
{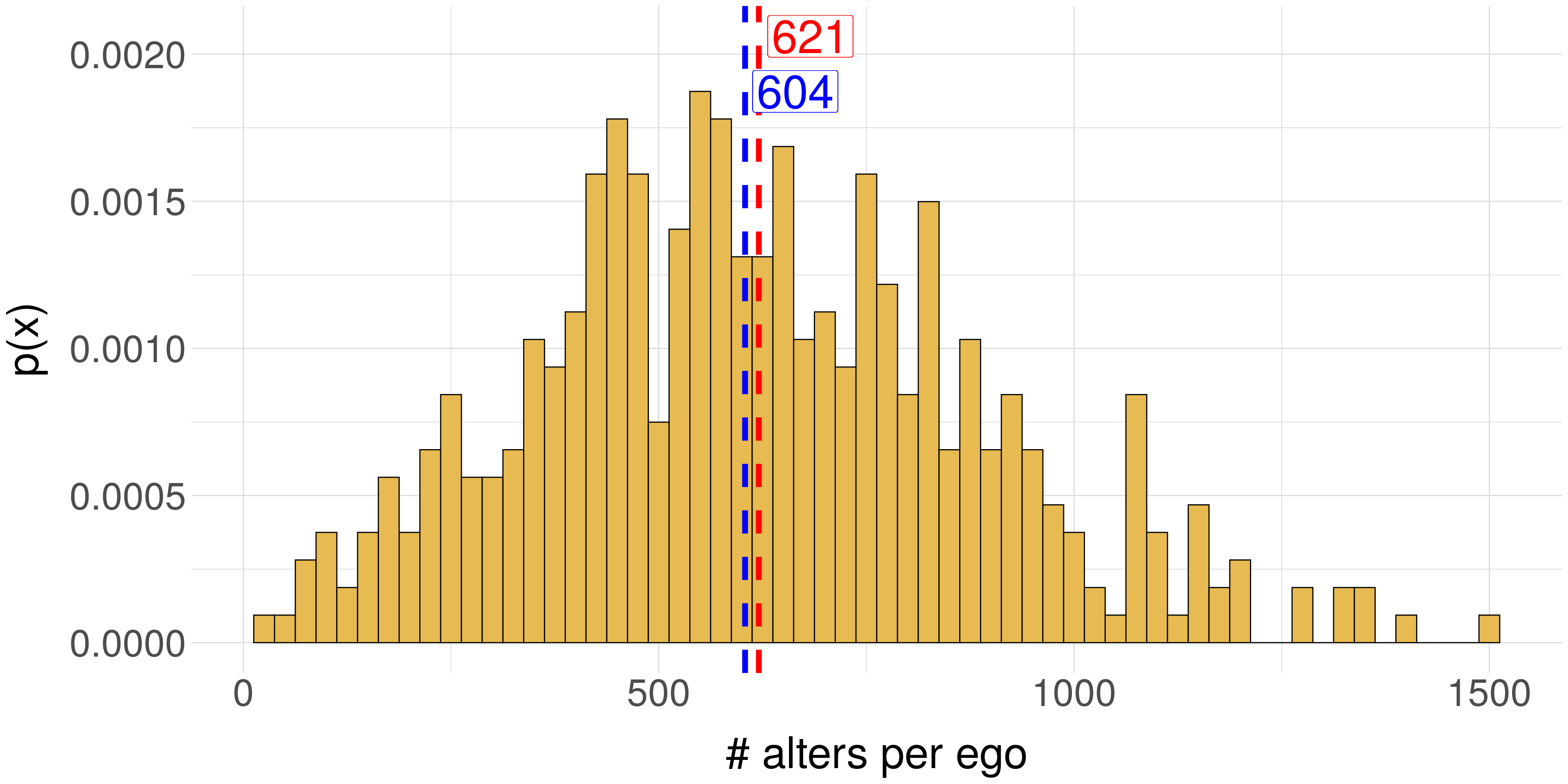}}
\hfill
\subfloat[Brasil
\label{fig_appendix:totalsize_BrazilianJournalists}]
{\includegraphics[width=0.28\textwidth]
{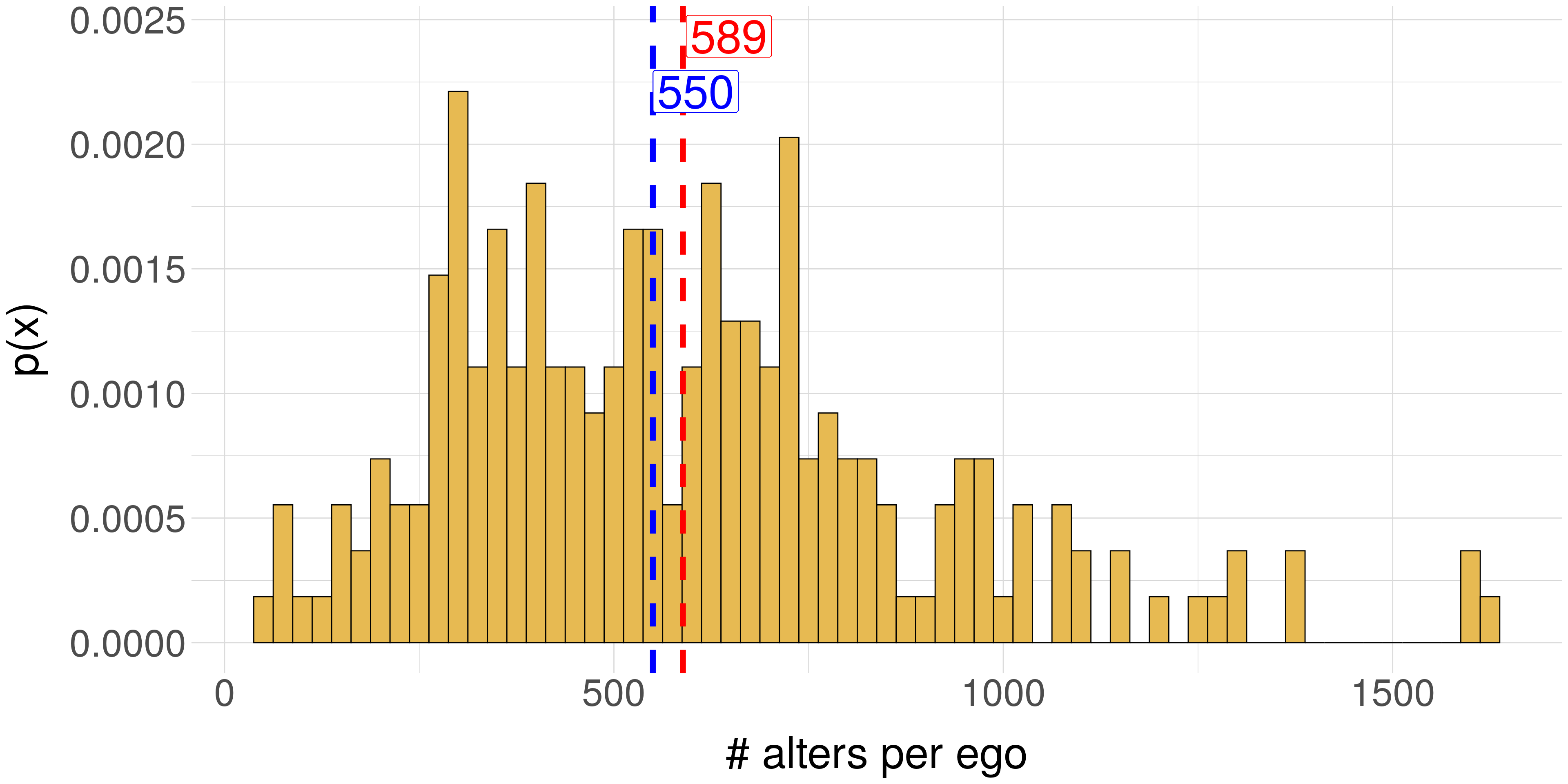}}
\hfill
\subfloat[Japan
\label{fig_appendix:totalsize_JapaneseJournalists}]
{\includegraphics[width=0.28\textwidth]
{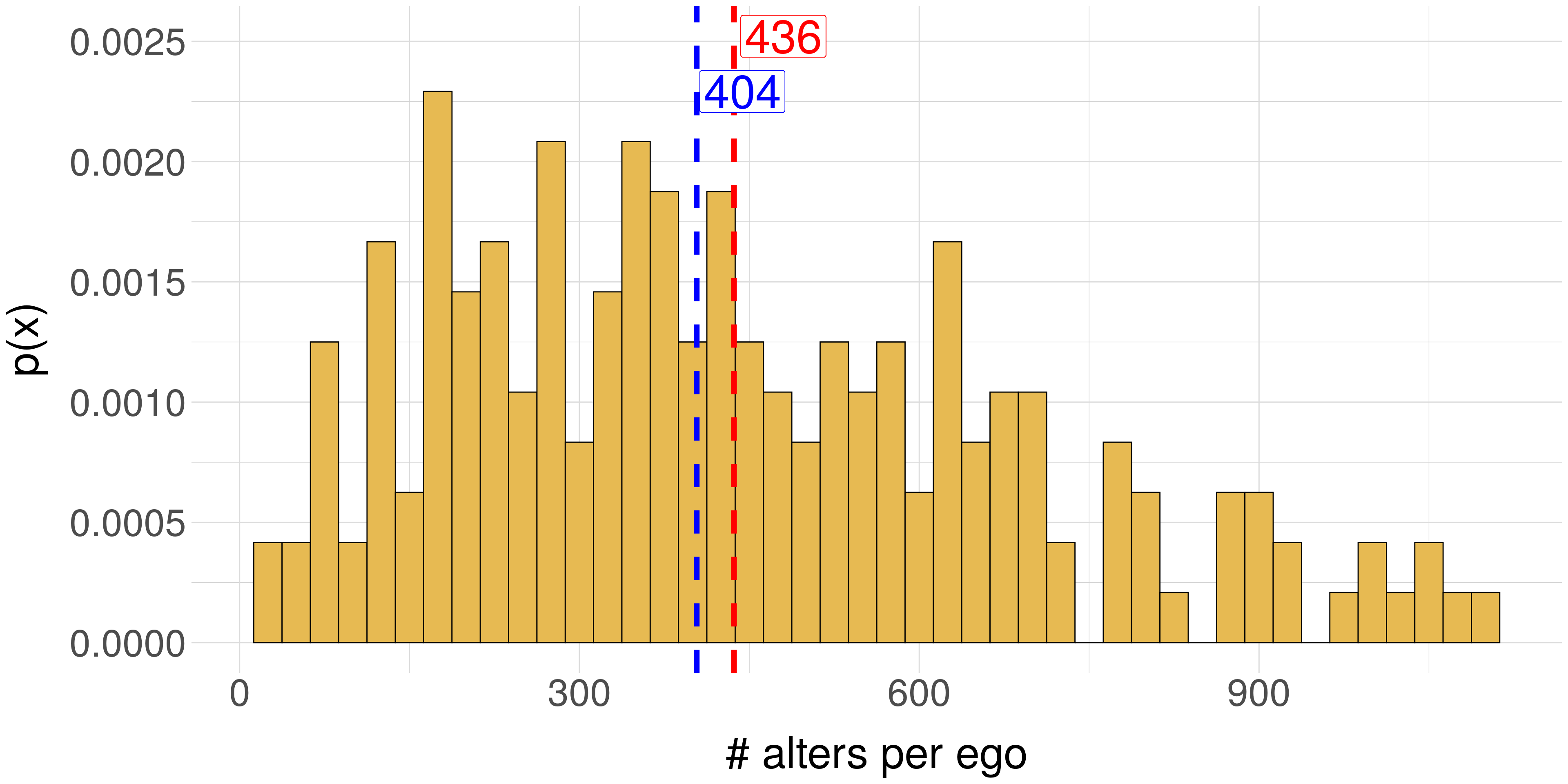}}
\hfill
\subfloat[Turkey
\label{fig_appendix:totalsize_TrukishJournalists}]
{\includegraphics[width=0.28\textwidth]
{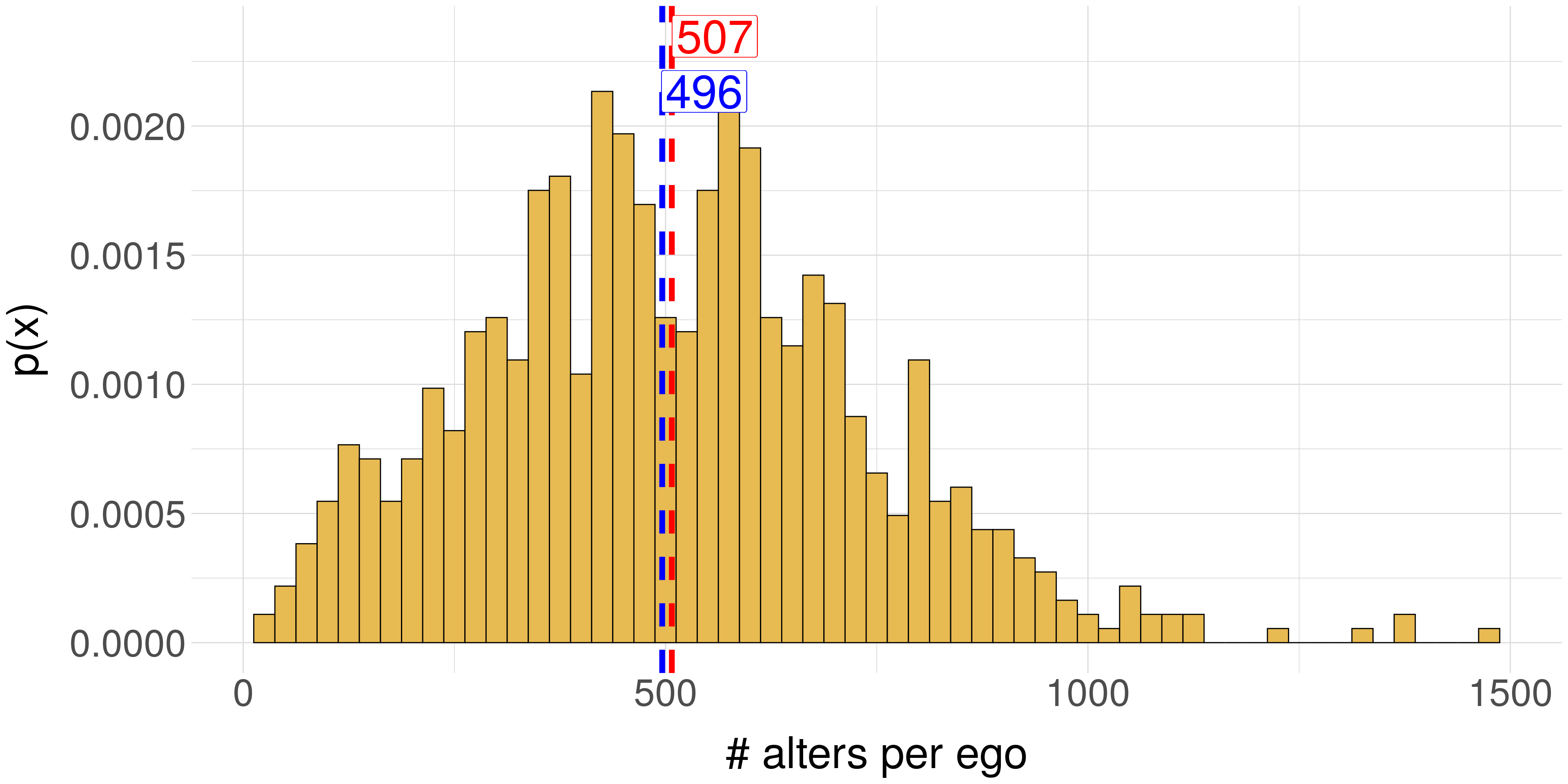}}
\hfill
\subfloat[UK
\label{fig_appendix:totalsize_BritishJournalists}]
{\includegraphics[width=0.28\textwidth]
{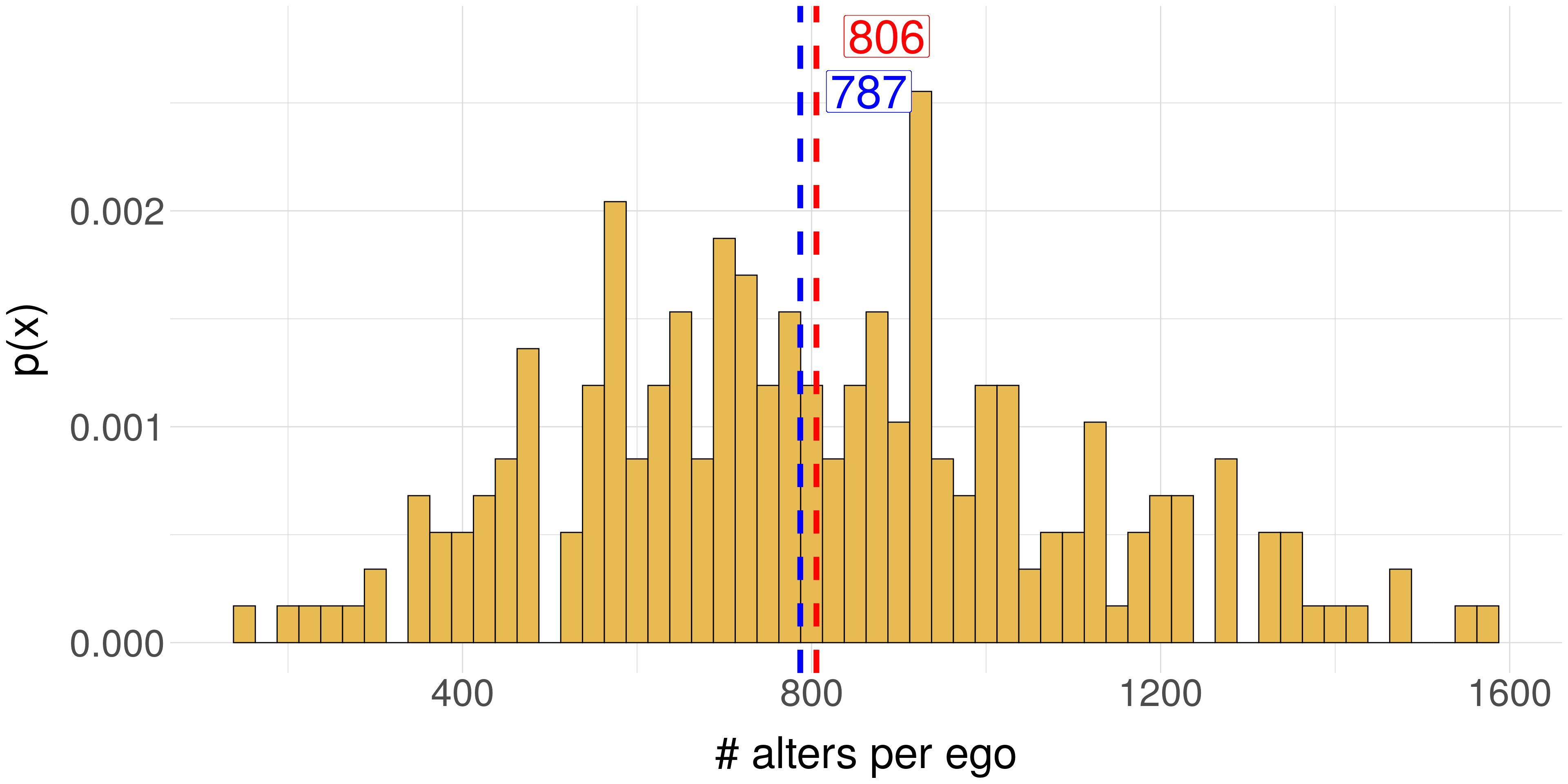}}
\hfill
\subfloat[Denmark
\label{fig_appendix:totalsize_DanishJournalists}]
{\includegraphics[width=0.28\textwidth]
{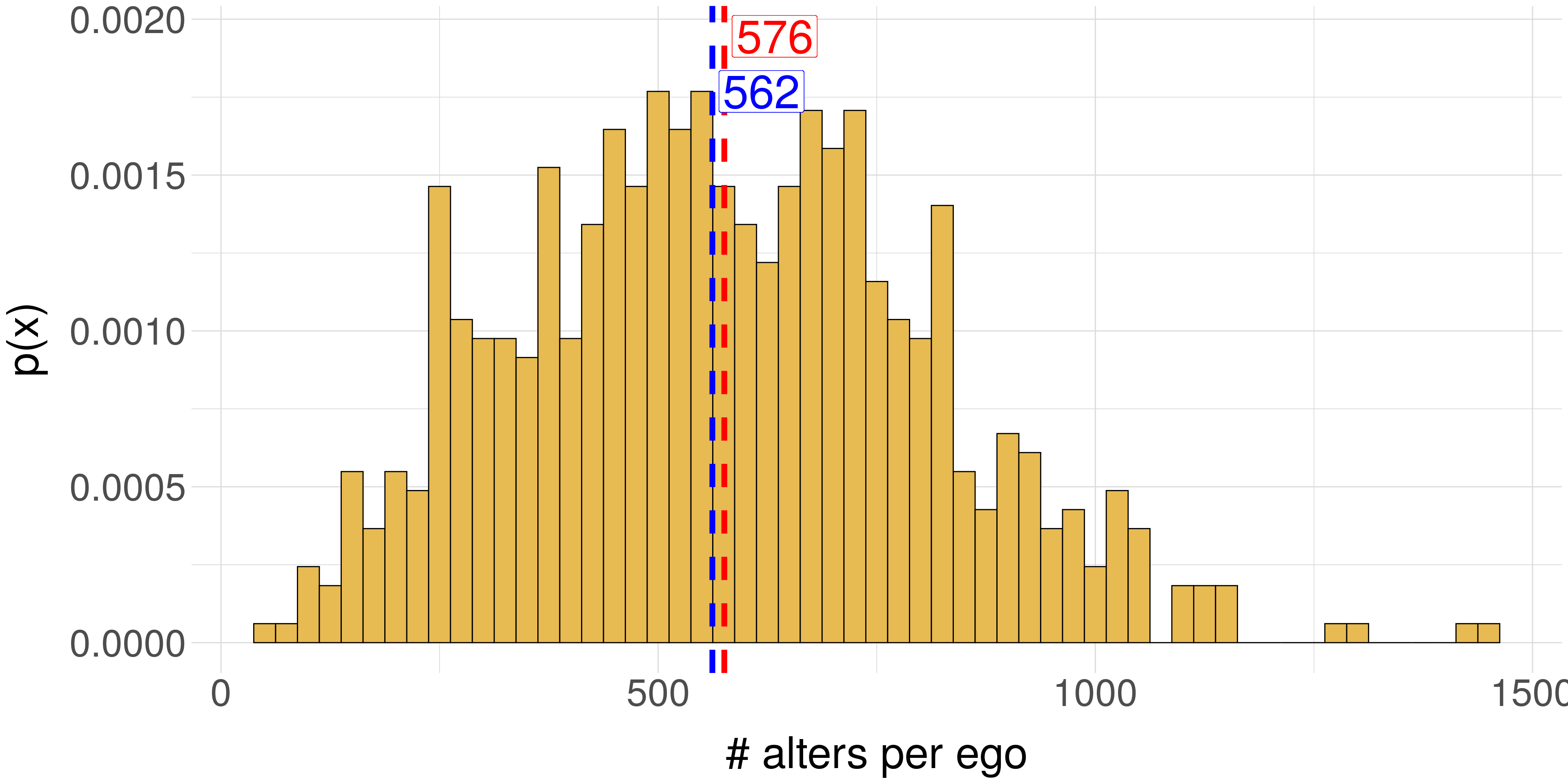}}
\hfill
\subfloat[Finland
\label{fig_appendix:totalsize_FinnishJournalists}]
{\includegraphics[width=0.28\textwidth]
{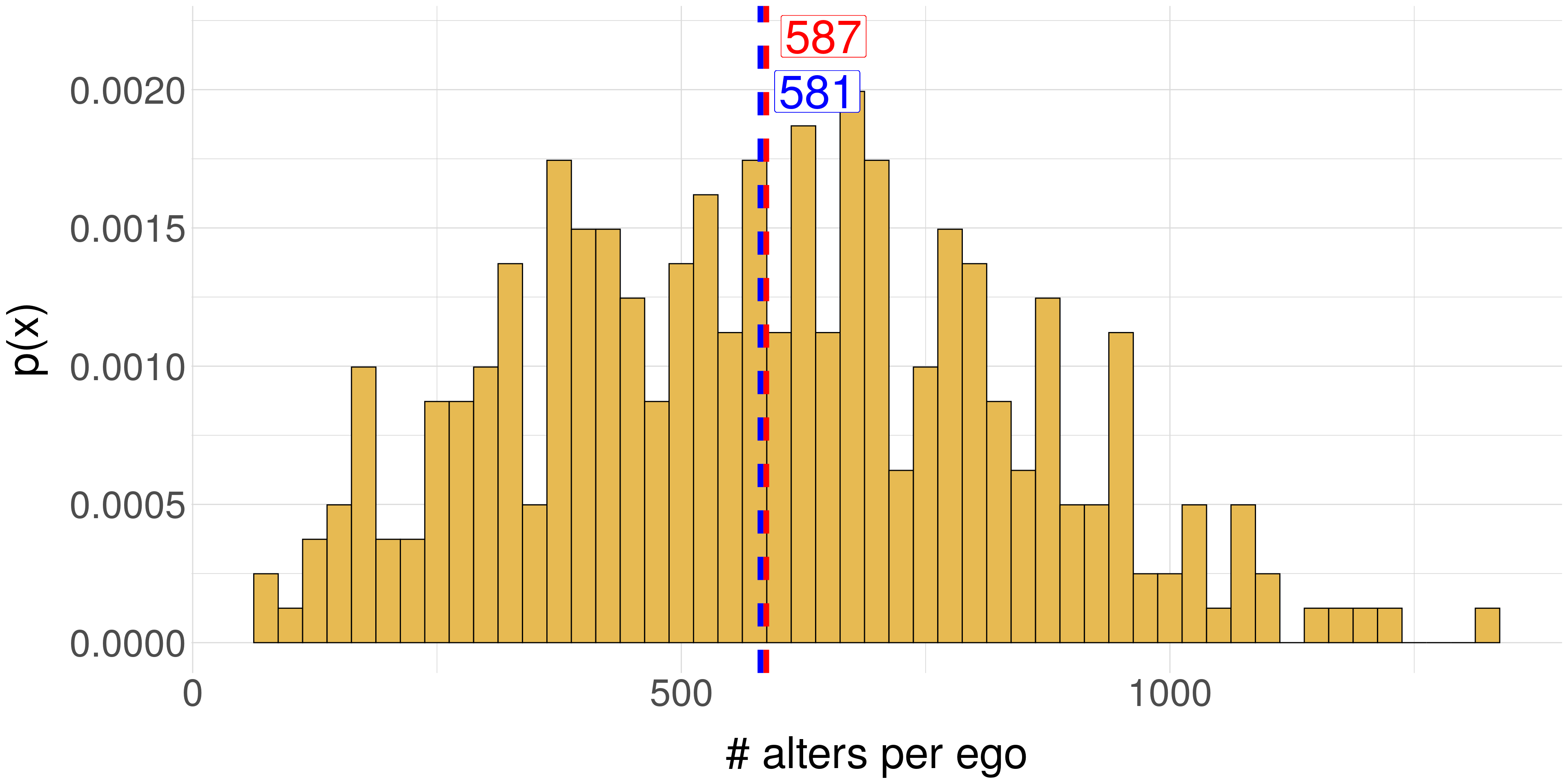}}
\hfill
\subfloat[Norway
\label{fig_appendix:totalsize_NorwegianJournalists}]
{\includegraphics[width=0.28\textwidth]
{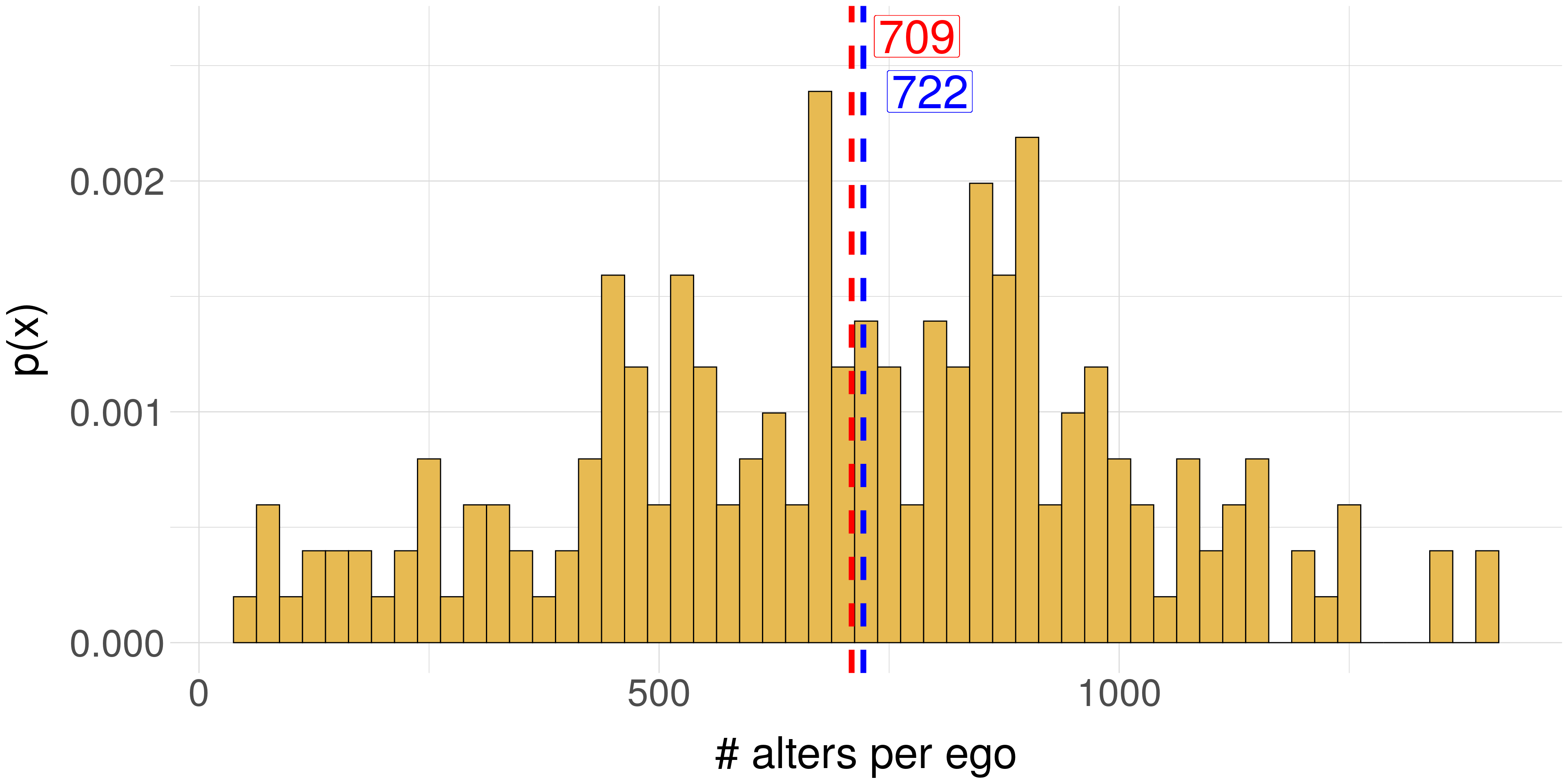}}
\hfill
\subfloat[Sweden
\label{fig_appendix:totalsize_SwedishJournalists}]
{\includegraphics[width=0.28\textwidth]
{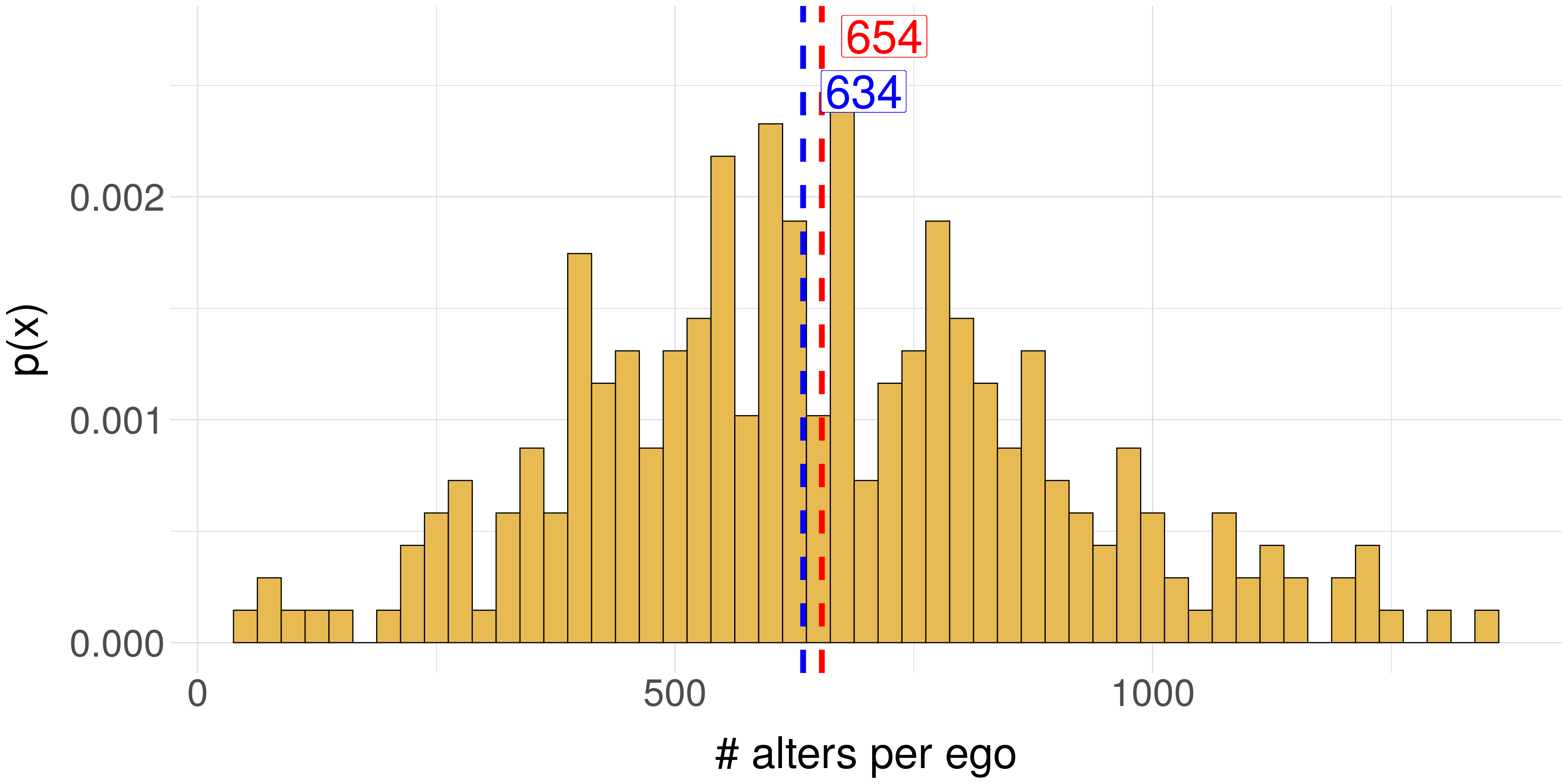}}
\hfill
\subfloat[Greece
\label{fig_appendix:totalsize_GreekJournalists}]
{\includegraphics[width=0.28\textwidth]
{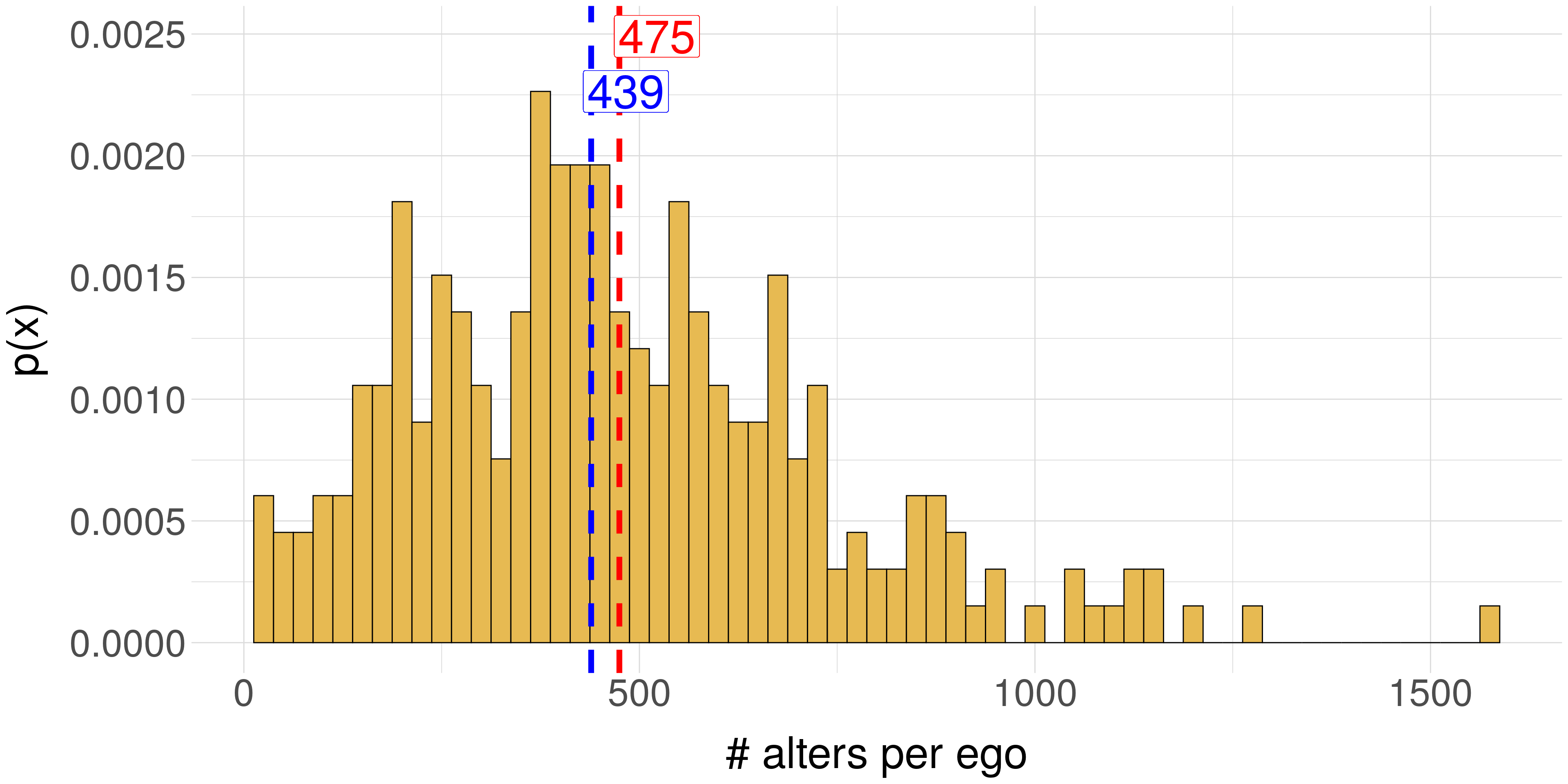}}
\hfill
\subfloat[Italy
\label{fig_appendix:totalsize_ItalianJournalists}]
{\includegraphics[width=0.28\textwidth]
{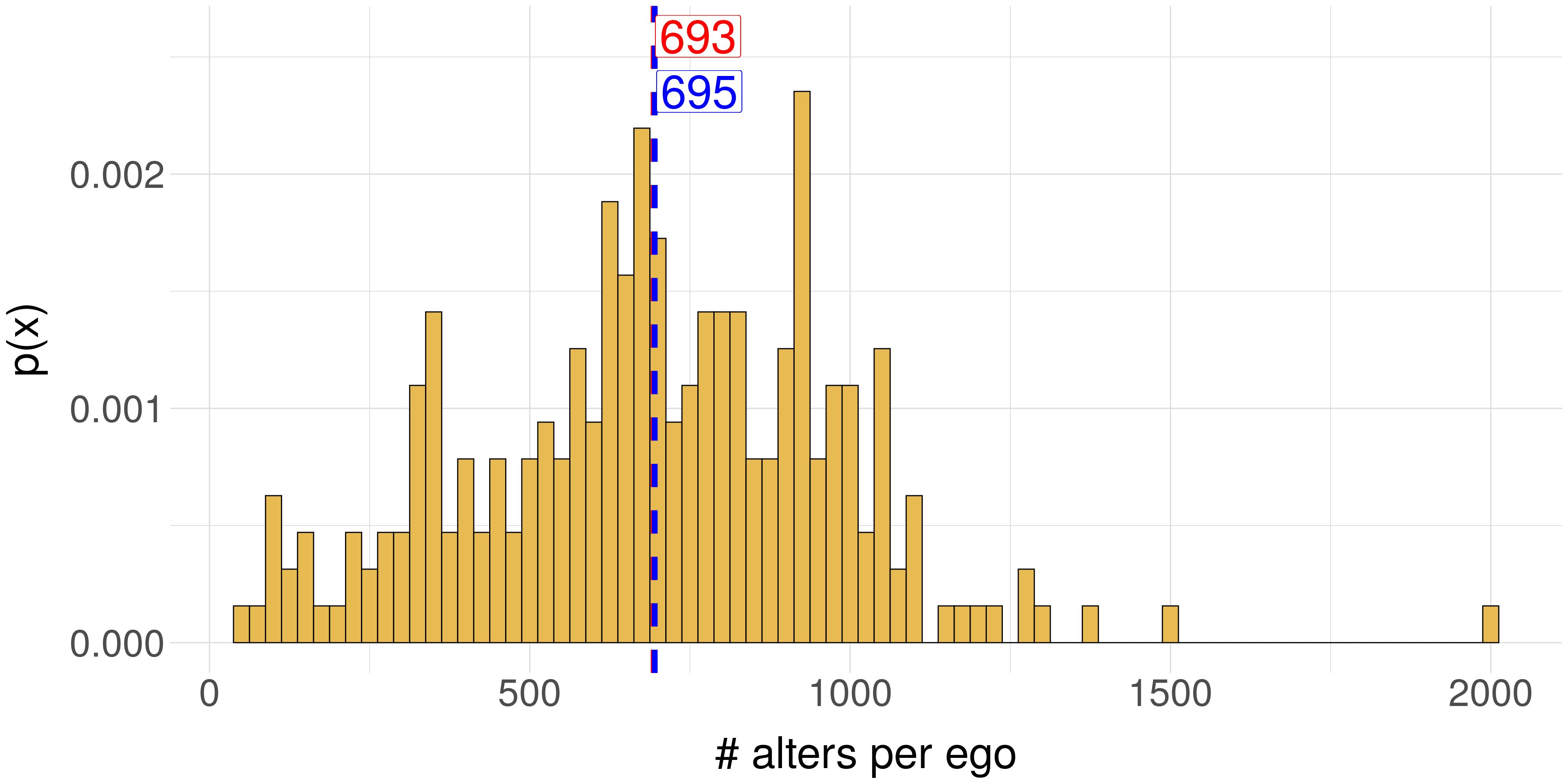}}
\hfill
\subfloat[Spain
\label{fig_appendix:totalsize_SpanishJournalists}]
{\includegraphics[width=0.28\textwidth]
{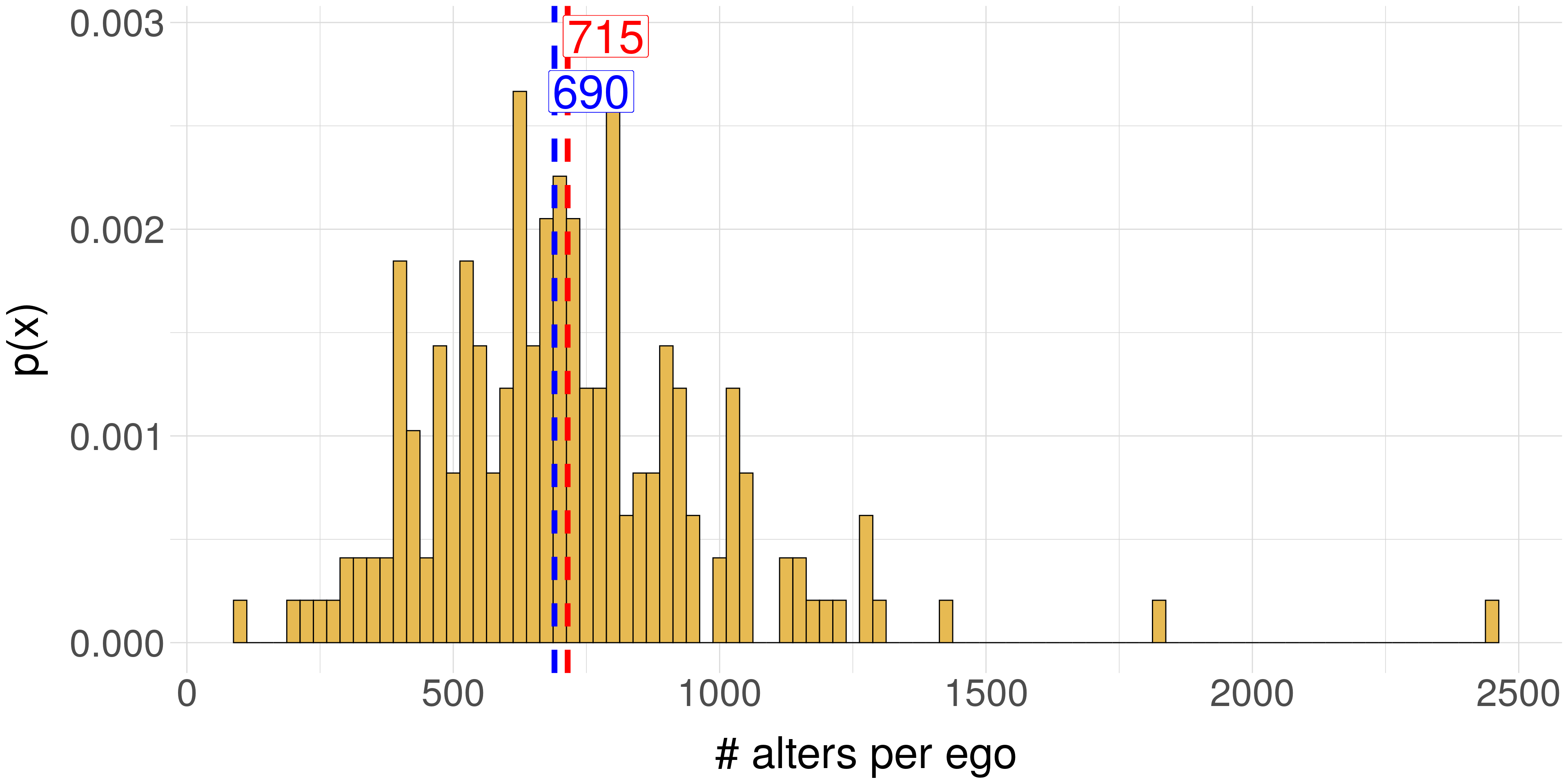}}
\hfill
\subfloat[France
\label{fig_appendix:totalsize_FrenchJournalists}]
{\includegraphics[width=0.28\textwidth]
{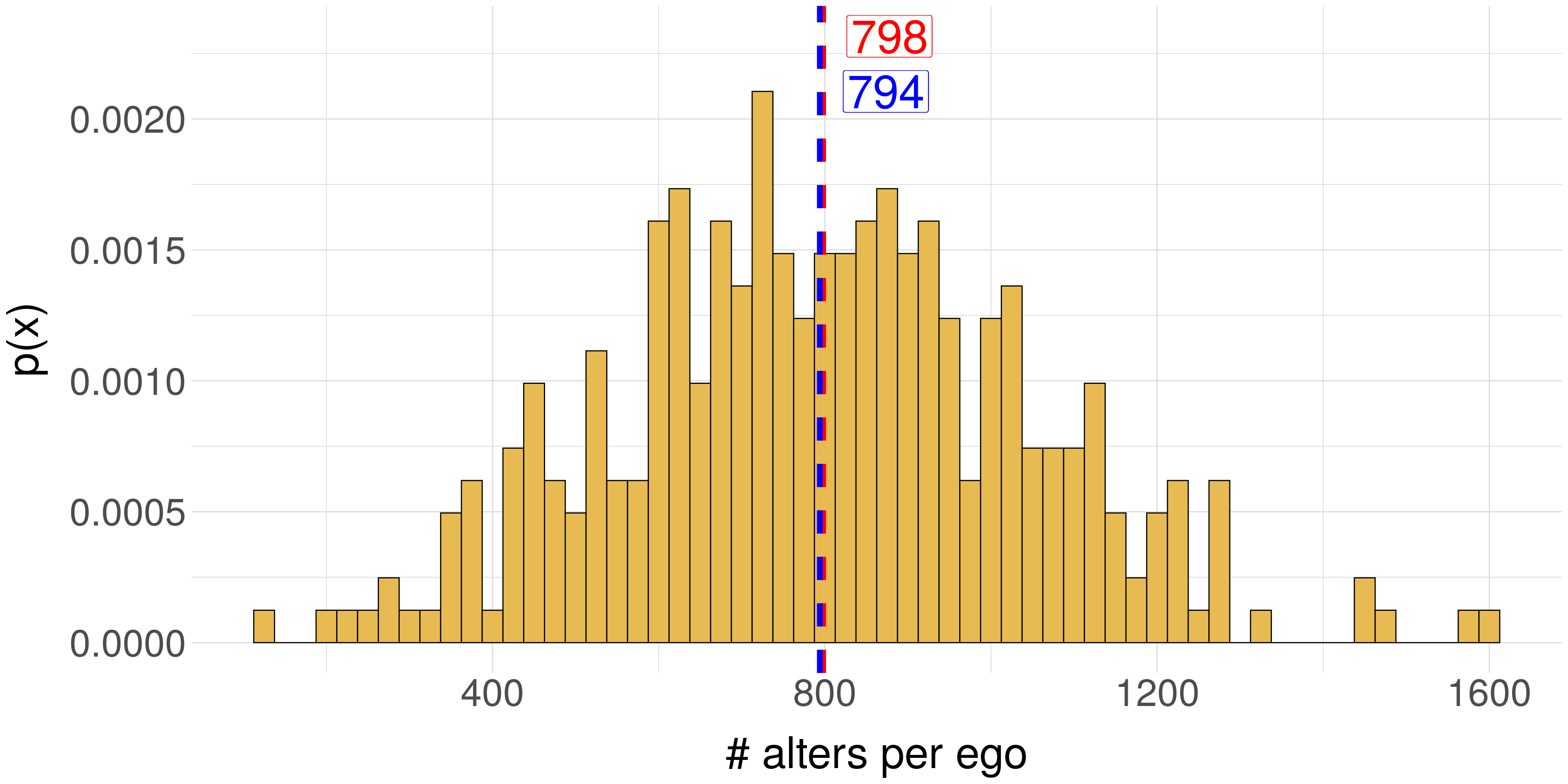}}
\hfill
\subfloat[Germany
\label{fig_appendix:totalsize_GermanJournalists}]
{\includegraphics[width=0.28\textwidth]
{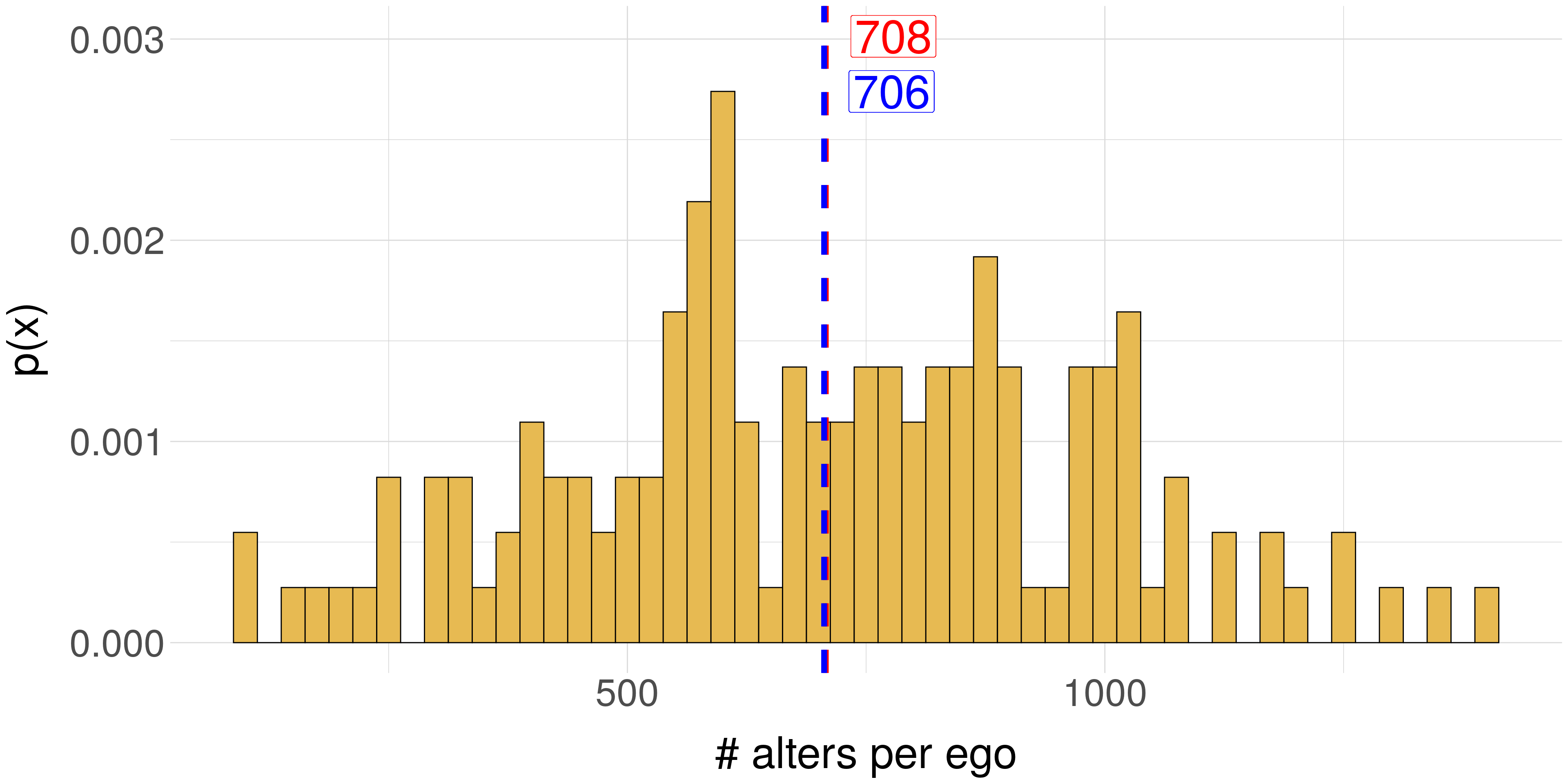}}
\hfill
\subfloat[Netherland
\label{fig_appendix:totalsize_NetherlanderJournalists}]
{\includegraphics[width=0.28\textwidth]
{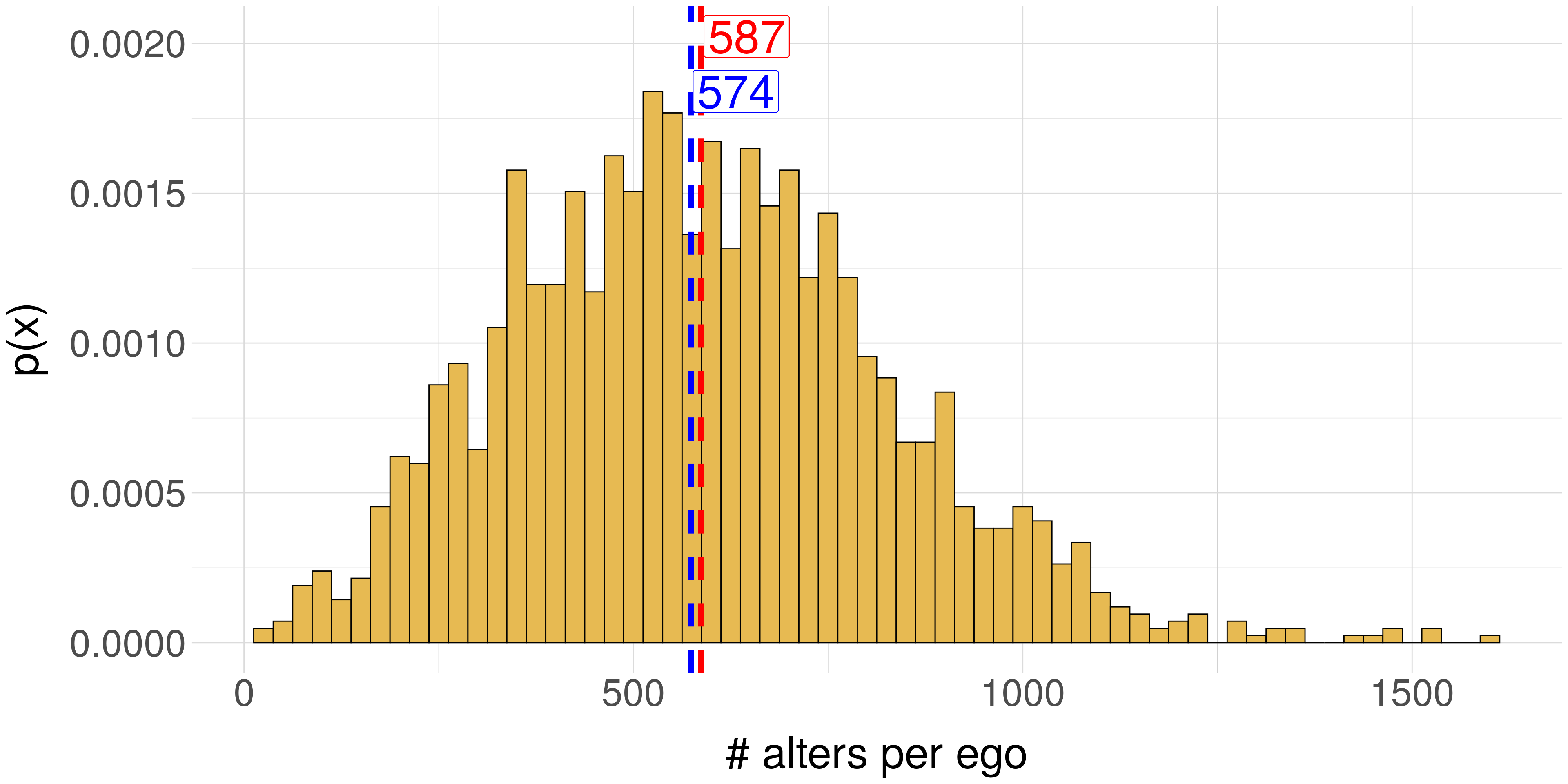}}
\hspace{1pt}
\subfloat[Australia
\label{fig_appendix:totalsize_AustralianJournalists}]
{\includegraphics[width=0.28\textwidth]
{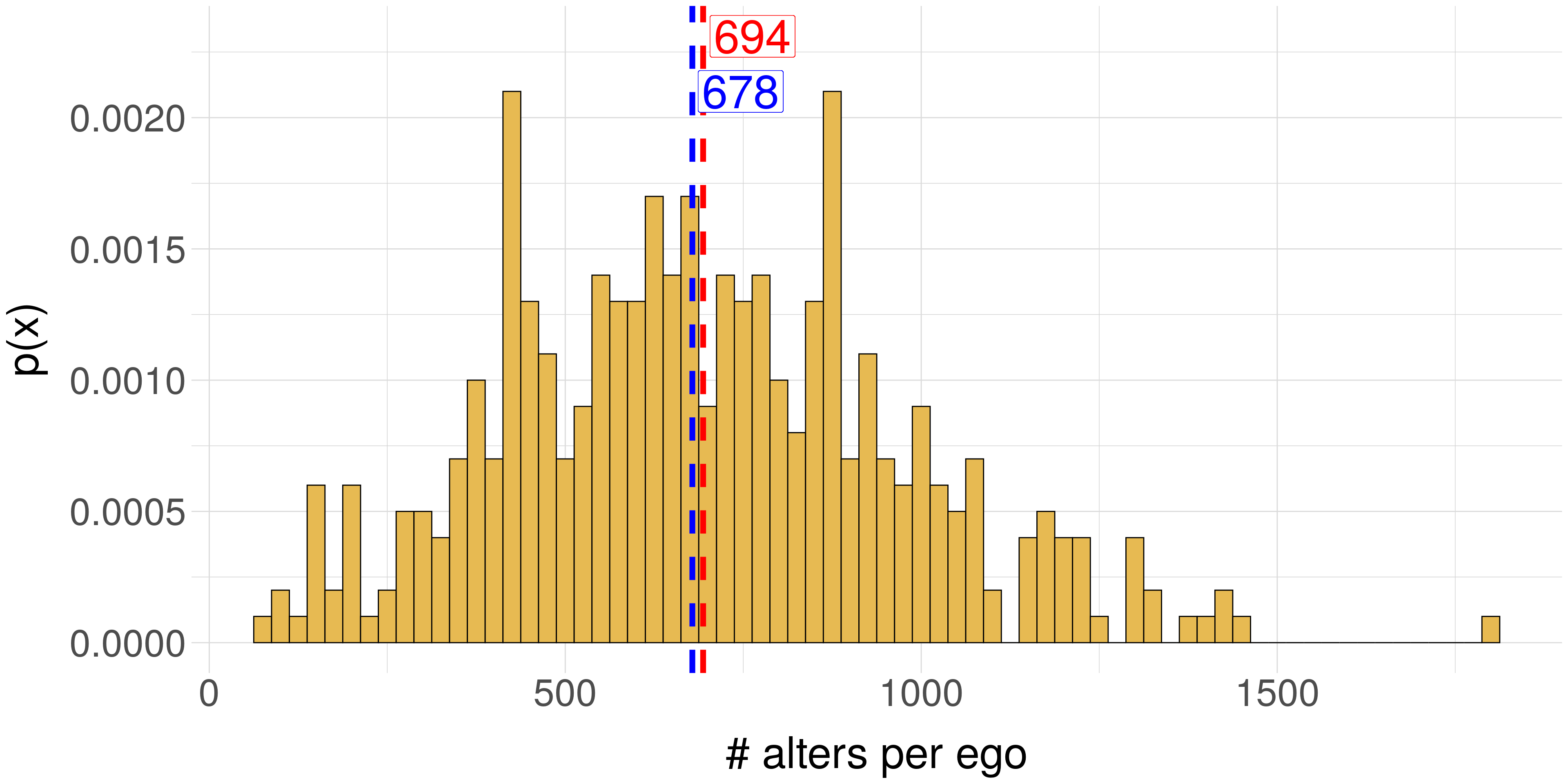}}
\end{center}
\end{adjustbox}
\caption{Distribution of ego network size, per country}
\label{fig_appendix:totalsize}
\end{figure}



\begin{figure}[!h]
\begin{adjustbox}{minipage=\linewidth}
\begin{center}
\subfloat[USA
\label{fig_appendix:activenet_size_AmericanJournalists}]
{\includegraphics[width=0.27\textwidth]
{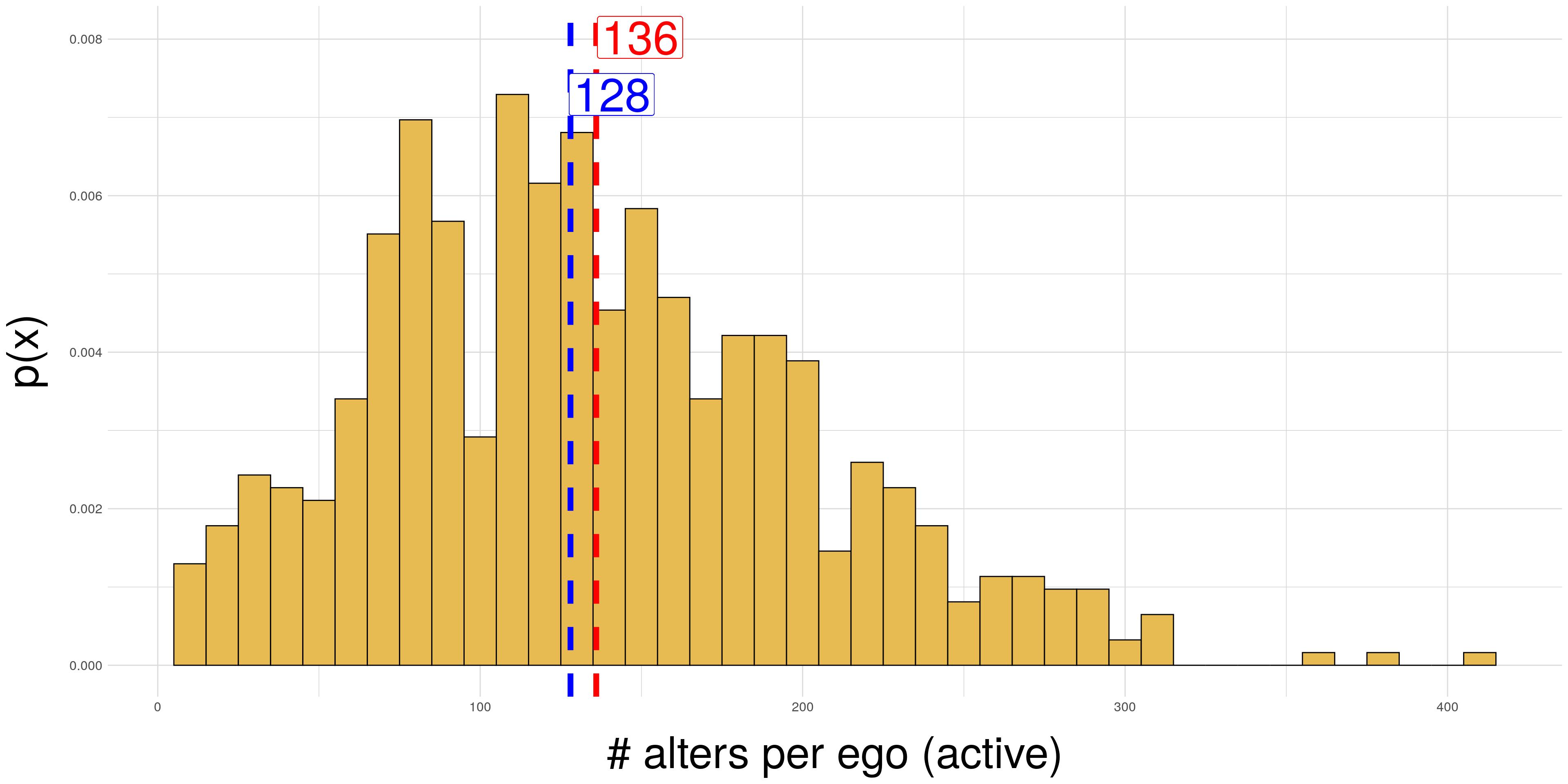}}
\hfill
\subfloat[Canada
\label{fig_appendix:activenet_size_CanadianJournalists}]
{\includegraphics[width=0.27\textwidth]
{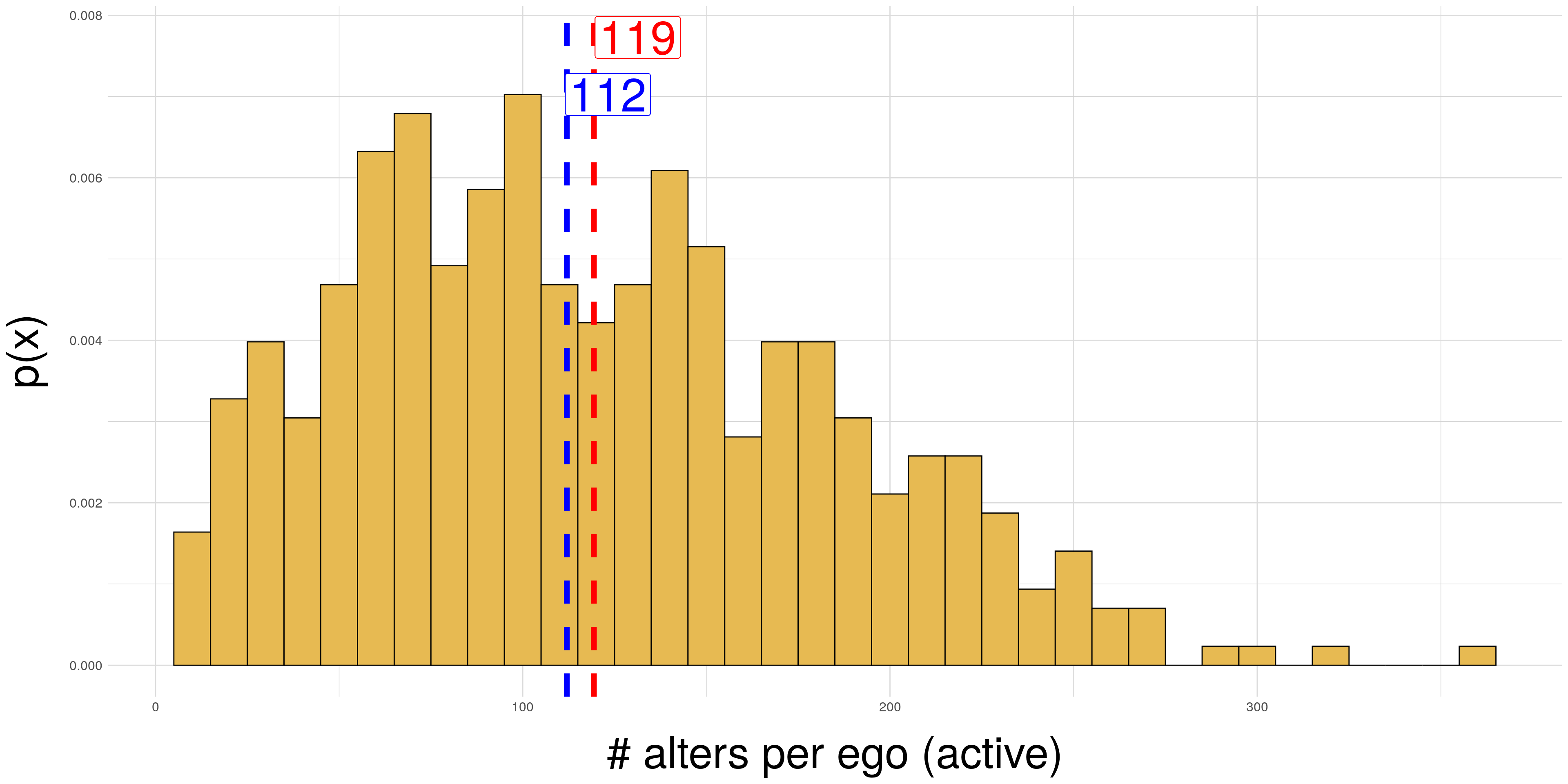}}
\hfill
\subfloat[Brasil
\label{fig_appendix:activenet_size_BrazilianJournalists}]
{\includegraphics[width=0.27\textwidth]
{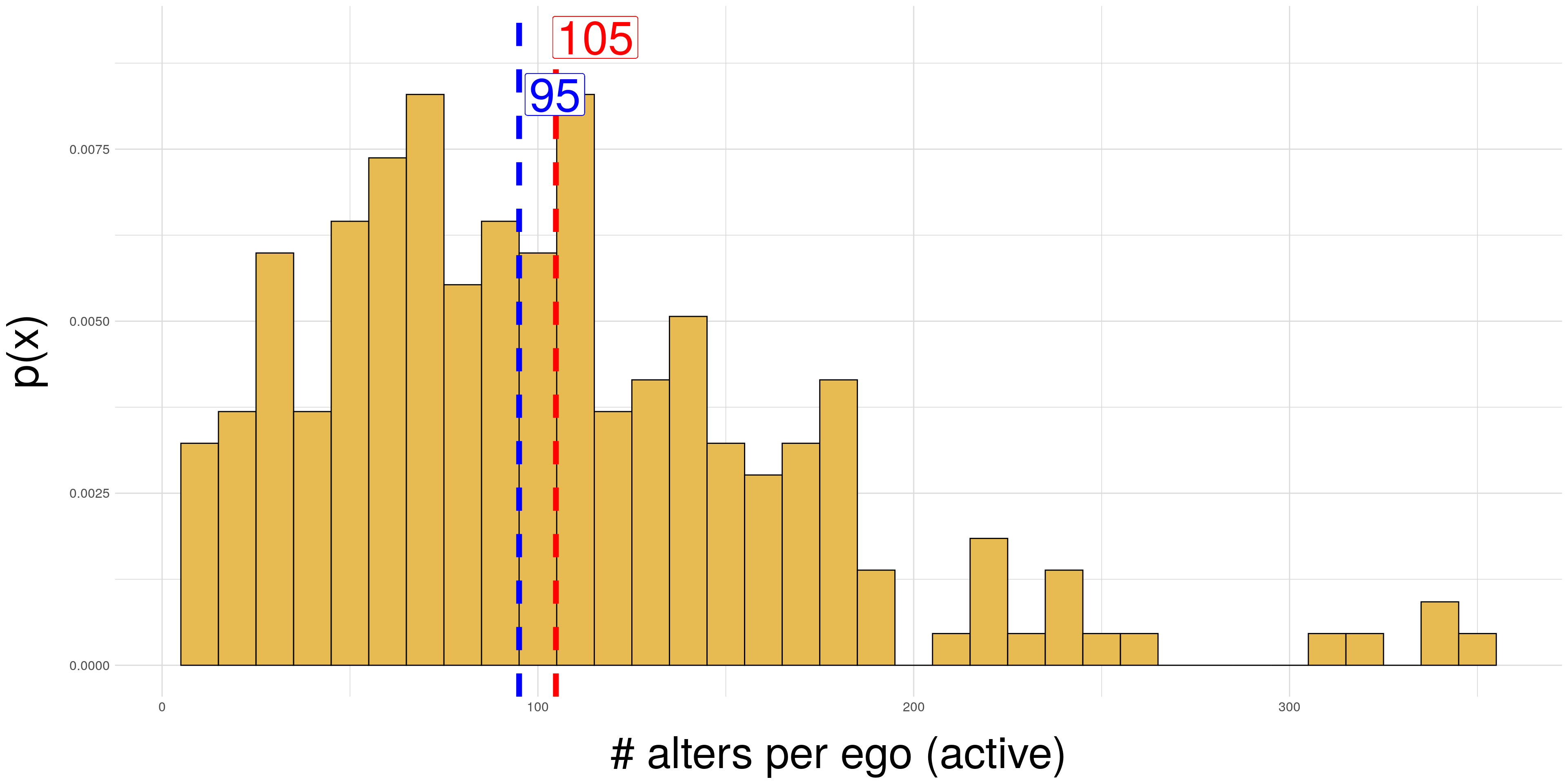}}
\hfill
\subfloat[Japan
\label{fig_appendix:activenet_size_JapaneseJournalists}]
{\includegraphics[width=0.27\textwidth]
{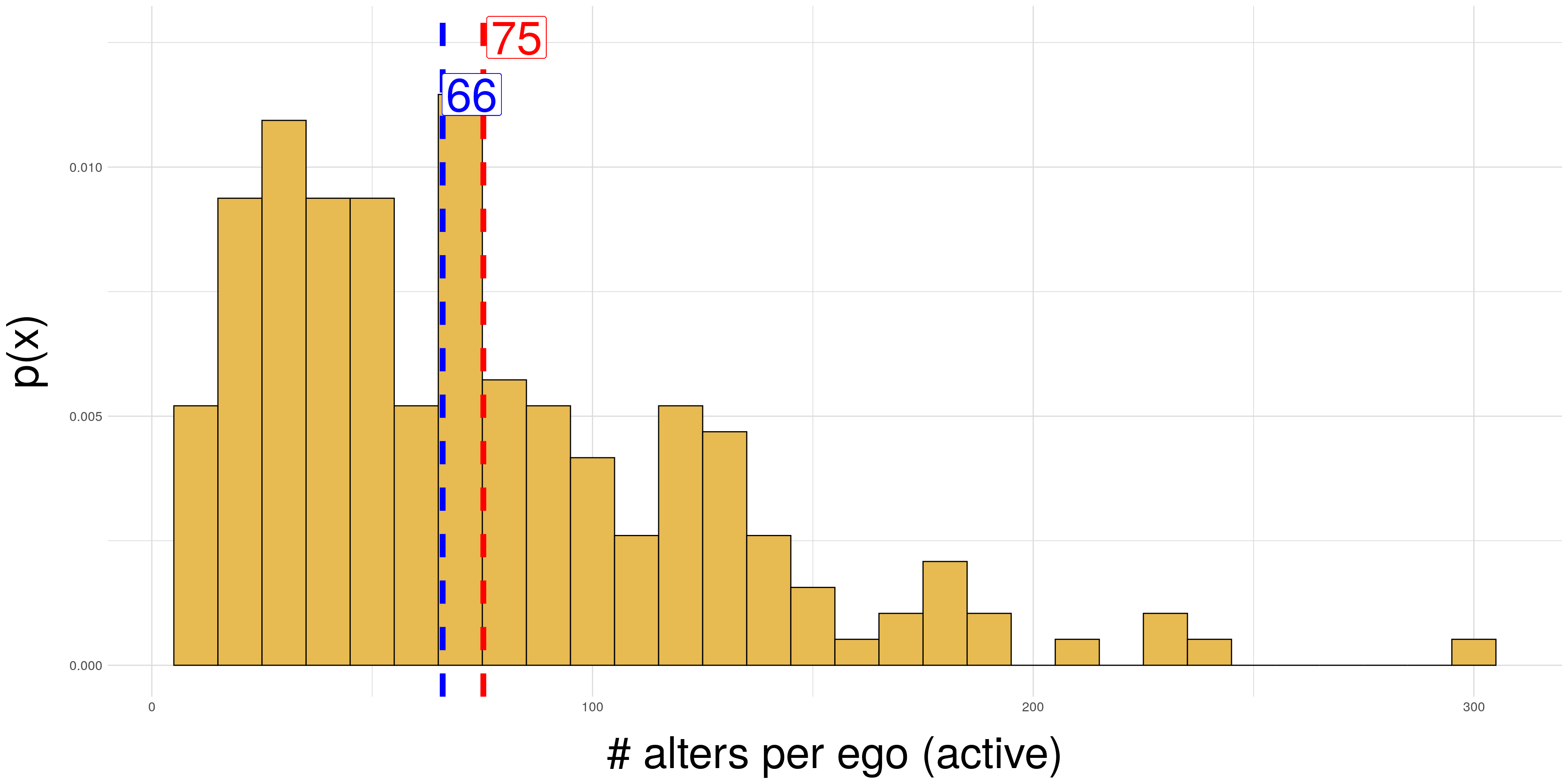}}
\hfill
\subfloat[Turkey
\label{fig_appendix:activenet_size_TrukishJournalists}]
{\includegraphics[width=0.27\textwidth]
{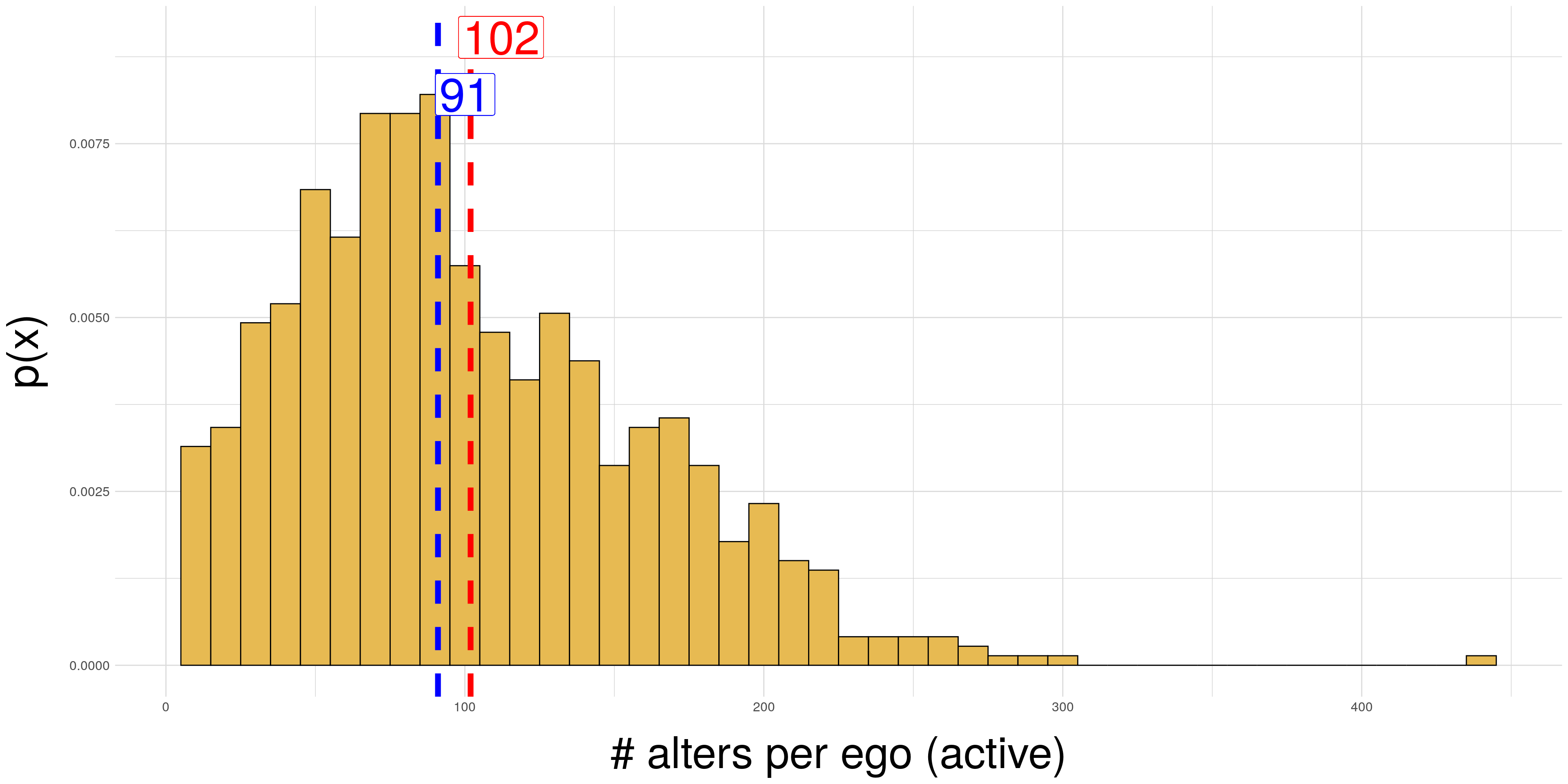}}
\hfill
\subfloat[UK
\label{fig_appendix:activenet_size_BritishJournalists}]
{\includegraphics[width=0.27\textwidth]
{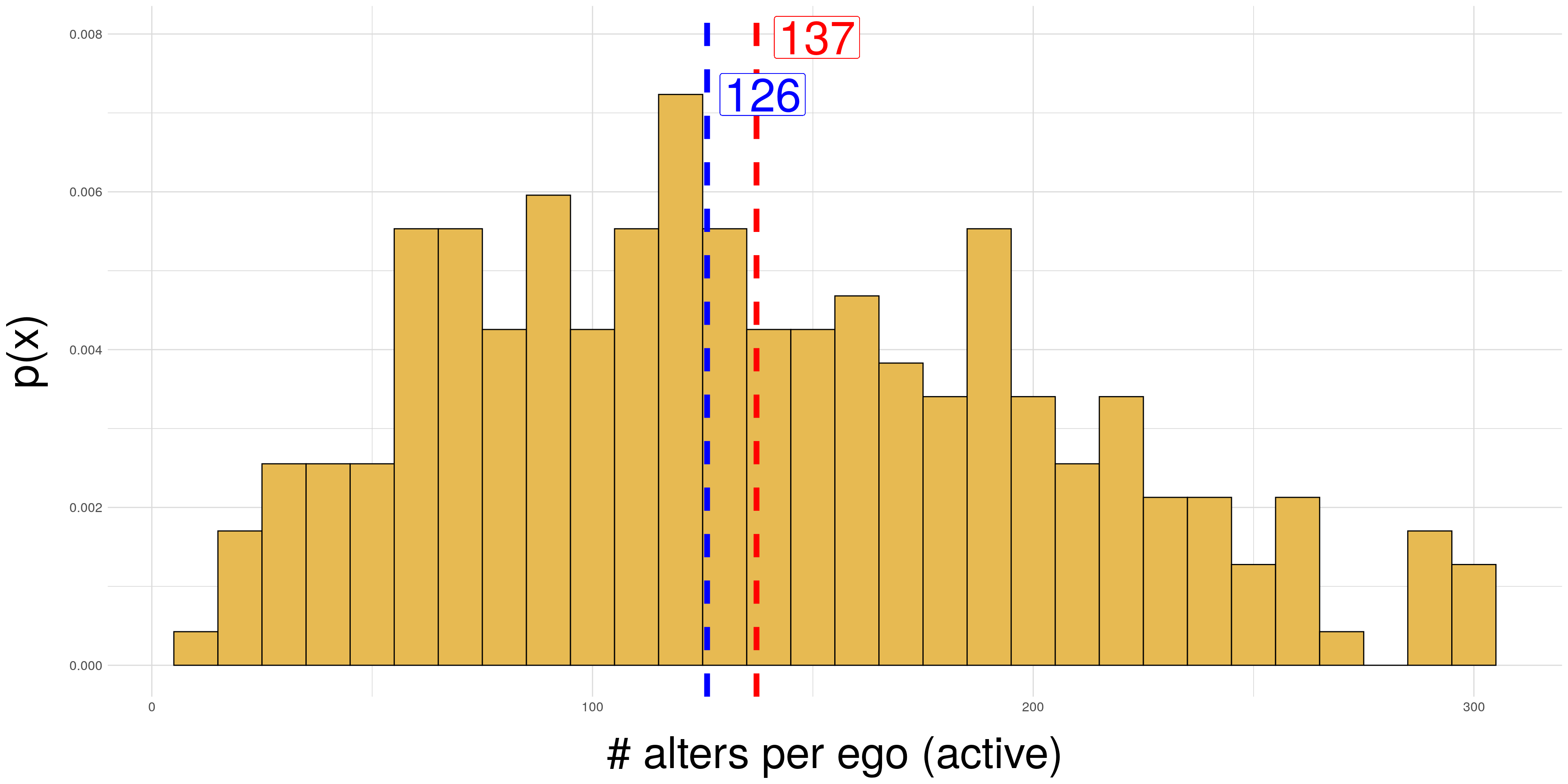}}
\hfill
\subfloat[Denmark
\label{fig_appendix:activenet_size_DanishJournalists}]
{\includegraphics[width=0.27\textwidth]
{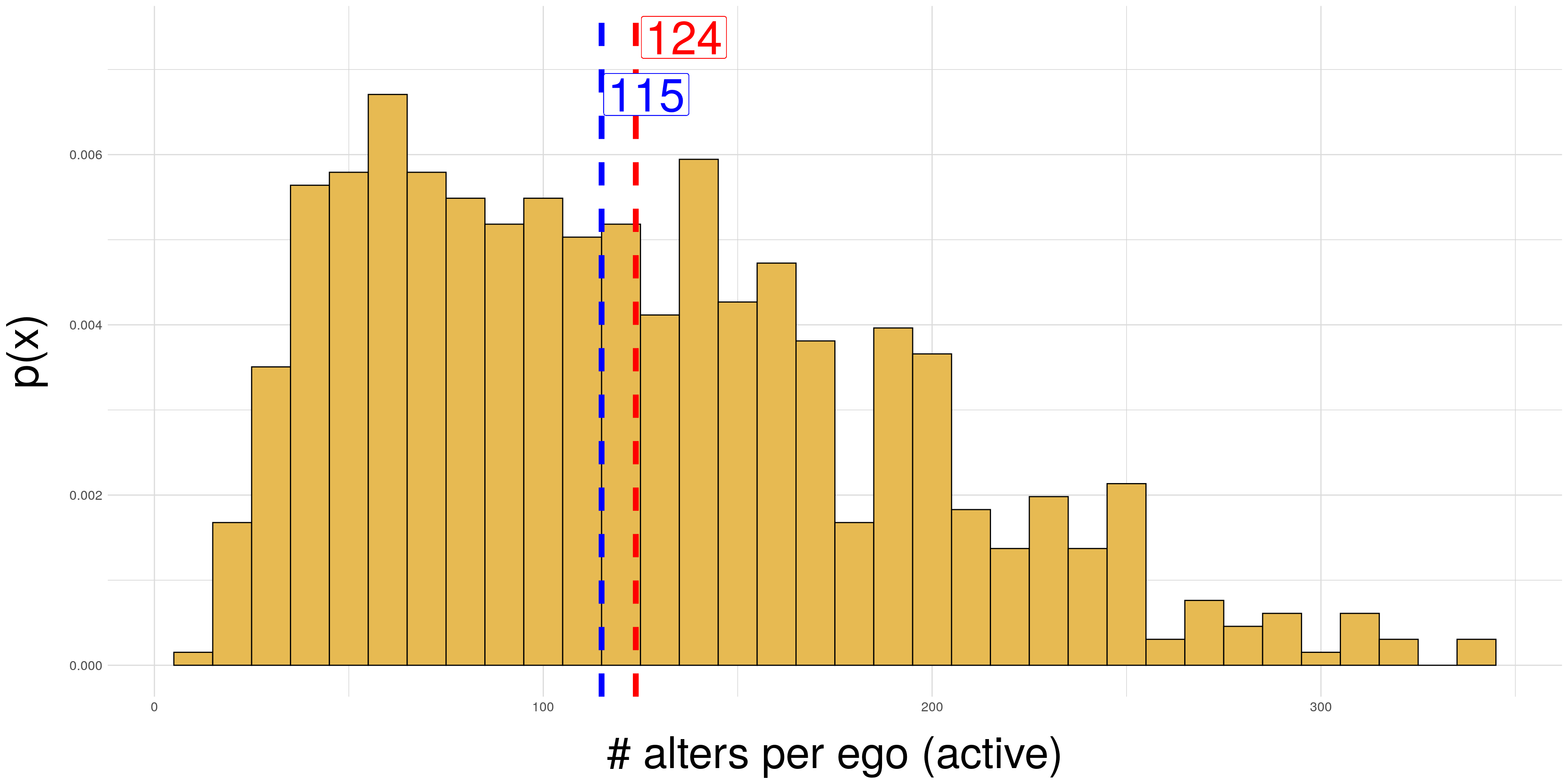}}
\hfill
\subfloat[Finland
\label{fig_appendix:activenet_size_FinnishJournalists}]
{\includegraphics[width=0.27\textwidth]
{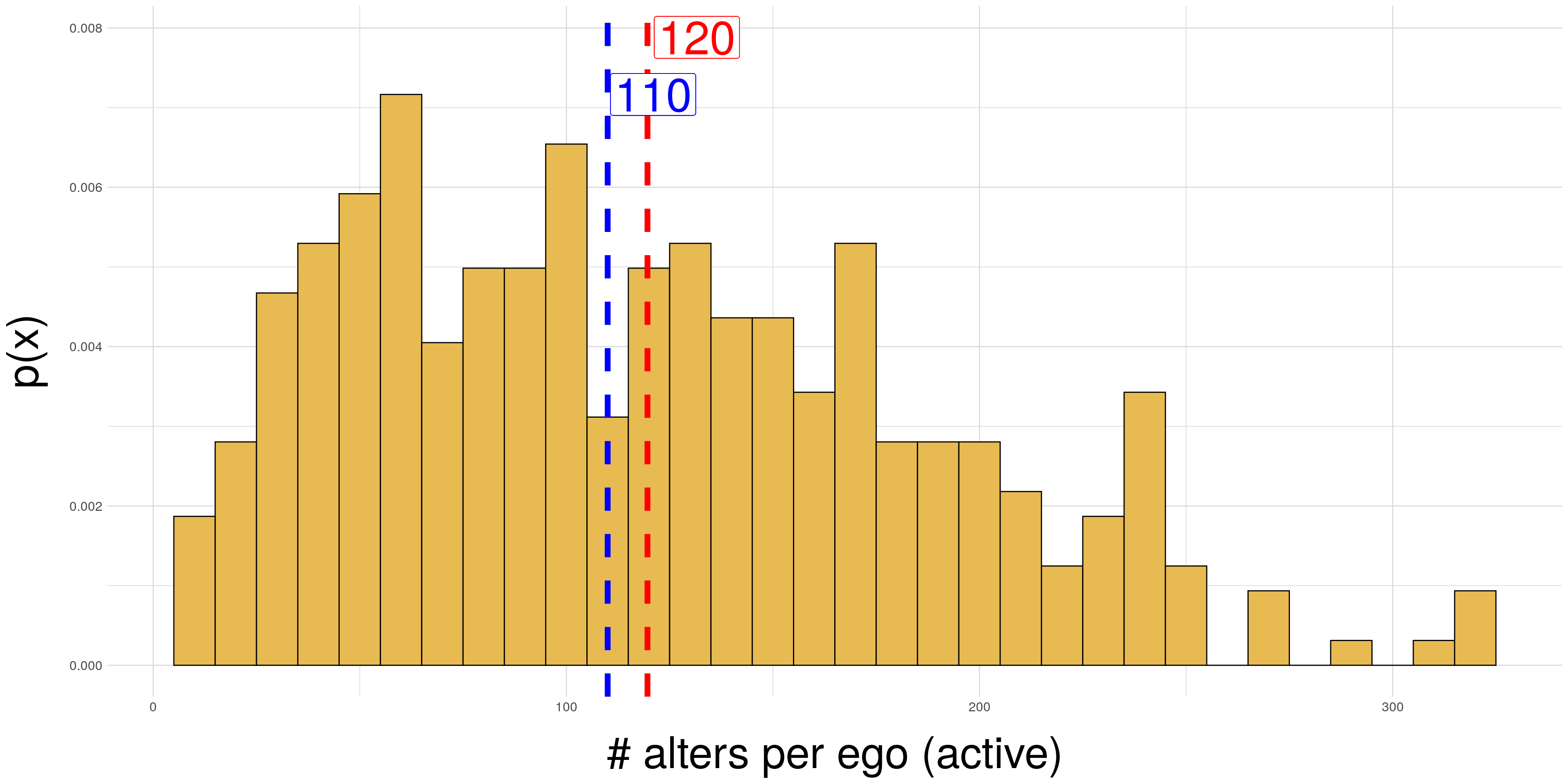}}
\hfill
\subfloat[Norway
\label{fig_appendix:activenet_size_NorwegianJournalists}]
{\includegraphics[width=0.27\textwidth]
{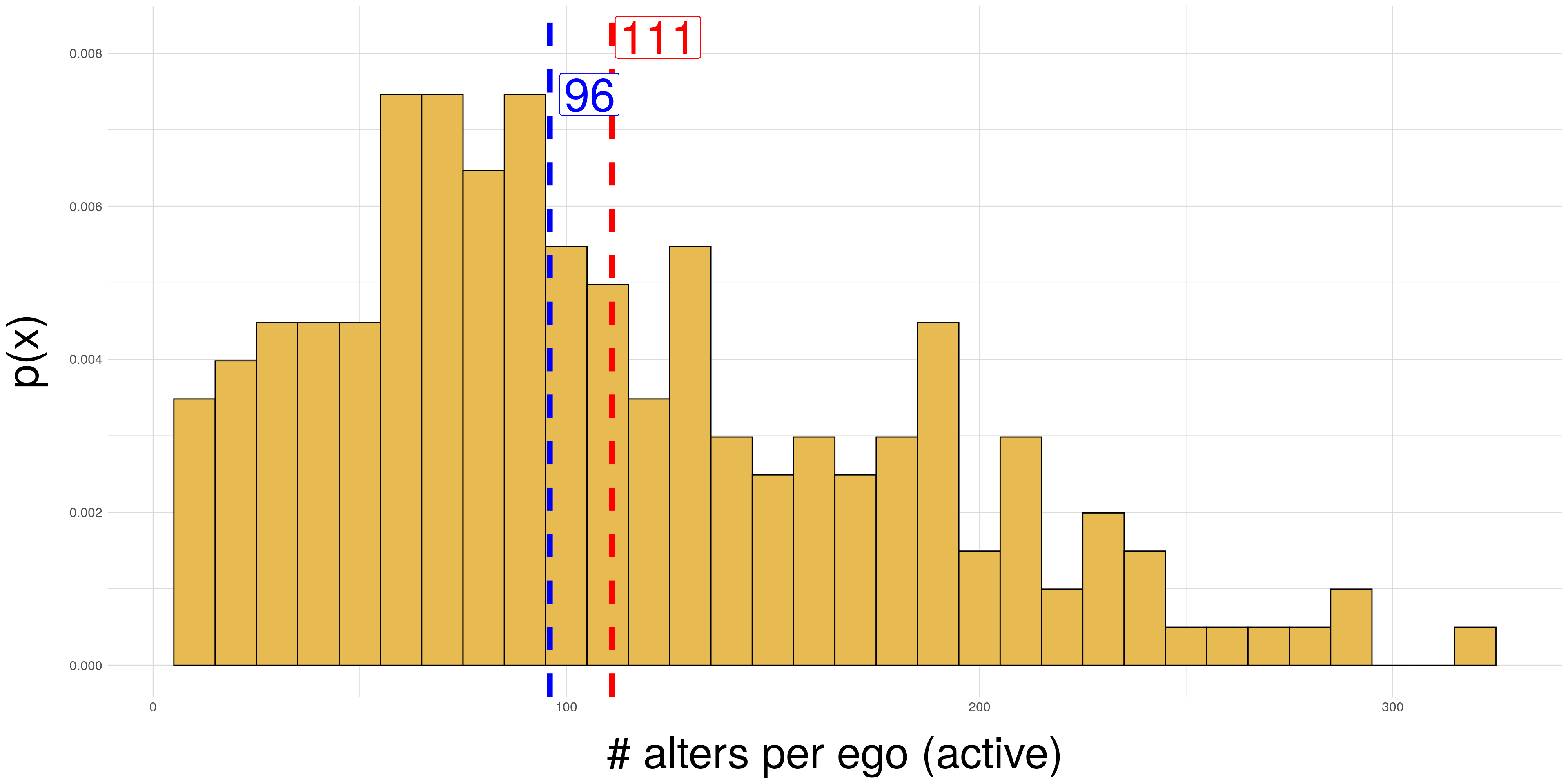}}
\hfill
\subfloat[Sweden
\label{fig_appendix:activenet_size_SwedishJournalists}]
{\includegraphics[width=0.27\textwidth]
{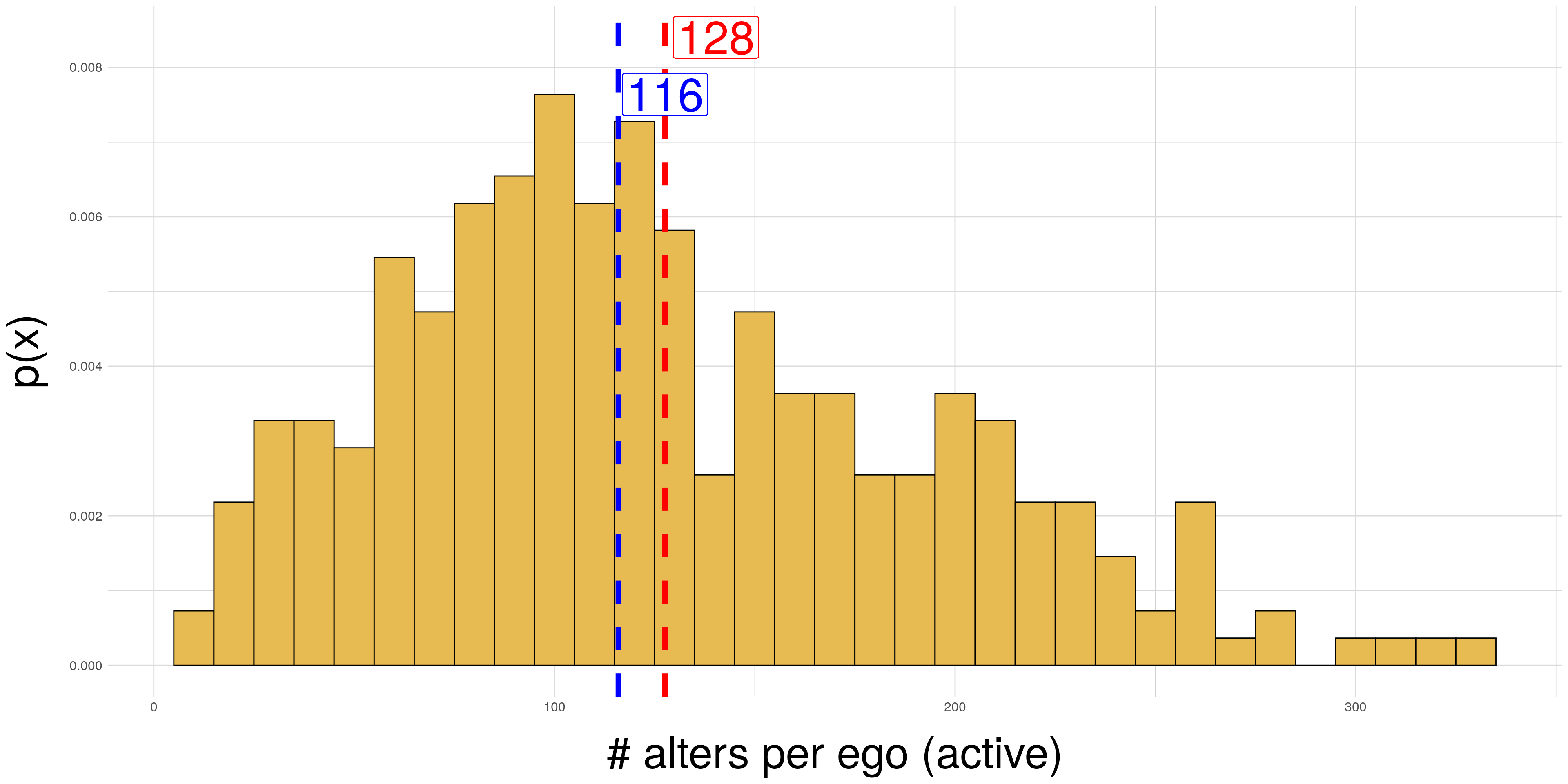}}
\hfill
\subfloat[Greece
\label{fig_appendix:activenet_size_GreekJournalists}]
{\includegraphics[width=0.27\textwidth]
{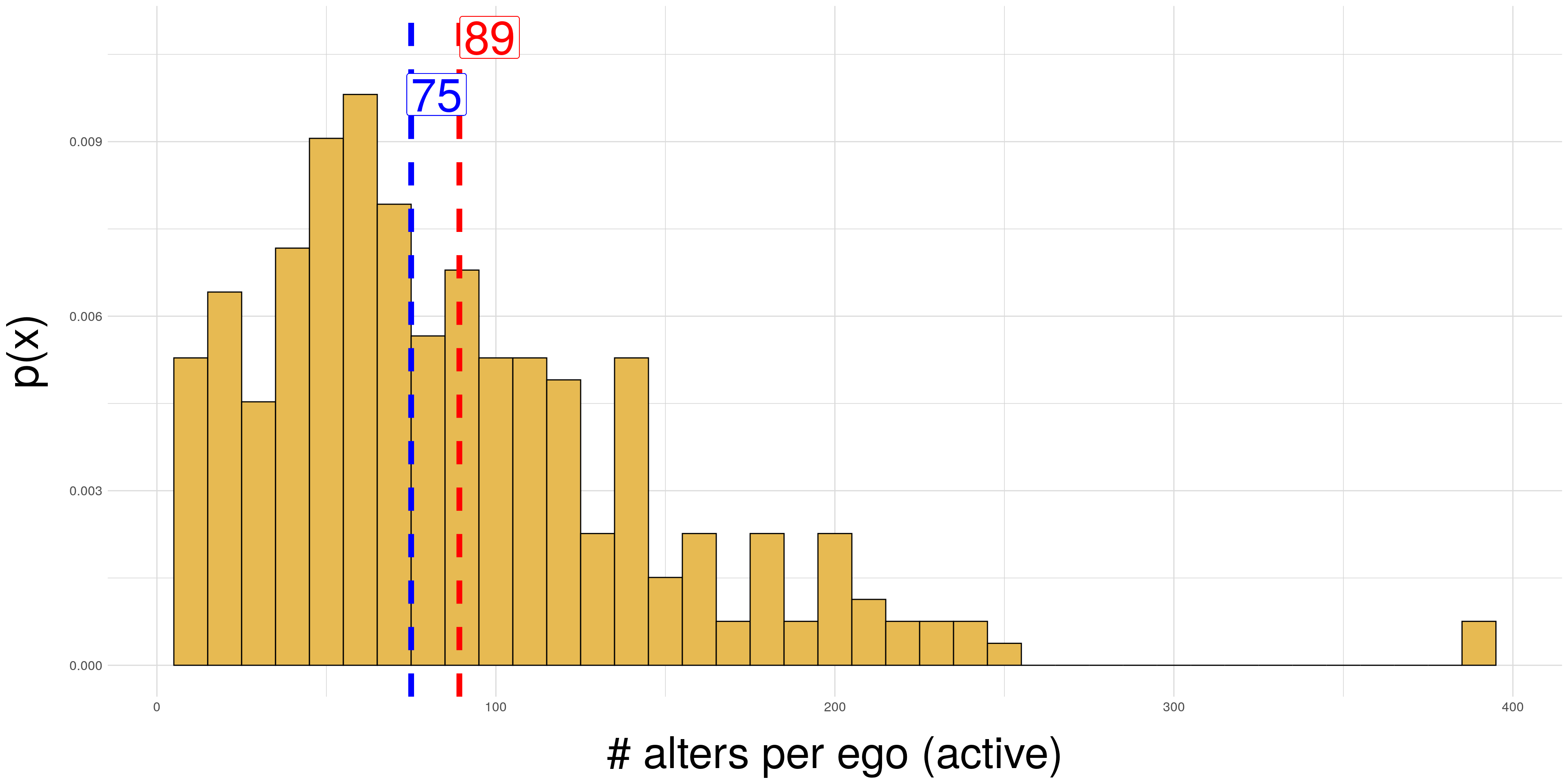}}
\hfill
\subfloat[Italy
\label{fig_appendix:activenet_size_ItalianJournalists}]
{\includegraphics[width=0.27\textwidth]
{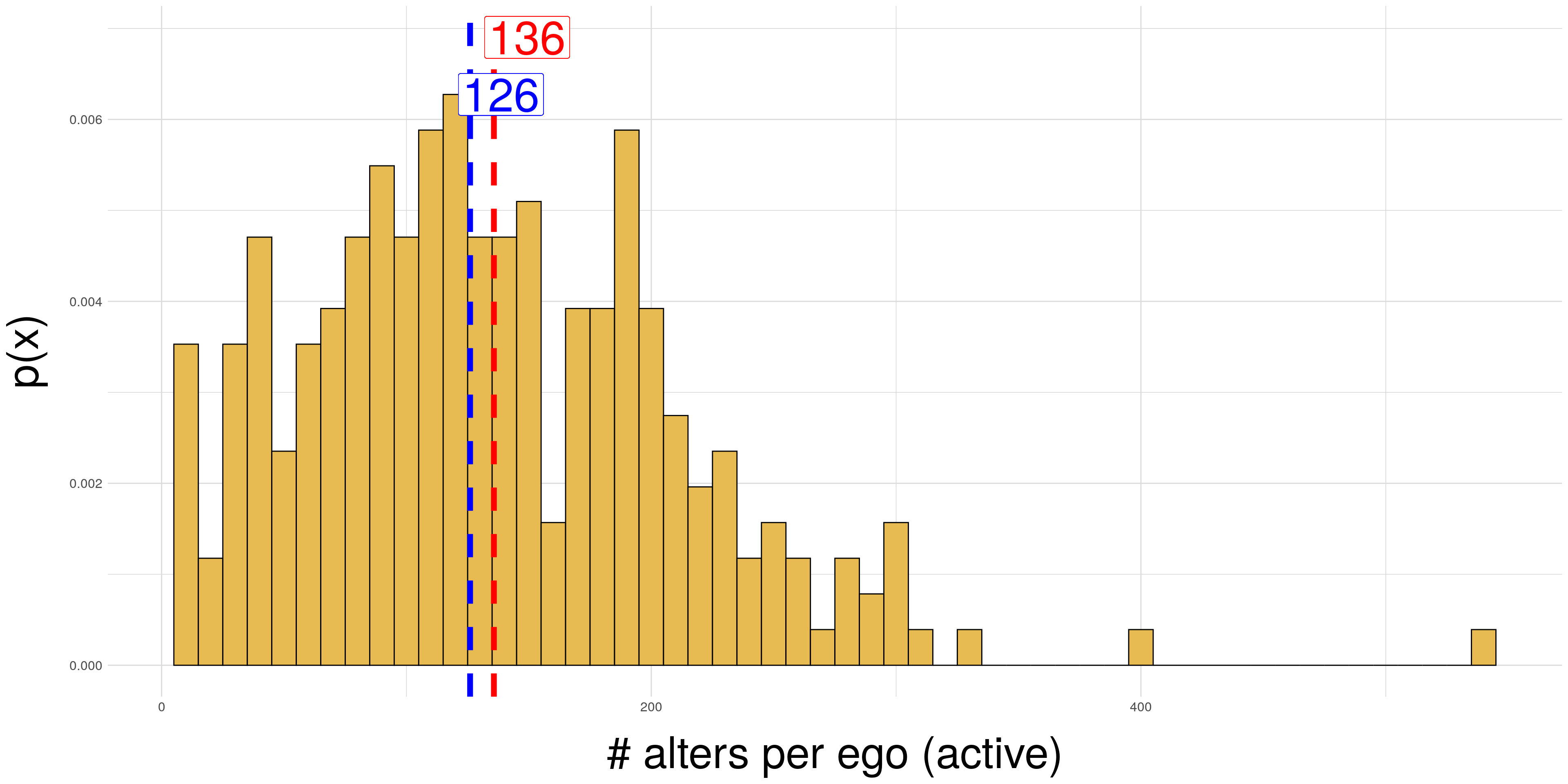}}
\hfill
\subfloat[Spain
\label{fig_appendix:activenet_size_SpanishJournalists}]
{\includegraphics[width=0.27\textwidth]
{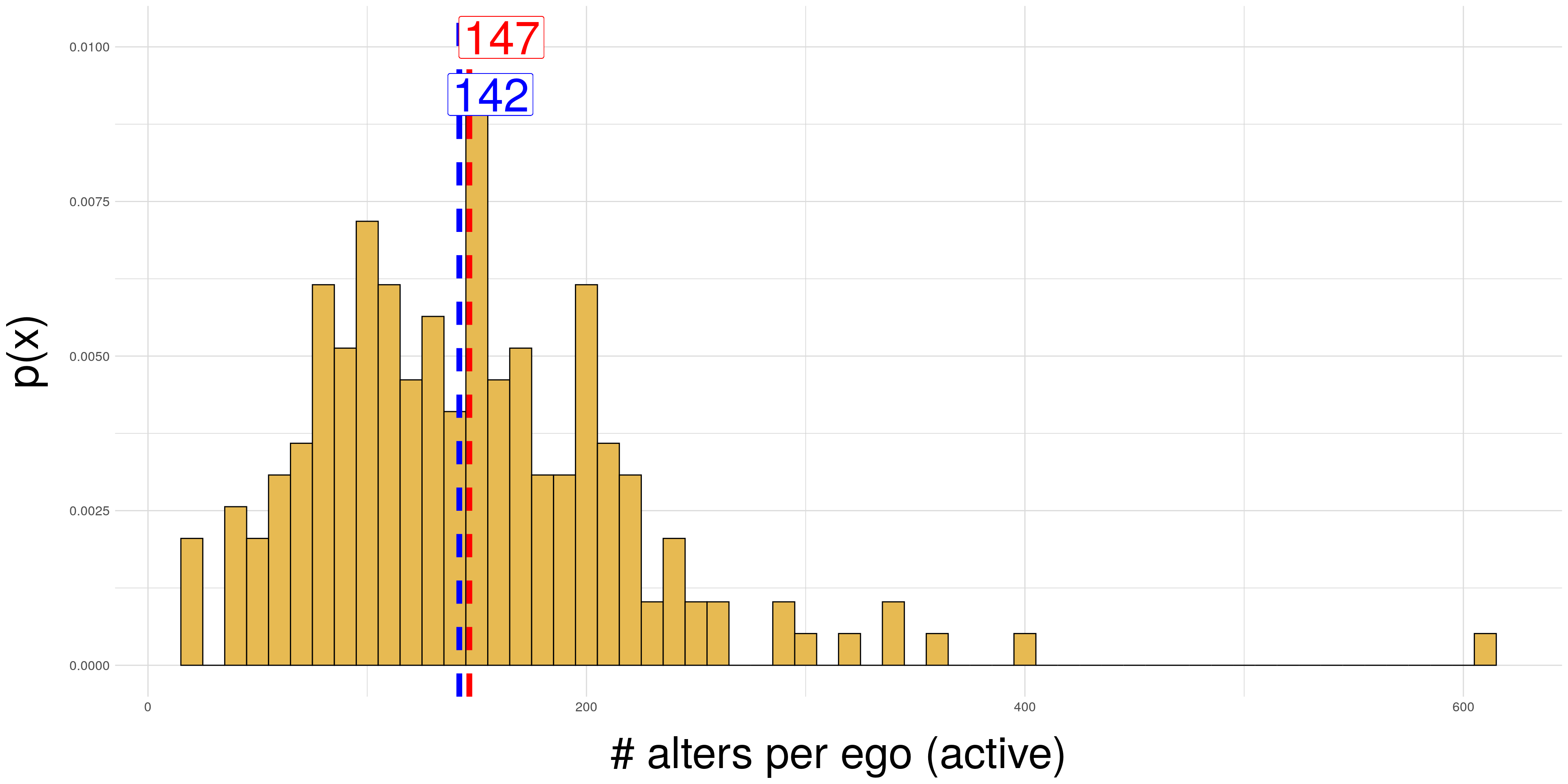}}
\hfill
\subfloat[France
\label{fig_appendix:activenet_size_FrenchJournalists}]
{\includegraphics[width=0.27\textwidth]
{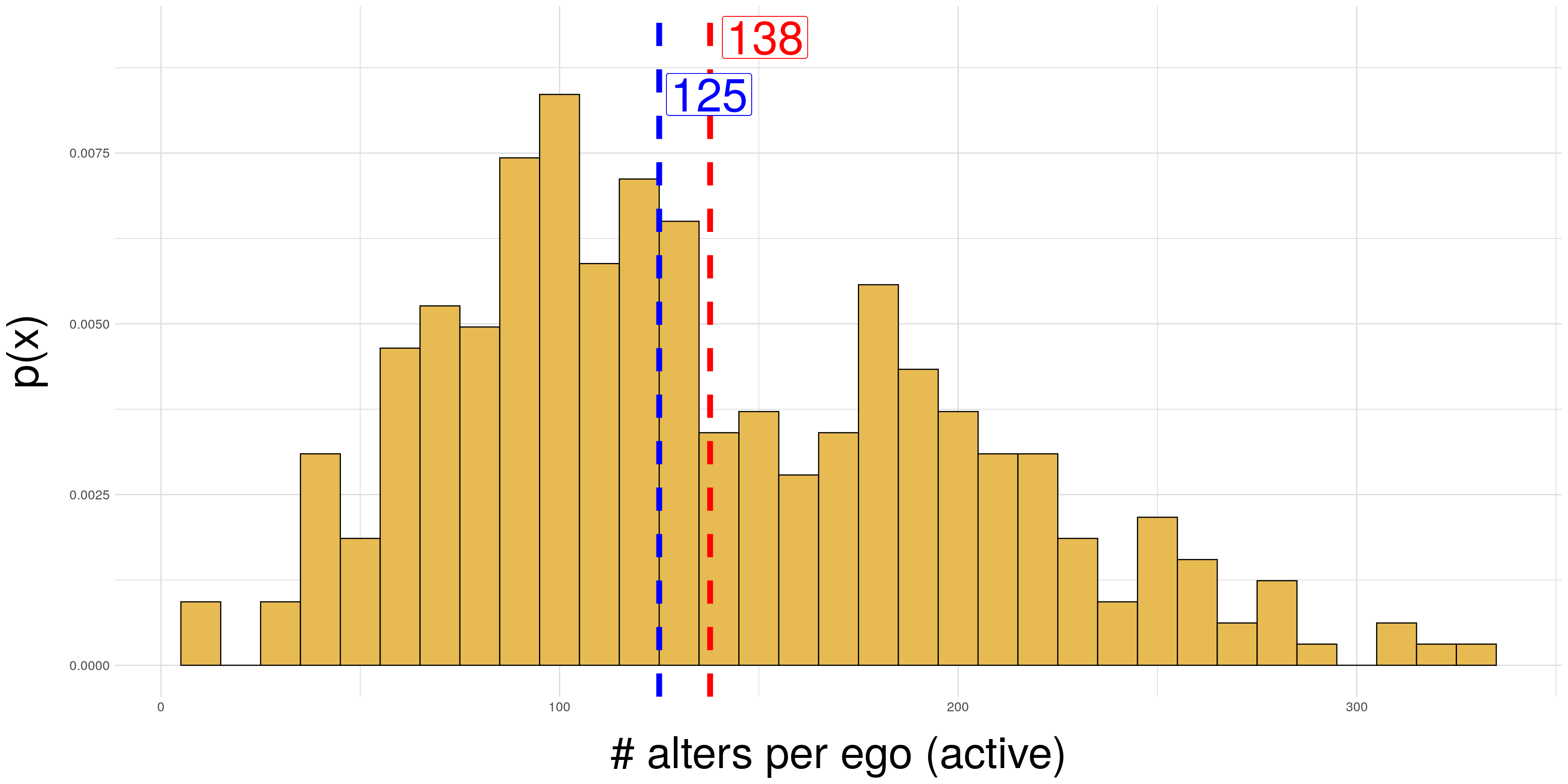}}
\hfill
\subfloat[Germany
\label{fig_appendix:activenet_size_GermanJournalists}]
{\includegraphics[width=0.27\textwidth]
{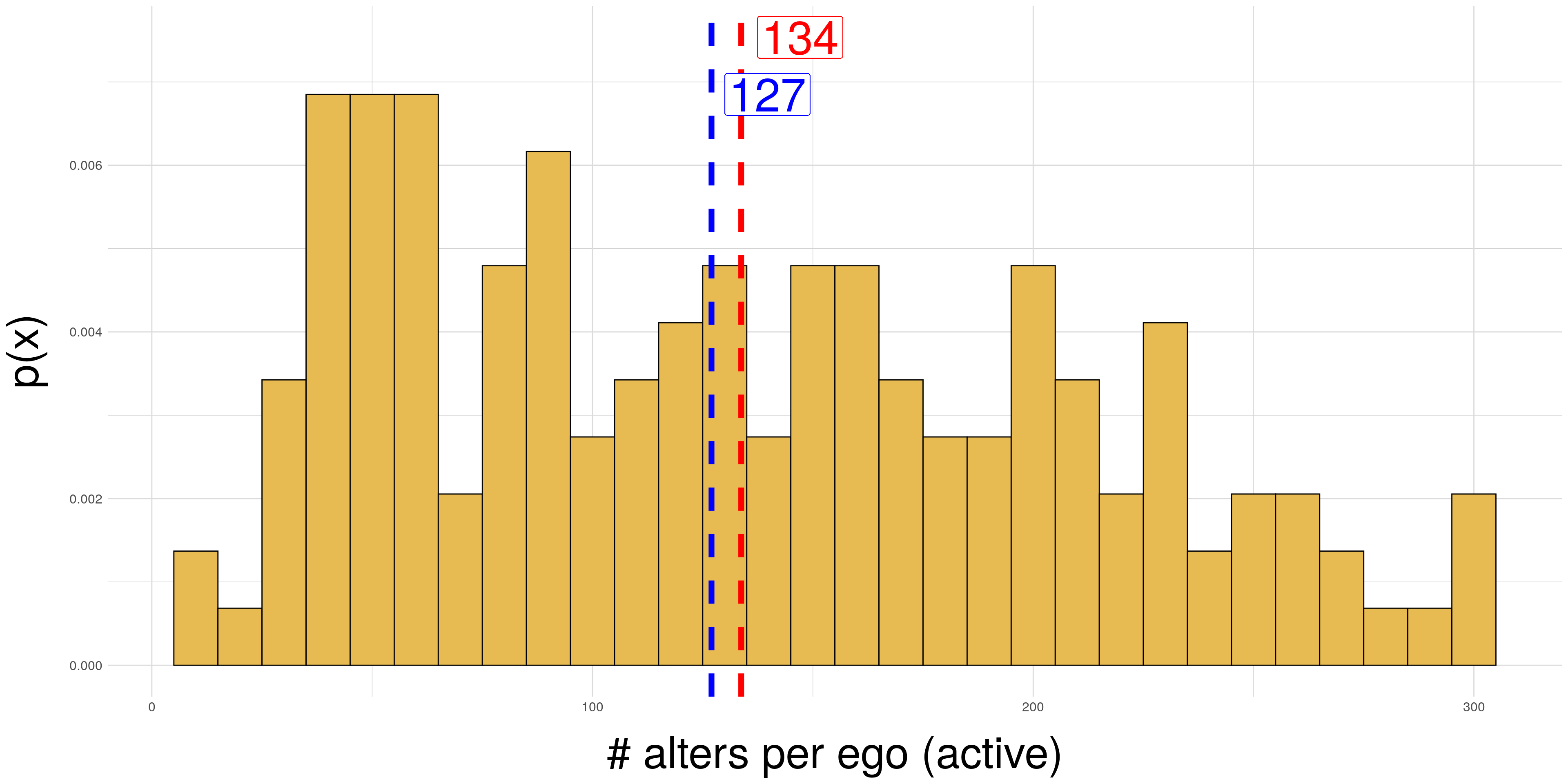}}
\hfill
\subfloat[Netherland
\label{fig_appendix:activenet_size_NetherlanderJournalists}]
{\includegraphics[width=0.27\textwidth]
{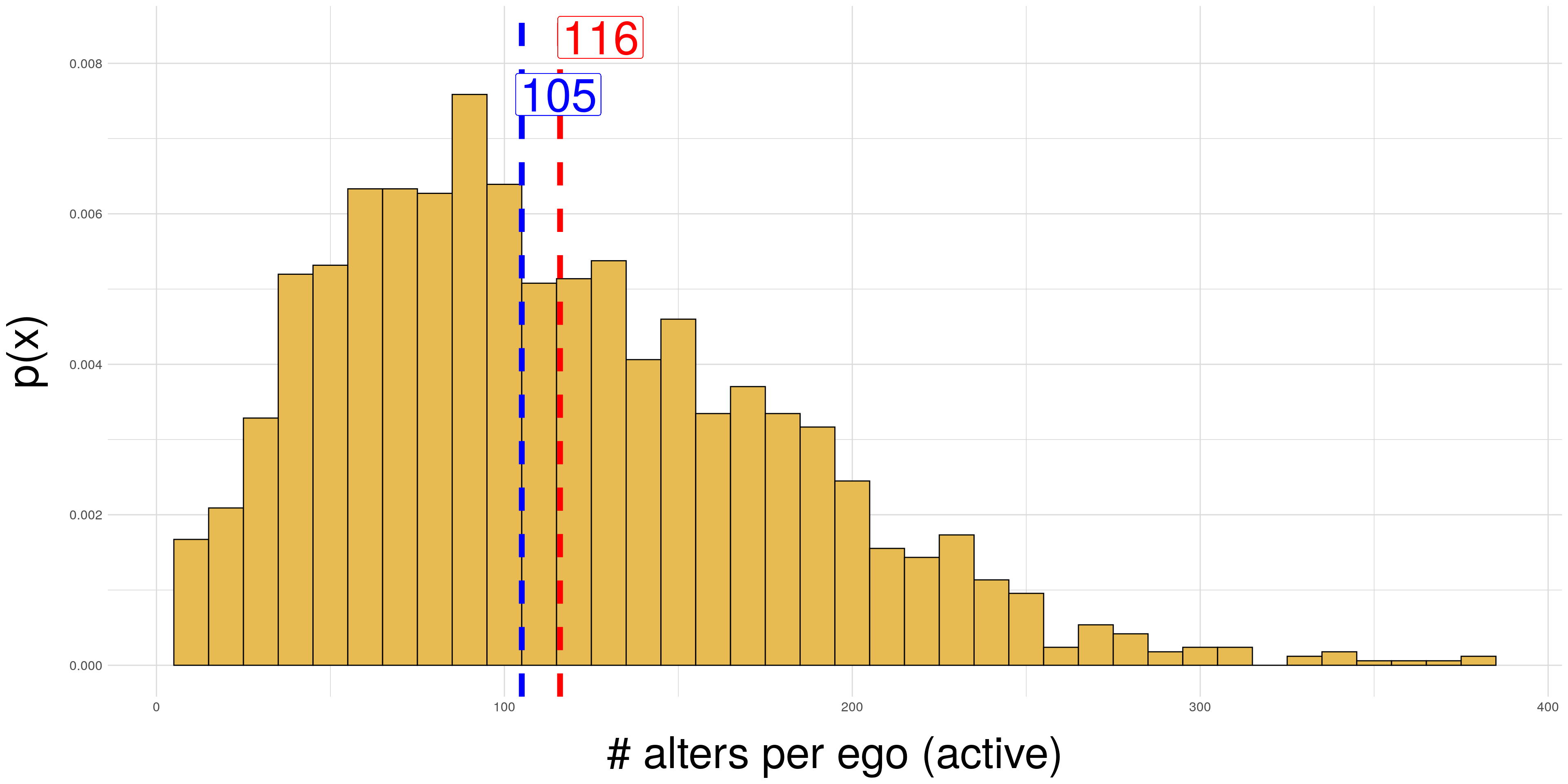}}
\hspace{1pt}
\subfloat[Australia
\label{fig_appendix:activenet_size_AustralianJournalists}]
{\includegraphics[width=0.27\textwidth]
{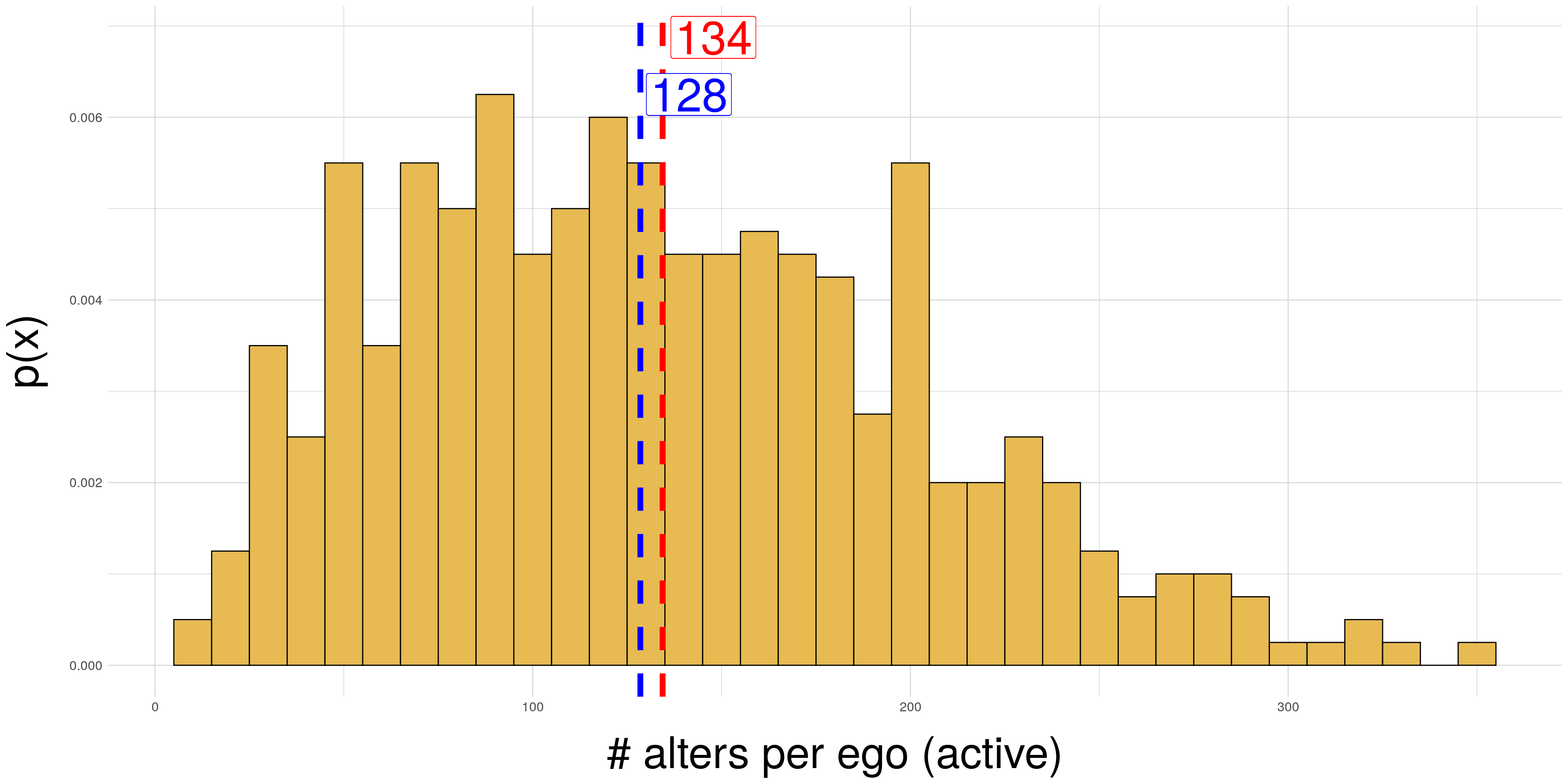}}
\end{center}
\end{adjustbox}
\caption{Distribution of active ego network size, per country}
\label{fig_appendix:activenet_size}
\end{figure}



\begin{figure}[!h]
\begin{adjustbox}{minipage=\linewidth}
\begin{center}
\subfloat[USA
\label{fig_appendix:optimal_circles_AmericanJournalists}]
{\includegraphics[width=0.27\textwidth]
{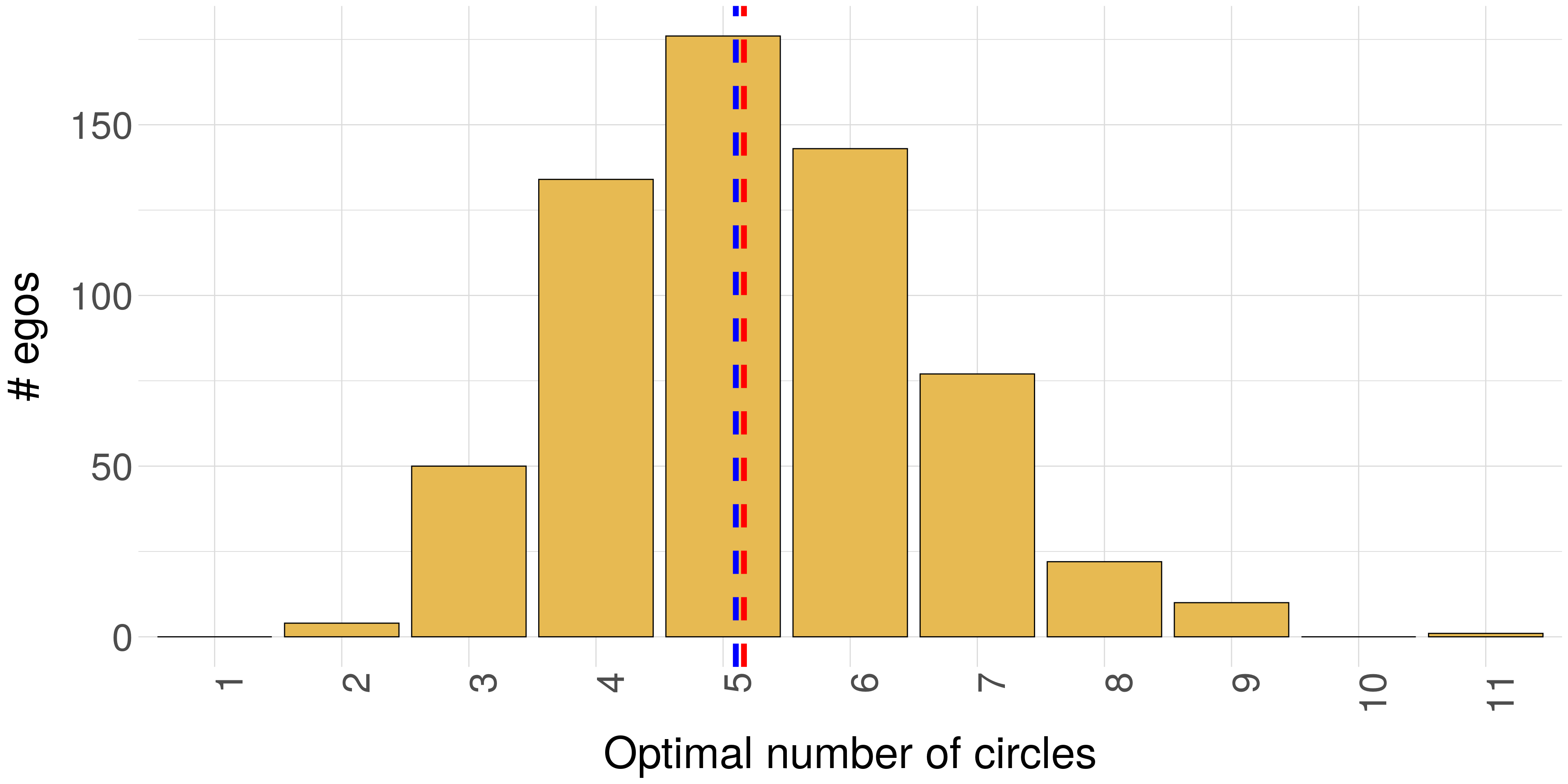}}
\hfill
\subfloat[Canada
\label{fig_appendix:optimal_circles_CanadianJournalists}]
{\includegraphics[width=0.27\textwidth]
{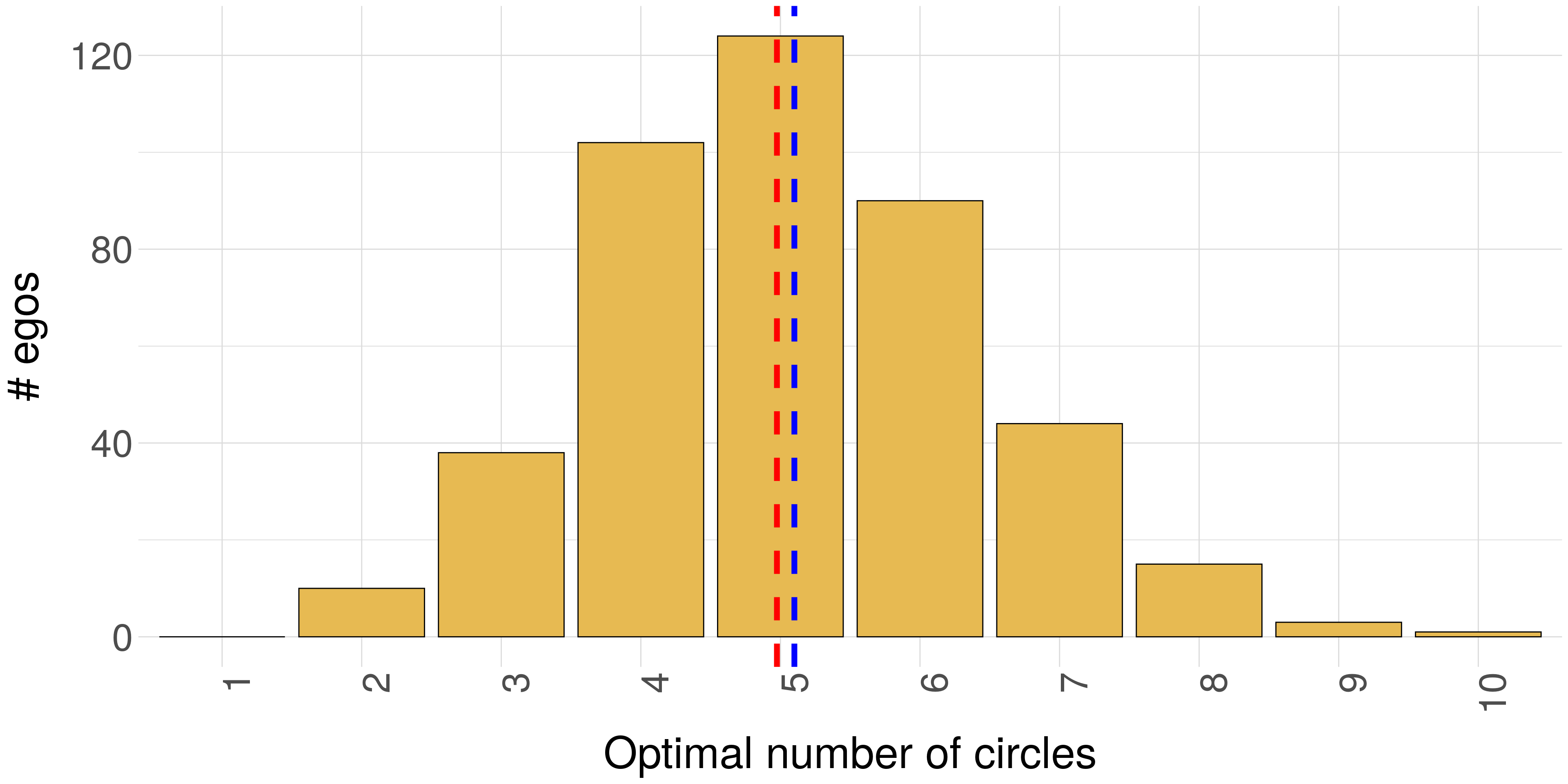}}
\hfill
\subfloat[Brasil
\label{fig_appendix:optimal_circles_BrazilianJournalists}]
{\includegraphics[width=0.27\textwidth]
{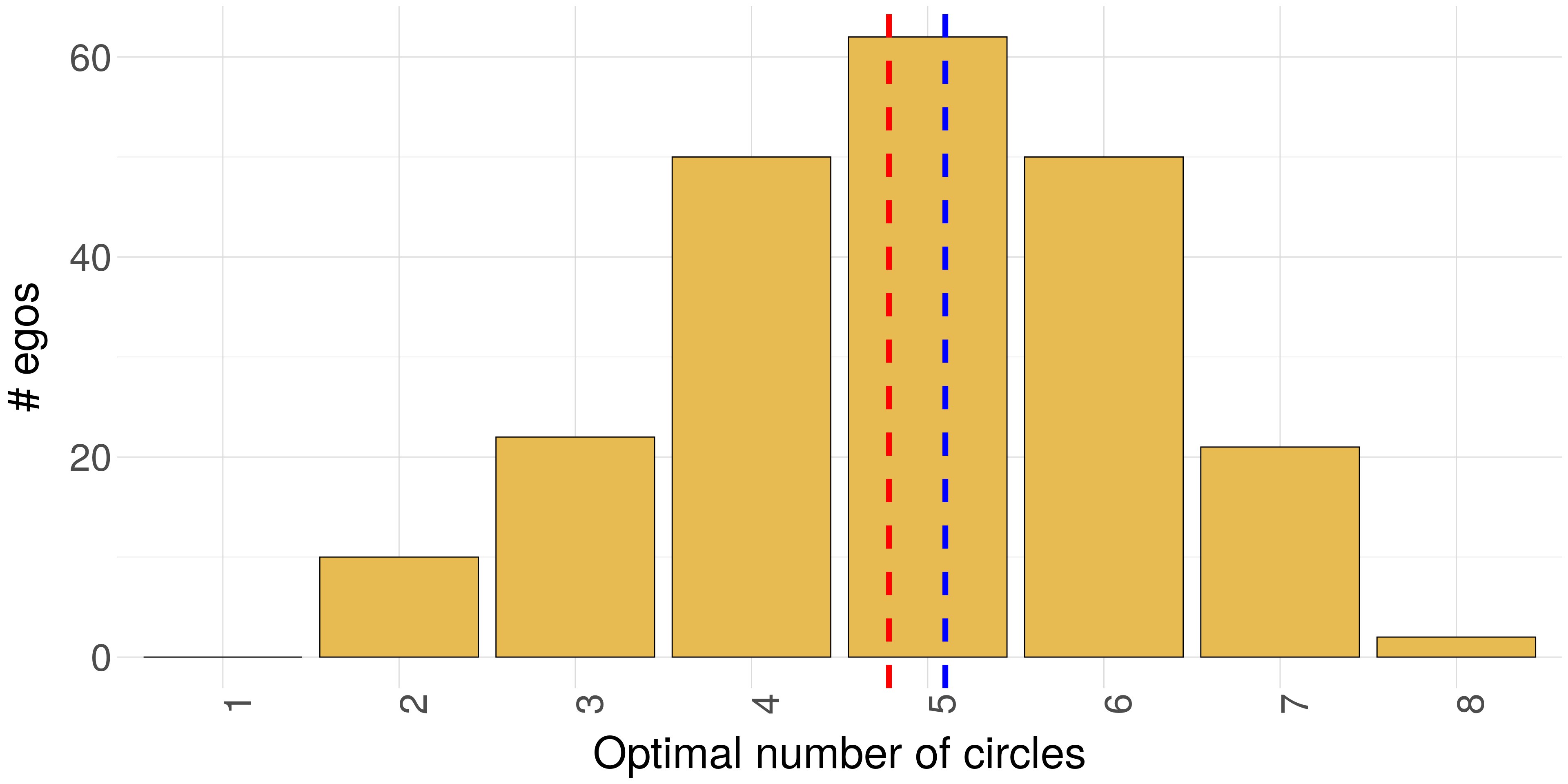}}
\hfill
\subfloat[Japan
\label{fig_appendix:optimal_circles_JapaneseJournalists}]
{\includegraphics[width=0.27\textwidth]
{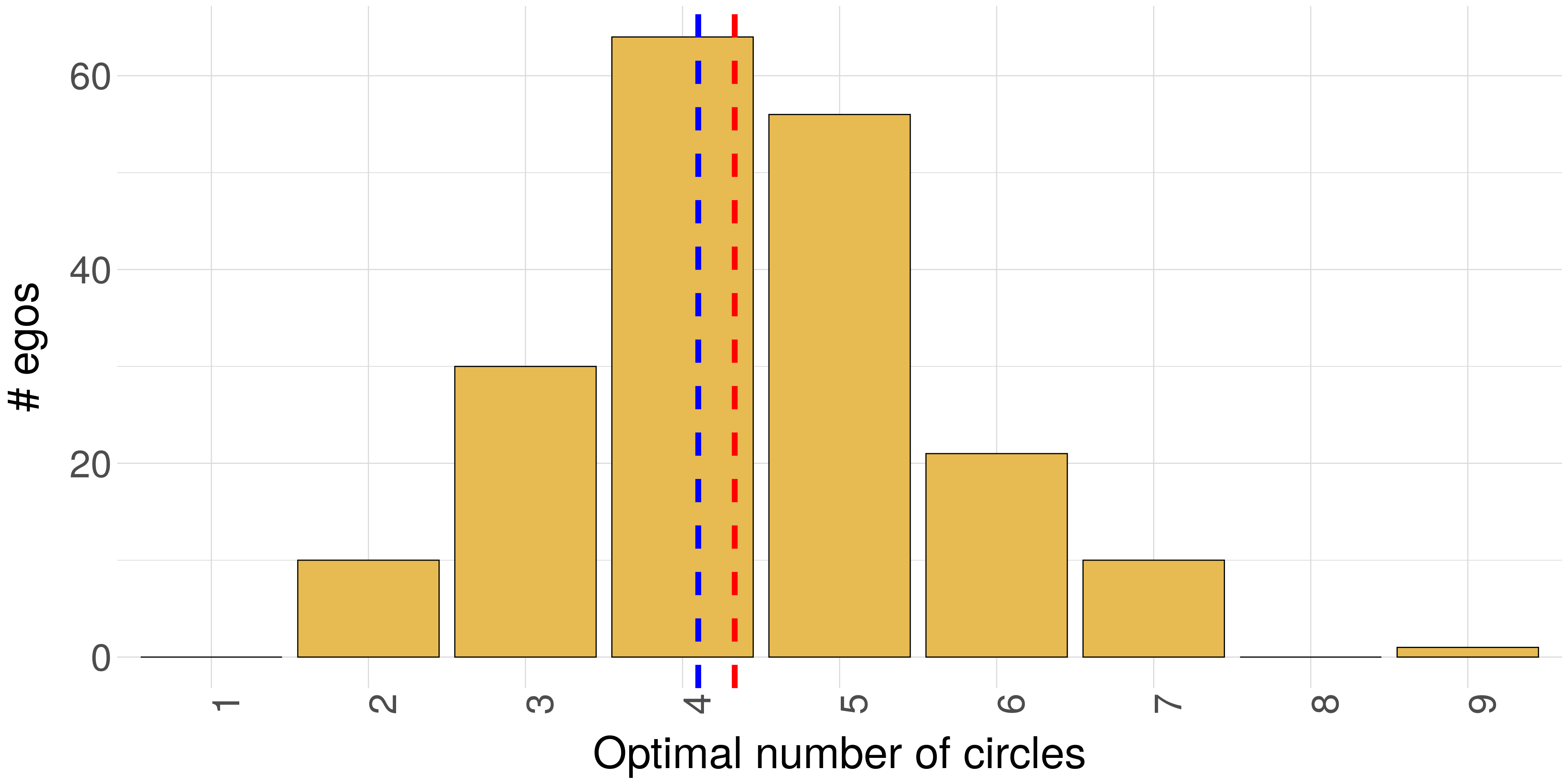}}
\hfill
\subfloat[Turkey
\label{fig_appendix:optimal_circles_TrukishJournalists}]
{\includegraphics[width=0.27\textwidth]
{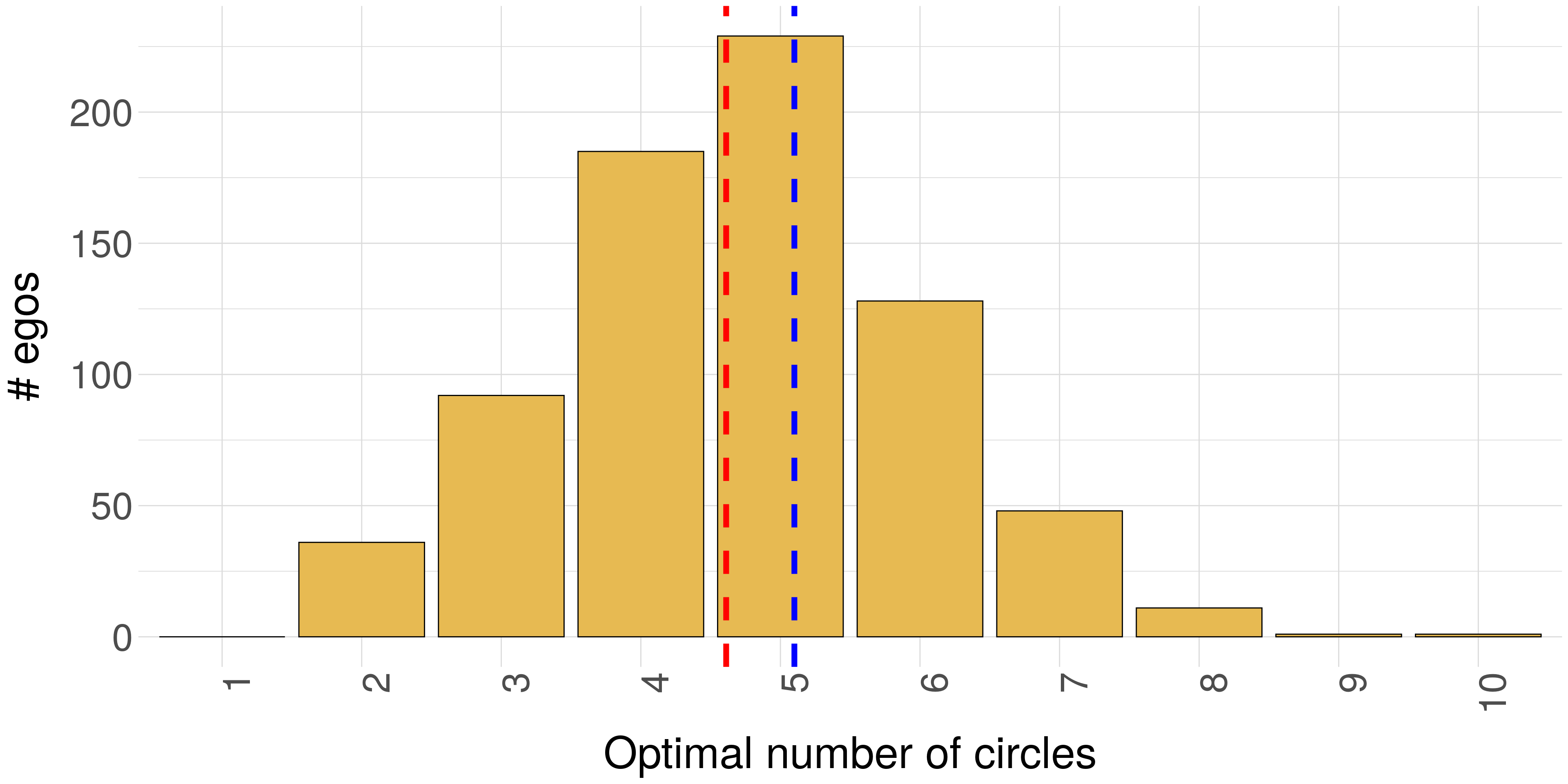}}
\hfill
\subfloat[UK
\label{fig_appendix:optimal_circles_BritishJournalists}]
{\includegraphics[width=0.27\textwidth]
{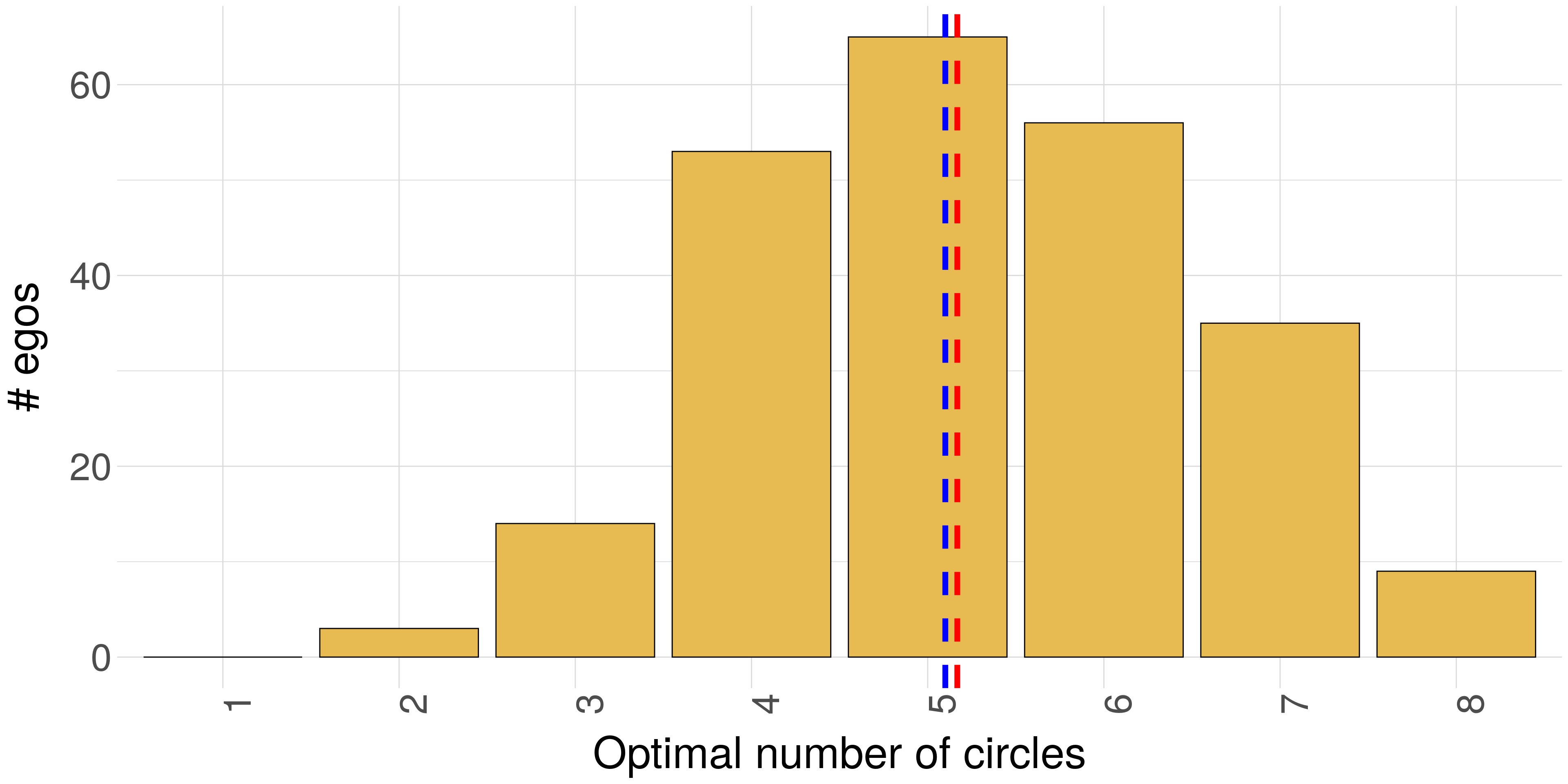}}
\hfill
\subfloat[Denmark
\label{fig_appendix:optimal_circles_DanishJournalists}]
{\includegraphics[width=0.27\textwidth]
{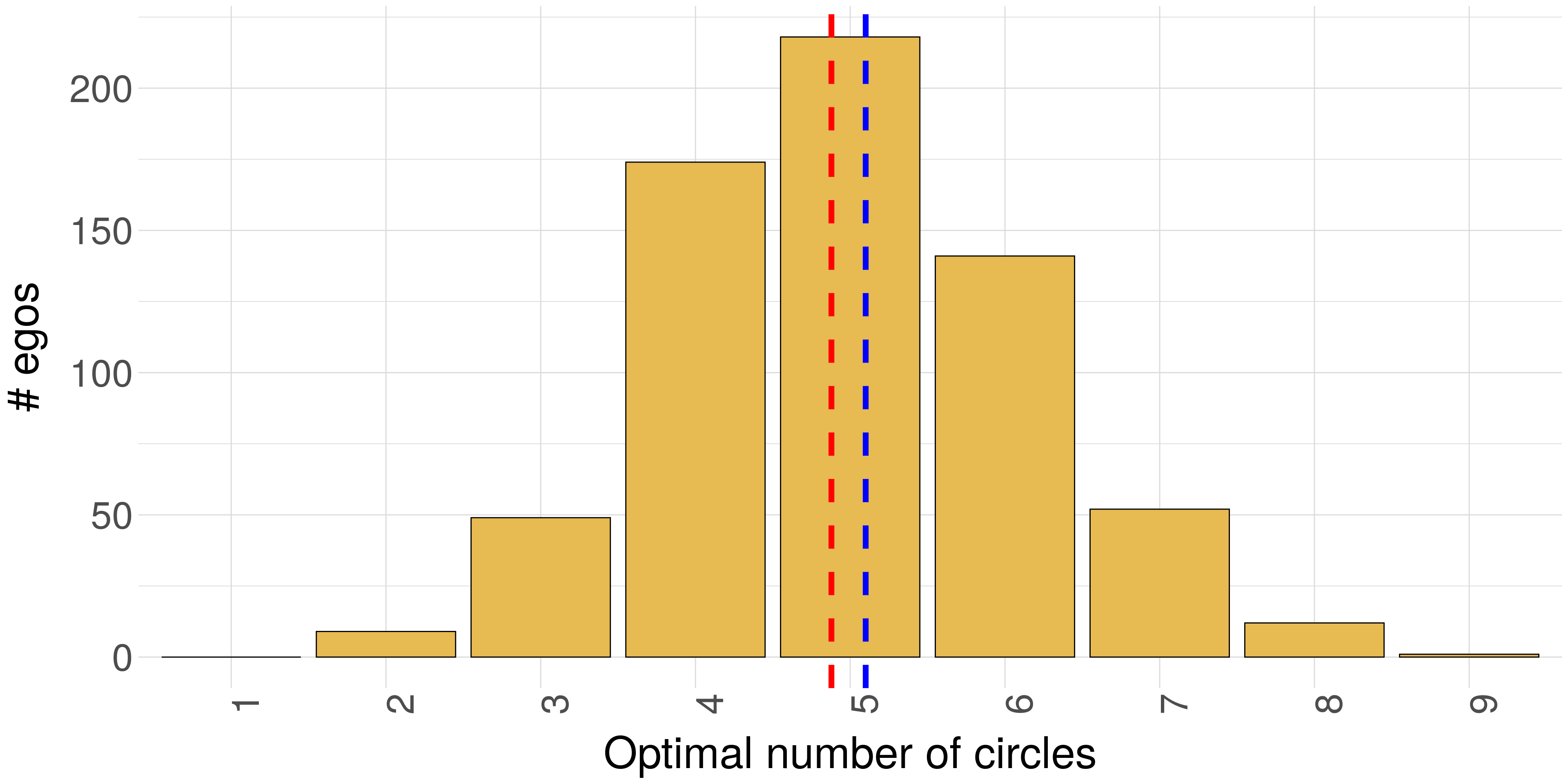}}
\hfill
\subfloat[Finland
\label{fig_appendix:optimal_circles_FinnishJournalists}]
{\includegraphics[width=0.27\textwidth]
{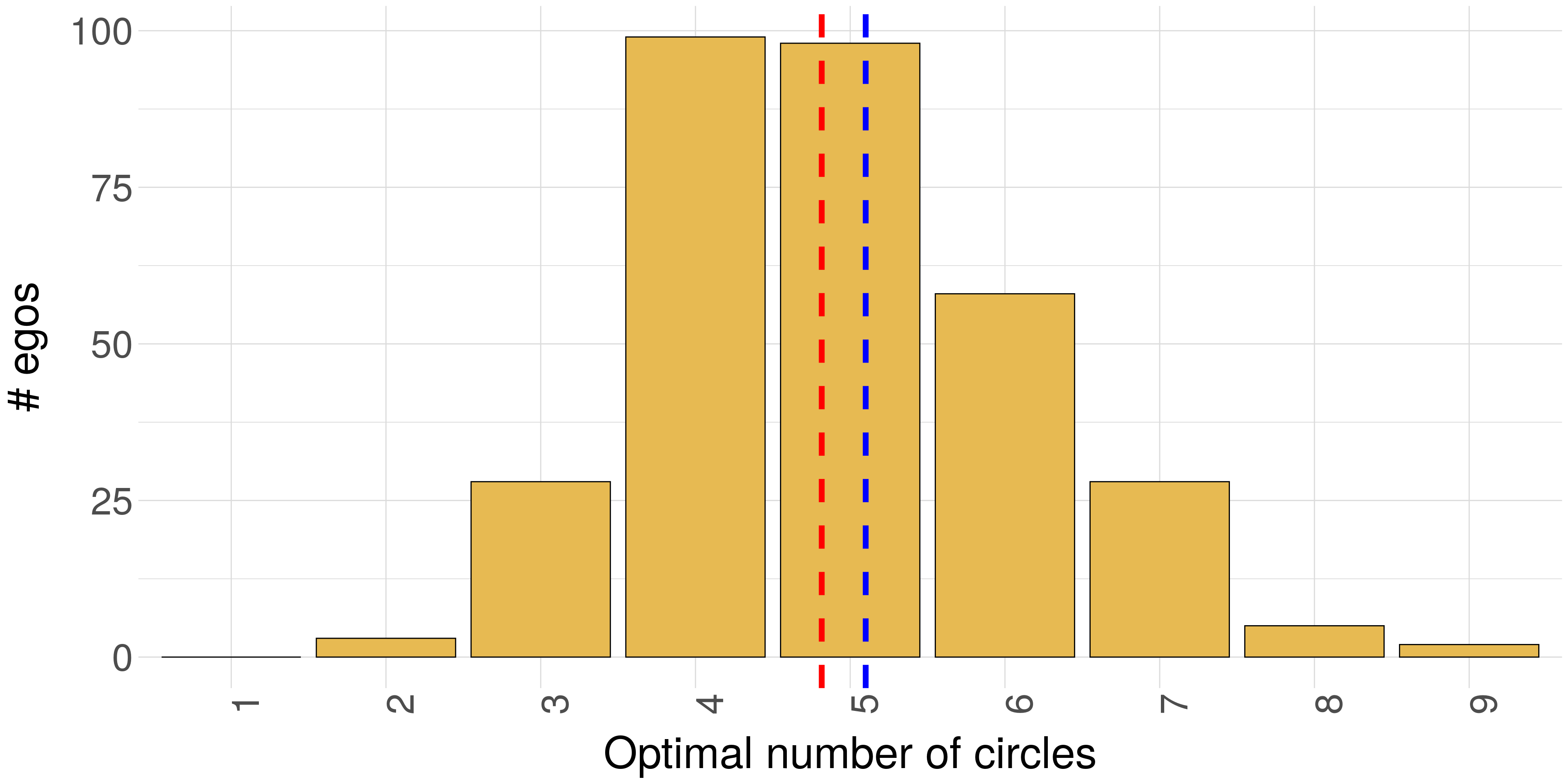}}
\hfill
\subfloat[Norway
\label{fig_appendix:optimal_circles_NorwegianJournalists}]
{\includegraphics[width=0.27\textwidth]
{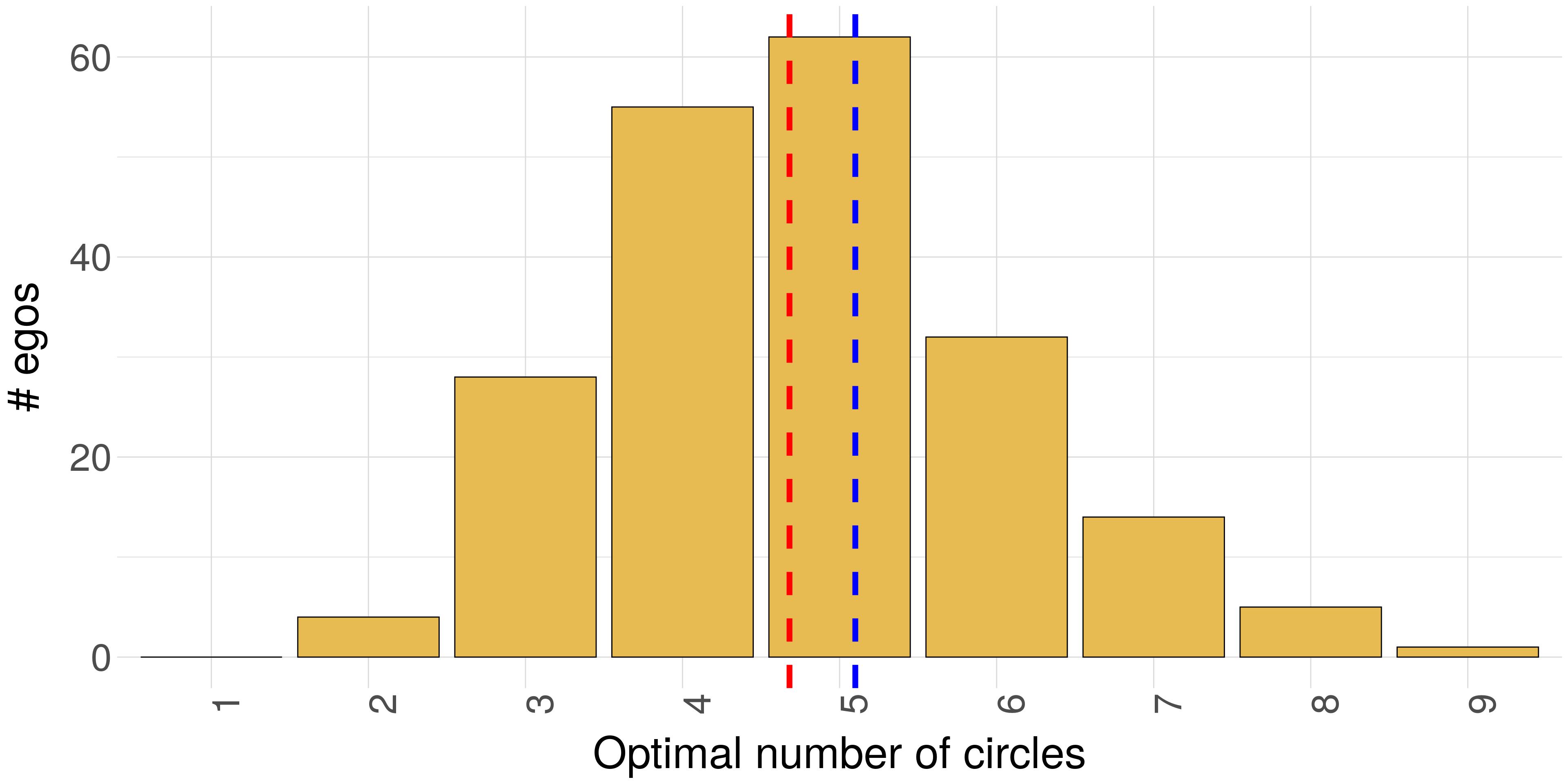}}
\hfill
\subfloat[Sweden
\label{fig_appendix:optimal_circles_SwedishJournalists}]
{\includegraphics[width=0.27\textwidth]
{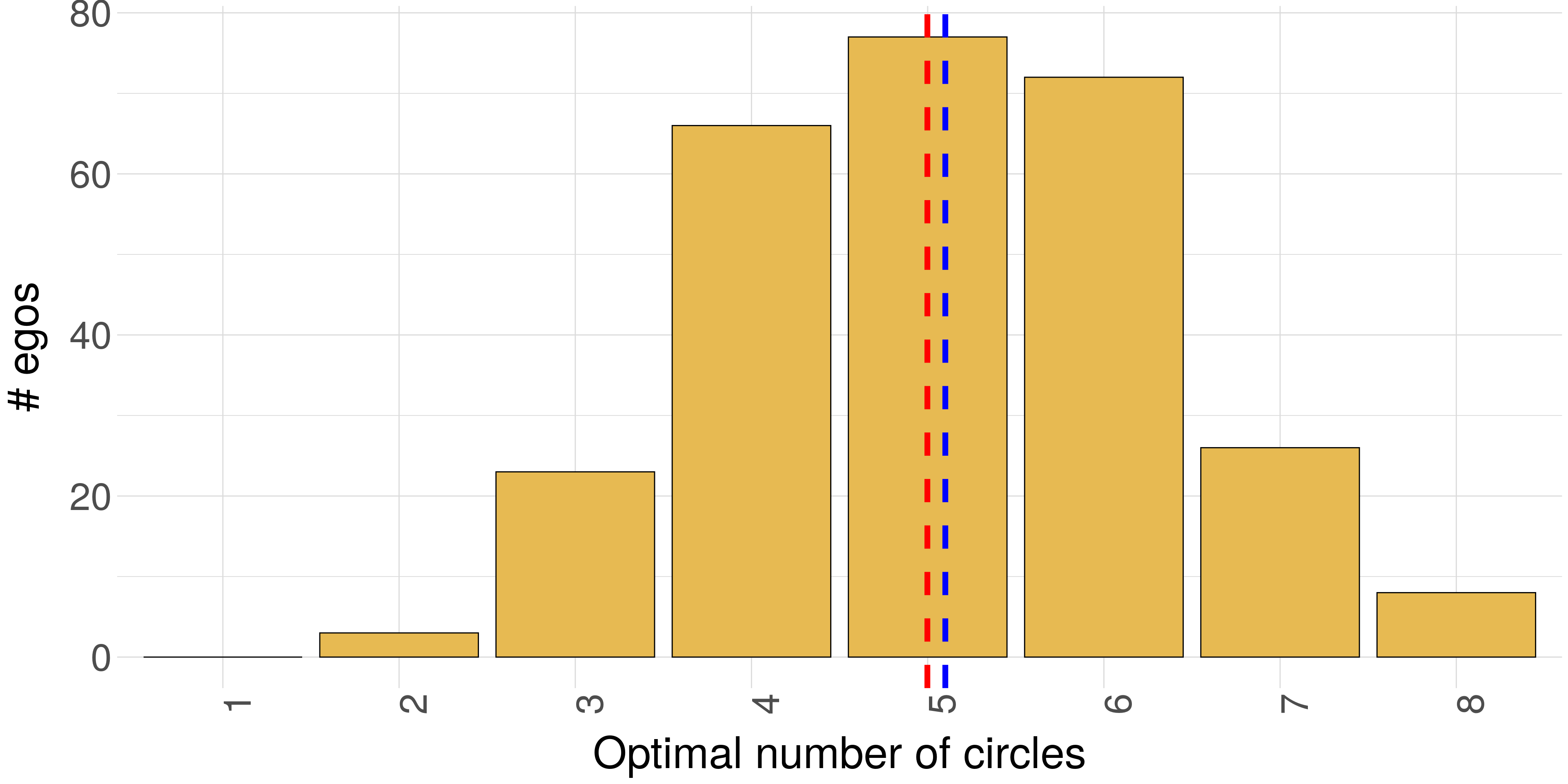}}
\hfill
\subfloat[Greece
\label{fig_appendix:optimal_circles_GreekJournalists}]
{\includegraphics[width=0.27\textwidth]
{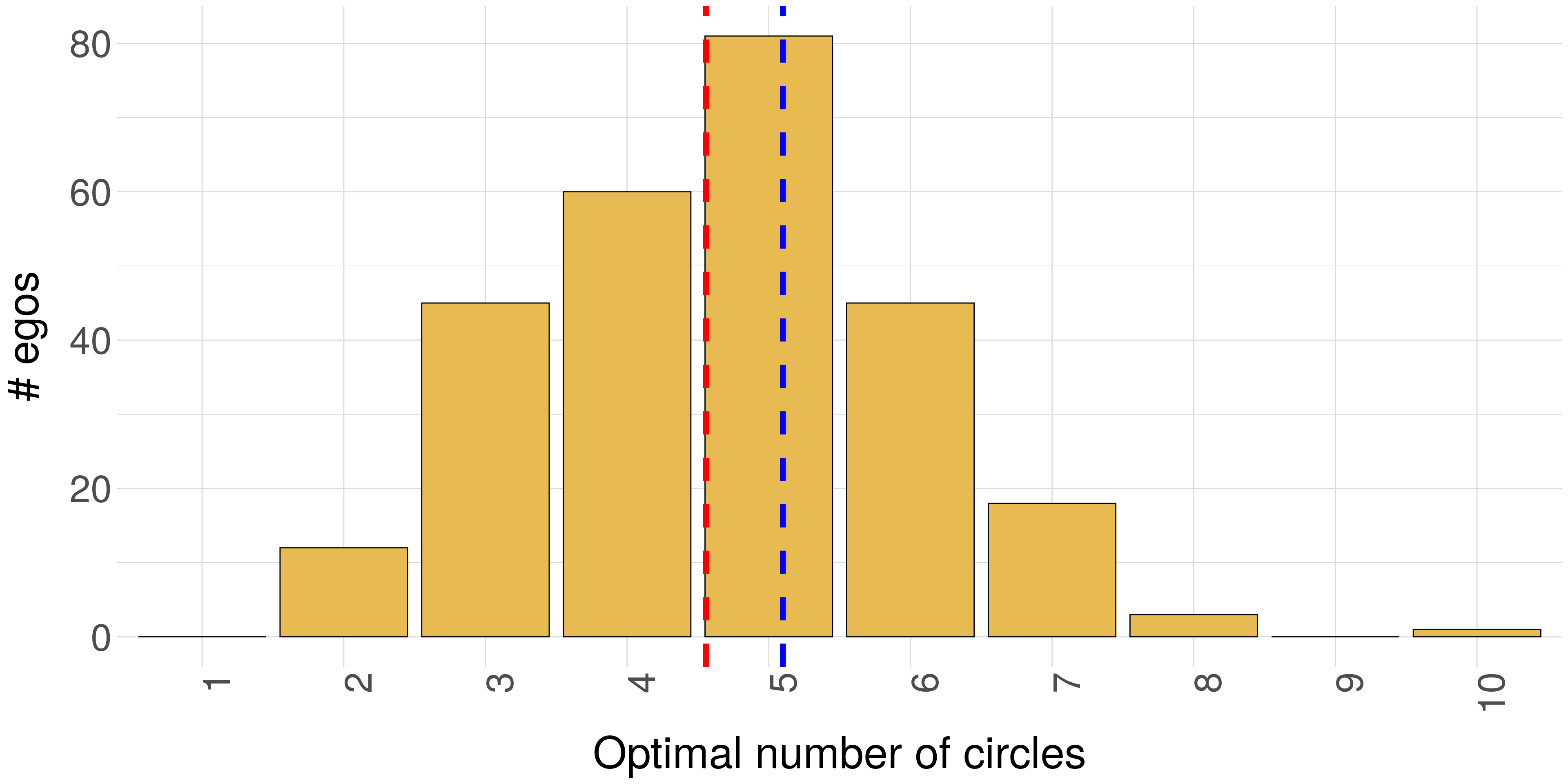}}
\hfill
\subfloat[Italy
\label{fig_appendix:optimal_circles_ItalianJournalists}]
{\includegraphics[width=0.27\textwidth]
{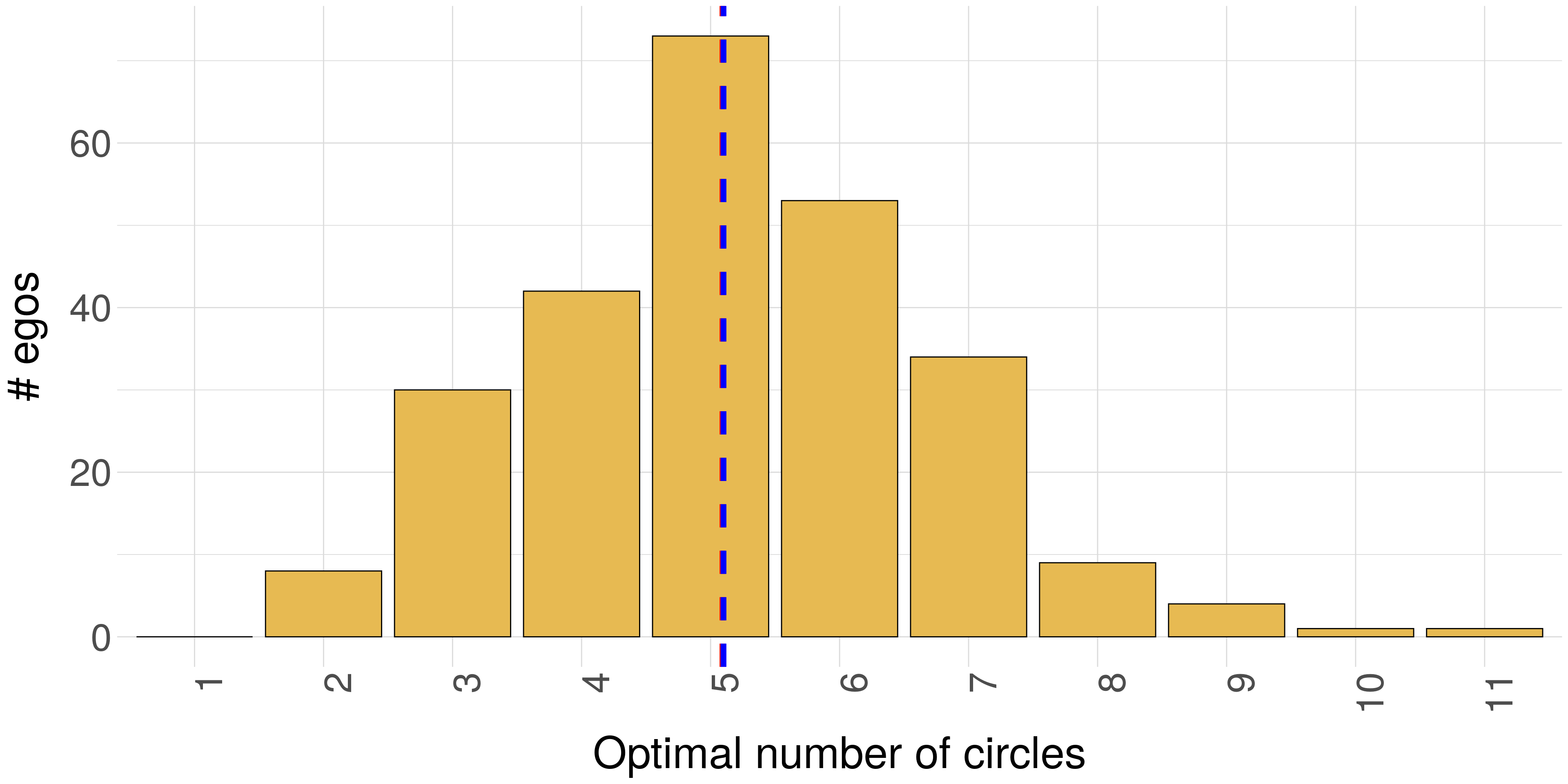}}
\hfill
\subfloat[Spain
\label{fig_appendix:optimal_circles_SpanishJournalists}]
{\includegraphics[width=0.27\textwidth]
{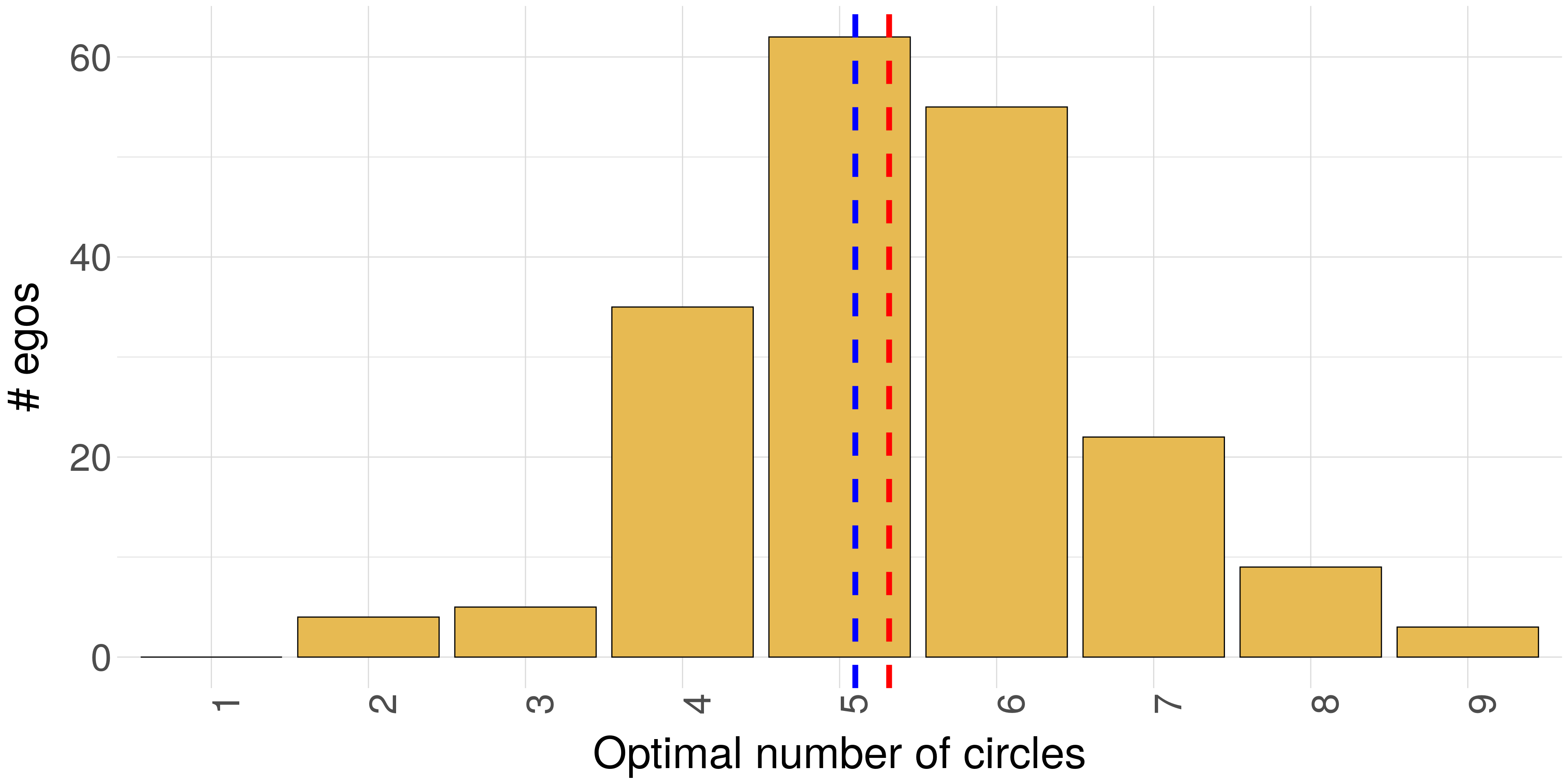}}
\hfill
\subfloat[France
\label{fig_appendix:optimal_circles_FrenchJournalists}]
{\includegraphics[width=0.27\textwidth]
{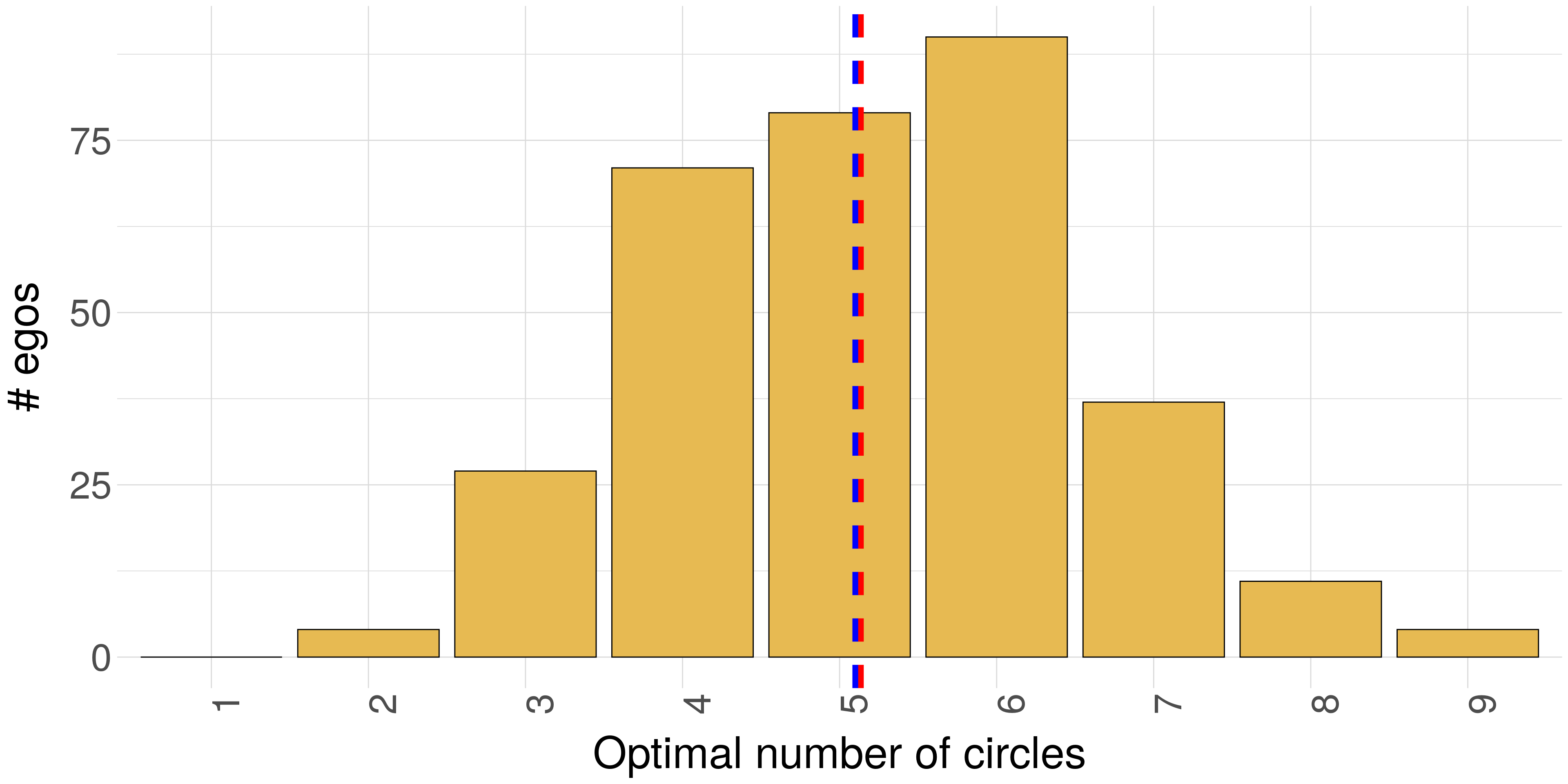}}
\hfill
\subfloat[Germany
\label{fig_appendix:optimal_circles_GermanJournalists}]
{\includegraphics[width=0.27\textwidth]
{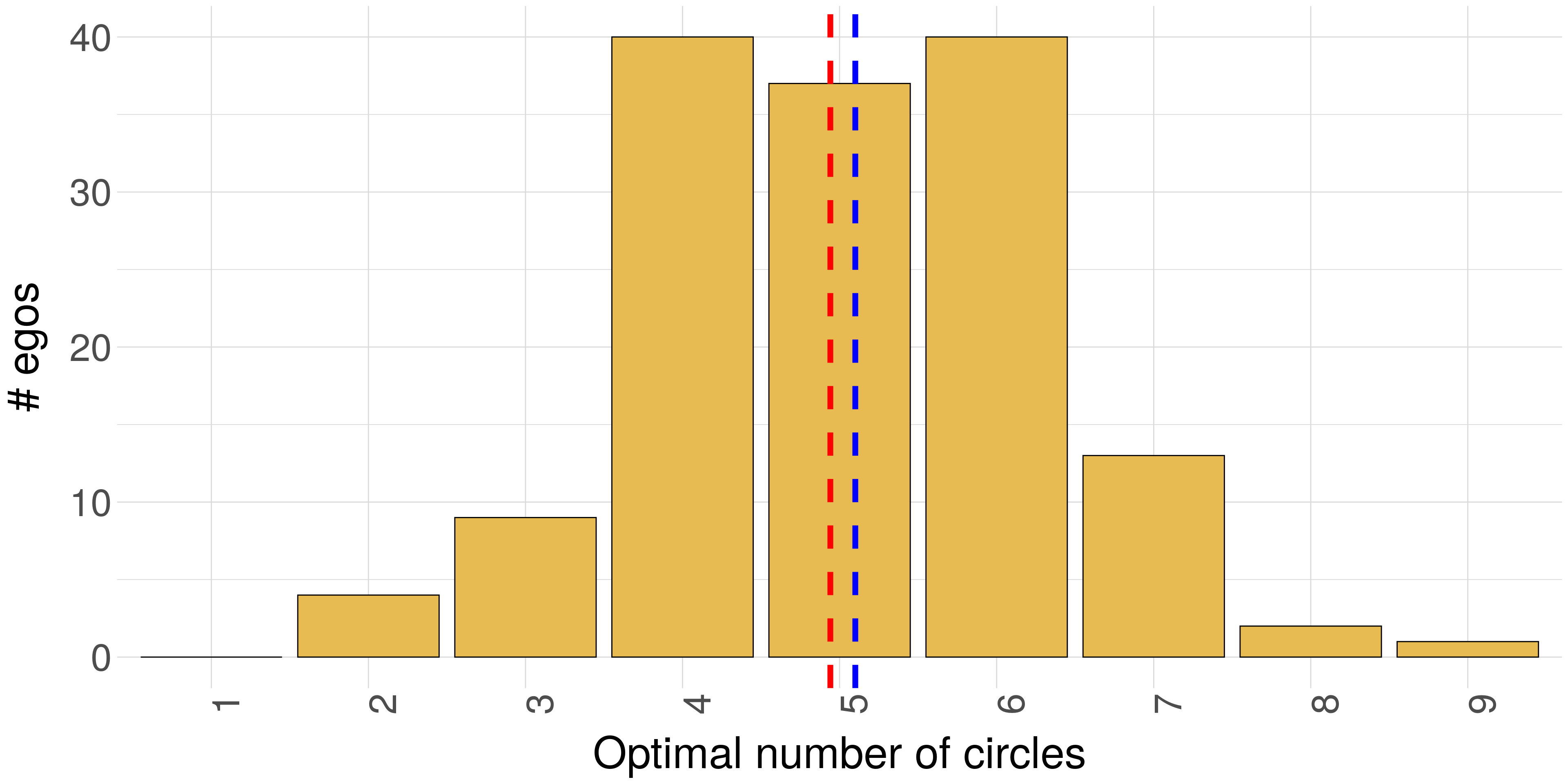}}
\hfill
\subfloat[Netherland
\label{fig_appendix:optimal_circles_NetherlanderJournalists}]
{\includegraphics[width=0.27\textwidth]
{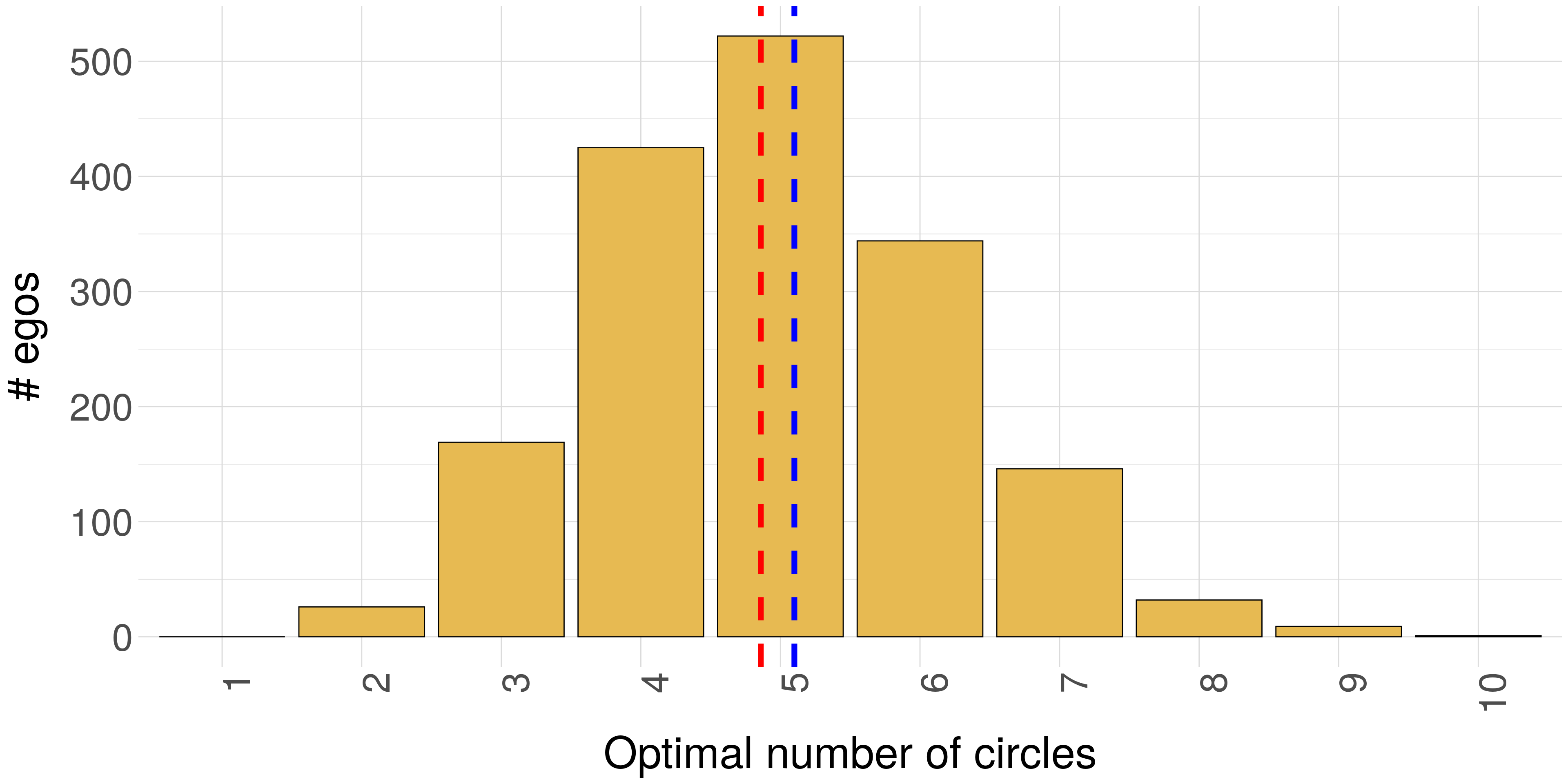}}
\hspace{1pt}
\subfloat[Australia
\label{fig_appendix:optimal_circles_AustralianJournalists}]
{\includegraphics[width=0.27\textwidth]
{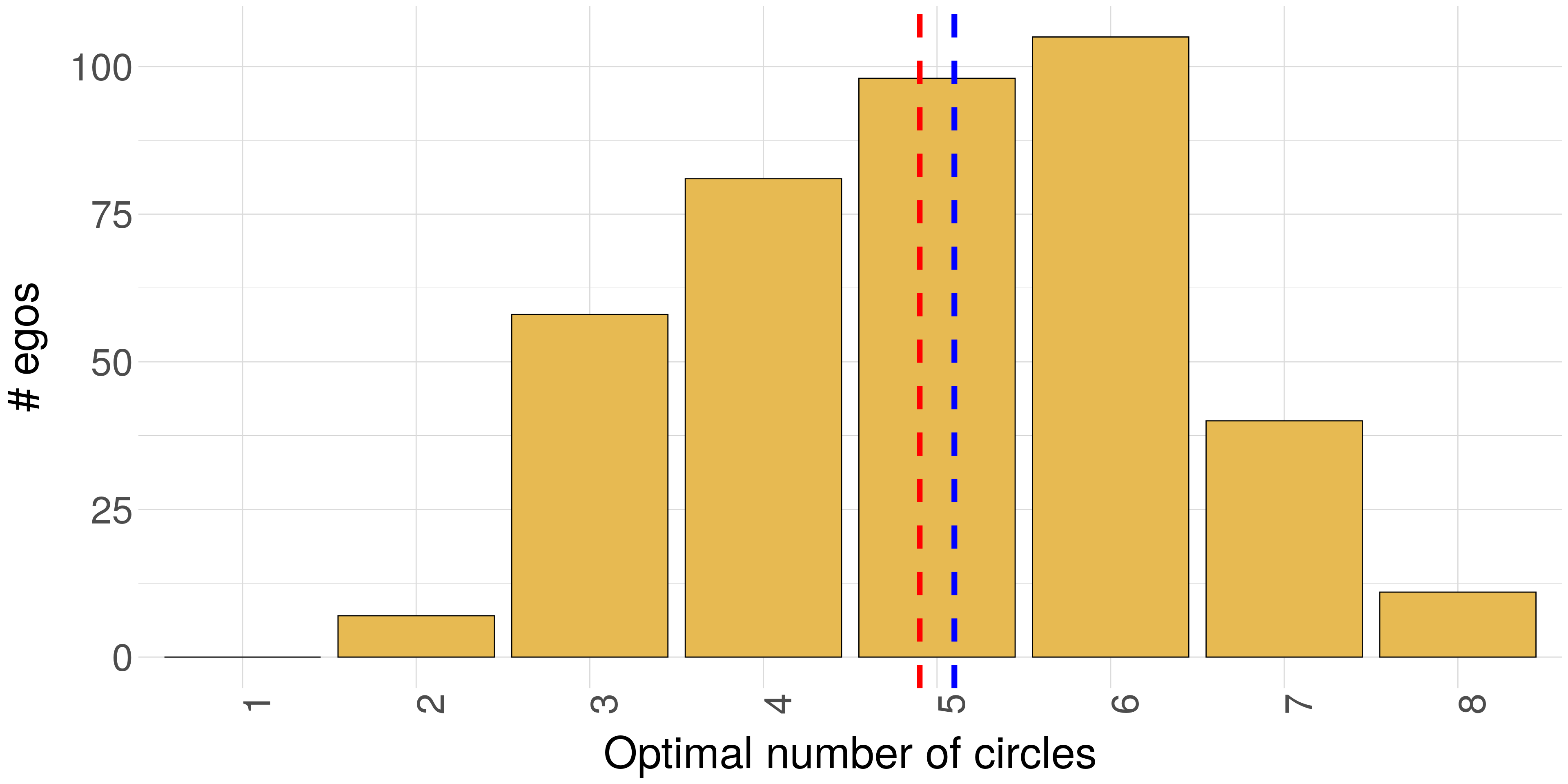}}
\end{center}
\end{adjustbox}
\caption{Distribution of the optimal circle number (red line: average, blue line: median), per country}
\label{fig_appendix:optimal_circles}
\end{figure}

\clearpage
 
\renewcommand\thefigure{\Alph{section}.\arabic{figure}} 
\setcounter{figure}{0}

\renewcommand\thetable{\Alph{section}.\arabic{table}} 
\setcounter{table}{0}

\section{Social tweets and hashtags - additional plots}
\label{appendix_hashtags_egonets}

\vspace{-20pt}

\begin{figure}[!h]
\begin{adjustbox}{minipage=\linewidth}
\begin{center}
\subfloat[USA
\label{fig_appendix:relations_hashtags_rings_AmericanJournalists}]
{\includegraphics[width=0.27\textwidth]
{./figures_new/relationships_hastags_rings/AmericanJournalists_ring_based_hashtag_act_relation.png}}
\hfill
\subfloat[Canada
\label{fig_appendix:relations_hashtags_rings_CanadianJournalists}]
{\includegraphics[width=0.27\textwidth]
{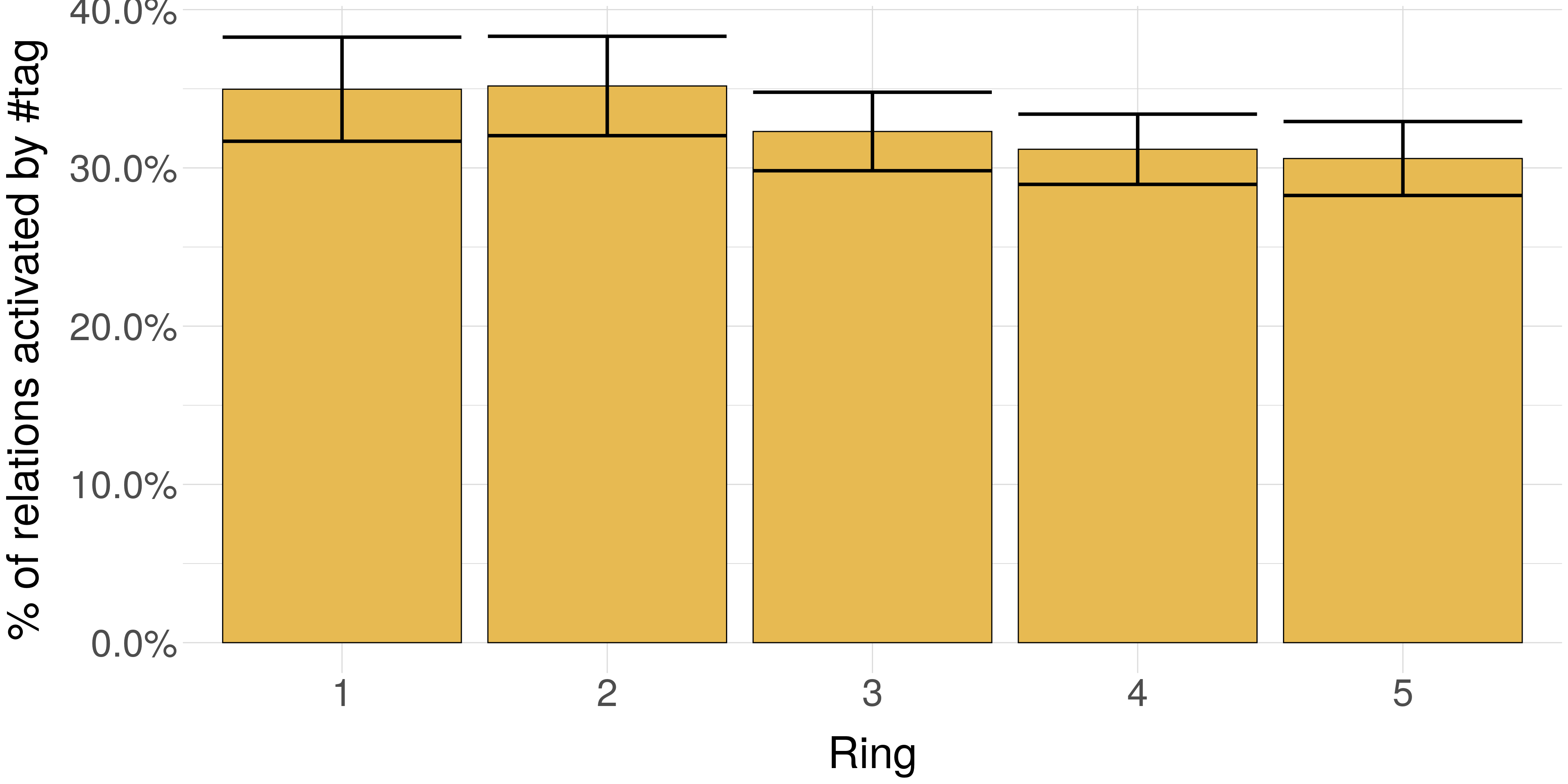}}
\hfill
\subfloat[Brasil
\label{fig_appendix:relations_hashtags_rings_BrazilianJournalists}]
{\includegraphics[width=0.27\textwidth]
{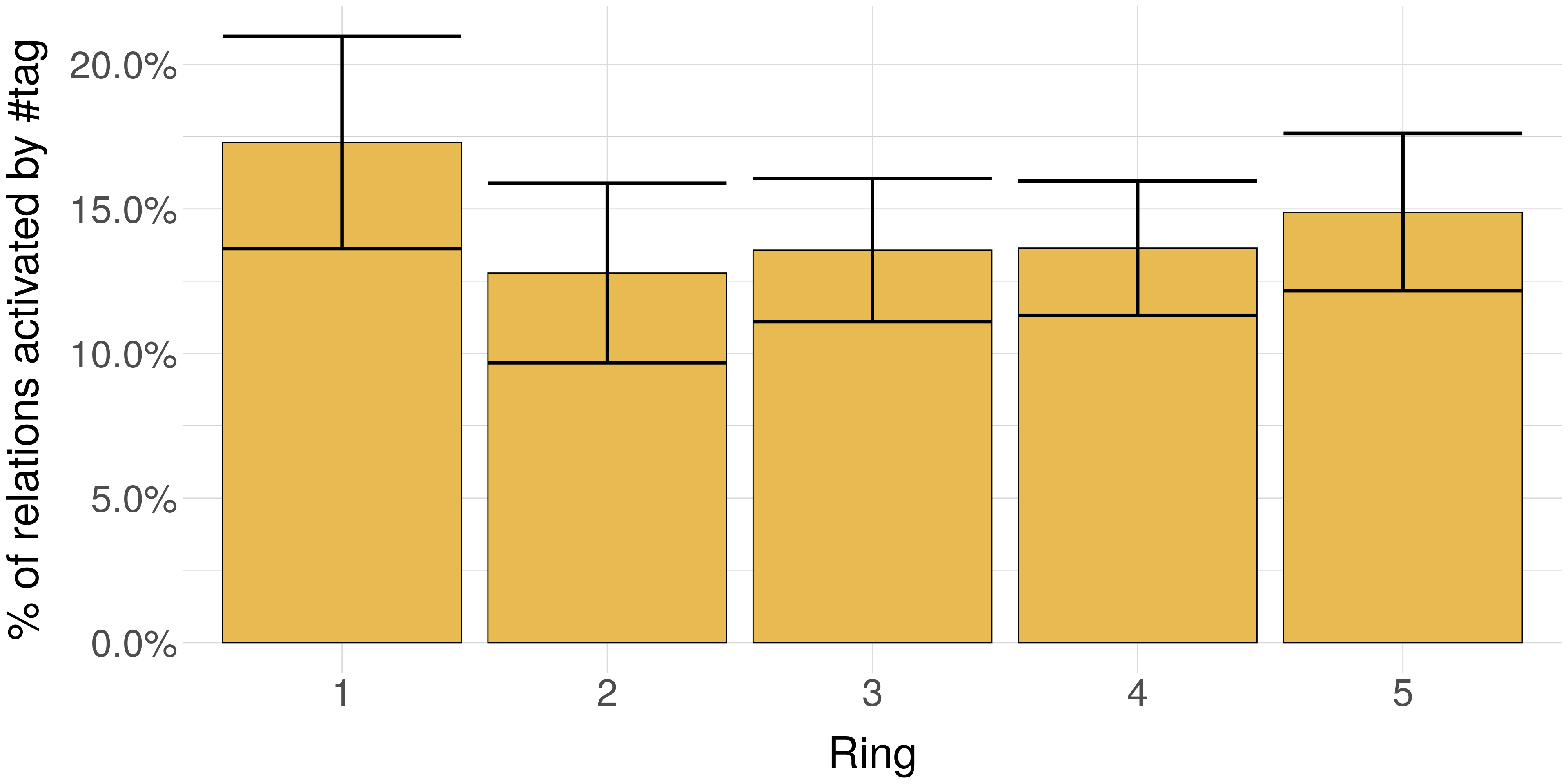}}
\hfill
\subfloat[Japan
\label{fig_appendix:relations_hashtags_rings_JapaneseJournalists}]
{\includegraphics[width=0.27\textwidth]
{./figures_new/relationships_hastags_rings/JapaneseJournalists_ring_based_hashtag_act_relation.png}}
\hfill
\subfloat[Turkey
\label{fig_appendix:relations_hashtags_rings_TrukishJournalists}]
{\includegraphics[width=0.27\textwidth]
{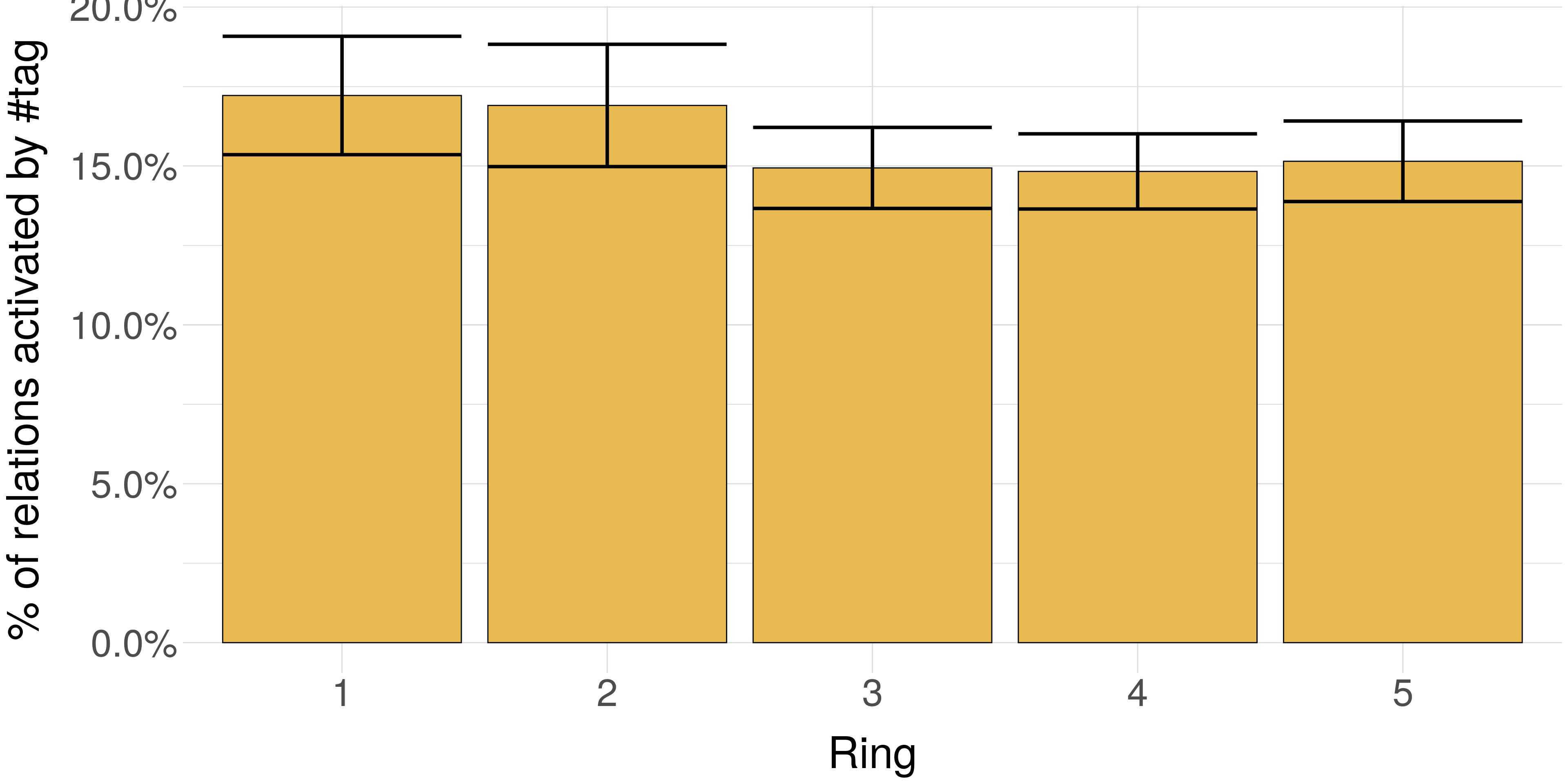}}
\hfill
\subfloat[UK
\label{fig_appendix:relations_hashtags_rings_BritishJournalists}]
{\includegraphics[width=0.27\textwidth]
{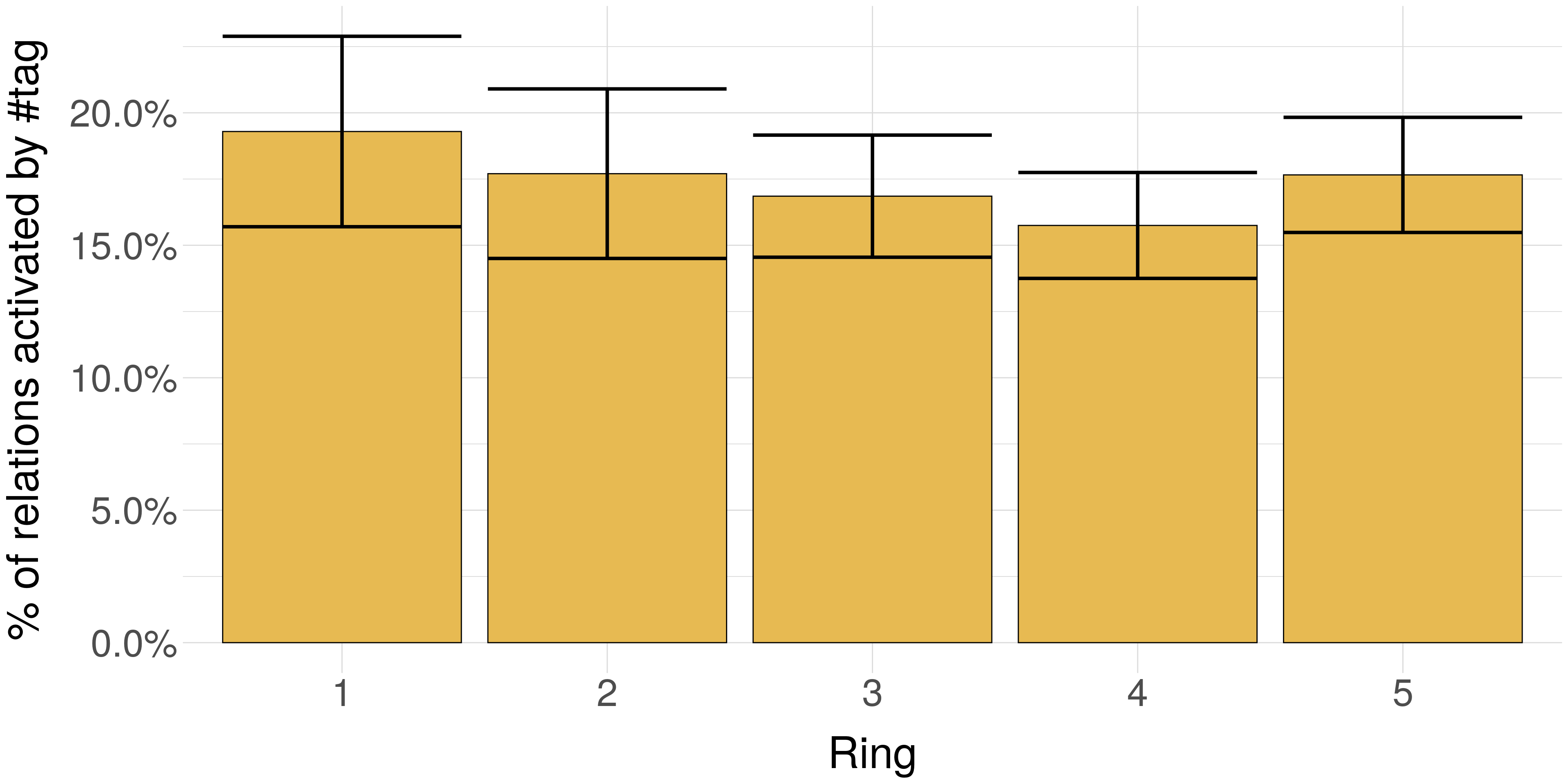}}
\hfill
\subfloat[Denmark
\label{fig_appendix:relations_hashtags_rings_DanishJournalists}]
{\includegraphics[width=0.27\textwidth]
{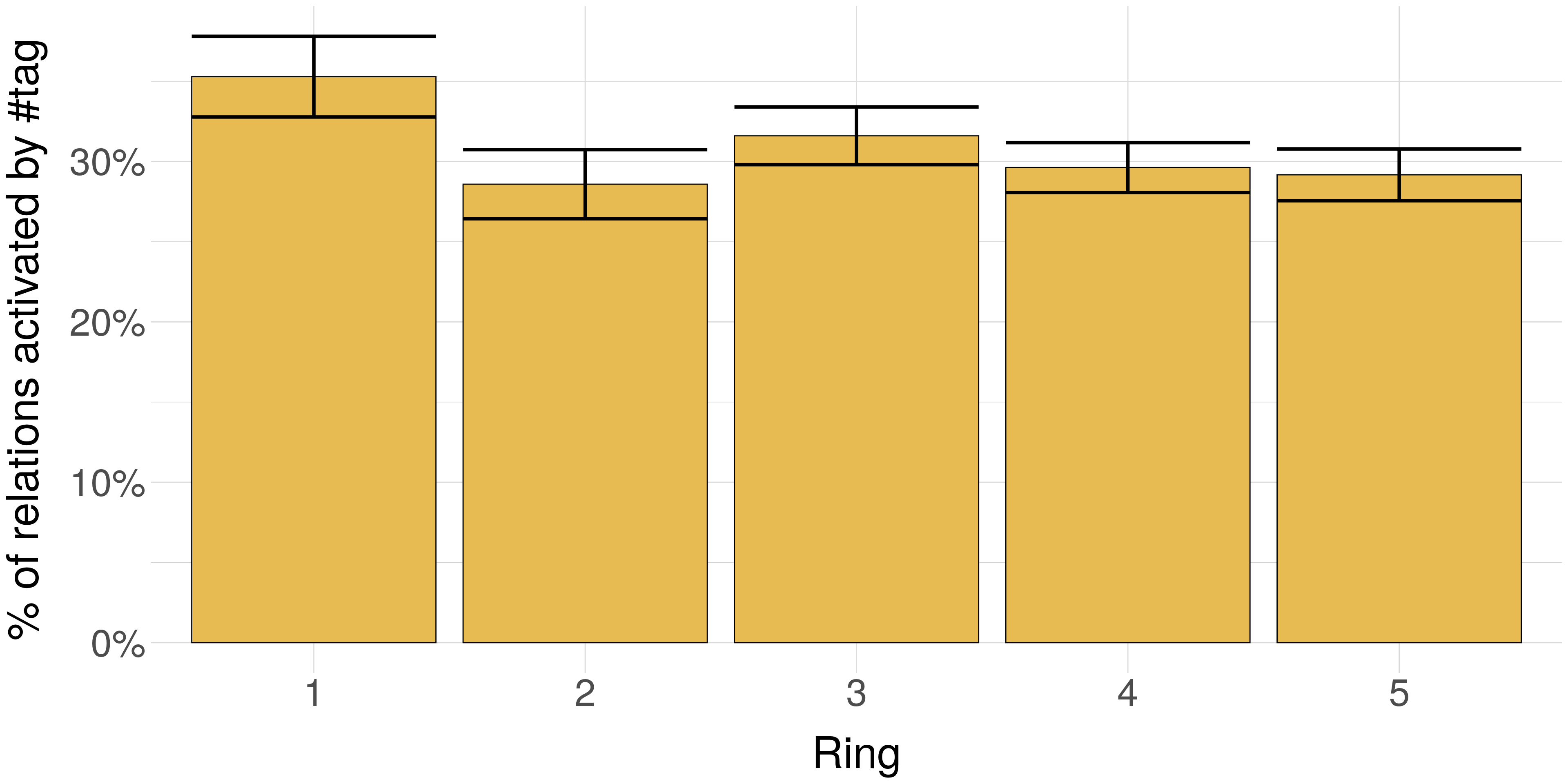}}
\hfill
\subfloat[Finland
\label{fig_appendix:relations_hashtags_rings_FinnishJournalists}]
{\includegraphics[width=0.27\textwidth]
{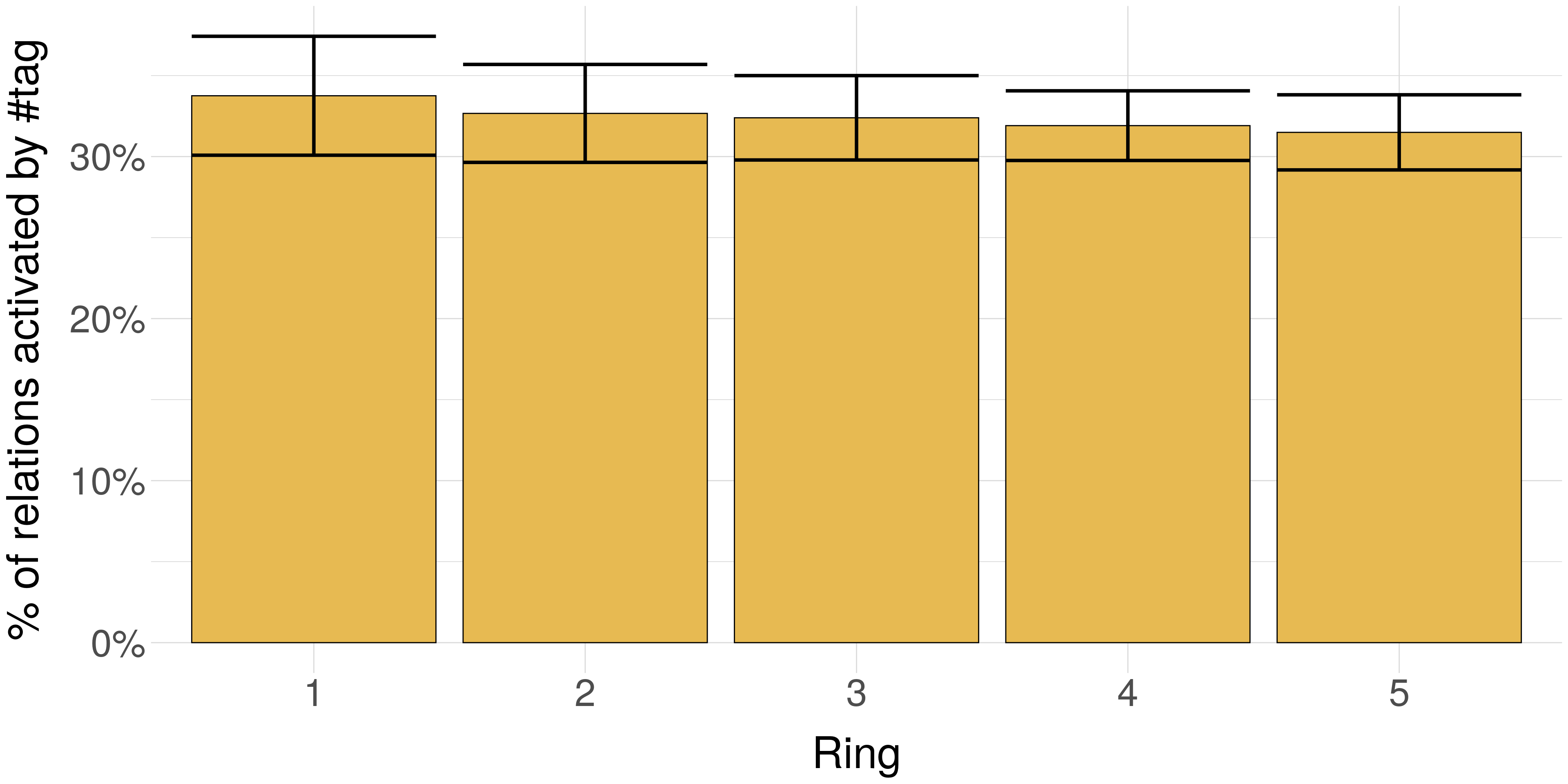}}
\hfill
\subfloat[Norway
\label{fig_appendix:relations_hashtags_rings_NorwegianJournalists}]
{\includegraphics[width=0.27\textwidth]
{./figures_new/relationships_hastags_rings/NorwegianJournalists_ring_based_hashtag_act_relation.png}}
\hfill
\subfloat[Sweden
\label{fig_appendix:relations_hashtags_rings_SwedishJournalists}]
{\includegraphics[width=0.27\textwidth]
{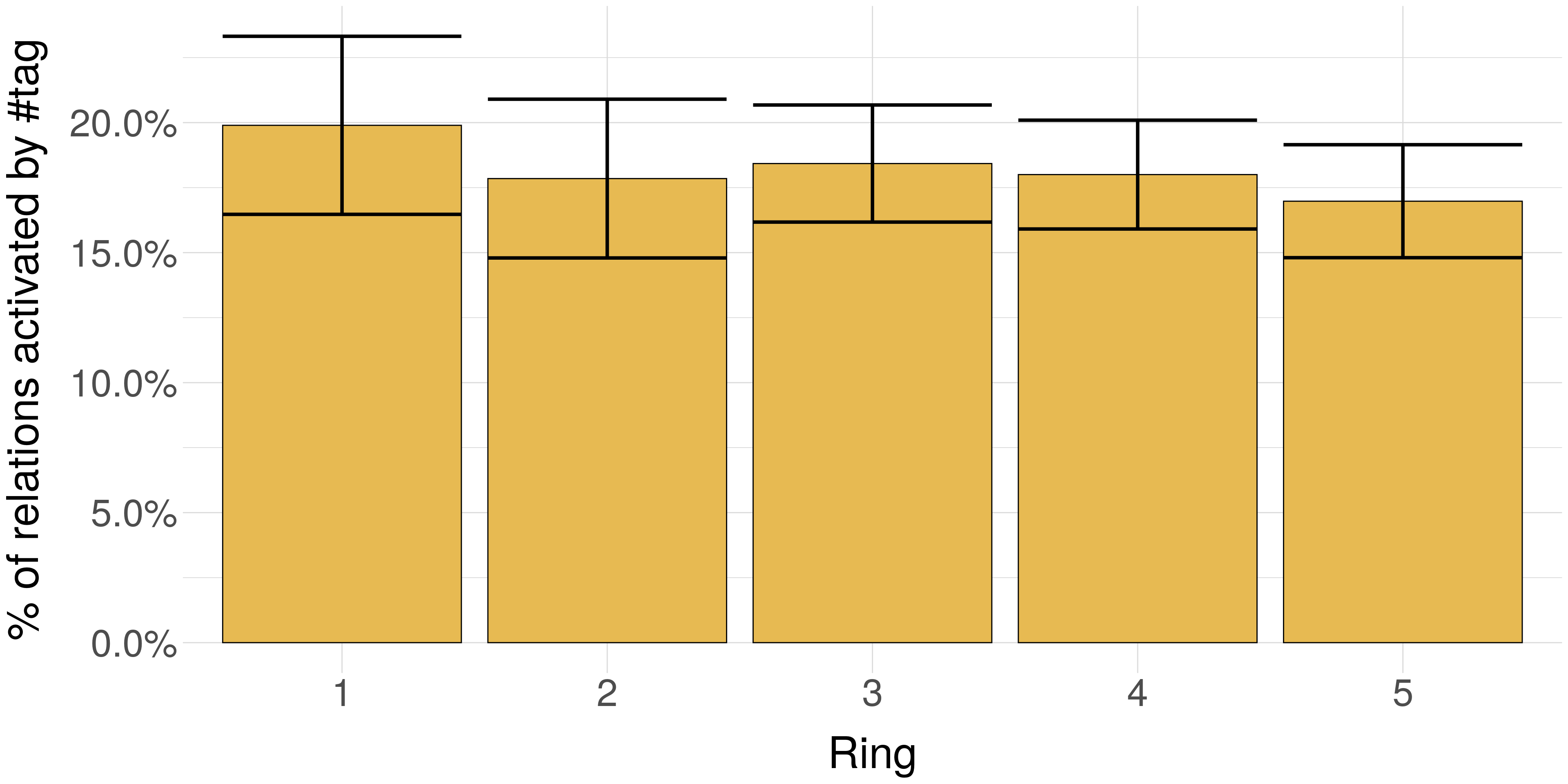}}
\hfill
\subfloat[Greece
\label{fig_appendix:relations_hashtags_rings_GreekJournalists}]
{\includegraphics[width=0.27\textwidth]
{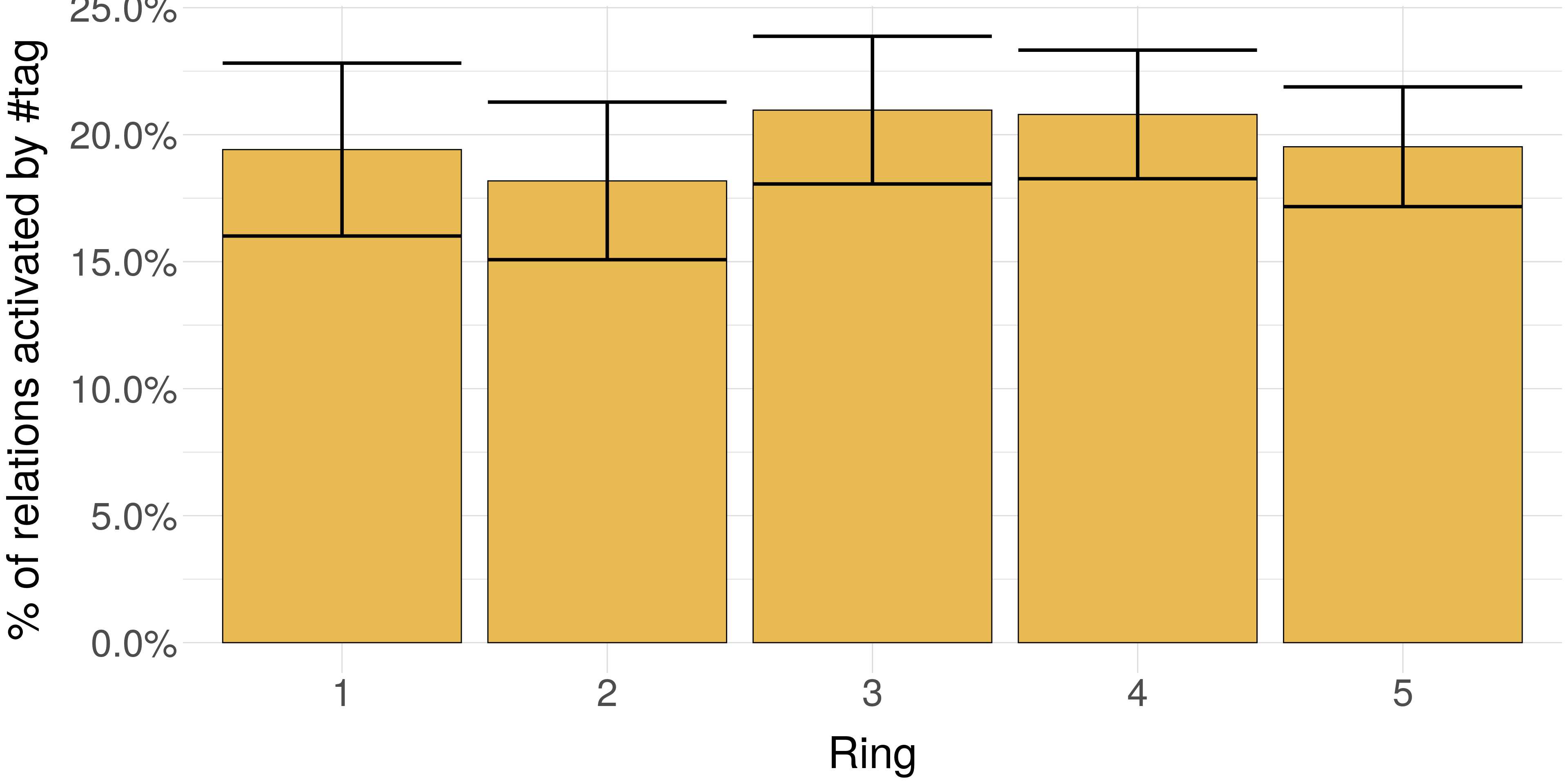}}
\hfill
\subfloat[Italy
\label{fig_appendix:relations_hashtags_rings_ItalianJournalists}]
{\includegraphics[width=0.27\textwidth]
{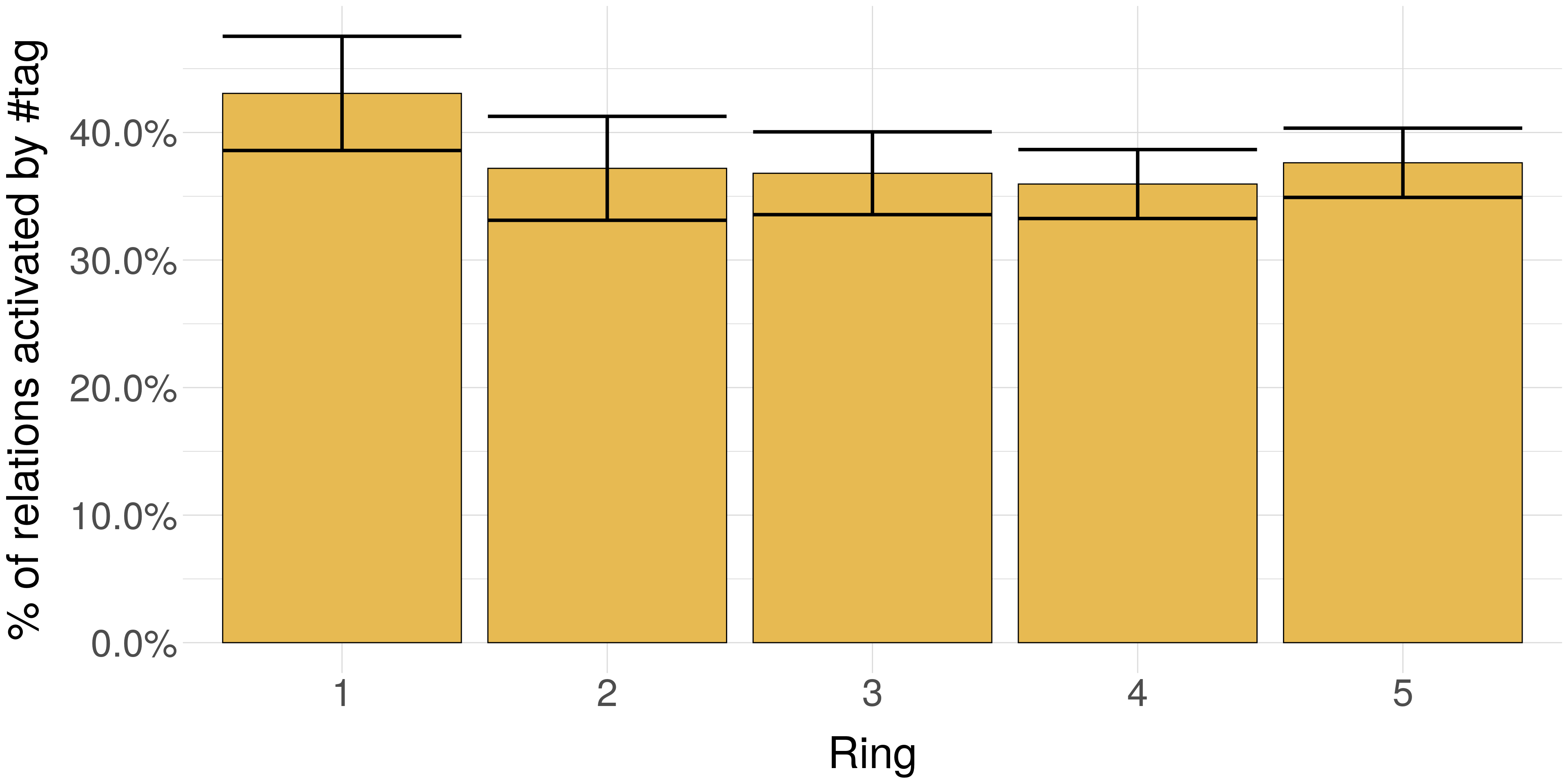}}
\hfill
\subfloat[Spain
\label{fig_appendix:relations_hashtags_rings_SpanishJournalists}]
{\includegraphics[width=0.27\textwidth]
{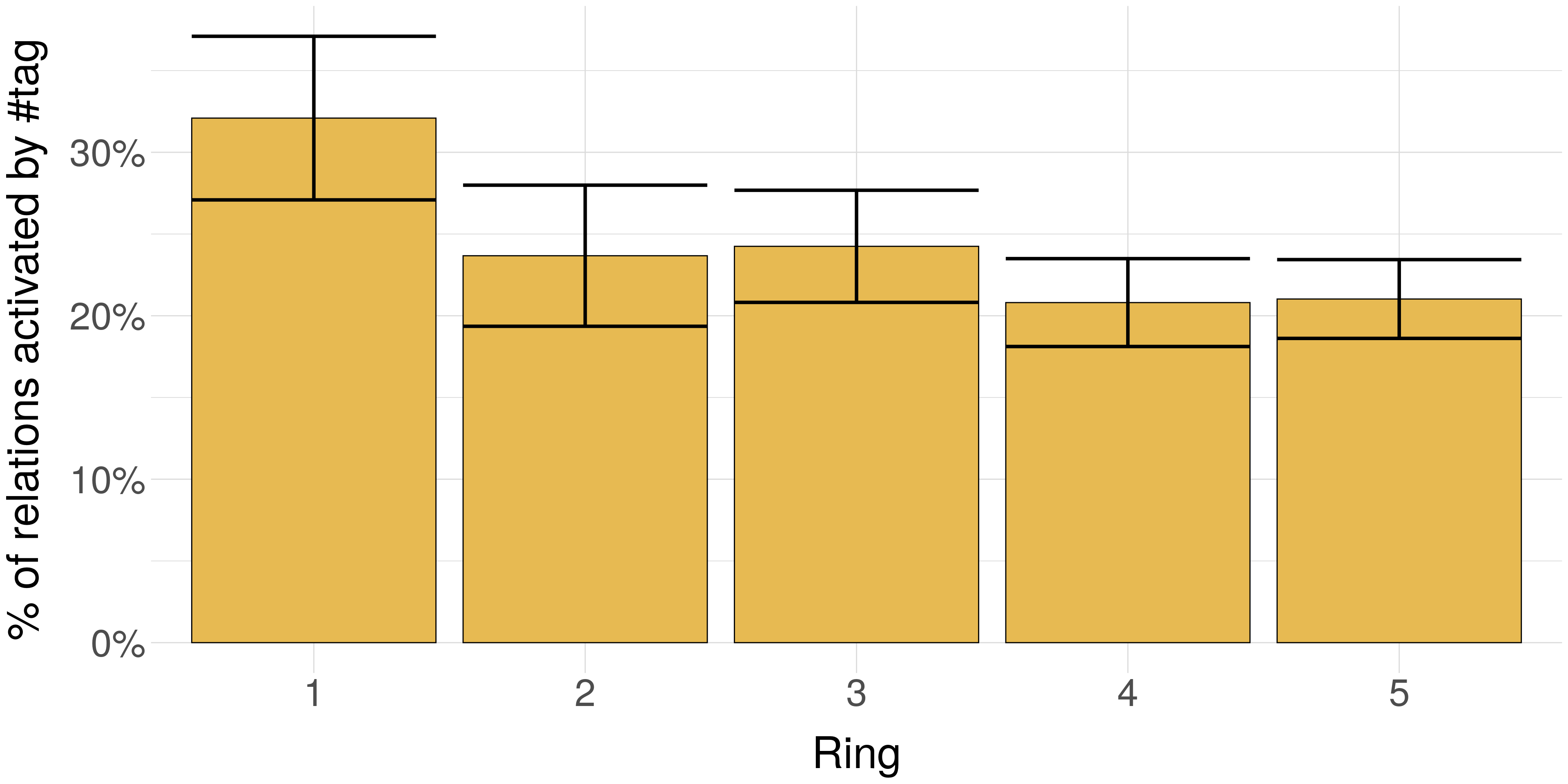}}
\hfill
\subfloat[France
\label{fig_appendix:relations_hashtags_rings_FrenchJournalists}]
{\includegraphics[width=0.27\textwidth]
{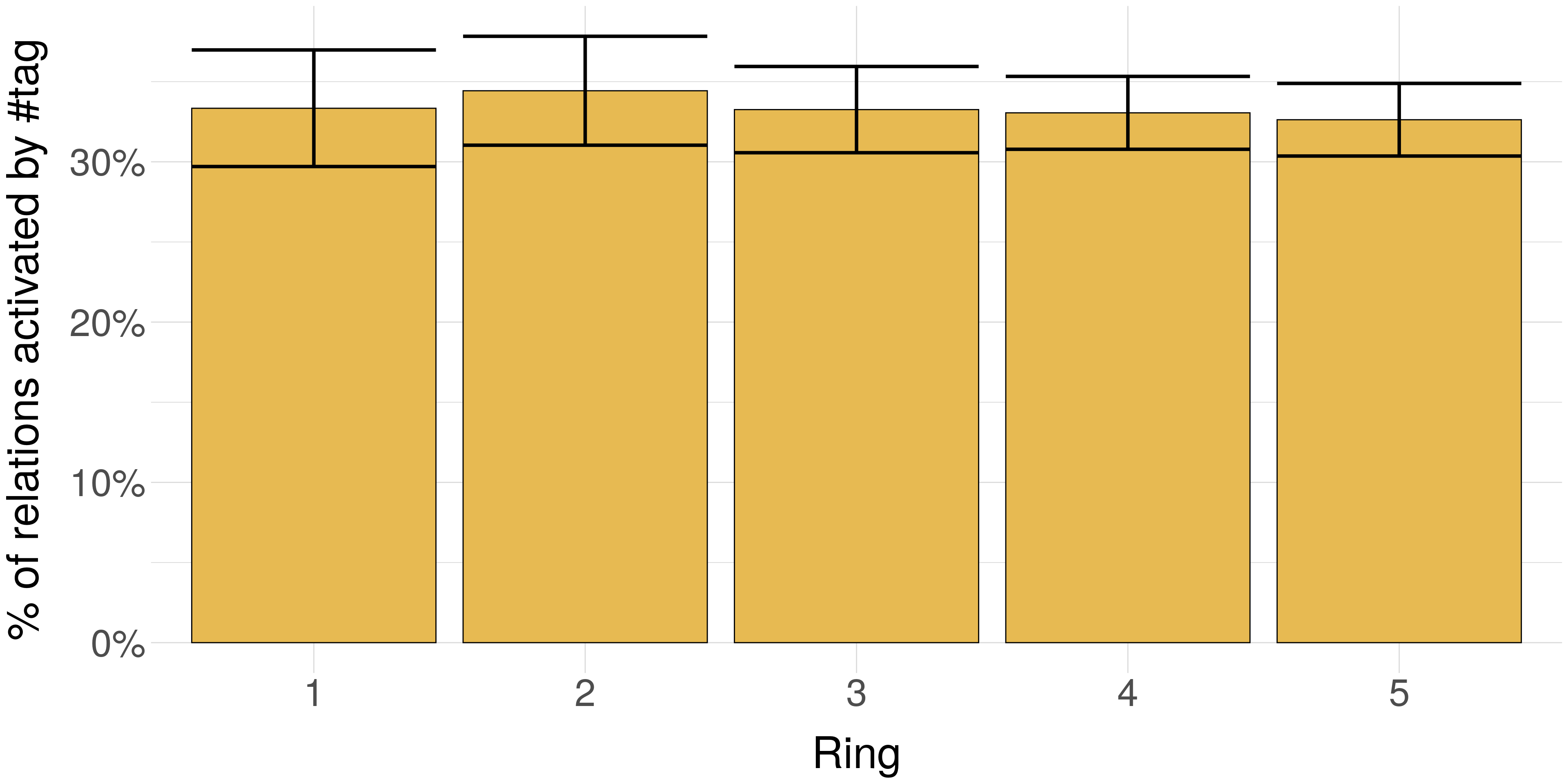}}
\hfill
\subfloat[Germany
\label{fig_appendix:relations_hashtags_rings_GermanJournalists}]
{\includegraphics[width=0.27\textwidth]
{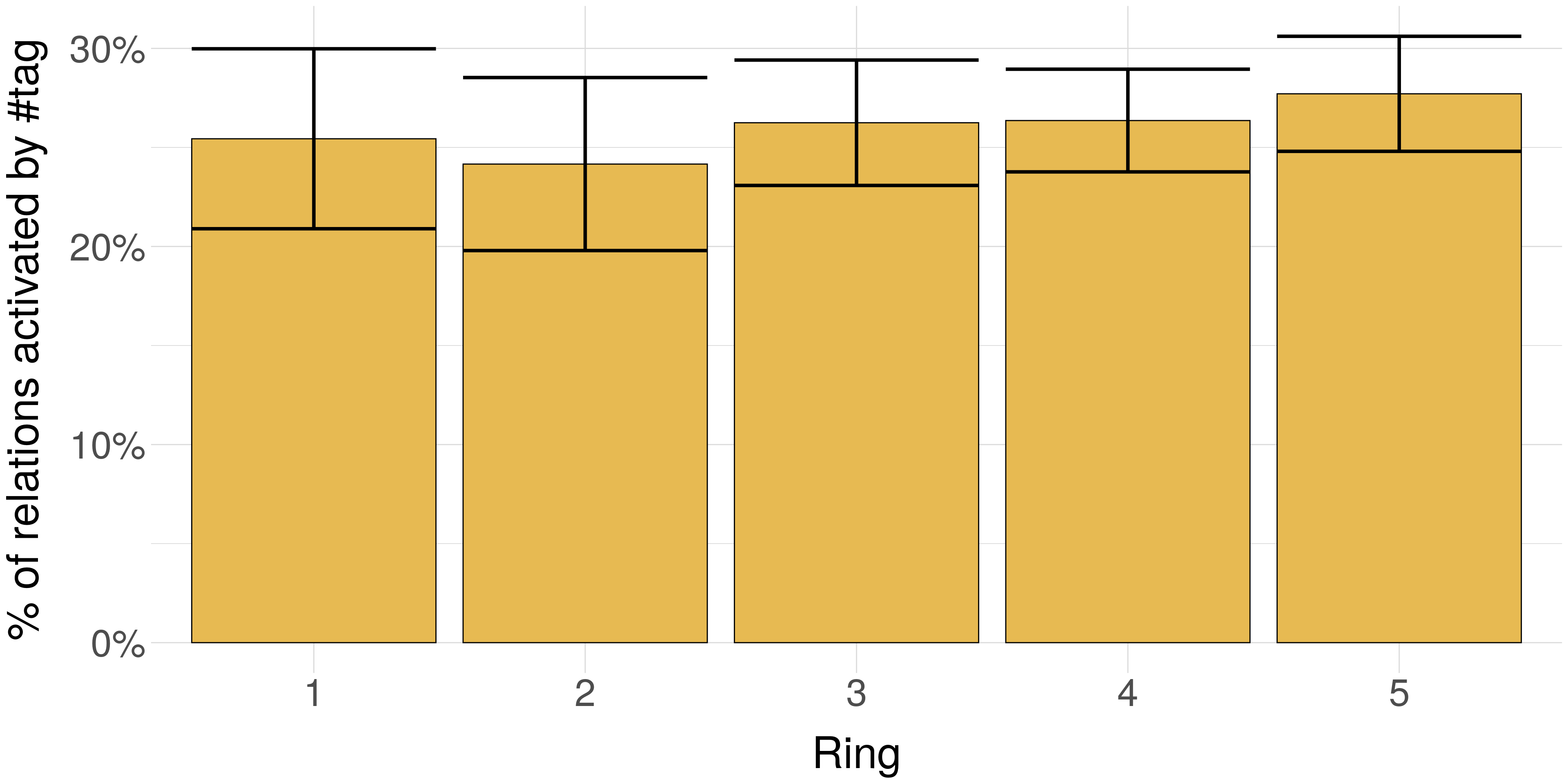}}
\hfill
\subfloat[Netherland
\label{fig_appendix:relations_hashtags_rings_NetherlanderJournalists}]
{\includegraphics[width=0.27\textwidth]
{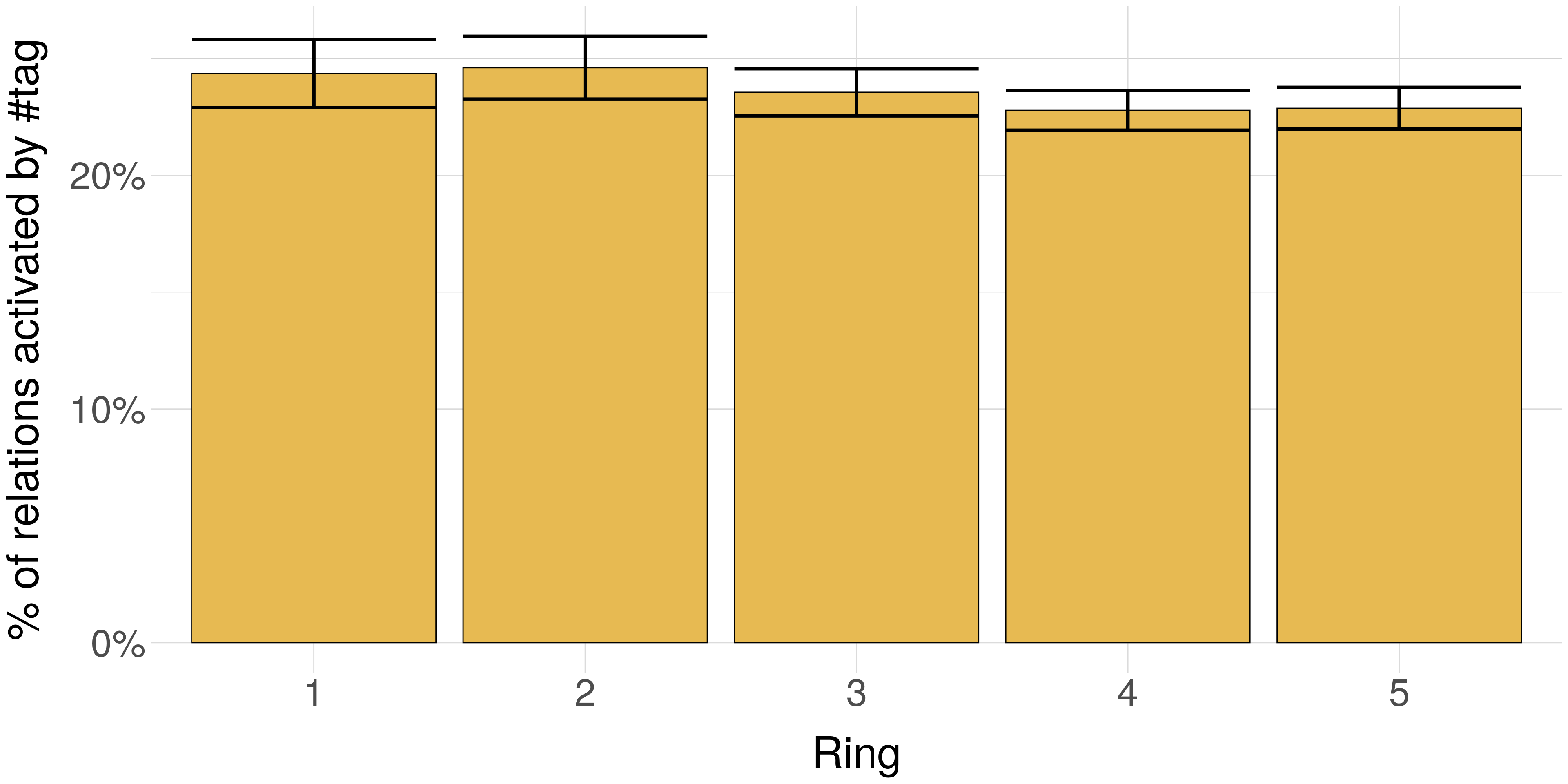}}
\hspace{1pt}
\subfloat[Australia
\label{fig_appendix:relations_hashtags_rings_AustralianJournalists}]
{\includegraphics[width=0.27\textwidth]
{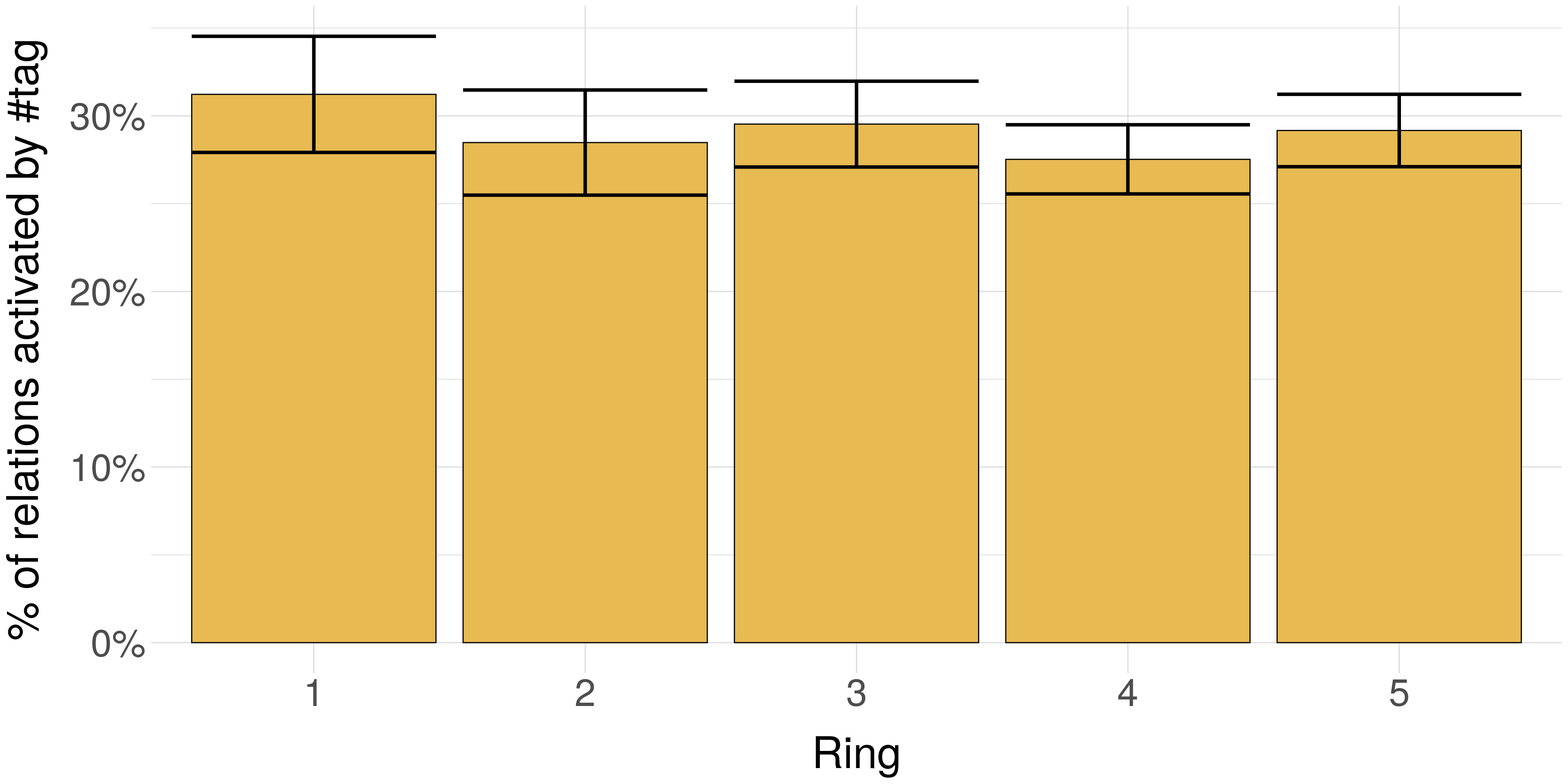}}
\end{center}
\end{adjustbox}
\caption{Average percentages of relationships activated by hashtags per ring, with confidence intervals}
\label{fig_appendix:relations_hashtags_rings}
\end{figure}


\begin{figure}[!h]
\begin{adjustbox}{minipage=\linewidth}
\begin{center}
\subfloat[USA
\label{fig_appendix:mean_hashtags_per_relation_AmericanJournalists}]
{\includegraphics[width=0.27\textwidth]
{./figures_new/mean_hashtag_per_rel/AmericanJournalists_avg_hashtag_per_ring_hashtag.png}}
\hfill
\subfloat[Canada
\label{fig_appendix:mean_hashtags_per_relation_CanadianJournalists}]
{\includegraphics[width=0.27\textwidth]
{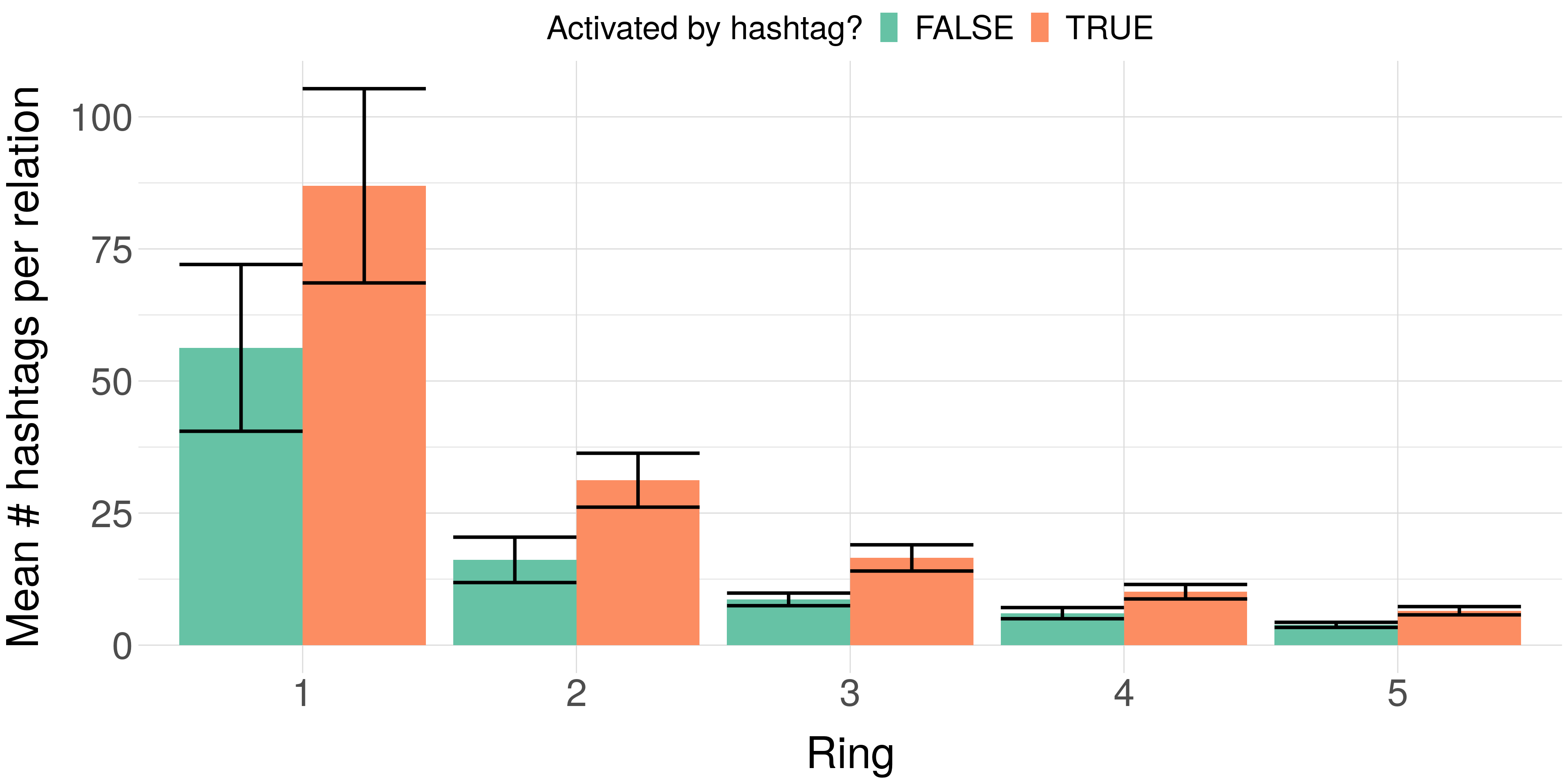}}
\hfill
\subfloat[Brasil
\label{fig_appendix:mean_hashtags_per_relation_BrazilianJournalists}]
{\includegraphics[width=0.27\textwidth]
{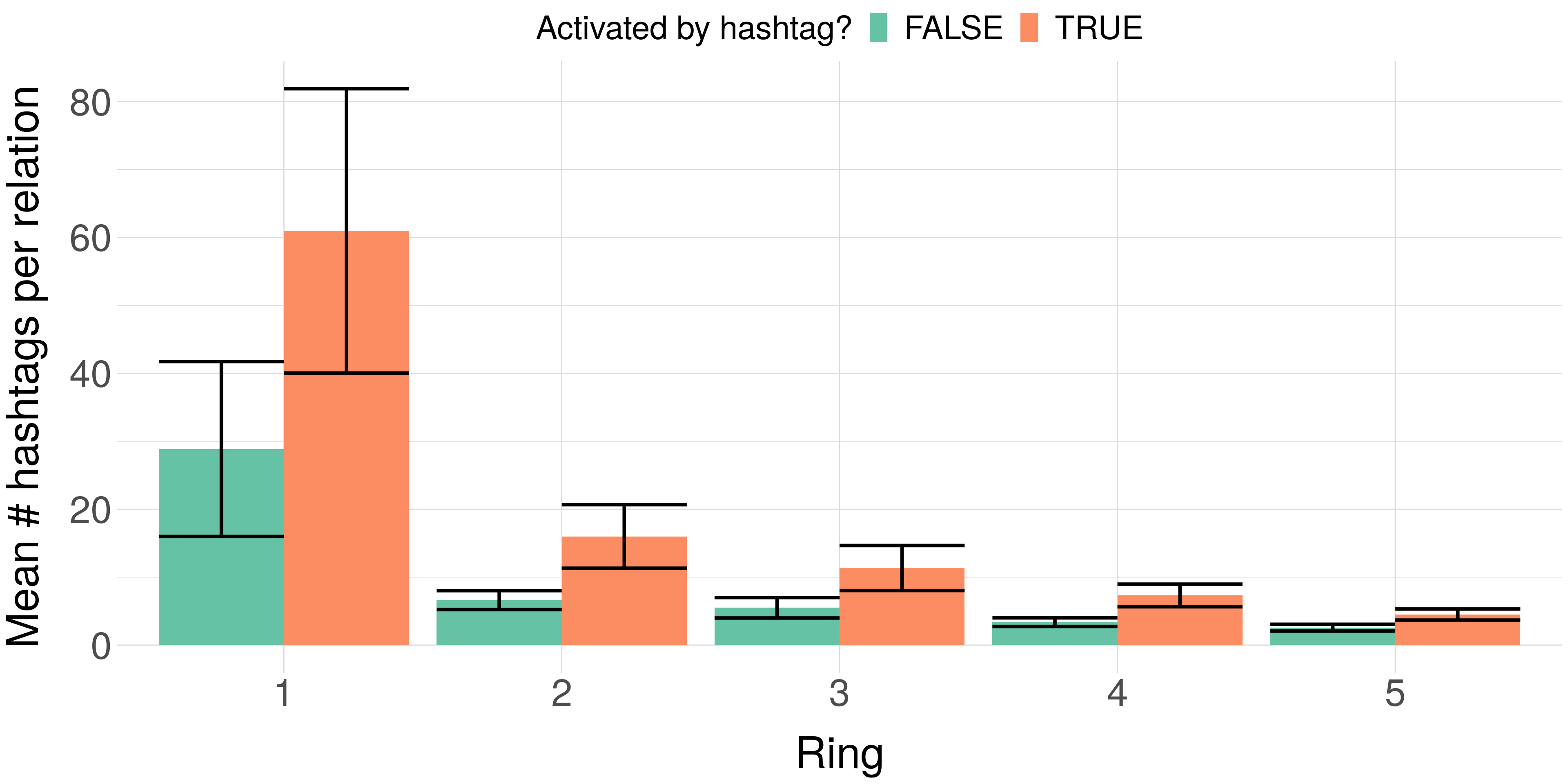}}
\hfill
\subfloat[Japan
\label{fig_appendix:mean_hashtags_per_relation_JapaneseJournalists}]
{\includegraphics[width=0.27\textwidth]
{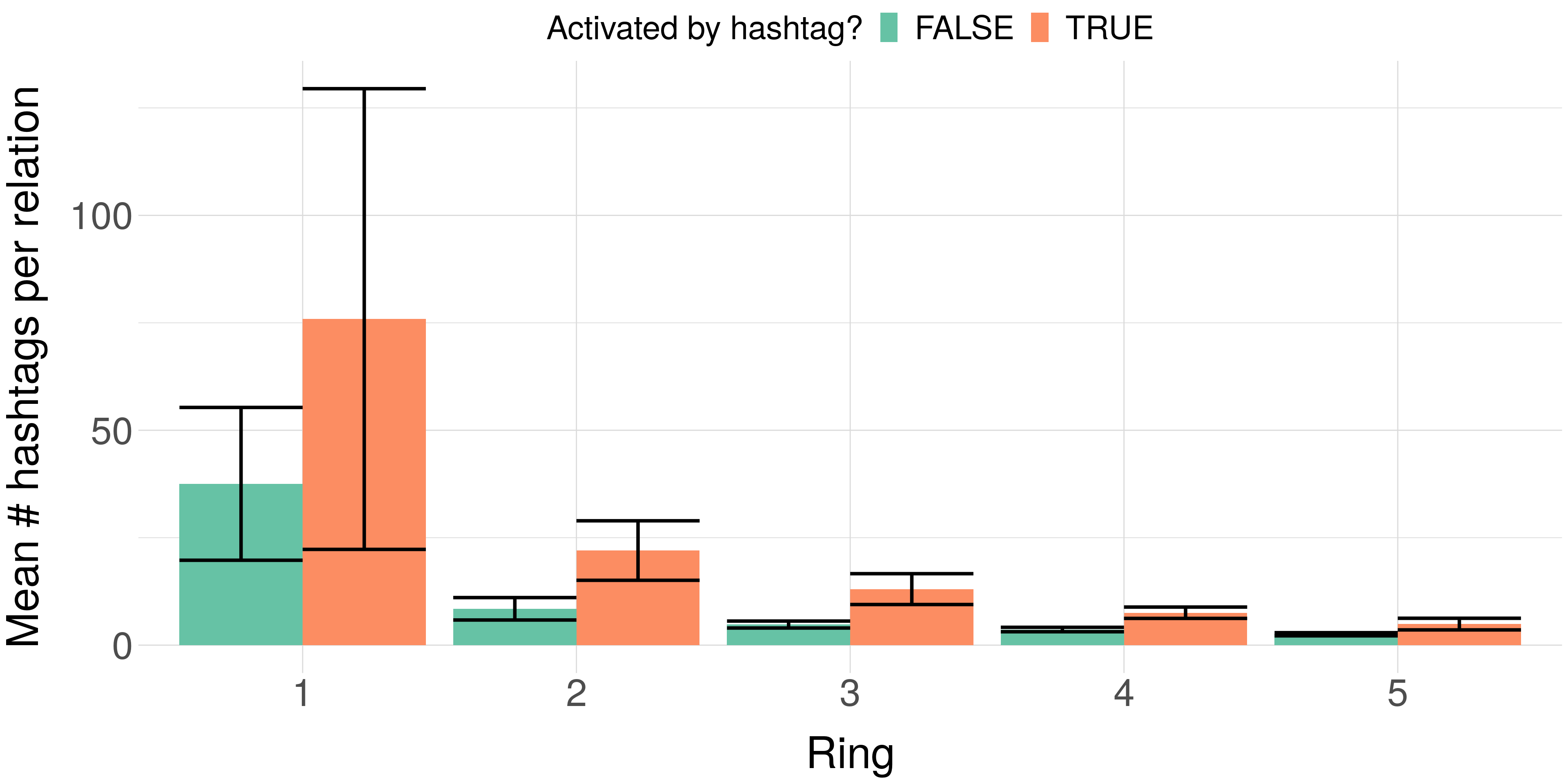}}
\hfill
\subfloat[Turkey
\label{fig_appendix:mean_hashtags_per_relation_TrukishJournalists}]
{\includegraphics[width=0.27\textwidth]
{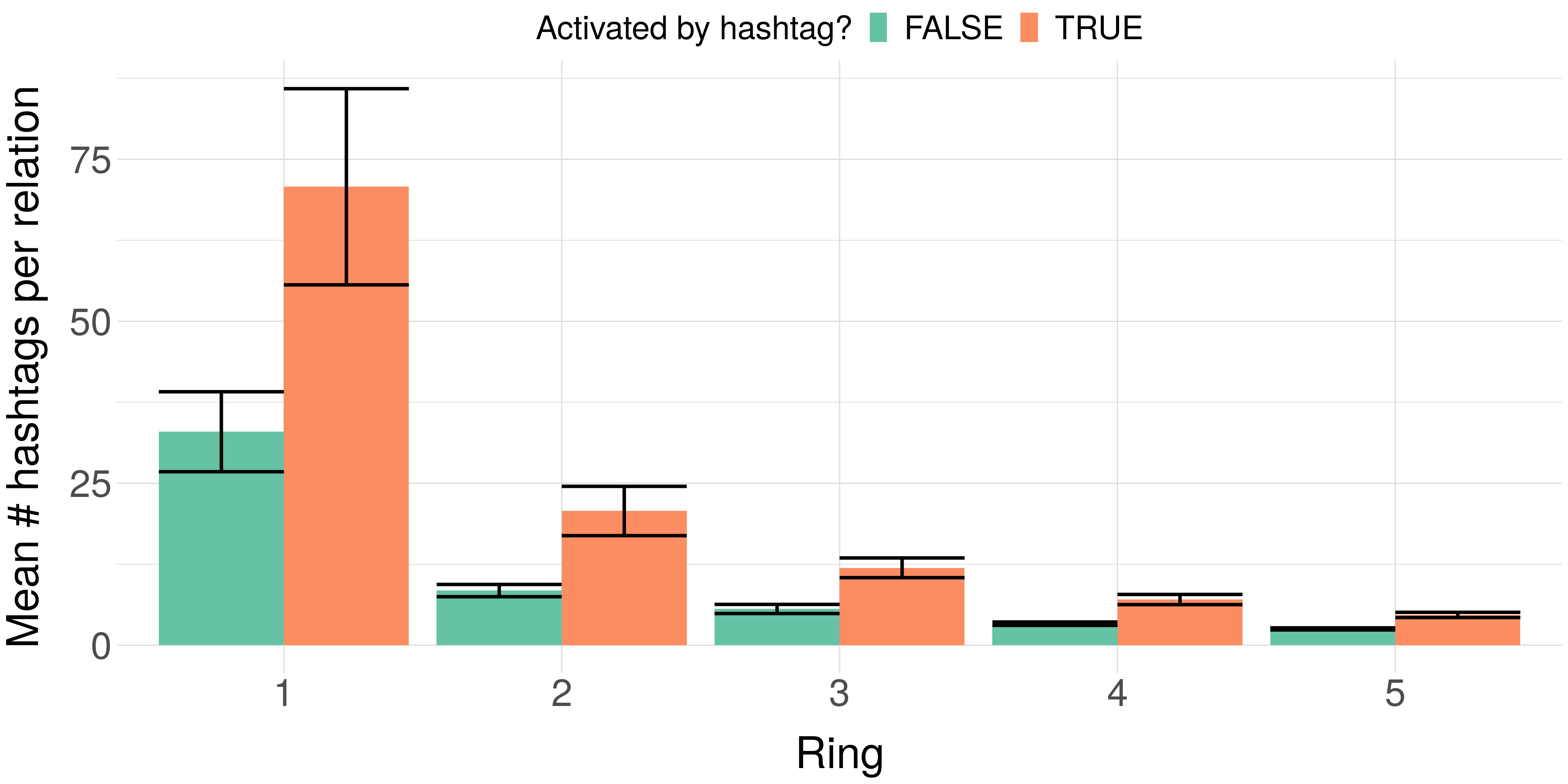}}
\hfill
\subfloat[UK
\label{fig_appendix:mean_hashtags_per_relation_BritishJournalists}]
{\includegraphics[width=0.27\textwidth]
{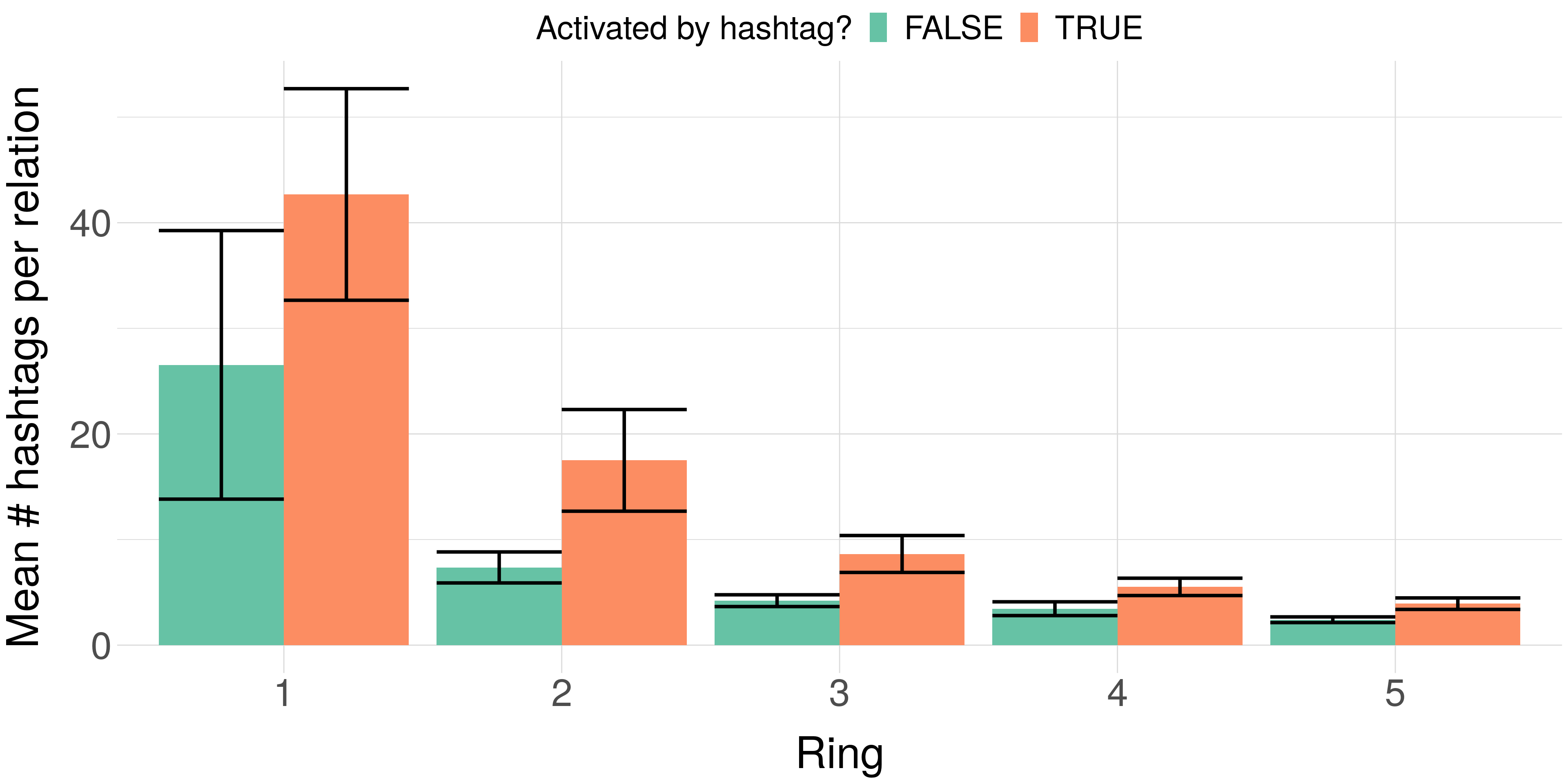}}
\hfill
\subfloat[Denmark
\label{fig_appendix:mean_hashtags_per_relation_DanishJournalists}]
{\includegraphics[width=0.27\textwidth]
{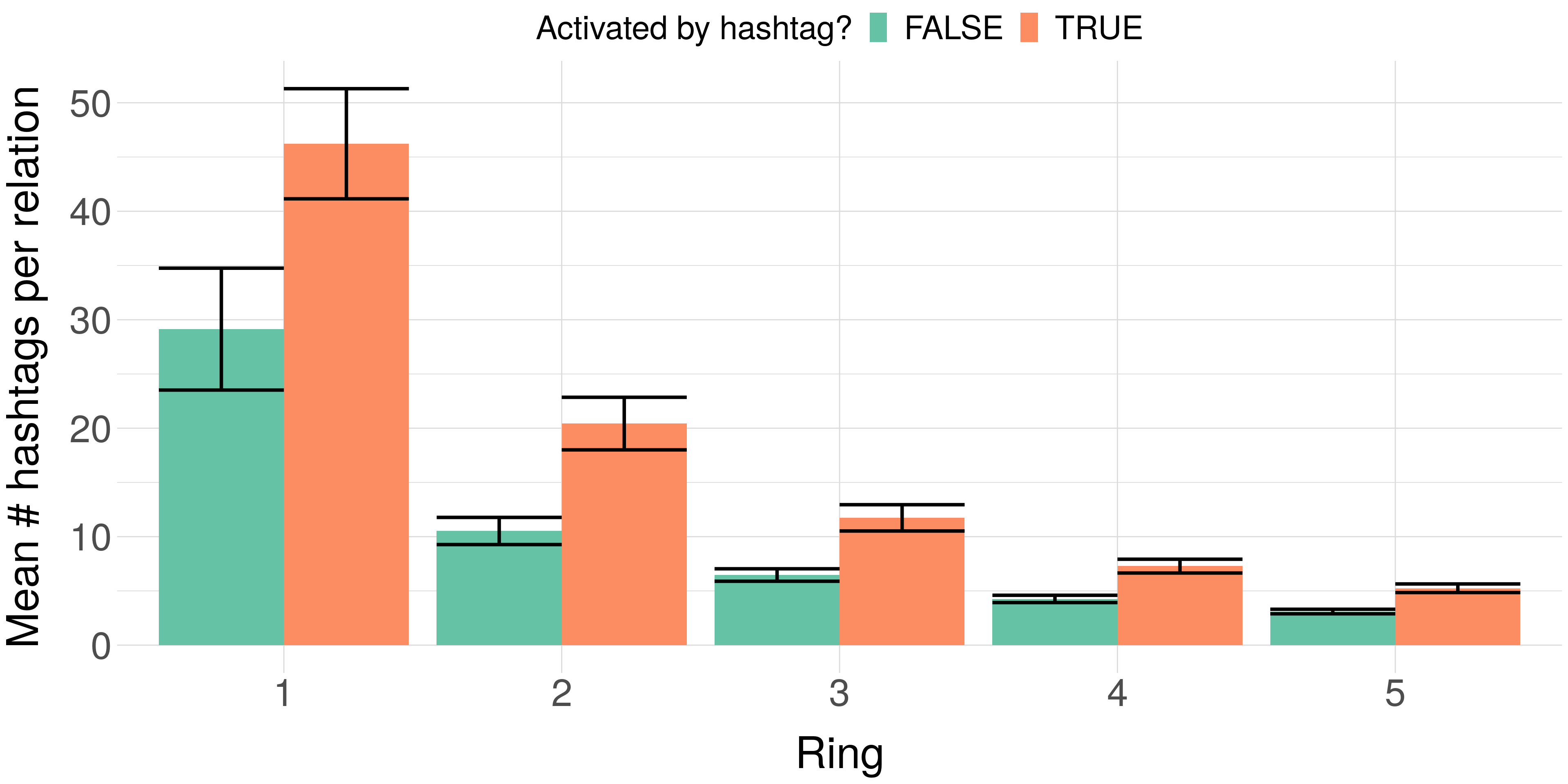}}
\hfill
\subfloat[Finland
\label{fig_appendix:mean_hashtags_per_relation_FinnishJournalists}]
{\includegraphics[width=0.27\textwidth]
{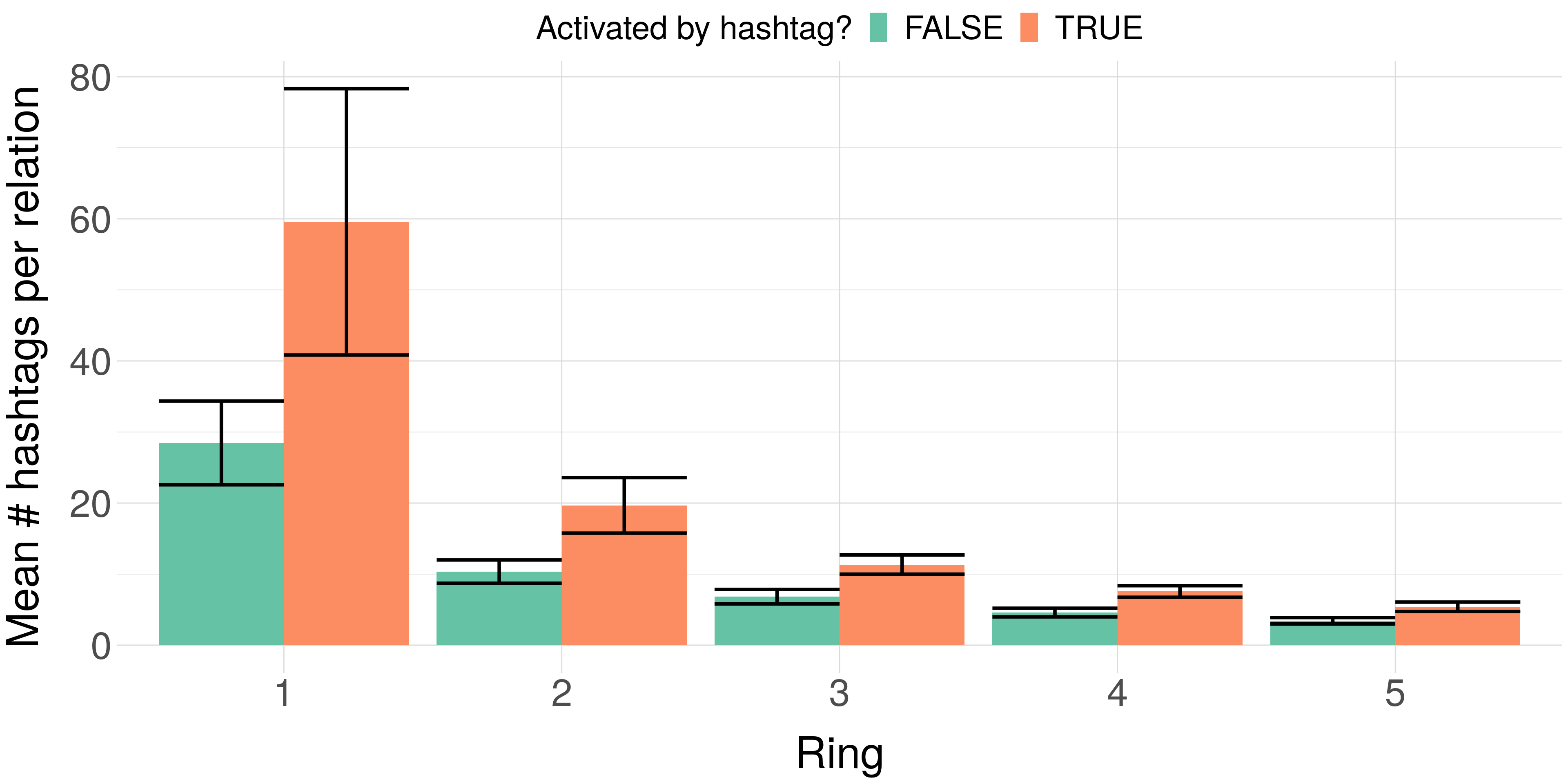}}
\hfill
\subfloat[Norway
\label{fig_appendix:mean_hashtags_per_relation_NorwegianJournalists}]
{\includegraphics[width=0.27\textwidth]
{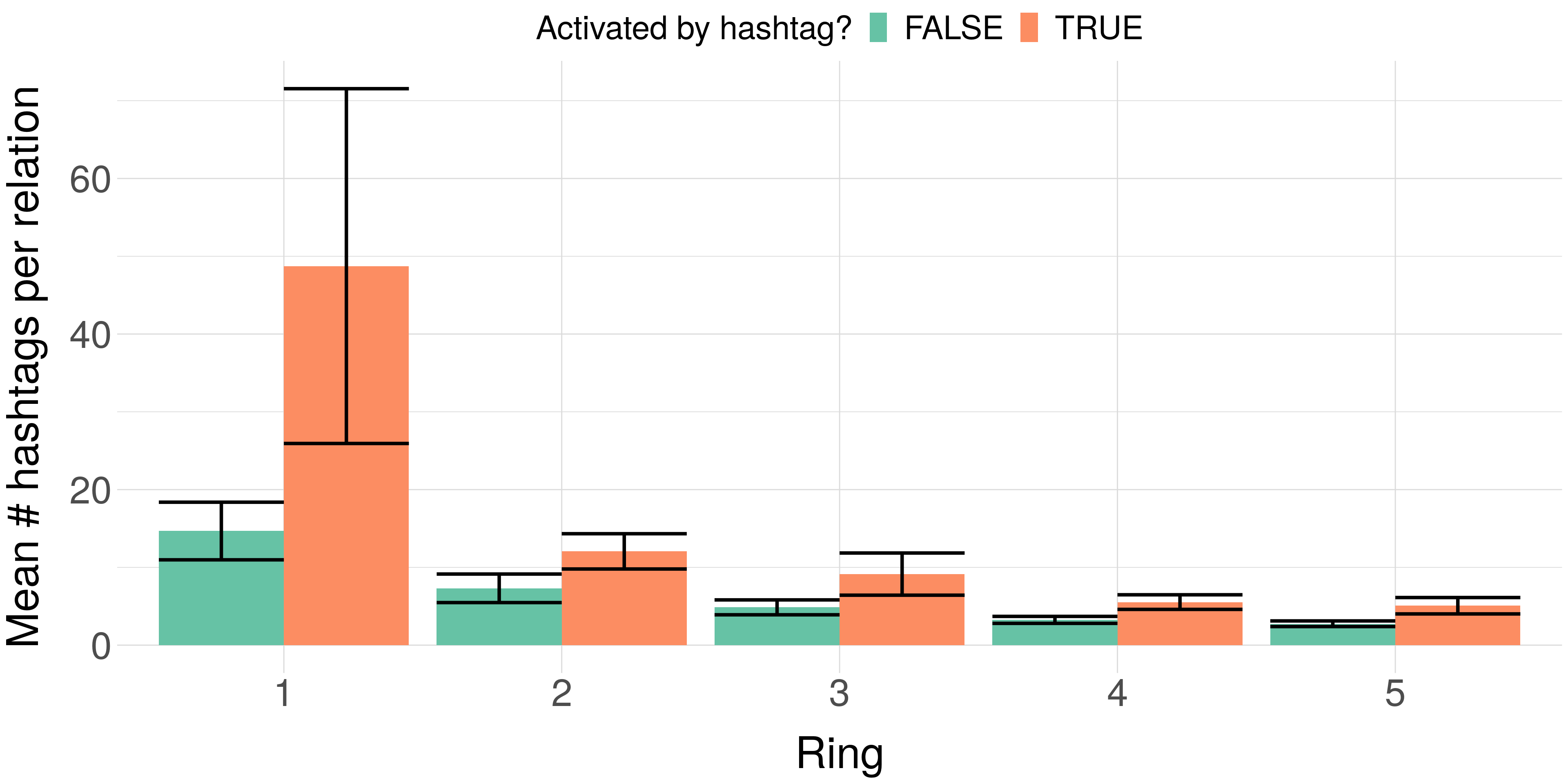}}
\hfill
\subfloat[Sweden
\label{fig_appendix:mean_hashtags_per_relation_SwedishJournalists}]
{\includegraphics[width=0.27\textwidth]
{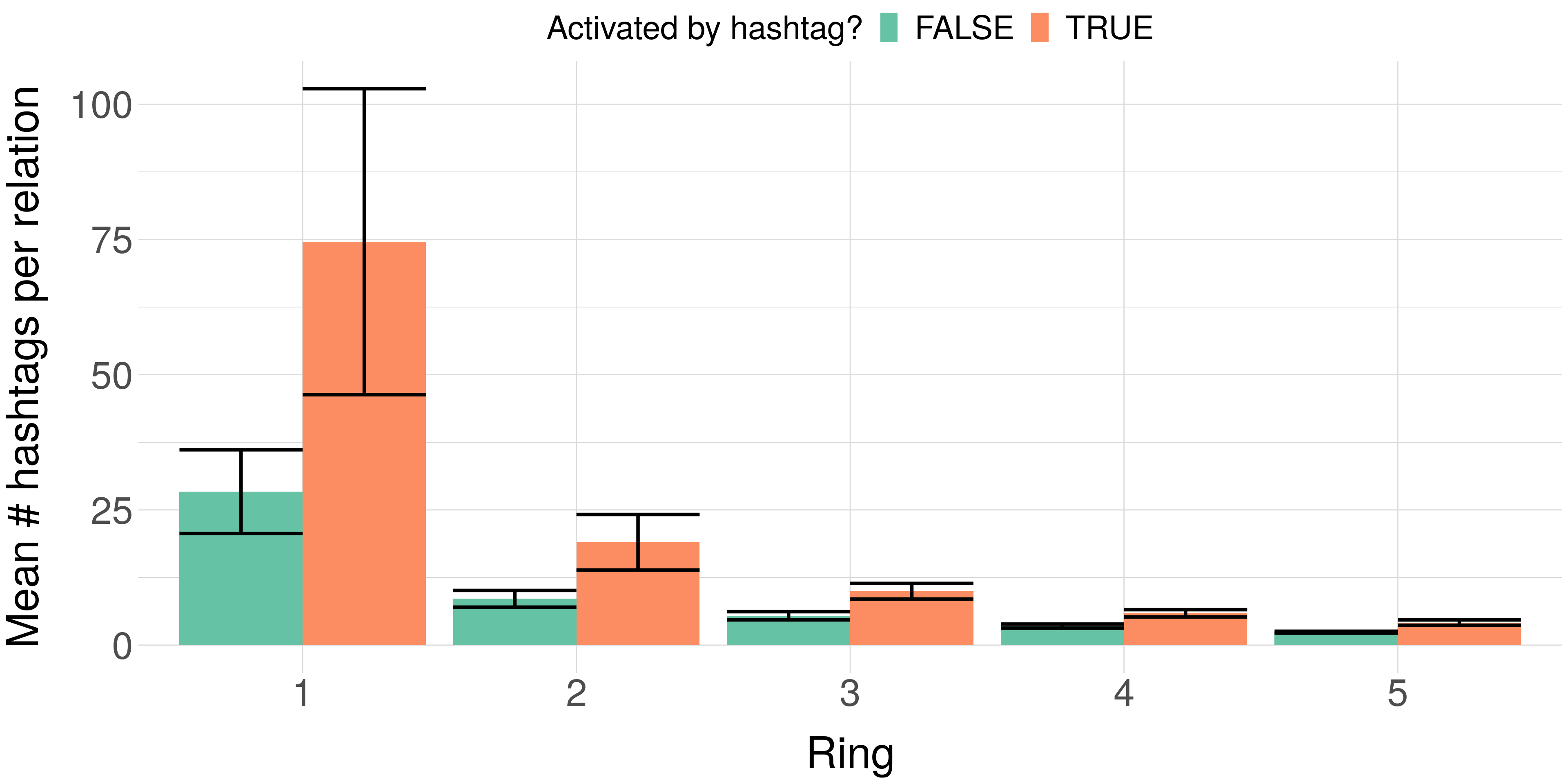}}
\hfill
\subfloat[Greece
\label{fig_appendix:mean_hashtags_per_relation_GreekJournalists}]
{\includegraphics[width=0.27\textwidth]
{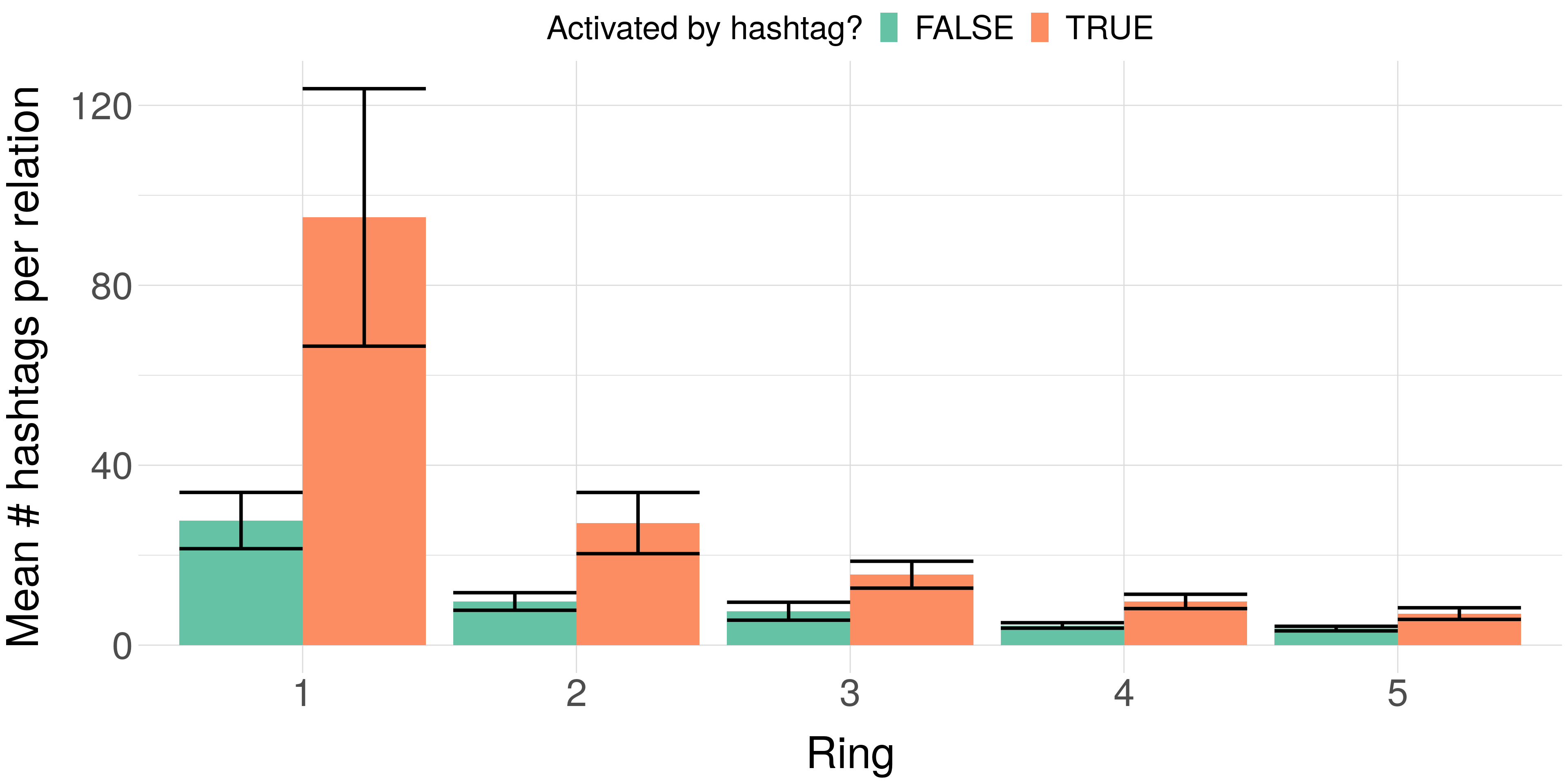}}
\hfill
\subfloat[Italy
\label{fig_appendix:mean_hashtags_per_relation_ItalianJournalists}]
{\includegraphics[width=0.27\textwidth]
{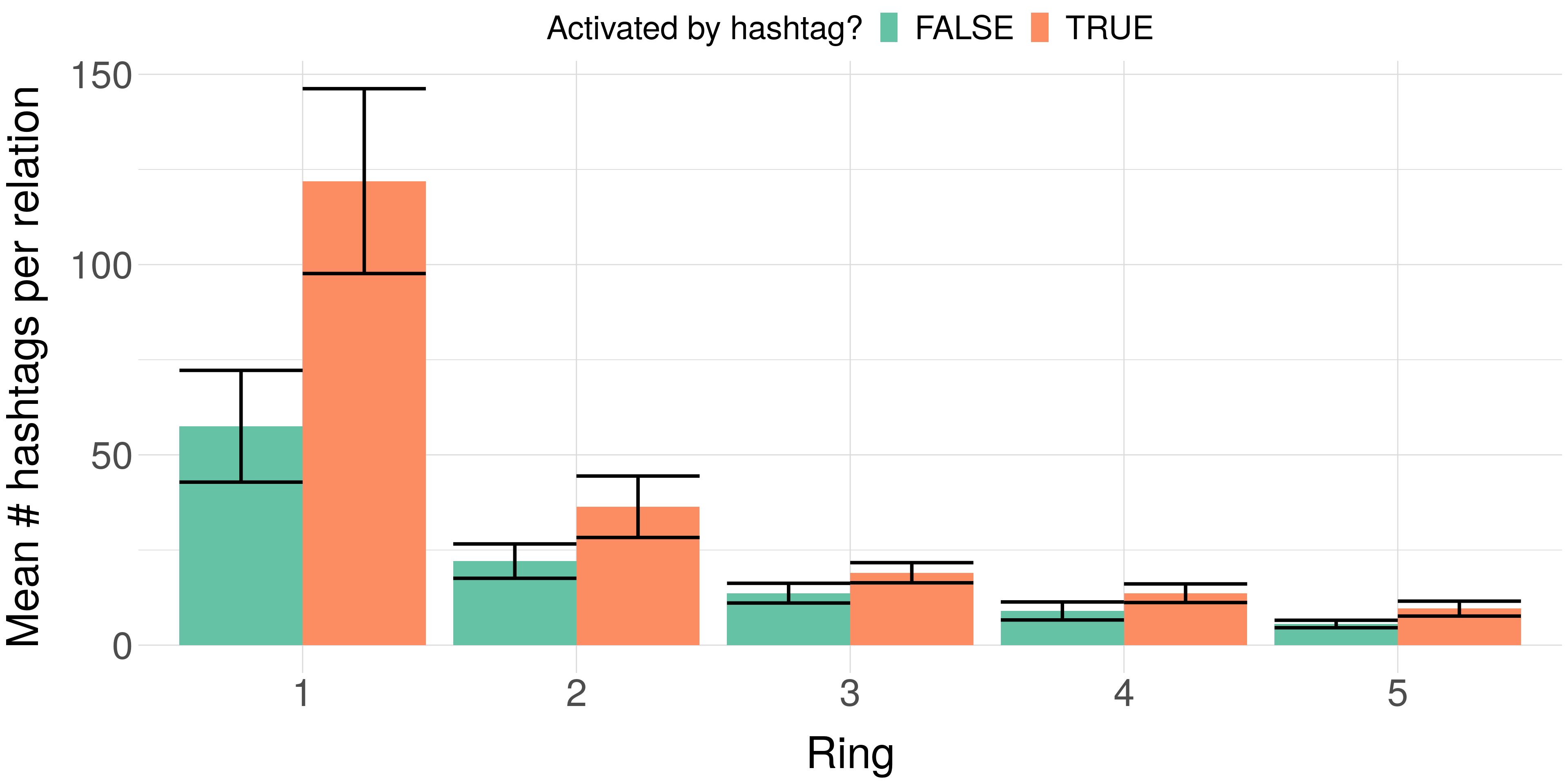}}
\hfill
\subfloat[Spain
\label{fig_appendix:mean_hashtags_per_relation_SpanishJournalists}]
{\includegraphics[width=0.27\textwidth]
{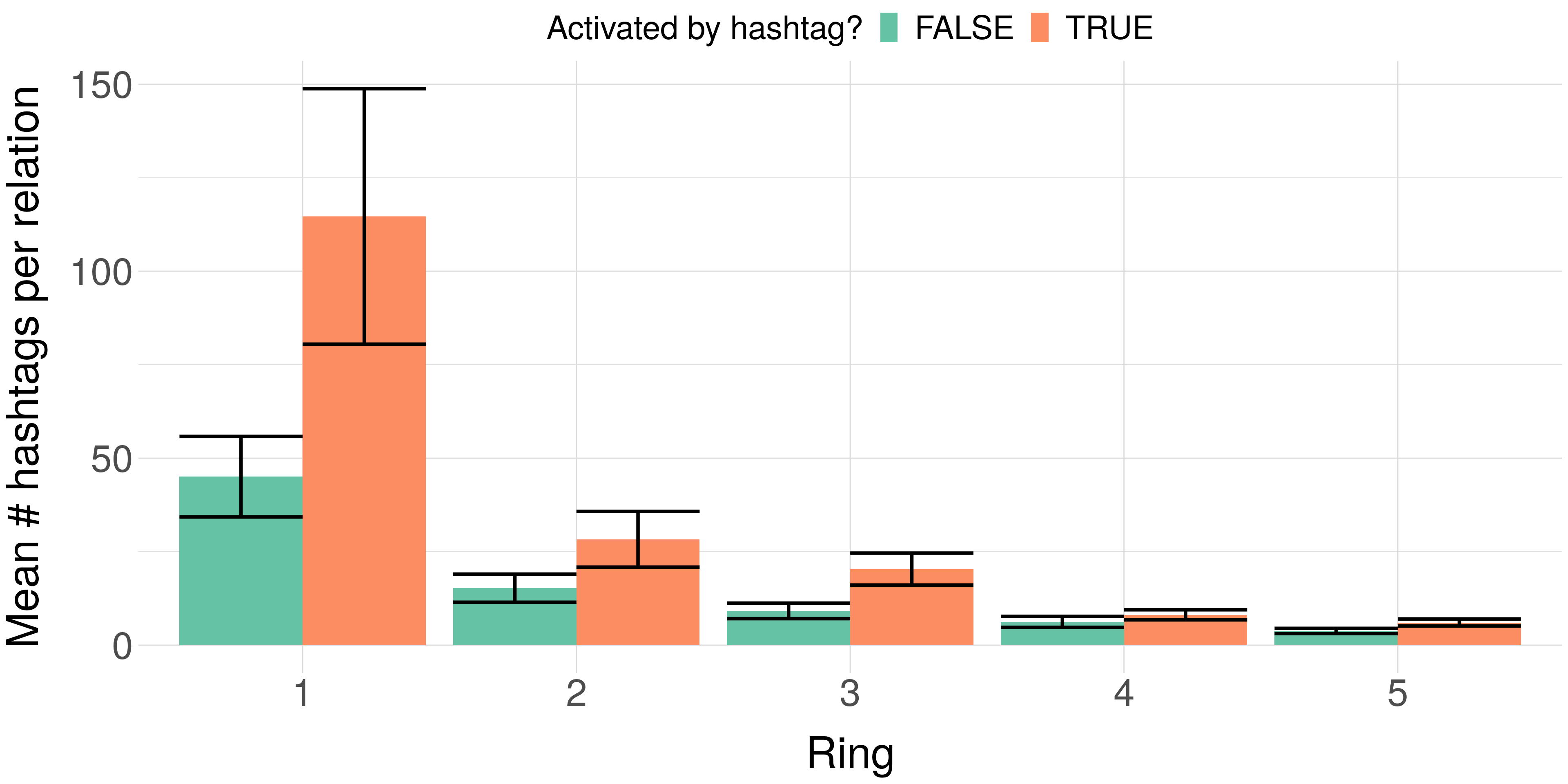}}
\hfill
\subfloat[France
\label{fig_appendix:mean_hashtags_per_relation_FrenchJournalists}]
{\includegraphics[width=0.27\textwidth]
{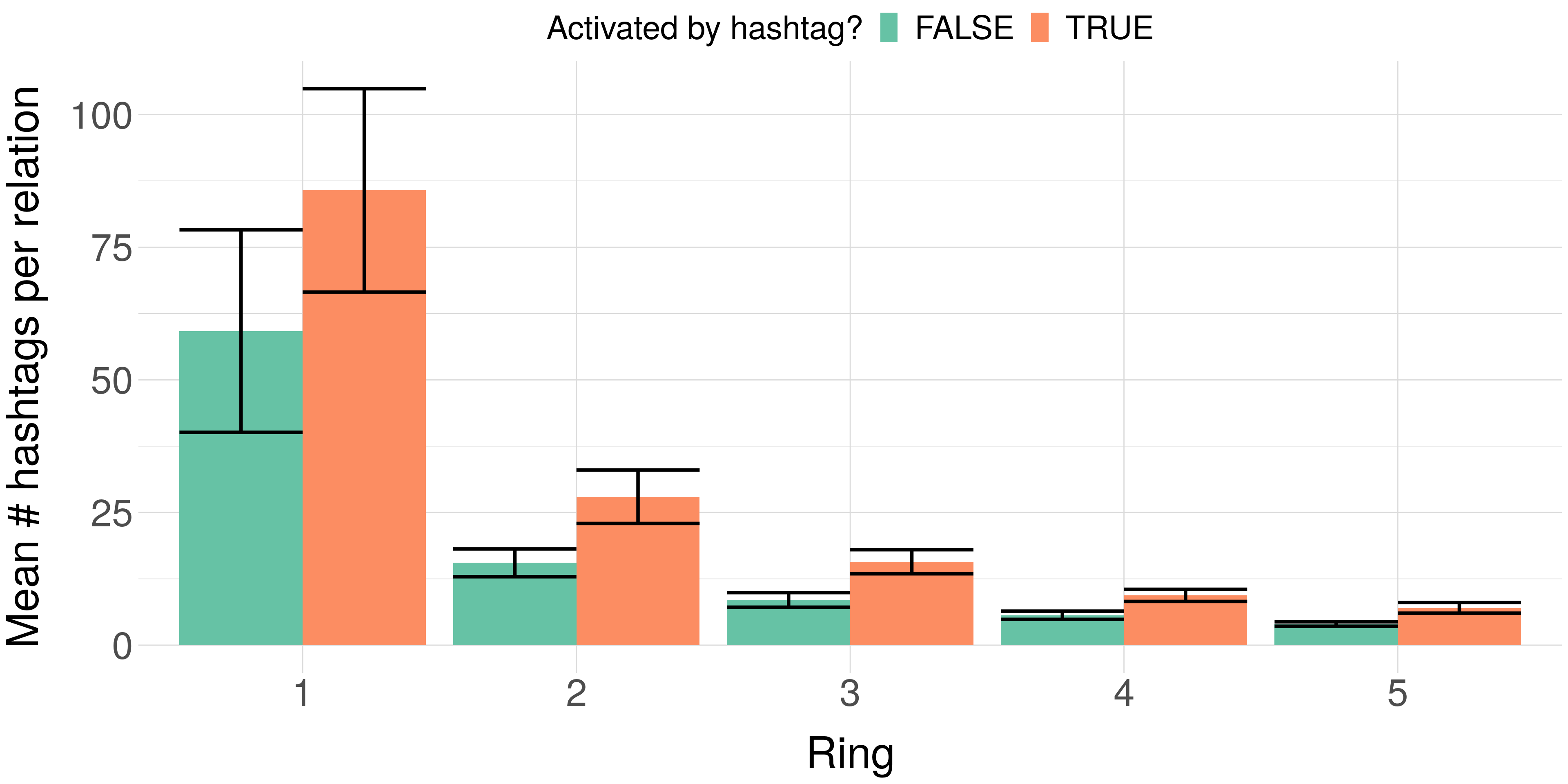}}
\hfill
\subfloat[Germany
\label{fig_appendix:mean_hashtags_per_relation_GermanJournalists}]
{\includegraphics[width=0.27\textwidth]
{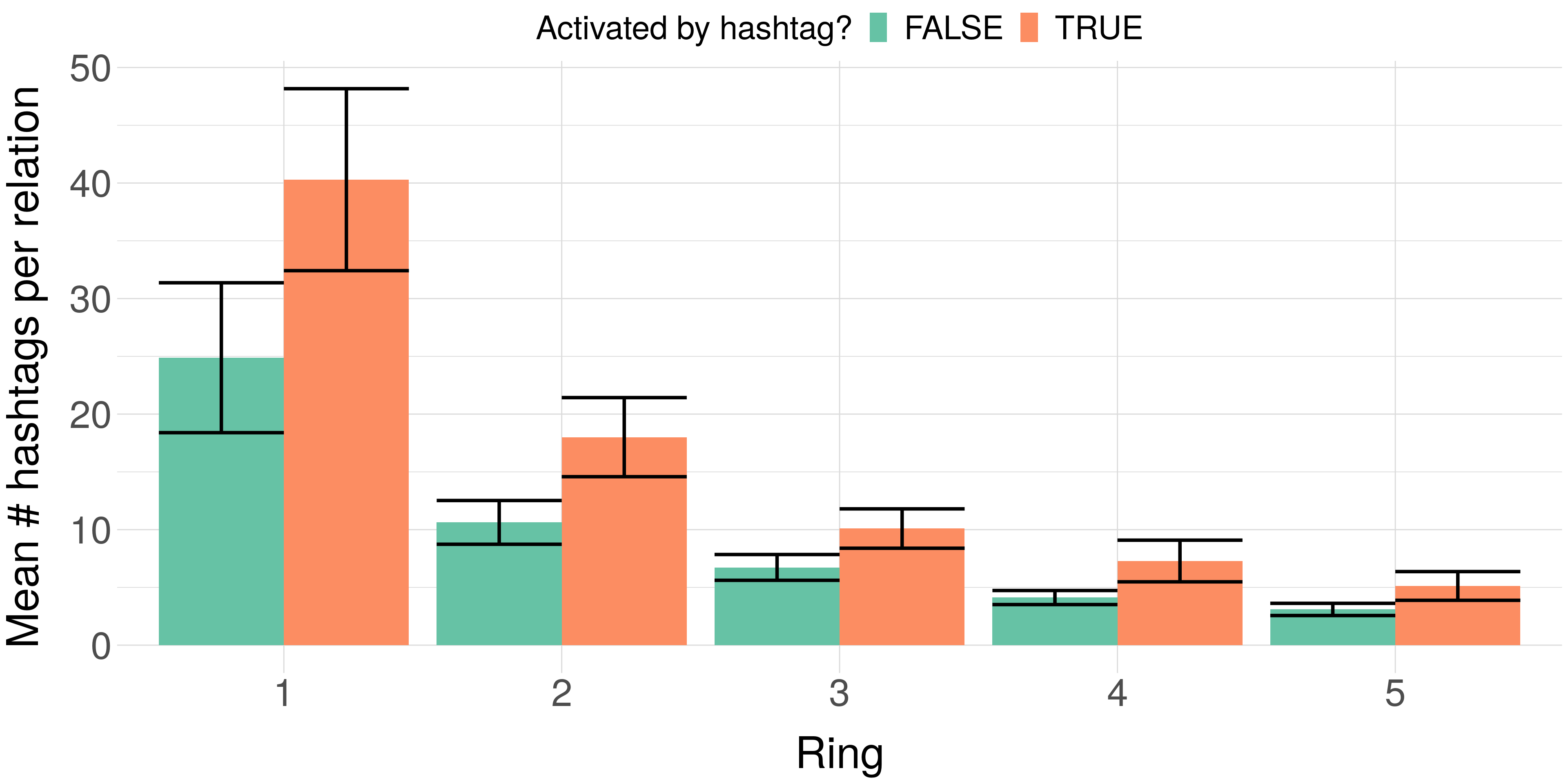}}
\hfill
\subfloat[Netherland
\label{fig_appendix:mean_hashtags_per_relation_NetherlanderJournalists}]
{\includegraphics[width=0.27\textwidth]
{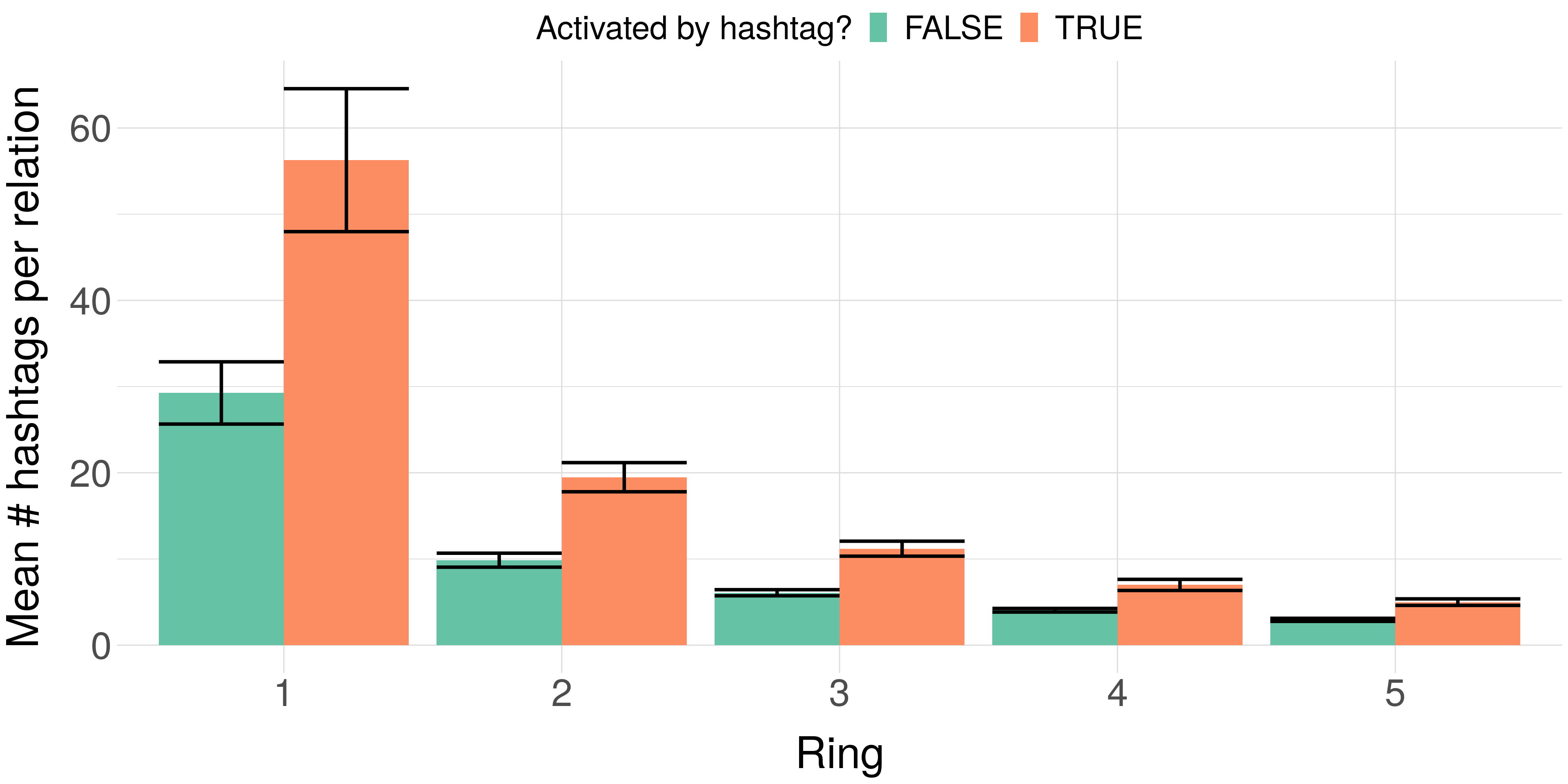}}
\hspace{1pt}
\subfloat[Australia
\label{fig_appendix:mean_hashtags_per_relation_AustralianJournalists}]
{\includegraphics[width=0.27\textwidth]
{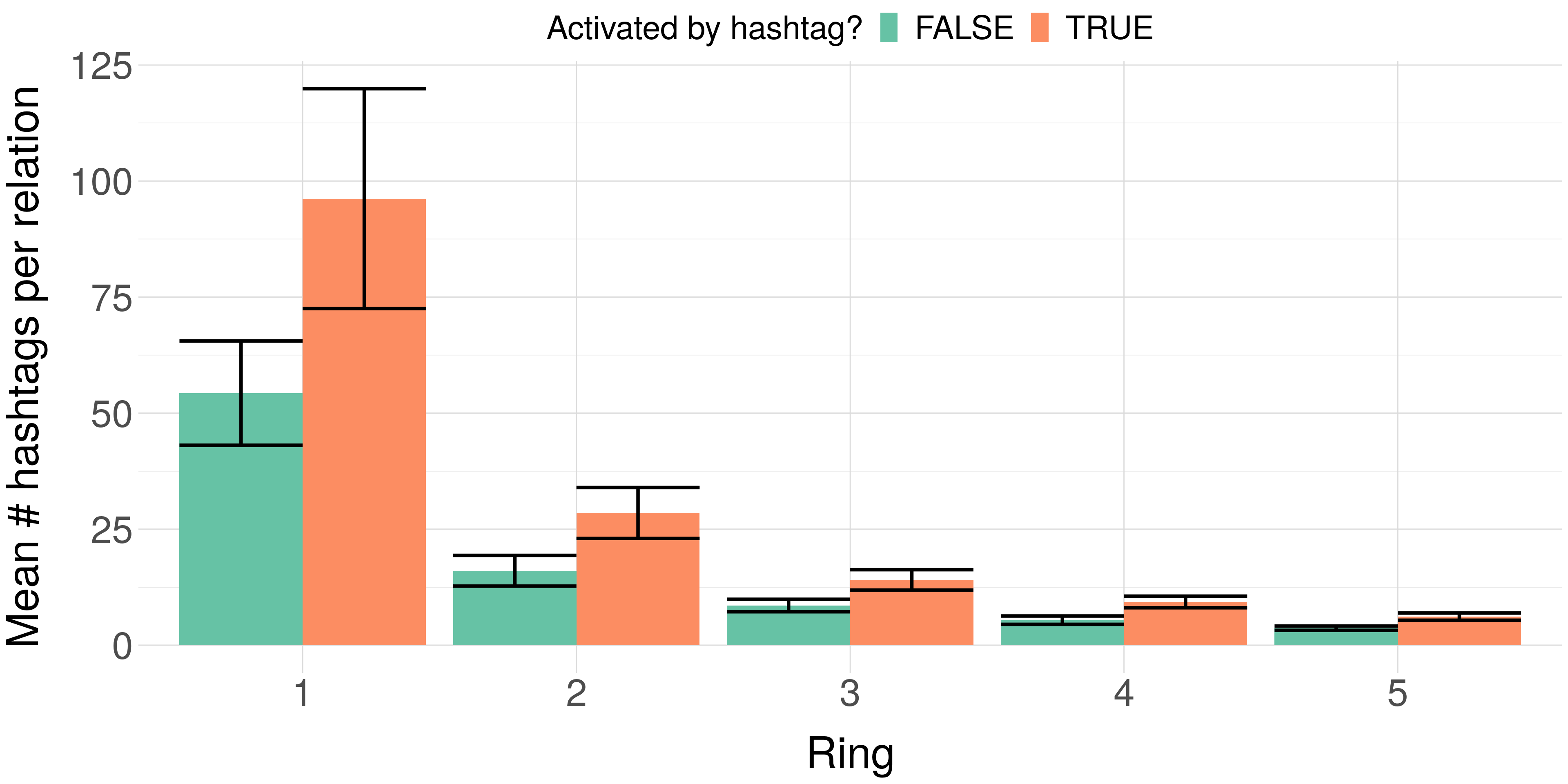}}
\end{center}
\end{adjustbox}
\caption{Average number of hashtags per relationship per ring, with confidence intervals}
\label{fig_appendix:mean_hashtags_per_relation}
\end{figure}


\begin{figure}[!h]
\begin{adjustbox}{minipage=\linewidth}
\begin{center}
\subfloat[USA
\label{fig_appendix:contact_freq_rings_hashtags_AmericanJournalists}]
{\includegraphics[width=0.27\textwidth]
{./figures_new/contact_freq_ring_hashtag/AmericanJournalists_avg_contact_freq_by_ring_hashtag.png}}
\hfill
\subfloat[Canada
\label{fig_appendix:contact_freq_rings_hashtags_CanadianJournalists}]
{\includegraphics[width=0.27\textwidth]
{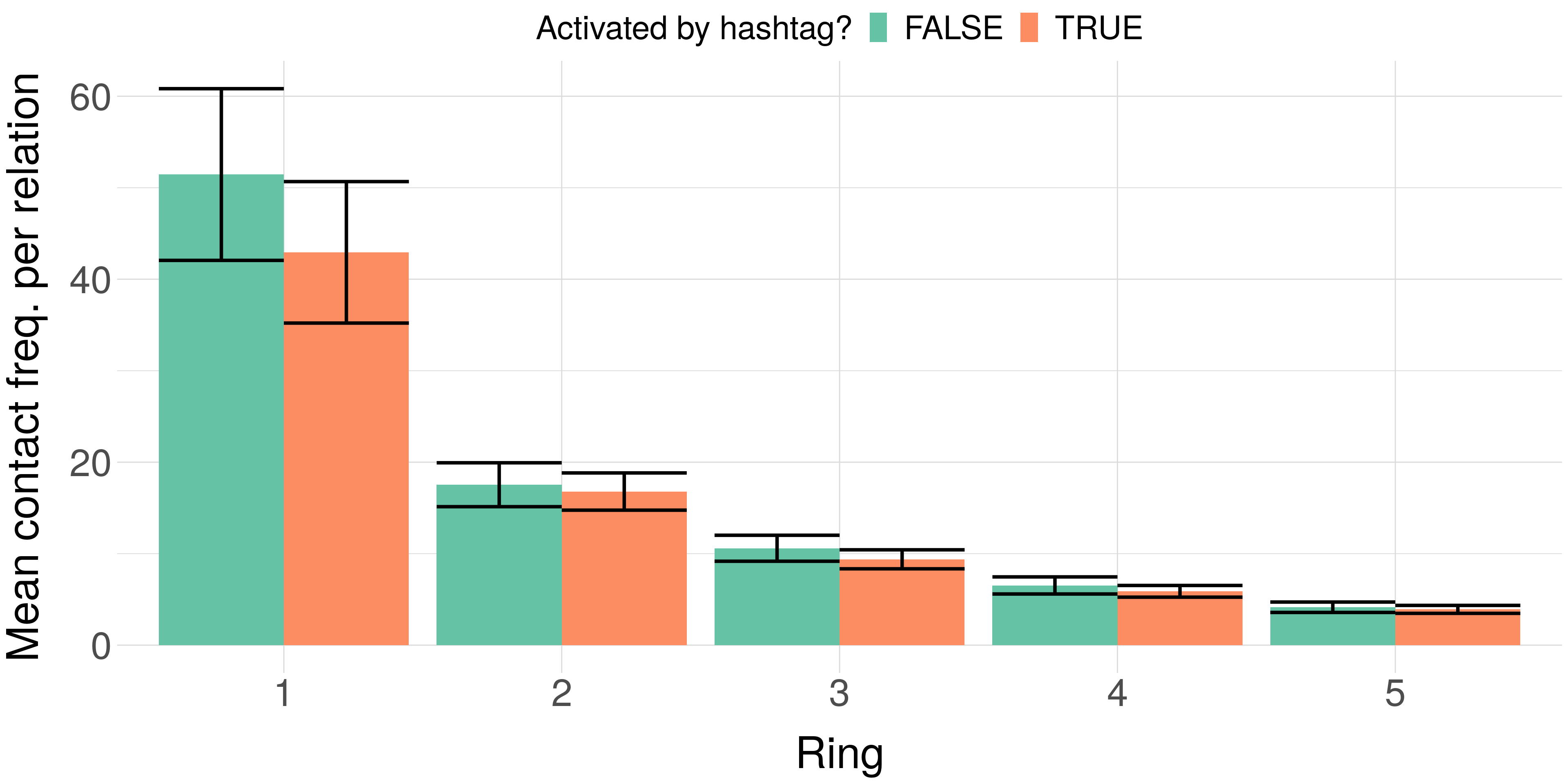}}
\hfill
\subfloat[Brasil
\label{fig_appendix:contact_freq_rings_hashtags_BrazilianJournalists}]
{\includegraphics[width=0.27\textwidth]
{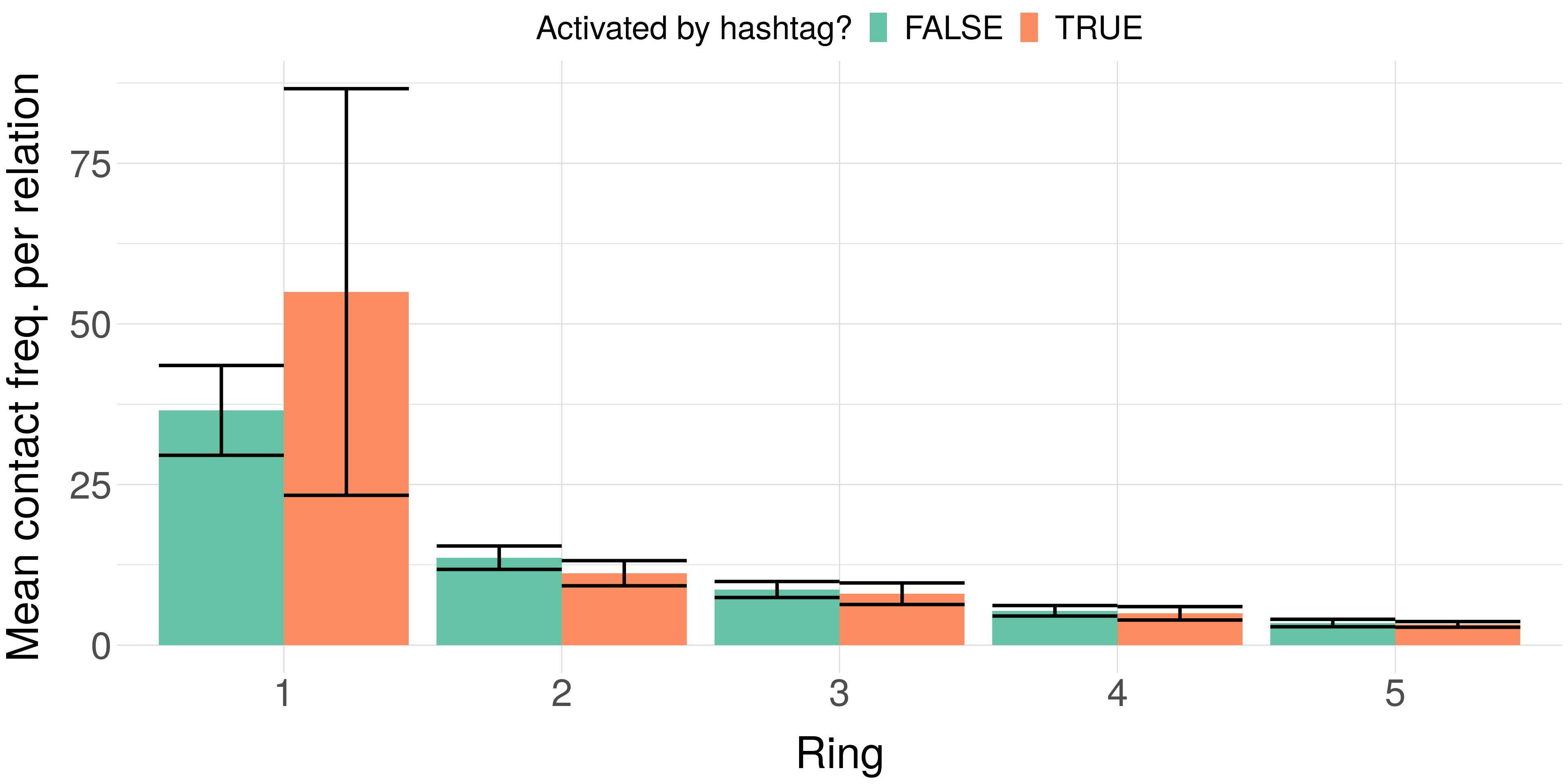}}
\hfill
\subfloat[Japan
\label{fig_appendix:contact_freq_rings_hashtags_JapaneseJournalists}]
{\includegraphics[width=0.27\textwidth]
{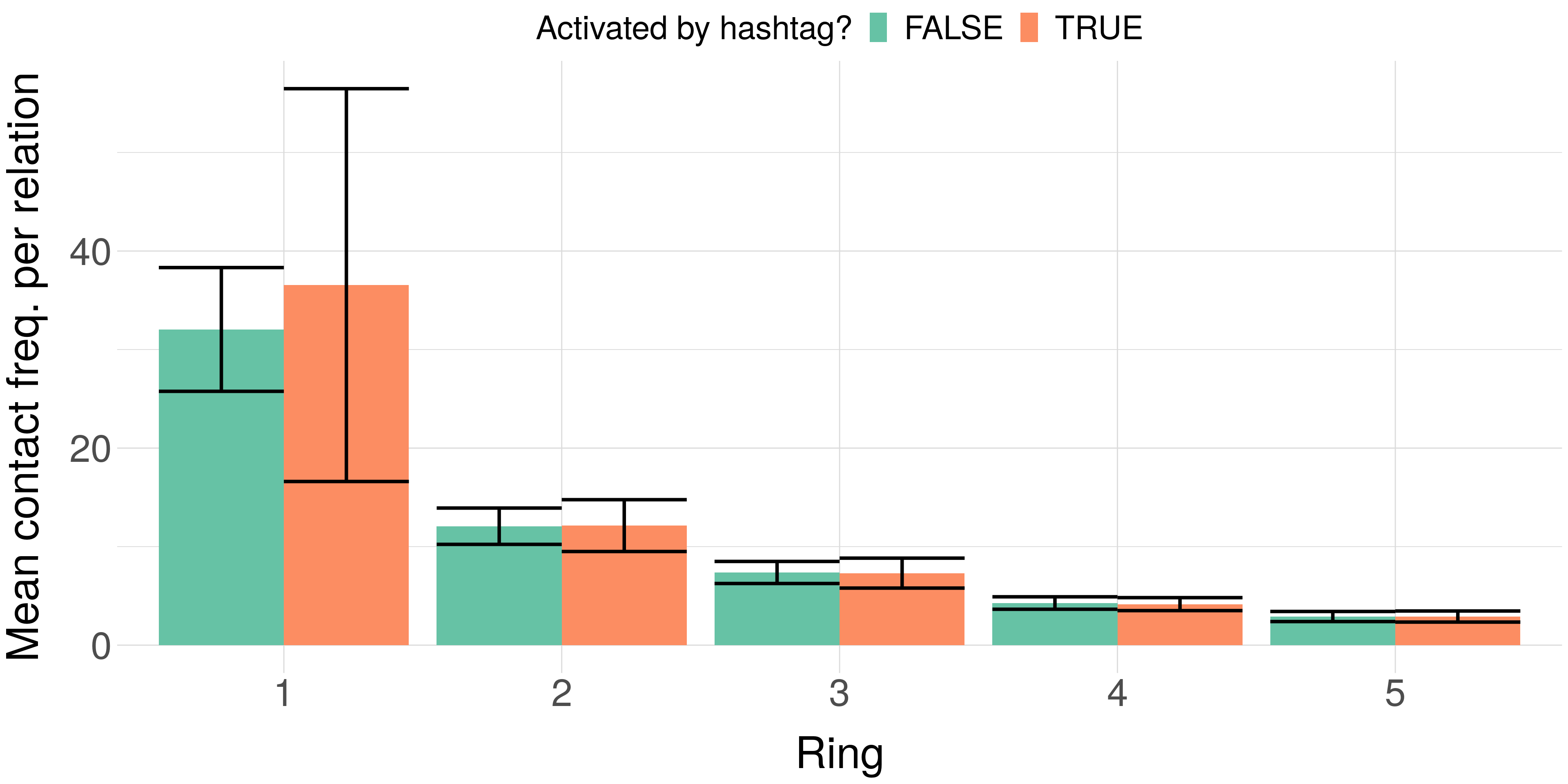}}
\hfill
\subfloat[Turkey
\label{fig_appendix:contact_freq_rings_hashtags_TrukishJournalists}]
{\includegraphics[width=0.27\textwidth]
{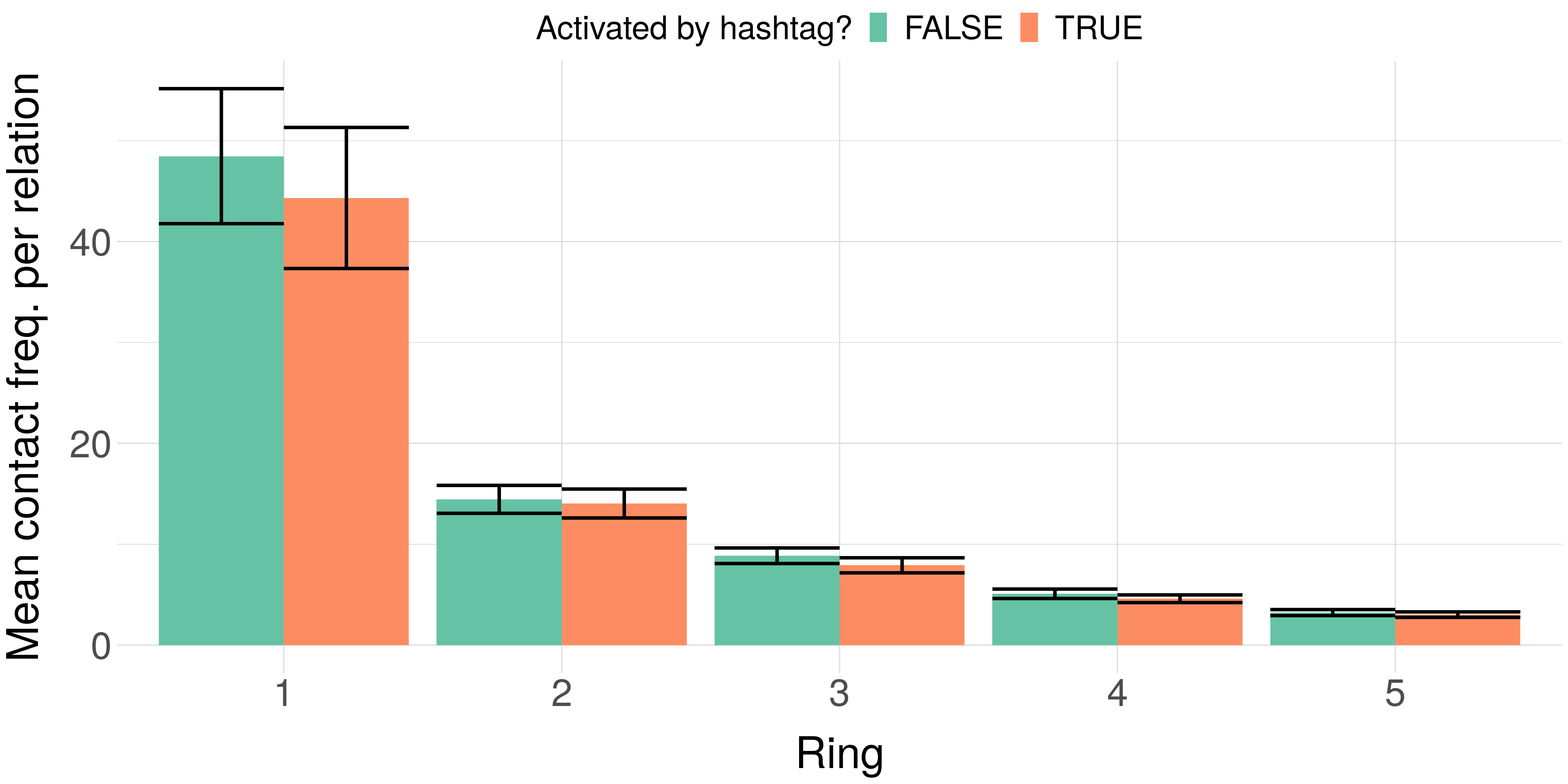}}
\hfill
\subfloat[UK
\label{fig_appendix:contact_freq_rings_hashtags_BritishJournalists}]
{\includegraphics[width=0.27\textwidth]
{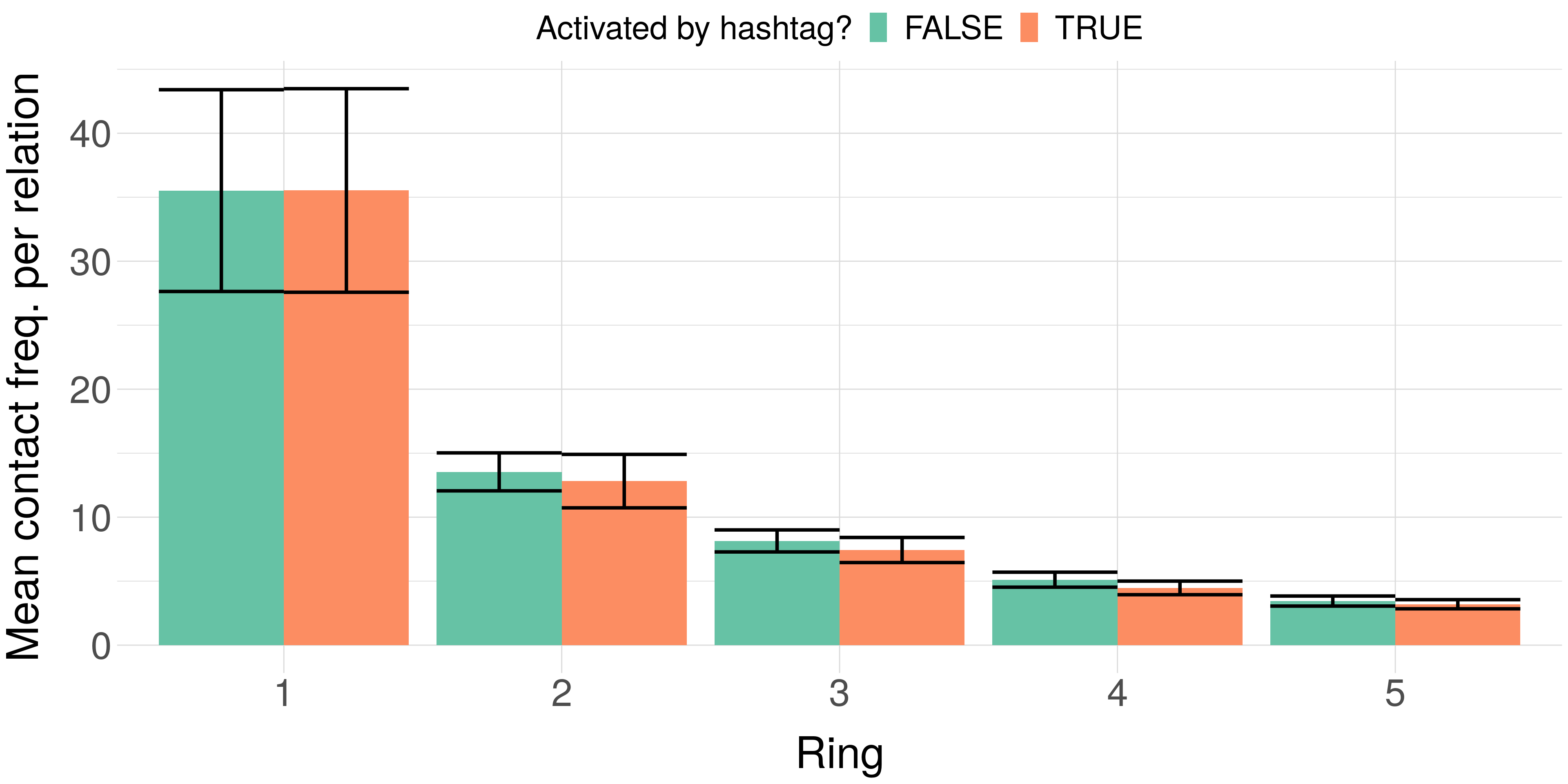}}
\hfill
\subfloat[Denmark
\label{fig_appendix:contact_freq_rings_hashtags_DanishJournalists}]
{\includegraphics[width=0.27\textwidth]
{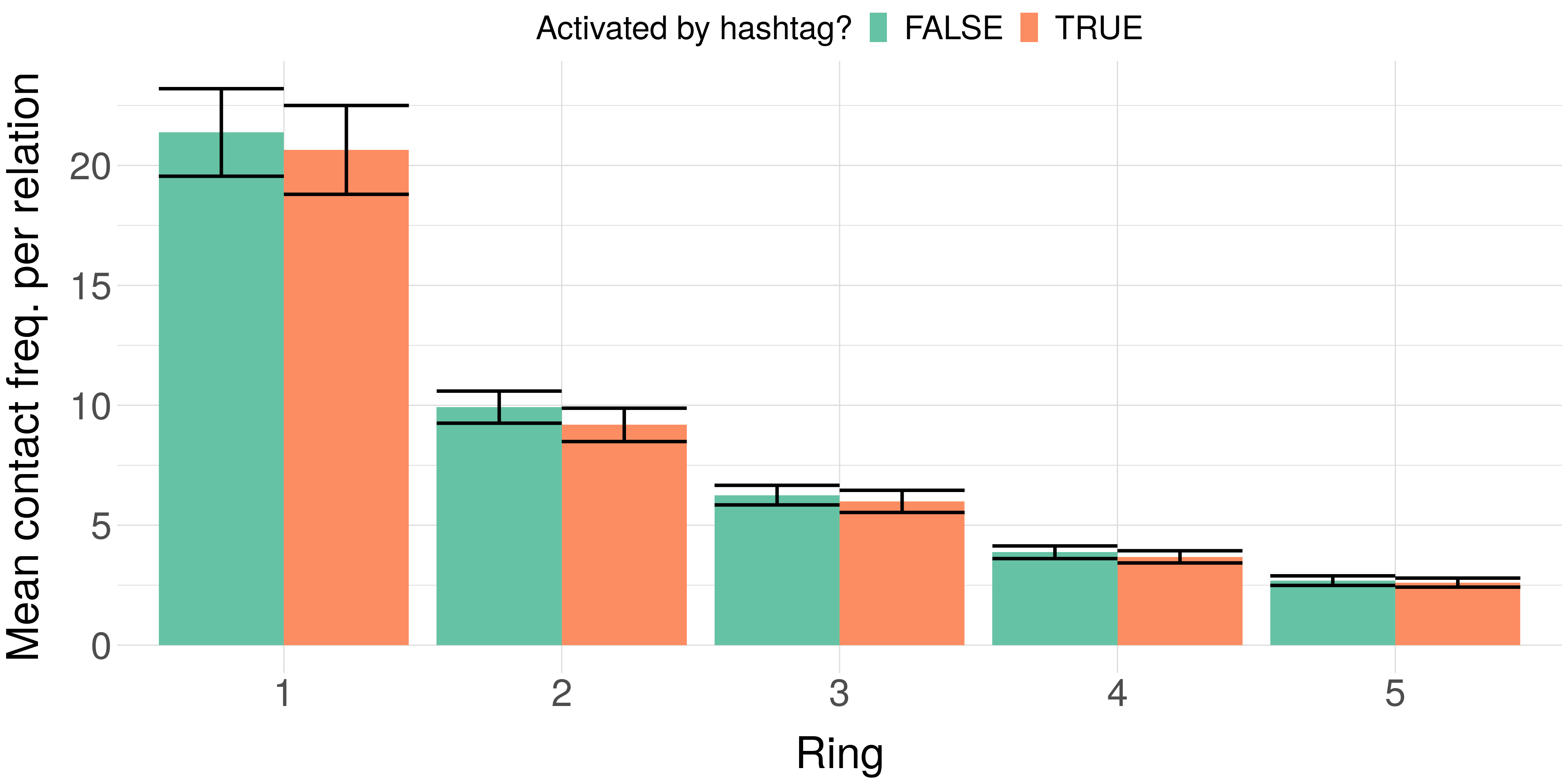}}
\hfill
\subfloat[Finland
\label{fig_appendix:contact_freq_rings_hashtags_FinnishJournalists}]
{\includegraphics[width=0.27\textwidth]
{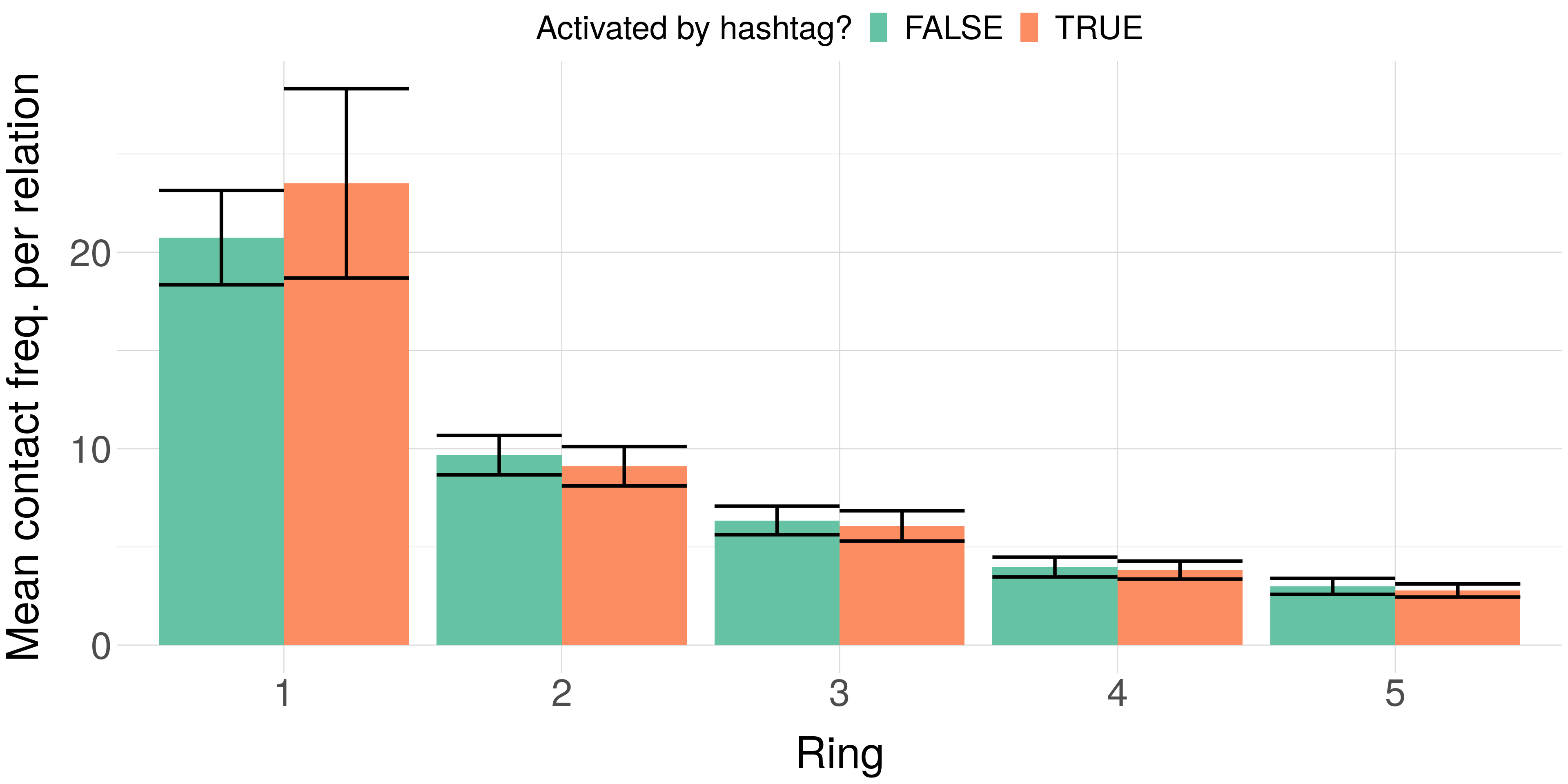}}
\hfill
\subfloat[Norway
\label{fig_appendix:contact_freq_rings_hashtags_NorwegianJournalists}]
{\includegraphics[width=0.27\textwidth]
{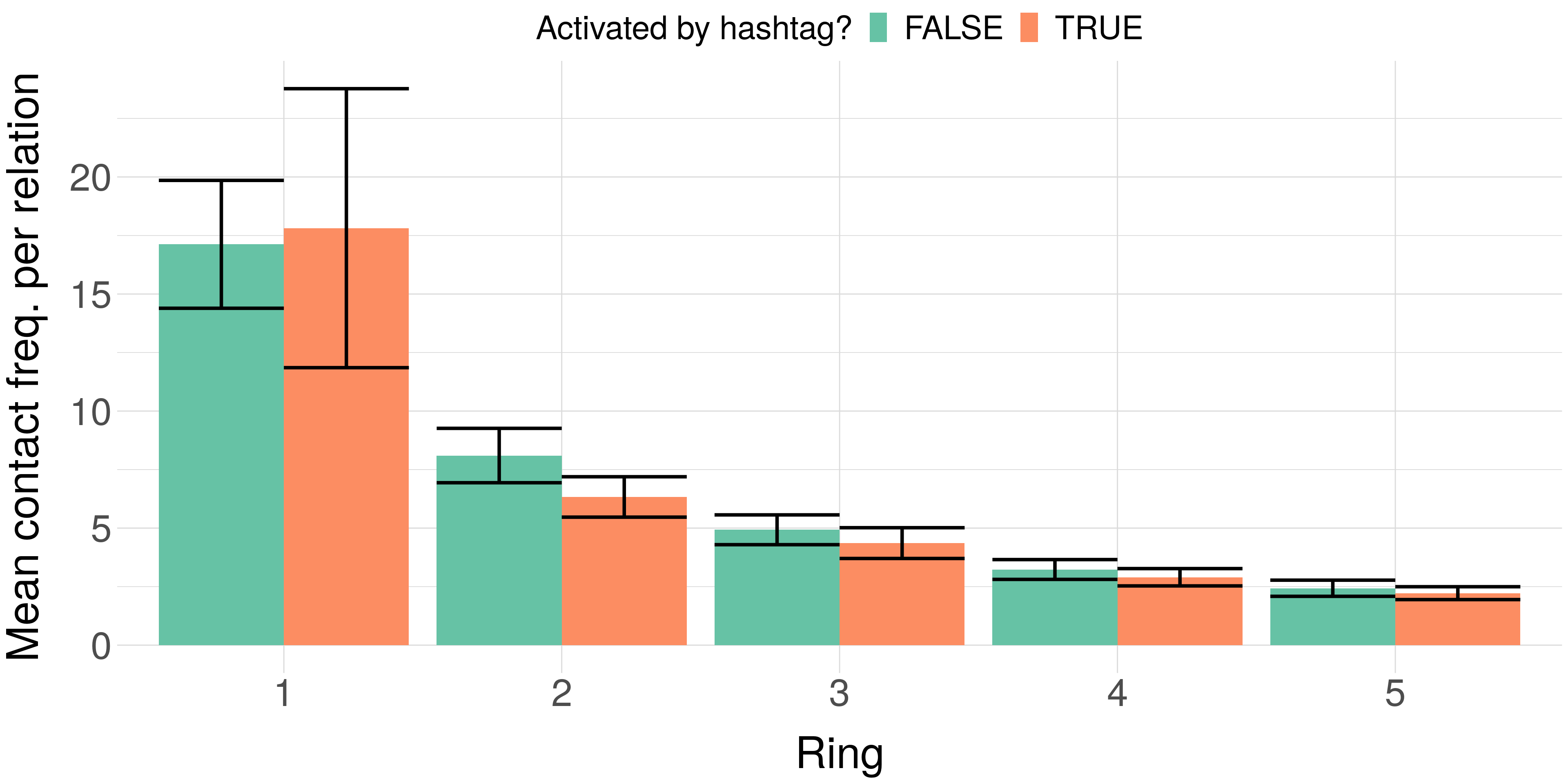}}
\hfill
\subfloat[Sweden
\label{fig_appendix:contact_freq_rings_hashtags_SwedishJournalists}]
{\includegraphics[width=0.27\textwidth]
{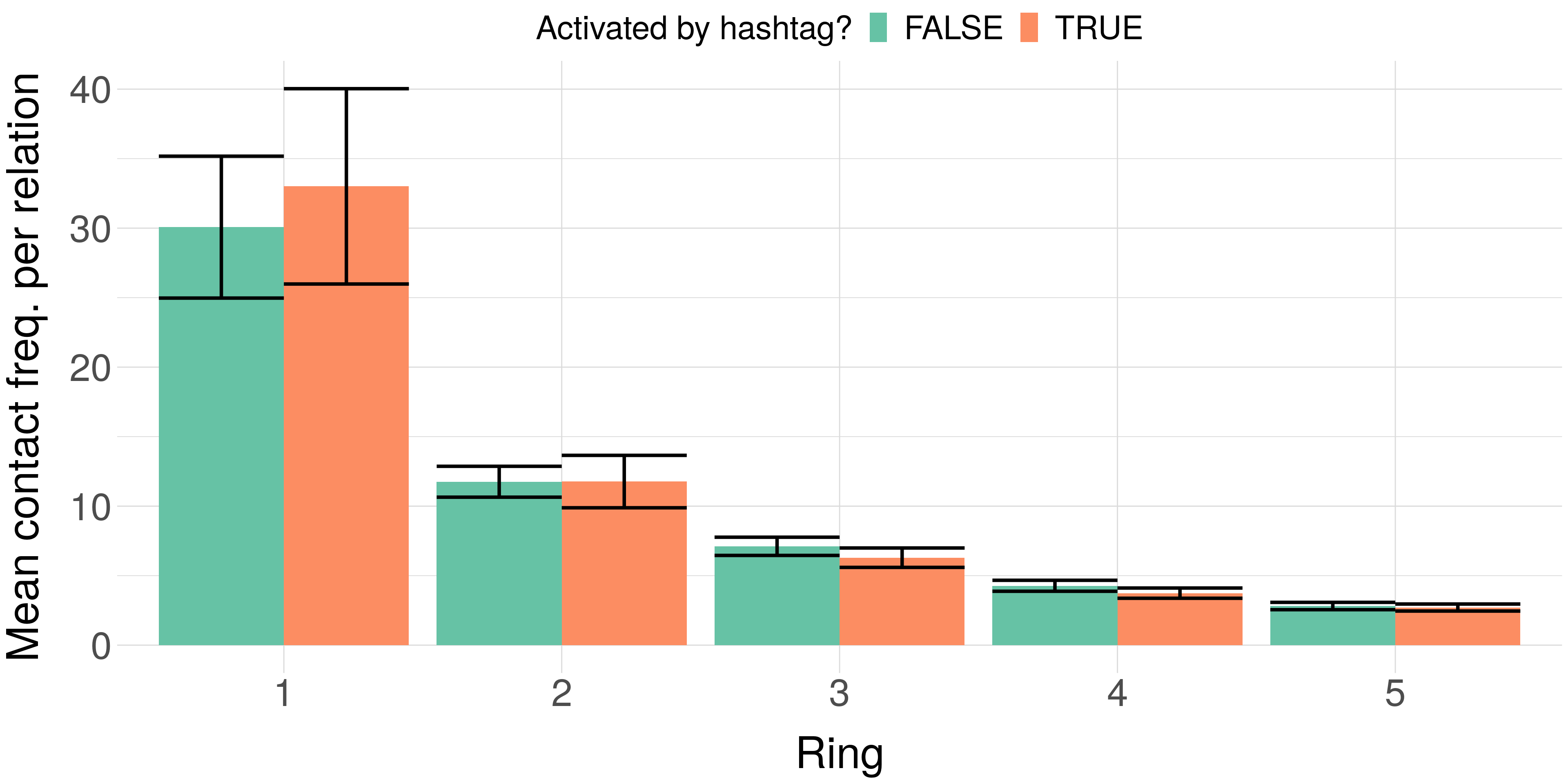}}
\hfill
\subfloat[Greece
\label{fig_appendix:contact_freq_rings_hashtags_GreekJournalists}]
{\includegraphics[width=0.27\textwidth]
{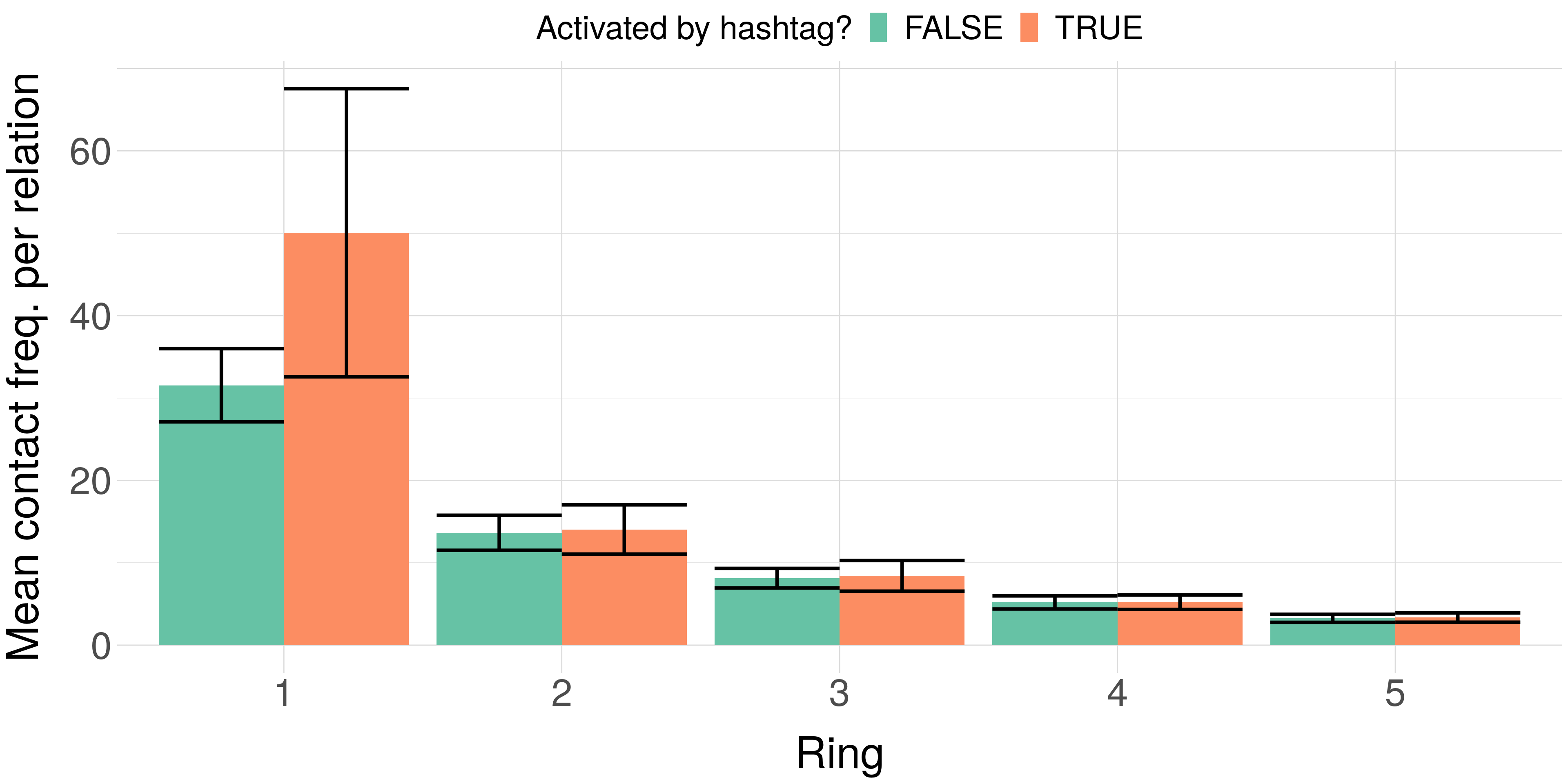}}
\hfill
\subfloat[Italy
\label{fig_appendix:contact_freq_rings_hashtags_ItalianJournalists}]
{\includegraphics[width=0.27\textwidth]
{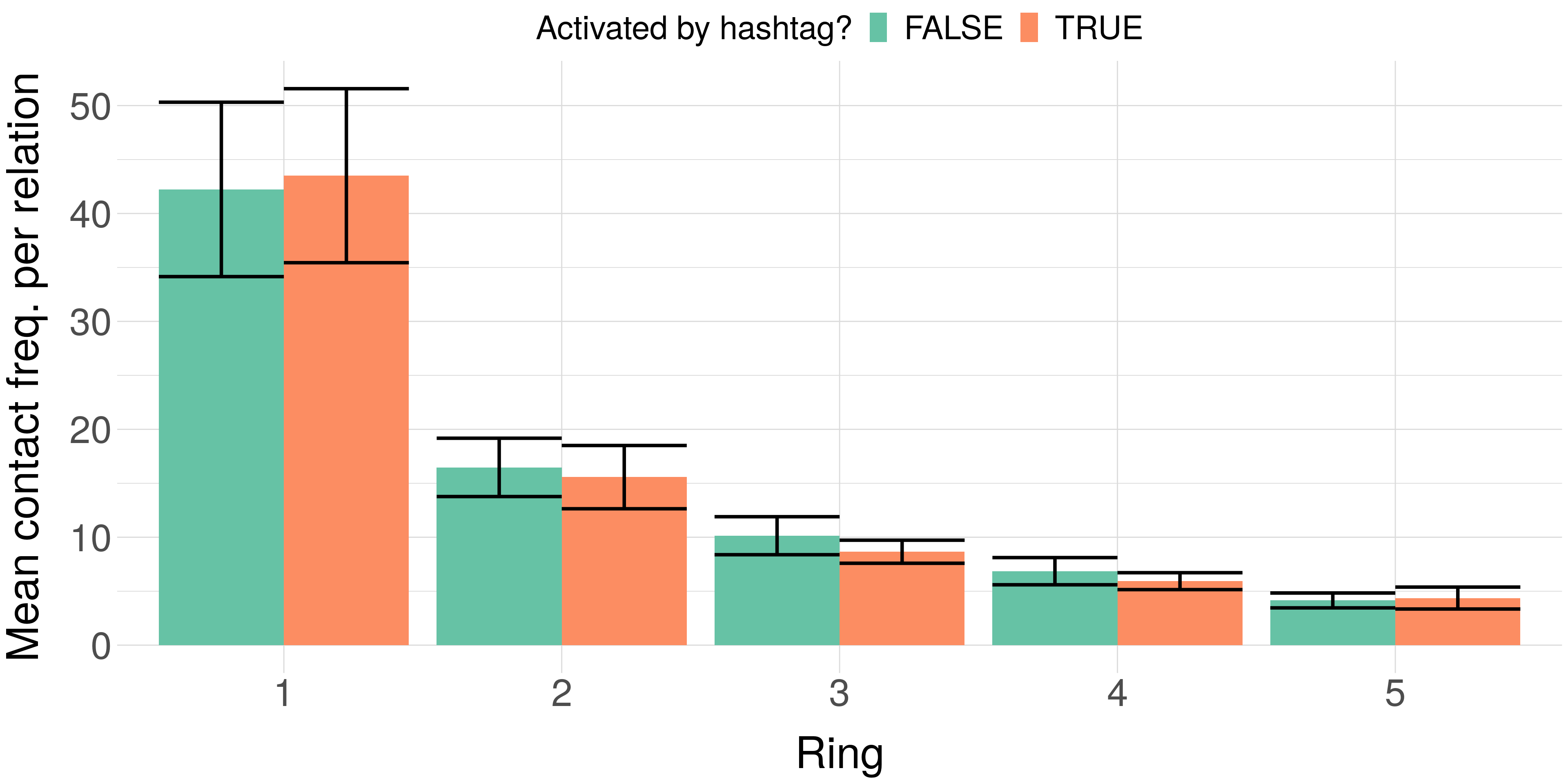}}
\hfill
\subfloat[Spain
\label{fig_appendix:contact_freq_rings_hashtags_SpanishJournalists}]
{\includegraphics[width=0.27\textwidth]
{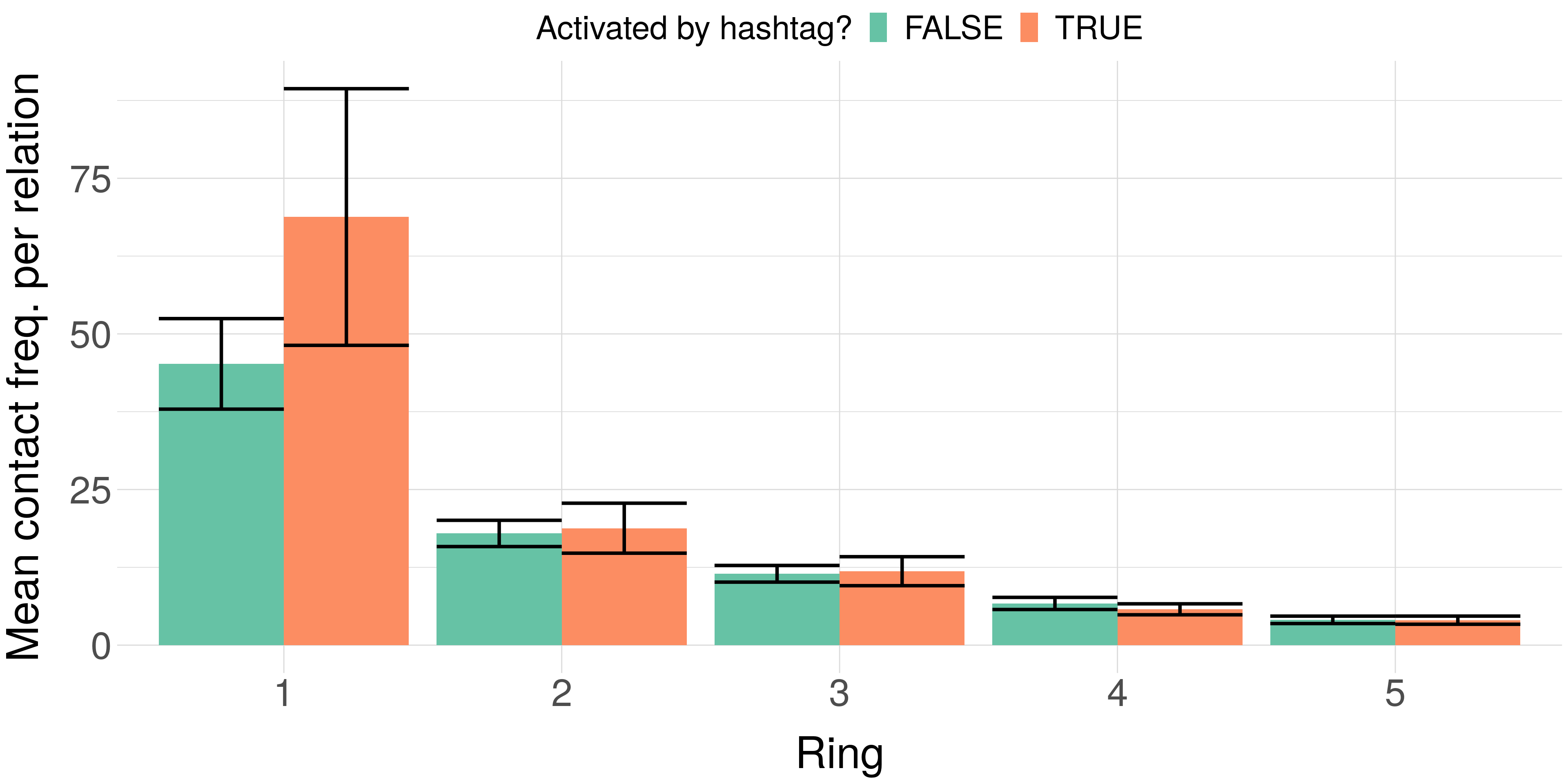}}
\hfill
\subfloat[France
\label{fig_appendix:contact_freq_rings_hashtags_FrenchJournalists}]
{\includegraphics[width=0.27\textwidth]
{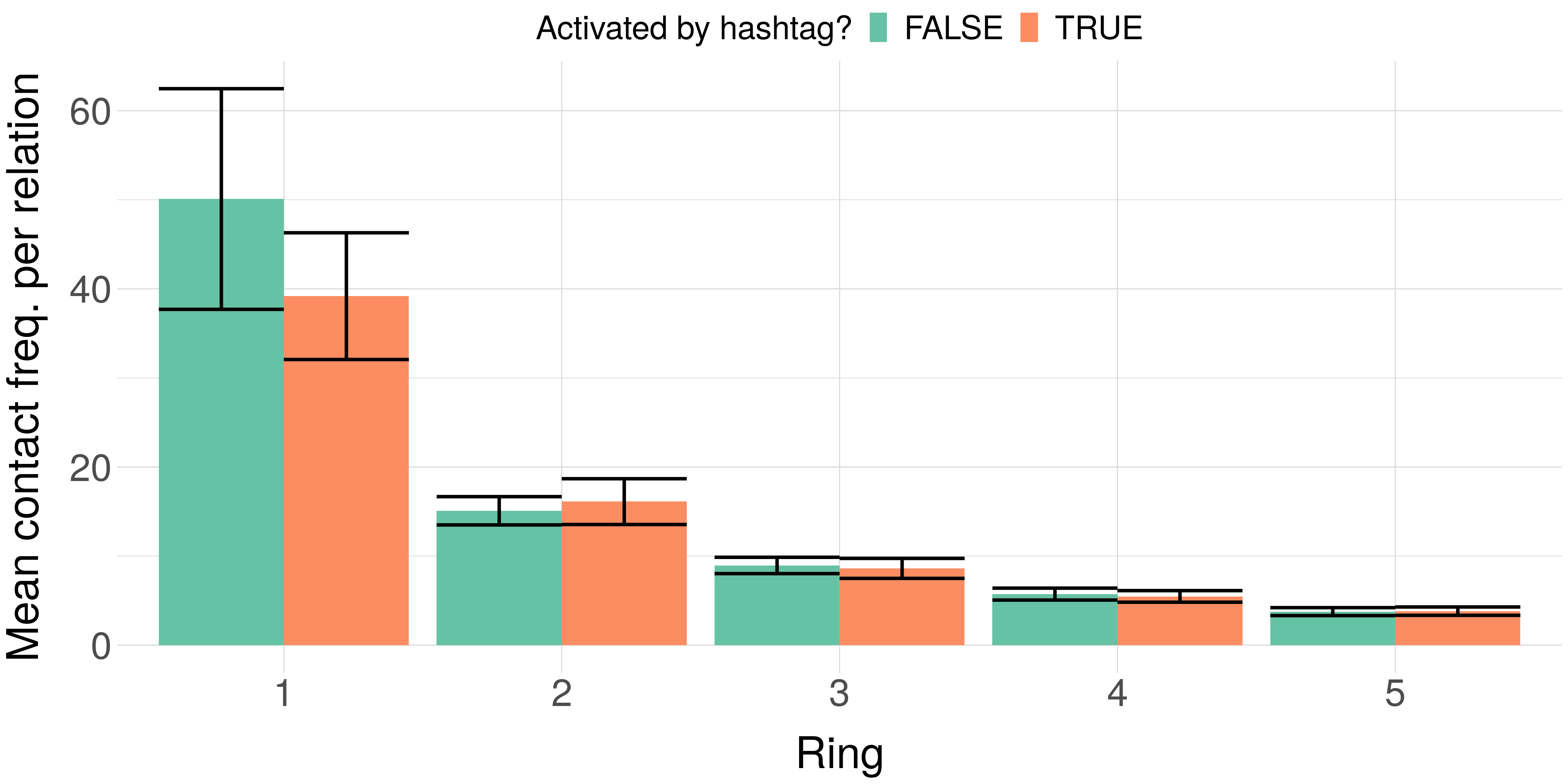}}
\hfill
\subfloat[Germany
\label{fig_appendix:contact_freq_rings_hashtags_GermanJournalists}]
{\includegraphics[width=0.27\textwidth]
{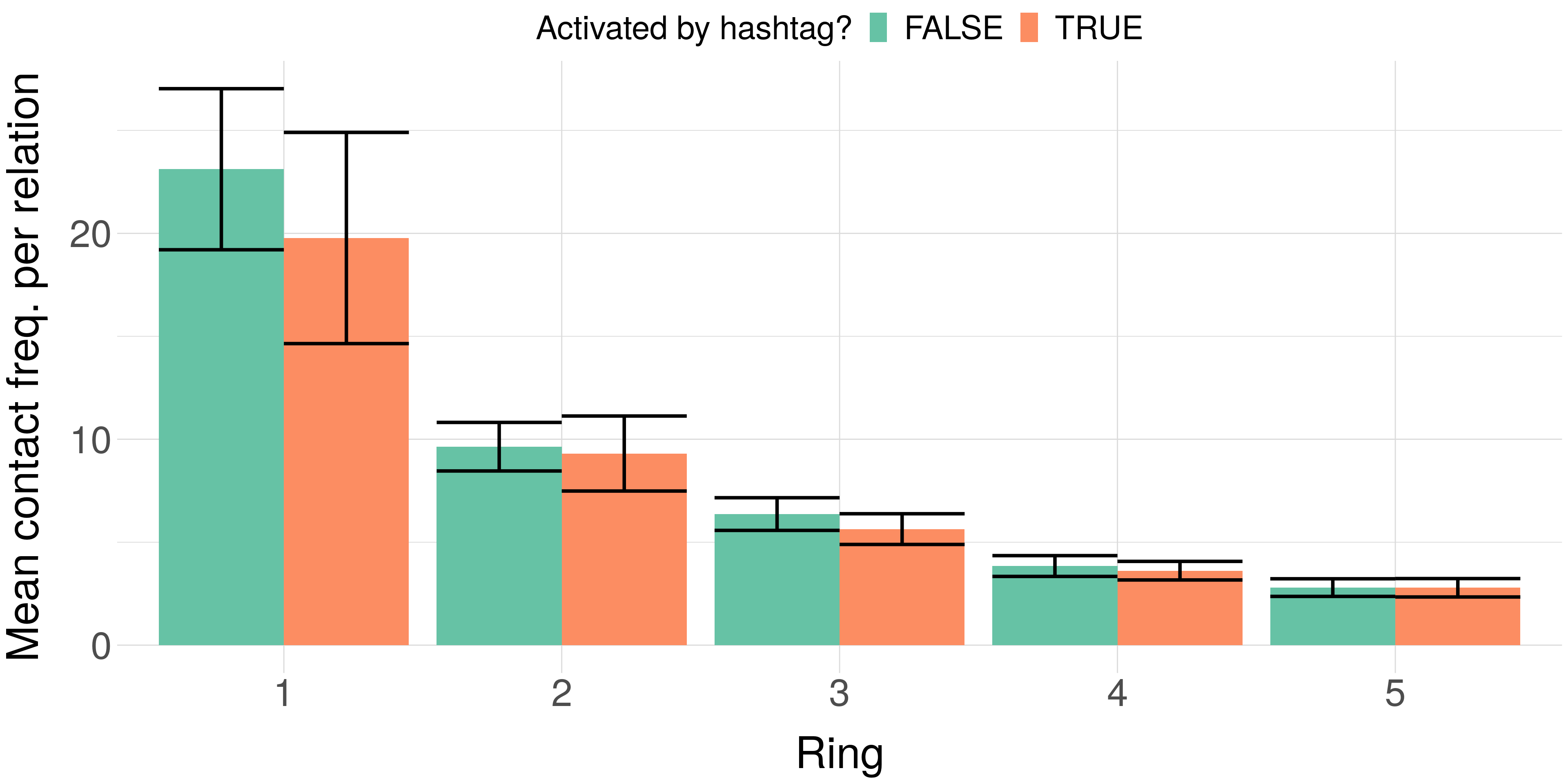}}
\hfill
\subfloat[Netherland
\label{fig_appendix:contact_freq_rings_hashtags_NetherlanderJournalists}]
{\includegraphics[width=0.27\textwidth]
{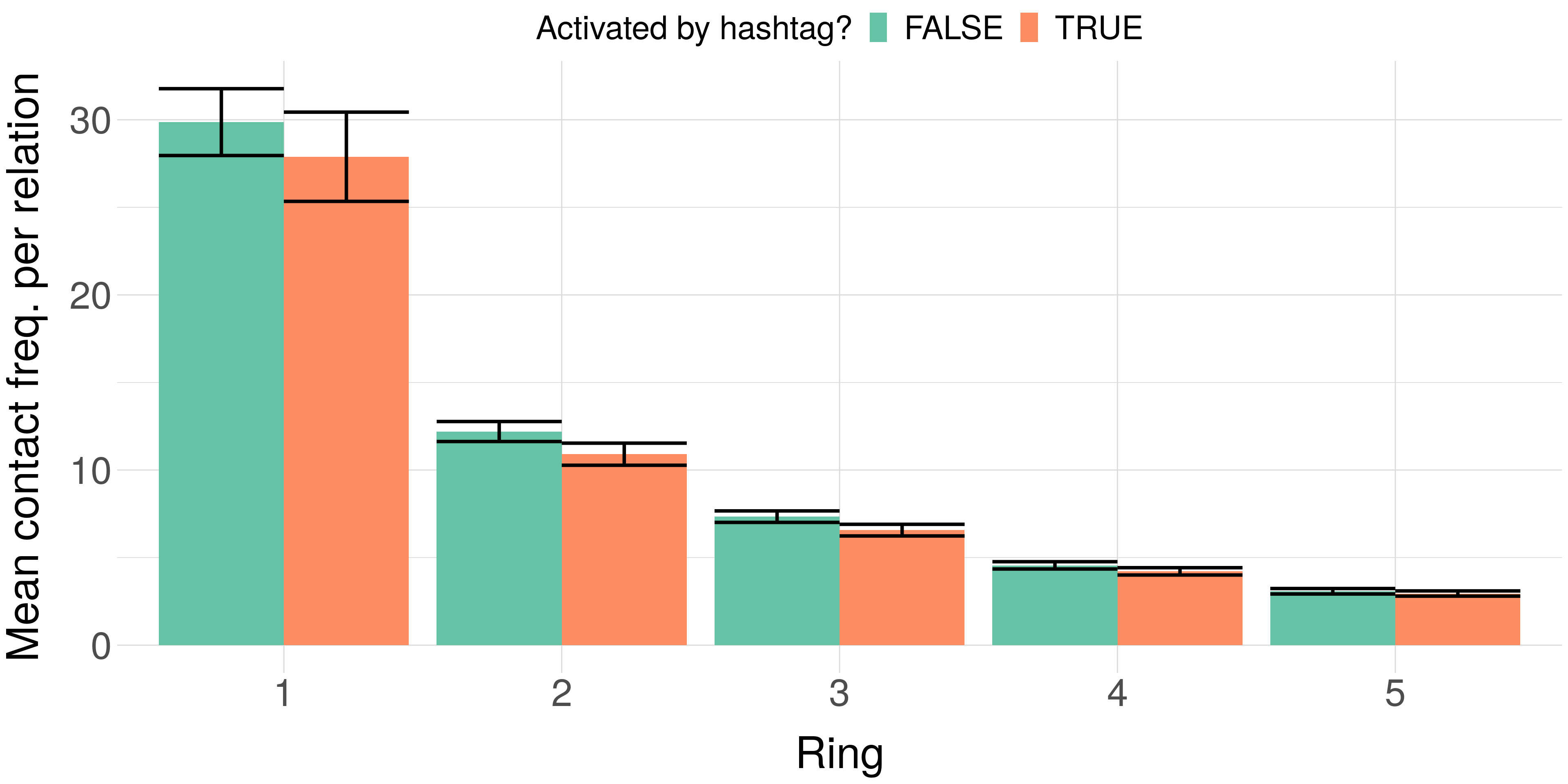}}
\hspace{1pt}
\subfloat[Australia
\label{fig_appendix:contact_freq_rings_hashtags_AustralianJournalists}]
{\includegraphics[width=0.27\textwidth]
{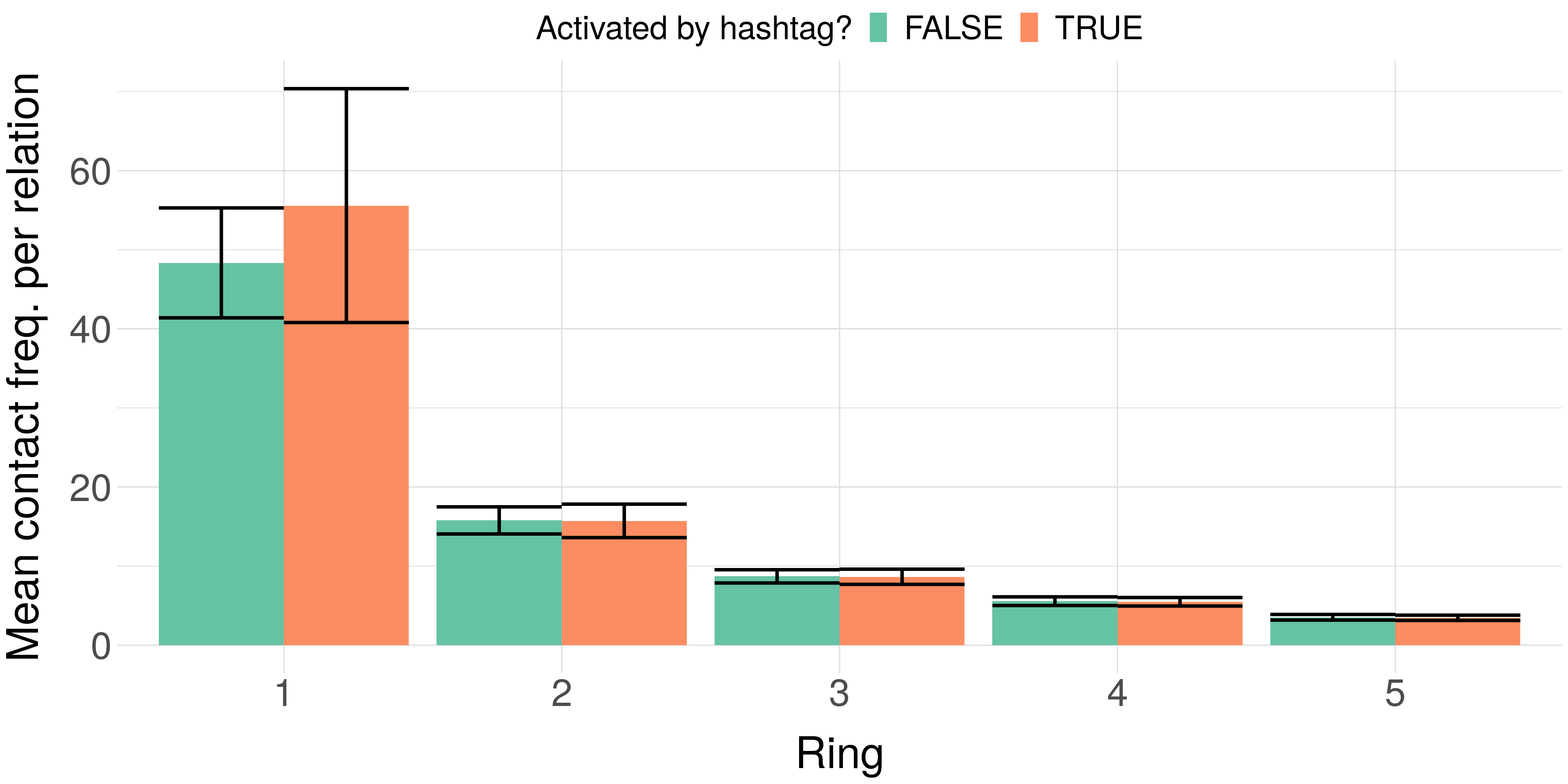}}
\end{center}
\end{adjustbox}
\caption{Average contact frequency per relationship per ring, with confidence intervals}
\label{fig_appendix:contact_freq_rings_hashtags}
\end{figure}